\documentclass{WileyMSP-template}
\usepackage{amsmath}
\usepackage{mathrsfs}
\usepackage{booktabs}
\usepackage{caption}
\usepackage{float}
\usepackage{placeins}
\usepackage{pdflscape}
\usepackage{pdfpages}
\usepackage{ragged2e}
\begin{document}

\pagestyle{fancy}
\rhead{}

\title{\parbox{\textwidth}{\hyphenpenalty=10000 \exhyphenpenalty=10000 Physical properties of compact star-like systems harboring traversable wormholes: Effects of chaotic magnetic fields and anisotropic matter}}

\maketitle


\author{Muhammad Lawrence Pattersons$^{1,2}$*}
\author{Freddy Permana Zen$^{1,2}$}
\author{Hadyan Luthfan Prihadi$^{3}$}\\
\author{Muhammad F. A. R. Sakti$^{4,5,6}$}



\begin{affiliations}
\setlength{\parskip}{0.5em}
$^1$Theoretical High Energy Physics Group, Department of Physics, Institut Teknologi Bandung\\
Jl. Ganesha 10, Bandung 40132,
Indonesia

$^2$Indonesia Center for Theoretical and Mathematical Physics (ICTMP), Institut Teknologi Bandung\\
Jl. Ganesha 10, Bandung 40132,
Indonesia

$^3$Research Center for Quantum Physics, National Research and Innovation Agency (BRIN)\\
South Tangerang 15314,
Indonesia

$^4$High Energy Physics Theory Group, Department of Physics, Faculty of Science,\\ Chulalongkorn University\\
Bangkok 10330,
Thailand

$^5$Department of Physics and Astronomy, University of Waterloo\\
Waterloo, Ontario, N2L 3G1, Canada

$^6$Perimeter Institute for Theoretical Physics\\
Waterloo, Ontario, N2L 2Y5, Canada

*Email Address: m.pattersons@proton.me

\end{affiliations}



\justifying

\begin{abstract}

We formulate exotic compact objects in form of compact star-like systems harboring traversable wormhole (CSSTW) with chaotic magnetic field and pressure anisotropy of the neutron fluid. The wormhole is traversable regardless of whether anisotropy of the neutron fluid and/or magnetic fields are included. For magnetized configurations, the resulting CSSTW can become extremely massive, with ADM masses exceeding $8\,M_\odot$, and can exhibit large surface redshifts exceeding $\mathcal{Z} \simeq 1.5$. The system can also reach the ultracompact regime, which allows us to calculate echo time that might be produced by the systems. Echo times of considered CSSTW vary depending on the chaotic magnetic field configuration and fluid anisotropy. For non-magnetized configurations, the gravitational-wave echo time is of the order of $10^{-2}-10^{-1}$ ms. For the magnetized configurations, however, it ranges from the order of $10^{-1}$ $\mu$s $-10^{-1}$ ms Moreover, we derive an explicit expression for the echo time as a function of uniform magnetic field. The resulting relation shows that the echo time decreases as the magnetic field strength increases. The stability analysis of radial perturbations shows that the ultracompact non-magneized CSSTW are prone to be unstable. For magnetized CSSTW, all considered configurations with a surface magnetic field of $B_s=10^{15}$~G are stable.

\end{abstract}

\begingroup
\setlength{\parskip}{0.5em}

\section{Introduction}\label{intro}
	
Wormholes are intriguing constructs that serve as tunnels connecting distant spacetime points~\cite{Dai2020,Turimov2025,Bronnikov2023,Cataldo2017,Yousaf2025,Prat2015,Dai2019,Godani2019}. Both distinct spacetime domains are linked through a throat-like geometry. Originally, wormholes emerged as exact solutions of the Einstein field equation~\cite{Zangeneh2025}. 
Einstein and Rosen~\cite{Einstein1935} found a solution bearing a resemblance to a wormhole, which was identified as a ``bridge'' that connects two identical physical spaces. This work gave rise to the term ``Einstein–Rosen bridge''. The term ``wormhole'' was later introduced by Ref.~\cite{Misner1957}. It is worth noting that the putative wormhole in the Einstein–Rosen bridge is not traversable, since it is colocated with the blackhole’s singularity. The surface of the blackhole is the event horizon, which cannot be a wormhole~\cite{Radhakrishnan2024}.
	
	The formalism of traversable wormholes was introduced by Morris and Thorne~\cite{Morris1988} in their seminal 1988 paper. In constructing traversable wormholes, the null energy condition (NEC) must be violated by the energy-momentum tensor (EMT)~\cite{Lobo2016}. It should be noted that there are three other energy conditions: the weak energy condition (WEC), the strong energy condition (SEC), and the dominant energy condition (DEC)~\cite{Radhakrishnan2024}. The violation of NEC implies that all the energy conditions are not validated~\cite{Moraes2019}. In general~relativity~(GR), the violation of NEC typically requires the presence of exotic matter~\cite{Kokubu2020}. Many authors~\cite{Chianese2017,Mustafa2020,AzregAinou2015,Kuhfittig2006,Bhattacharya2019,Sreekumar2025} have included the exotic matter in their studies on wormholes. Furthermore, Ref.~\cite{Lobo2016} provided a comprehensive review of wormhole physics. Refs.~\cite{Falco1,Falco2,Falco3,Falco4,Falco5,Falco6} worked on the development of possible astrophysical techniques to detect the wormholes.
	
	Some authors have highlighted the potential existence of wormholes under extreme astrophysical conditions. Ref.~\cite{Bhar2016} investigated the possibility of sustaining static, spherically symmetric traversable wormhole geometries in GR, supported by dark energy and admitting conformal motion. Ref.~\cite{Biswas2024} proposed that wormholes could mimic supermassive black holes at galactic centers. Refs.~\cite{Hao1,Hao2} work on wormhole-plus-black-hole systems. Other extreme astrophysical environments where wormholes are hypothesized to exist include compact stars~\cite{Dzhunushaliev2011,Dzhunushaliev2012,Dzhunushaliev2013,Dzhunushaliev2014,Aringazin2015,Dzhunushaliev2015,Dzhunushaliev2016,Dzhunushaliev2023,Nojiri2024,Pattersons2026mpla}. However, in the works of Refs.~\cite{Dzhunushaliev2011,Dzhunushaliev2012,Dzhunushaliev2013,Dzhunushaliev2014,Aringazin2015,Dzhunushaliev2015,Dzhunushaliev2016,Dzhunushaliev2023}, the existence of hybrid systems composed of wormholes and compact stars is sustained by the presence of ghosts, which leads to undesirable consequences: both mixed star-plus-wormhole systems and pure wormholes are unstable~\cite{Nojiri2024}. Ref.~\cite{Nojiri2024} eliminated the ghosts by imposing the constraints given by the additional Lagrange multipliers.
	
	It has to be noted that although the model proposed by Ref.~\cite{Nojiri2024} has successfully constructed a formalism of ghost-free systems, it did not take into account pressure anisotropy of the fluid that may arise. It is widely known that the matter inside compact stars like neutron stars (NSs) may be anisotropic, meaning that the radial and tangential pressures differ~\cite{Ghosh:2025lma}. Various factors can lead to such pressure anisotropy; including boson condensation, the presence of a solid core, different types of phase transitions, strong magnetic or electric fields, and the influence of modified spacetime geometry~\cite{Pattersons2021}. In line with these physical considerations, several studies on compact stars~\cite{Pattersons2021,Sulaksono2014,Sulaksono2015,Silva2015,Setiawan2017,Lopes2019,Rizaldy2024,Posada2024,Becerra2024,Pattersons2025,Yusmantoro2025,Sakti2016,Becerra2024b,Becerra2025,Torres2019,Rahmansyah2020} have incorporated anisotropic pressure. Anisotropy plays an important role in the mass calculation of compact stars. For instance, Ref.~\cite{Pattersons2025} shows that anisotropy can enhance the mass of NSs, allowing them to reach the mass range of the secondary compact object in GW190814, which has been reported by Ref.~\cite{Abbott2019} and hypothesized as the heaviest NS ever observed~\cite{Fattoyev2020}. The review of anisotropic compact stars can be referred to Ref.~\cite{Kumar2022}.
	
	Another aspect that is not taken into account in the model presented by Ref.~\cite{Nojiri2024} is the presence of magnetic fields. The magnetic fields on NSs are typically in the range of $10^{12}$–$10^{15}$ G~\cite{Lander2009}. 
	One framework for describing magnetic fields in NSs is through chaotic magnetic fields, which has been investigated by Ref.~\cite{Lopes2015}. In their work, they also introduced an ansatz in which the magnetic field is coupled to the energy density of the star. An important advantage of adopting a chaotic magnetic field is the elimination of another anisotropy induced by the magnetic field, thereby simplifying the mathematical formulation. In addition to this simplification, the use of chaotic magnetic fields allows one to obtain a magnetic pressure that is independent of both direction and the choice of coordinate system, thus restoring the proper thermodynamic concept of pressure and ensuring consistency with field theory~\cite{Lopes2015,Pattersons2024a}. Several studies~\cite{Pattersons2024a,Wu2017,Lopes2020,Pattersons2024b} have subsequently incorporated chaotic magnetic fields in their analyses.
	
	Our aim in this work is to extend the star-plus-wormhole system constructed by Ref.~\cite{Nojiri2024} by incorporating the anisotropy of the neutron fluid and the effects of chaotic magnetic fields that may arise in such a system, \textcolor{black}{thereby constructing compact star-like systems harboring traversable wormhole (CSSTW)}. Other distinctions of our work from previously studied star-plus-wormhole configurations constructed by Refs.~\cite{Dzhunushaliev2011,Dzhunushaliev2012,Dzhunushaliev2013,Dzhunushaliev2014,Aringazin2015,Dzhunushaliev2015,Dzhunushaliev2016,Dzhunushaliev2023,Nojiri2024} is the inclusion of the calculations of surface redshift and the gravitational-wave echo time.
	
	
	The paper is organized as follows. In Sect.~\ref{anisotropicmagnetic}, we introduce the formalisms of the anisotropic fluid and the chaotic magnetic field. Sec.~\ref{GRcoupled} presents the formulation of GR coupled with two scalar fields, and demonstrates the elimination of ghosts. In this section, we also analyze the energy conditions and the travesability of the wormholes. In Sec.~\ref{nonmagnetized}, we construct a model of non-magnetized CSSTW that consider anisotropy. Sec. \ref{magnetized} shows the model of anisotropic magnetized CSSTW. Note that, in both Secs. \ref{nonmagnetized} and \ref{magnetized}, we compare the obtained CSSTW with ordinary NSs; and also present the calculation results of the surface redshift and the echo time that may arise in such systems. \textcolor{black}{The stability analysis of the systems related to radial perturbation is given in Sec.~\ref{stabilityanalysissss}}.
	Finally, Sec.~\ref{conclusion} is devoted to the conclusions of our study.

\section{Anisotropic fluid and chaotic magnetic field}
	\label{anisotropicmagnetic}
	To make this section self-contained, in Subsect.~\ref{anisotropy} we introduce the anisotropic fluid model employed in this study, while the formulation and ansatz of the chaotic magnetic field are given in Subsect.~\ref{magnetic}.
	\subsection{Model of anisotropy}
	\label{anisotropy}
	It is important to consider the EMT of matter with anisotropic pressure, given by~\cite{Pretel2023}  
	\begin{eqnarray}
		T_{\text{matter}\,\mu\nu} = (\rho + p_t)u_\mu u_\nu + p_t\,g_{\mu\nu} + \sigma\,\xi_\mu \xi_\nu,
		\label{EMT}
	\end{eqnarray}
	where $\rho$ is the energy density, $p_t$ is the tangential pressure, and $\sigma = p_r - p_t$ denotes the anisotropy term, where $p_r$ is the radial pressure. At this stage, we have to emphasize that $\rho$, $p_r$, and $p_t$ belong to matter. In Eq. (\ref{EMT}), $u^\mu$ is the four-velocity of the fluid, satisfying the normalization condition $u^\mu u_\mu = -1$, 
	and $\xi^\mu$ is a unit radial four-vector, satisfying $\xi^\mu \xi_\mu = 1$. \textcolor{black}{The vector $\xi_\mu$ is physically an auxiliary quantity that parametrizes the anisotropic stresses in the neutron fluid. Its components are fully determined by the geometry and matter profiles, in such a way so that $p_r=\xi^\mu\xi^\nu T_{\text{matter}\:\mu\nu}$, and $p_t=\varsigma^\mu\varsigma^\nu T_{\text{matter}\:\mu\nu}$, where $\varsigma^\mu$ is a unit spacelike vector (orthogonal to $u^\mu$ and $\xi^\mu$)~\cite{anisoadd}.}
	
	By choosing $\xi_\mu=(0,\sqrt{g_{rr}},\:0,\:0)$, for static spherically symmetric configuration, we can easily find
	\begin{eqnarray}
		T_{\text{matter}\:\mu\nu}=
		\begin{pmatrix}
			-g_{tt}\:\rho & 0 & 0 & 0 \\
			0 & g_{rr}\:p_r & 0 & 0 \\
			0 & 0 & g_{\theta\theta}\:p_t & 0 \\
			0 & 0 & 0 & g_{\varphi\varphi}\:p_t
		\end{pmatrix}.
		\label{EMTmatrix}
	\end{eqnarray}
	
	In this work, we consider Cosenza-Herrera-Esculpi-Witten (CHEW) model~\cite{Cosenza1981} in the realization of the CSSTW with anisotropic neutron fluid.
	
	\textcolor{black}{For a static, spherically symmetric spacetime, the metric describing the interior of an ordinary NS is given by the interior Schwarzschild metric, i.e.,~\cite{Pattersons:2026hnq}}
	\begin{eqnarray}
		\textcolor{black}{ds^2=-e^{2\nu(r)}dt^2+e^{2\lambda(r)}+r^2(d\theta^2+\sin^2\theta d\phi^2),}
	\end{eqnarray}
	\textcolor{black}{where $\lambda(r)=\left(1-\frac{2m(r)}{r}\right)^{-1}$, and $m(r)$ denotes mass of the NSs as a function of $r$. Using the EMT shown in Eq. (\ref{EMT}), Tolman-Oppenheimer-Volkoff (TOV) equation for anisotropic NSs reads}
	\begin{eqnarray}
		\textcolor{black}{\frac{dp_r}{dr}=-\frac{d\nu}{dr}(\rho+p_r)-\frac{2\sigma}{r}.}
	\end{eqnarray}
	
	\textcolor{black}{Cosenza et al.~\cite{Cosenza1981} considered that $\sigma/r$ can be written as}
	\begin{eqnarray}
		\textcolor{black}{\frac{\sigma}{r}=f(p_r,r)(\rho+p_r),}
	\end{eqnarray}
	\textcolor{black}{where $f(p_r,r)$ is an arbitrary function. Motivated to obtain a minimal modification of the TOV equation, Cosenza et al.~\cite{Cosenza1981} assumed $f(p_r,r)=\frac{1-h}{2}\frac{d\nu}{dr}$, where $h$ is a constant. On the other words, \textcolor{black}{the model is entirely ad hoc}, and it is not assumed for any specific physical reason but only because it transforms TOV equation into the simple form, i.e.}
	\begin{eqnarray}
		\textcolor{black}{\frac{dp_r}{dr}=-h\frac{d\nu}{dr}(\rho+p_r)}.
	\end{eqnarray}
	\textcolor{black}{Finally, the expression of the anisotropy term can also be written as~\cite{Rahmansyah2020}}
	\begin{eqnarray}
		\sigma=-\frac{r}{2}\frac{1-h}{h}\frac{dp_r}{dr},\label{CHEW}
	\end{eqnarray}
	where $h$ denotes the free parameter of anisotropy. The system becomes isotropic when we set $h=1$. \textcolor{black}{It is worth noting that the parameter $h$ is not introduced based on any specific physical mechanism. Therefore, it does not represent any particular physical process in NSs. Instead, it is merely a phenomenological parameter to conveniently model the anisotropic NS configuration.}
	\subsection{Chaotic magnetic fields}
	\label{magnetic}
	Consider a magnetic field $\mathcal{B}$ oriented along the $z$-axis. The corresponding stress tensor can then be written in the form: diag$(\frac{\mathcal{B}^2}{8\pi};\:\frac{\mathcal{B}^2}{8\pi};\:-\frac{\mathcal{B}^2}{8\pi})$, which is non-identical \cite{Lopes2015,Pattersons2024a}. However, Ref.~\cite{Zeldovich1971} argued that the effect of a magnetic field can be described in terms of pressure only in the case of a small-scale chaotic field. In such a case, the pressure associated with the magnetic field, $p_{\mathcal{B}}$, is shown to be consistent with field theory. Thus, $p_\mathcal{B}$ now writes~\cite{Lopes2015,Pattersons2024a,Zeldovich1971}
	\begin{eqnarray}
		p_\mathcal{B}=\frac{1}{3}<T^{\:\:\:j}_{\mathcal{B}\:j}>\:=\frac{1}{3}\left(\frac{\mathcal{B}^2}{8\pi}+\frac{\mathcal{B}^2}{8\pi}-\frac{\mathcal{B}^2}{8\pi}\right).
	\end{eqnarray}
	Here, $T^{\:\:\:j}_{\mathcal{B}\:j}$ is the spatial component of the magnetic EMT.

The magnetic energy density $\rho_\mathcal{B}$ and magnetic pressure $p_\mathcal{B}$ can now conveniently be written as
	\begin{eqnarray}
		\rho_\mathcal{B}=\frac{\mathcal{B}^2}{8\pi},\:\:\:\:\:
		p_\mathcal{B}=\frac{\mathcal{B}^2}{24\pi}.
	\end{eqnarray}
	
	It should be noted that the standard treatment of magnetic fields in NSs, which leads to the appearance of anisotropy even though the original matter of the NSs is isotropic~\cite{Bordbar:2022}, complicates the mathematical formulation and prevents the stellar configuration from remaining spherically symmetric \cite{Mallick2014}.
	
	According to Ref.~\cite{Lopes2015}, the chaotic magnetic field approximation adopted here is physically motivated by the expectation that magnetic fields generated during the turbulent core-collapse supernova phase are highly disoriented. In this framework, the magnetic pressure is assumed to be statistically isotropic, so that no global preferred direction exists and anisotropic magnetic stresses associated with ordered field configurations are effectively averaged out. This allows the magnetic contribution to be incorporated consistently within the spherically symmetric formalism through an effective isotropic pressure. A limitation of the model is that it is expected that the magnetic field evolves and becomes at least partially oriented. 
	In this work, the chaotic magnetic field is employed as an approximation to calculate the influence of the magnetic field in NS properties within the spherically symmetric TOV formalism. \textcolor{black}{Thus, the magnetic field model employed in this paper is purely a simplified framework to investigate additional magnetic-field effects on energy density and pressure of the system.}
	
	\textcolor{black}{Note that an ordered magnetic field breaks the spherical symmetry of NSs and may deform them into an oblate configuration, as discussed by Refs.~\cite{Mallick2014,Rizaldy2018a}. In this case, the spacetime metric must be expanded up to the quadrupole order, which naturally introduces stellar deformation. The detailed formalism can be found in Ref.~\cite{Mallick2014}.}
	
	Another important point to note is that the exact mathematical profile of the magnetic field inside NSs remains unknown. Ref.~\cite{Lopes2015} proposed an ansatz in which the magnetic field is coupled to the matter energy density $\rho$, i.e.
	\begin{eqnarray}
		\mathcal{B}=B_s+B_0\left(\frac{\rho}{\rho_0}\right)^\eta,
	\end{eqnarray}
	where $B_s$ denotes the magnetic field at the surface of the NS, $B_0$ represents the magnetic field expected at the core. For the CSSTW, $B_0$ can be interpreted as the magnetic field expected at the wormhole’s throat. Moreover, $\rho_0$ denotes the central energy density of the non-magnetized NS configuration (without the wormhole) corresponding to the maximum mass, and $\eta$ is an arbitrary positive constant. It is worth noting that choosing $\eta$ to be very close to zero would make the magnetic field profile resemble a constant magnetic field. To simplify the calculation of the radius of the wormhole throat in Sec.~\ref{magnetized}, we set $\eta = 1$ throughout this work. Moreover, the magnetic field parameters that are varied in our analysis are both $B_0$ and $B_s$.
	
	\textcolor{black}{It is important to note that the internal magnetic field profile of NSs is still subject to significant uncertainties. For this reason, it is customary to introduce phenomenological but physically motivated ansatz for the magnetic field profile. Different prescriptions have been proposed in the literature, including magnetic field profiles coupled to the baryon number density, as in Ref.~\cite{Mallick2014}, as well as profiles coupled to the energy density.}
	
	\textcolor{black}{In the latter approach, i.e. the ansatz proposed by Ref.~\cite{Lopes2015}, the energy density can enter directly into the TOV equations governing the macroscopic structure of ordinary NSs. The parameter $\rho_0$ serves as a reference energy density scale, while the parameter $\eta$ controls the strength of the magnetic field. This choice provides a convenient and physically motivated parametrization for exploring the impact of strong magnetic fields on NS properties.}

\section{GR coupled with two scalar fields and a magnetic field}
	\label{GRcoupled}
	For completeness, Subsect.~\ref{einsteinfieldequation} contains the derivation of the field equations from the action, Subsect.~\ref{elimination} addresses the procedure for eliminating the ghosts, and Subsect.~\ref{energyconditions} analyzes the energy conditions and traversability of the wormholes.
	\subsection{Einstein field equations}
	\label{einsteinfieldequation}
	The action of GR coupled with 2 scalar fields and magnetic field is given by
\begin{eqnarray}
			S_{\text{GR}\phi\chi\mathcal{B}} &=& \int d^4x \sqrt{-g} \left[ \frac{\mathcal{R}}{2\kappa^2} 
			- \frac{1}{2} A(\phi,\chi) \partial_\mu \phi \partial^\mu \phi 
			- B(\phi,\chi) \partial_\mu \phi \partial^\mu \chi  \right. \nonumber \\
			&& \left. - \frac{1}{2} C(\phi,\chi) \partial_\mu \chi \partial^\mu \chi 
			- V(\phi,\chi) + \mathcal{L}_{\text{matter}} + \mathcal{L}_{\mathcal{B}} \right].
			\label{action}
		\end{eqnarray}
		Here $\mathcal{R}$ denotes the Ricci scalar, $\kappa^2=8\pi G c^{-4}$, where $G$ is the universal constant of gravitation, and $c$ denotes the speed of light; $A(\phi,\chi)$, $B(\phi,\chi)$, and $C(\phi,\chi)$ are arbitrary functions; $V(\phi,\chi)$ denotes the scalar fields potential; and $\mathcal{L}_{\text{matter}}$ is the Lagrangian of matter, and $\mathcal{L}_{\mathcal{B}}$ is the Lagrangian of magnetic field. Throughout the paper, we use $G=c=1$.
		
		By varying the action in Eq. (\ref{action}) with respect to the metric $g_{\mu\nu}$, we can obtain
		\begin{eqnarray}
			0 &=& \frac{1}{2\kappa^2} \left( - \mathcal{R}_{\mu\nu} + \frac{1}{2} g_{\mu\nu} \mathcal{R} \right) \nonumber \\
			&& + \frac{1}{2} g_{\mu\nu} \left[ -\frac{1}{2} A(\phi, \chi) \partial_\rho \phi \partial^\rho \phi - B(\phi, \chi) \partial_\rho \phi \partial^\rho \chi - \frac{1}{2} C(\phi, \chi) \partial_\rho \chi \partial^\rho \chi - V(\phi, \chi) \right] \nonumber \\
			&& + \frac{1}{2} \left[ A(\phi, \chi) \partial_\mu \phi \partial_\nu \phi + B(\phi, \chi)(\partial_\mu \phi \partial_\nu \chi + \partial_\nu \phi \partial_\mu \chi) + C(\phi, \chi) \partial_\mu \chi \partial_\nu \chi \right] \nonumber\\
			&& + \frac{1}{2}\left(T_{\text{matter}\, \mu\nu} + T_{\mathcal{B}\, \mu\nu}\right).
			\label{field}
		\end{eqnarray}
		Here, $T_{\text{matter}\,\mu\nu}$ denotes the EMT of the matter sector, while $T_{\mathcal{B}\,\mu\nu}$ represents the EMT of the magnetic field. The Greek indices run from 0 to 3.
		
		Furthermore, by varying the action with respect to the fields $\phi$ and $\chi$, one obtains
		\begin{align}
			\frac{1}{2} A_{\phi} \partial_{\mu} \phi \, \partial^{\mu} \phi 
			+ A \nabla^{\nu} \partial_{\mu} \phi 
			+ A_{\chi} \partial_{\mu} \phi \, \partial^{\mu} \chi 
			+ \left( B_{\chi} - \frac{1}{2} C_{\phi} \right) \partial_{\mu} \chi \, \partial^{\mu} \chi 
			+ B \nabla^{\mu} \partial_{\mu} \chi 
			- V_{\phi}=0, \\
			\left( -\frac{1}{2} A_{\chi} + B_{\phi} \right) \partial_{\mu} \phi \, \partial^{\mu} \phi
			+ B \nabla^{\mu} \partial_{\mu} \phi 
			+ \frac{1}{2} C_{\chi} \partial_{\mu} \chi \, \partial^{\mu} \chi 
			+ C \nabla^{\mu} \partial_{\mu} \chi 
			+ C_{\phi} \partial_{\mu} \phi \, \partial^{\mu} \chi 
			- V_{\chi}=0. \label{fieldscalar}
		\end{align}

Here, $A_\phi$ is the derivative of $A(\phi,\chi)$ with respect to $\phi$, etc.
	\subsection{Elimination of the ghosts}
	\label{elimination}
	First, it is important to note that the elimination of ghosts presented in this subsection follows the procedure outlined by Ref.~\cite{Nojiri2024}. Now consider a metric of a general spherically symmetric and time-dependent spacetime, given by
	\begin{align}
		ds^2 =& -e^{2\nu(t,r)} dt^2  +e^{2\lambda(t,r)} dr^2 +r^2 (d\theta^2 + \sin^2 \theta \, d\varphi^2).
		\label{generalmetric}
	\end{align}
	Let us now assume the ansatz
	\begin{align}
		\phi = t, \quad \chi = r.
		\label{fieldansatz}
	\end{align}
	A detailed justification that this ansatz does not lead to any loss of generality can be found in Refs.~\cite{Nojiri2021,NojiriNashed2024,Nojiri2024NPB}.
	
	According to Refs.~\cite{Nojiri2024,Nojiri2021,NojiriNashed2024,Nojiri2024NPB,Nojiri2023}, the functions $A(\phi,\chi)$ and/or $C(\phi,\chi)$ are often negative, which implies that $\phi$ and/or $\chi$ behave as ghosts. To eliminate the ghosts, appropriate constraints must be introduced by adding Lagrange multipliers \textcolor{black}{$\Lambda_\phi$} and \textcolor{black}{$\Lambda_\chi$}. This results in the addition of an additional action \textcolor{black}{$S_\Lambda$}, which reads \cite{Nojiri2024,NojiriNashed2024}
	\begin{align}
		\textcolor{black}{S_\Lambda} = \int d^4x\sqrt{-g} \big[ & \textcolor{black}{\Lambda_\phi} \big(e^{-2\nu(t=\phi,r=\chi)}\partial_\mu\phi\partial^\mu\phi + 1\big) \nonumber \\
		& + \textcolor{black}{\Lambda_\chi} \big(e^{-2\lambda(t=\phi,r=\chi)}\partial_\mu\chi\partial^\mu\chi - 1\big) \big].\label{actionlambda}
	\end{align}
	
	Variations of $S_\Lambda$ with respect to $\Lambda_\phi$ and $\Lambda_x$ give the following constraints:
	\begin{align}
		0 &= e^{-2\nu(t=\phi,r=\chi)}\partial_\mu\phi\partial^\mu\phi + 1, \label{constraint1} \\
		0 &= e^{-2\lambda(t=\phi,r=\chi)}\partial_\mu\chi\partial^\mu\chi - 1.\label{constraint2}
	\end{align}
	
	The fluctuations of $\phi$ and $\xi$ can be written as \cite{Nojiri2024,NojiriNashed2024,Nojiri2024NPB}
	\begin{eqnarray}
		\phi=t+\delta\phi,\:\:\:\ \chi=r+\delta\chi.
	\end{eqnarray}
	By using Eqs. (\ref{constraint1}) and (\ref{constraint2}), one can obtain
	\begin{eqnarray}
		\partial_t(\delta\phi)=\partial_r(\delta r)=0,
	\end{eqnarray}
	which means that both $\delta\phi$ and $\delta\chi$ vanish in the whole spacetime if we set the initial condition $\delta\phi=0$ and the boundary condition $\delta\chi\rightarrow0$ when $r\rightarrow \infty$, and we are left with $\delta\phi=\delta\chi=0$ \cite{Nojiri2024,NojiriNashed2024,Nojiri2024NPB}.
	
	It is worth noting that, as pointed out by Ref.~\cite{Nojiri2024}, even within the model defined by the modified action $S_{GR\phi\chi\mathcal{B}} + \textcolor{black}{S_\Lambda}$, the choice $\textcolor{black}{\Lambda_\phi = \Lambda_\chi = 0}$ consistently satisfies the field equations. Consequently, any solution of Eq.~(\ref{field}) derived from the original action (\ref{action}) also remains a solution in the modified model with action $S_{GR\phi\chi\mathcal{B}} + \textcolor{black}{S_\Lambda}$. It is also worth noting that the constraints given by Eqs.~(\ref{constraint1}) and (\ref{constraint2}) generalize the mimetic constraint introduced by Ref.~\cite{Chamseddine2013}, where nondynamical dark matter effectively emerges. The review of mimetic gravity can be referred to Ref.~\cite{Vagnozzi}.
	
	It is important to clarify the role of the auxiliary scalar fields $\chi$ and $\phi$ introduced in this construction. These fields are not intended to represent independent propagating degrees of freedom, but rather serve as auxiliary variables that parametrize the geometry and generate an effective EMT through the structure of the action. 
	\textcolor{black}{In particular, once the ansatz $\chi=r$ and $\phi=t$ is imposed, the functions $A(\phi,\chi)$, $C(\phi,\chi)$, and the potential $V(\phi,\chi)$ become completely determined. Consequently, the constrained scalar sector gives rise to an nondynamical additional term in EMT that mimics the role of exotic matter required to support the wormhole geometry. It should be emphasized, however, that these fields are geometric ansatzes and should not be interpreted as an ordinary physical matter component.}
	
	\textcolor{black}{Note that the absence of ghosts does not imply full dynamical stability. A radial perturbation analysis is employed in Sec.~\ref{stabilityanalysissss} to investigate the stability of the systems against radial perturbations.}
	
	\subsection{Energy conditions and traversability}
	\label{energyconditions}
	First, we have to highlight that the total EMT in our model generally consist of contributions from ordinary matter, the magnetic field, and the scalar fields. In GR, the NEC, WEC, SEC, and DEC impose the following sequence of inequalities on the total EMT $T_{\text{tot}\,\mu\nu}$ \cite{Nojiri2024}:
	\begin{eqnarray*}
		T_{\text{tot}\:\mu\nu}\:k^\mu k^\nu \ge 0,\qquad
		T_{\text{tot}\:\mu\nu}\:V^\mu V^\nu \ge 0,&\\
		\left( T_{\text{tot}\:\mu\nu} - \frac{1}{2} g_{\mu\nu} T_\text{tot} \right) V^\mu V^\nu \ge 0,\qquad
		&T_{\text{tot}\:\mu\nu}\:V^\mu V^\nu \ge 0,&\\
		\text{and}\qquad T_{\text{tot}\:\mu\nu}\:V^\nu \qquad\text{is not spacelike,}&
	\end{eqnarray*}
	for any null vector $k^\mu$, $g_{\mu\nu}k^\mu k^\nu=0$, and for any timelike vector $V^\mu$, $g_{\mu\nu}V^\mu V^\nu<0$. For the system that we are considering in this work, generally, we have 
	$T_{\text{tot}\:\mu}^\nu=\text{diag}(-\rho_{\text{tot}}; p_{r,\text{tot}}; p_{t,\text{tot}};p_{t,\text{tot}})$. Now the energy conditions writes \cite{Nojiri2024,Nojiri2023}
	\begin{equation}
		\begin{aligned}
			\mathrm{NEC}: \qquad &
			\rho_{\rm tot}+p_{r,\rm tot}\geq 0,
			\qquad
			\rho_{\rm tot}+p_{t,\rm tot}\geq 0,
			\\[2mm]
			\mathrm{WEC}: \qquad &
			\rho_{\rm tot}+p_{r,\rm tot}\geq 0,
			\qquad
			\rho_{\rm tot}+p_{t,\rm tot}\geq 0,
			\\
			&
			\rho_{\rm tot}\geq 0,
			\\[2mm]
			\mathrm{SEC}: \qquad &
			\rho_{\rm tot}+p_{r,\rm tot}\geq 0,
			\\
			&
			\rho_{\rm tot}+p_{t,\rm tot}\geq 0,
			\qquad
			\rho_{\rm tot}+p_{r,\rm tot}+2p_{t,\rm tot}\geq 0,
			\\[2mm]
			\mathrm{DEC}: \qquad &
			\rho_{\rm tot}\geq 0,
			\qquad
			\rho_{\rm tot}\geq |p_{r,\rm tot}|,
			\qquad
			\rho_{\rm tot}\geq |p_{t,\rm tot}|.
		\end{aligned}
	\end{equation}
	
	To obtain the traversable wormhole, we need the violation of the NEC in the vicinity of a throat \cite{Zaslavskii2007}. It implies the violation of all energy conditions \cite{Nojiri2024}. Before analyzing the traversability of the wormhole, we must first determine the total energy density $\rho_{\text{tot}}$, the total radial pressure $p_{r,\text{tot}}$, and the total tangential pressure $p_{t,\text{tot}}$. Once again we have to highlight that these quantities consist of contributions from ordinary matter, the magnetic field, and the scalar fields. To obtain them, one must first derive the explicit expressions for the functions $A$, $B$, $C$, and $V$.
	
	Now we reconsider the metric given in Eq.~(\ref{generalmetric}) for a static, spherically symmetric stellar object harboring a wormhole, which reads
	\begin{eqnarray}
		ds^{2} &=& -e^{2\nu(r)} dt^{2} + e^{2\lambda(r)} dr^{2} + r^{2} ( d\theta^{2}+ \sin^{2}\theta\, d\varphi^{2} ).
		\label{metric}
	\end{eqnarray}
	\textcolor{black}{We employ an ansatz}
	\begin{eqnarray}
		e^{2\lambda(r)} = \left( 1 - \frac{r_{0}}{r} \right)^{-1},\label{elambdaansatz}
	\end{eqnarray}
	where \( r_{0} \) denotes the radius of the wormhole throat. Furthermore, for \( r \geq R_{s} \), where \( R_{s} \) is the surface radius, the exterior solution satisfies
	\begin{eqnarray}
		e^{2\nu(r)} = e^{-2\lambda(r)} = 1 - \frac{r_{0}}{r}.
		\label{exterior}
	\end{eqnarray}
	\textcolor{black}{The ansatzes shown by Eqs.~(\ref{elambdaansatz}) and (\ref{exterior}) have also been used in the ad hoc formalism in Ref.~\cite{Nojiri2024}.}
	
	
	For the CSSTW system, near the throat $r\sim r_0$, the metric function $e^{2\lambda(r)}$ behaves as~\cite{Nojiri2024}
	\begin{eqnarray}
		e^{2\lambda(r)}\sim \frac{r_0}{r-r_0}e^{2\lambda_0},
	\end{eqnarray}
	where $\lambda_0$ is a constant. The radial coordinate $r$ can be redefined as~\cite{Nojiri2024}
	\begin{eqnarray}
		l=2\sqrt{r_0(r-r_0)}.\label{coordinatelllll}
	\end{eqnarray}
	Near the throat, the metric becomes
	\begin{eqnarray}
		ds^2&\sim&-e^{2\nu(l)}dt^2+e^{2\lambda_0}dl^2+r_0^2(d\theta^2
        +\sin^2\theta d\varphi^2).\label{metricell}
	\end{eqnarray}
	
	From the ansatz given in Eq.~(\ref{elambdaansatz}), one immediately obtains $\lambda_0 = 0$.
	According to Ref.~\cite{Nojiri2024}, although the new radial coordinate $l$ is defined to be positive, the metric with the new radial coordinate $l$ admits an analytic continuation to the region $l < 0$.
	We assume that the region $l > 0$ corresponds to our spacetime, while the region $l < 0$ represents another spacetime, which is smoothly connected to ours through the wormhole.
	
	The EMT for anisotropic neutron fluid with chaotic magnetic field $\mathcal{T}_{\mu\nu}$ writes
		\begin{eqnarray}
			\mathcal{T}_{\mu\nu}&\equiv& T_{\text{matter}\:\mu\nu}+T_{\mathcal{B}\:\mu\nu}\nonumber\\
            &=&\begin{pmatrix}
				-\left(\rho + \dfrac{\mathcal{B}^{2}}{8\pi}\right) g_{tt} & 0 & 0 & 0 \\[6pt]
				0 & \left(p_{r} + \dfrac{\mathcal{B}^{2}}{24\pi}\right) g_{rr} & 0 & 0 \\[6pt]
				0 & 0 & \left(p_{t} + \dfrac{\mathcal{B}^{2}}{24\pi}\right) g_{\theta\theta} & 0 \\[6pt]
				0 & 0 & 0 & \left(p_{t} + \dfrac{\mathcal{B}^{2}}{24\pi}\right) g_{\varphi\varphi}
			\end{pmatrix}.\label{EMTmagnetized}
		\end{eqnarray}
		
		
		Now we can define
		\begin{eqnarray}
			\varrho \equiv\left(\rho + \dfrac{\mathcal{B}^{2}}{8\pi}\right),\:\:\:\:
			P_r\equiv\left(p_{r} + \dfrac{\mathcal{B}^{2}}{24\pi}\right),\:\:\:\:P_t\equiv\left(p_{t} + \dfrac{\mathcal{B}^{2}}{24\pi}\right),
		\end{eqnarray}
		where we still have $\sigma=p_r-p_t=P_r-P_t$.
		
		With the current EMT in our hands, and by using Eq. (\ref{field}), the $(t,t)$, $(r,r)$, $(\theta,\theta)$ components can be written as
		\begin{equation}
			-e^{2\nu}\left(-\frac{A}{2}e^{-2\nu}-\frac{C}{2}e^{-2\lambda}-V\right)+e^{2\nu}\varrho=\frac{e^{-2\lambda+2\nu}}{\kappa^2}\left(\frac{2\lambda'}{r}+\frac{e^{2\lambda}-1}{r^2}\right),\label{tt}
		\end{equation}
		\begin{equation}
			e^{2\lambda}\left(\frac{A}{2}e^{-2\nu}+\frac{C}{2}e^{-2\lambda}-V\right)+e^{2\lambda}P_r=\frac{1}{\kappa^2}\left(\frac{2\nu'}{r}-\frac{e^{2\lambda}-1}{r^2}\right),\label{rr}
		\end{equation}
		\begin{align}
			r^2\bigg(\frac{A}{2}e^{-2\nu}
			-\frac{C}{2}e^{-2\lambda}-V\bigg)
			+r^2P_t=\:&\frac{1}{\kappa^2}\big\{e^{-2\lambda}[r(\nu'-\lambda')+r^2\nu''+r^2(\nu'-\lambda')\nu']\big\}.\label{thetatheta}
		\end{align}
		With a little algebra, we obtain
		\begin{align}
			A &= \frac{e^{2\nu}}{\kappa^2} \left\{ e^{-2\lambda} \left[ \frac{\nu' + \lambda'}{r} + \nu'' + (\nu' - \lambda') \nu' + \frac{e^{2\lambda} - 1}{r^2} \right] \right\} - e^{2\nu} (\varrho + P_r - \sigma),\label{A} \\[10pt]
			C &= \frac{e^{2\lambda}}{\kappa^2} \left\{- e^{-2\lambda} \left[ -\frac{\nu' + \lambda'}{r} + \nu'' + (\nu' - \lambda') \nu' + \frac{e^{2\lambda} - 1}{r^2} \right] \right\}-e^{2\lambda}\sigma, \label{C}\\[10pt]
			V &= \frac{e^{-2\lambda}}{\kappa^2} \left( \frac{\lambda' - \nu'}{r} + \frac{e^{2\lambda} - 1}{r^2} \right) - \frac{1}{2} (\varrho - P_r).\label{V}
		\end{align}
	Note that the function $B$ appearing in Eq.~(\ref{action}) can be determined by allowing for a time-dependent metric: in that case the $(t,r)$ component of the field equations does not vanish but yields $B = \frac{2\dot{\lambda}}{\kappa^{2} r}$. Since in the present time-independent metric $\dot{\lambda}=0$, we therefore obtain $B=0$.
	
	From Eqs. (\ref{tt})-(\ref{thetatheta}), we have
	\begin{eqnarray}
		\textcolor{black}{\rho_{\text{tot}}= \frac{A}{2}e^{-2\nu}+\frac{C}{2}e^{-2\lambda}+V+\varrho,}\label{rhototmentah}
	\end{eqnarray}
	\begin{eqnarray}
		\textcolor{black}{p_{r,\text{tot}}=\frac{A}{2}e^{-2\nu}+\frac{C}{2}e^{-2\lambda}-V+P_r,}\label{prtotmentah}
	\end{eqnarray}
	\begin{eqnarray}
		\textcolor{black}{p_{t,\text{tot}}=\frac{A}{2}e^{-2\nu}
			-\frac{C}{2}e^{-2\lambda}-V+P_t.}\label{pttotmentah}
	\end{eqnarray}
	\textcolor{black}
	{After substituting the expressions of $A$, $C$, and $V$ from Eqs. (\ref{A})-(\ref{V}) into Eqs. (\ref{rhototmentah})-(\ref{pttotmentah}) we obtain}
	\begin{eqnarray}
		\textcolor{black}{\rho_{\text{tot}}=0}\label{rhotot},
	\end{eqnarray}
	\begin{eqnarray}
		\textcolor{black}{p_{r,\text{tot}}=\frac{2r(r-r_0)\nu'-r_0}{\kappa^2r^3},}\label{prtot}
	\end{eqnarray}
	\begin{eqnarray}
		\textcolor{black}{ p_{t,\text{tot}}=\frac{2r^2(r-r_0)(\nu''+\nu'^2)+r(2r-r_0)\nu'+r_0}{2\kappa^2 r^3}.}\label{pttot}
	\end{eqnarray}
	
	\textcolor{black}{From Eq.~(\ref{rhotot}), we see that the total energy density vanishes. This result arises from the exact cancellation between the neutron-fluid energy density and the energy densities of the scalar fields. Note that the scalar fields play role of the exotic matter supporting the wormhole geometry. Consequently, the mass cannot be defined through the volume integral of the total energy density. Instead, it is determined from the metric component $e^{2\lambda}=(1-r_0/r)^{-1}$, where $r_0$ plays the role of $2M$, with $M$ being the ADM mass. Therefore, throughout this work, we adopt this definition of the mass. In view of this result, the object considered here cannot be regarded as a NS in the astrophysical sense, even though it is composed of neutron fluid. Instead, it is more appropriately interpreted as a compact star-like system.}
	
	\textcolor{black}{It is important to emphasize that the absence of explicit anisotropic or magnetic-field terms in the final expressions for the energy density and pressures, shown by Eqs.~(\ref{rhotot})--(\ref{pttot}), does not imply that anisotropy or the magnetic field are physically absent from the system. 
		Rather, within the present effective mathematical expression, the total energy density, the total radial pressure, and the total tangential pressure admit the same functional form in terms of the metric functions, independently of whether anisotropy and/or magnetic fields are included in the matter sector.}
	
	\textcolor{black}{As a consequence, the traversability conditions are governed by the total EMT, and remain unchanged both in the presence and in the absence of anisotropy and magnetic fields. 
		In this sense, anisotropy and the magnetic field do not explicitly modify the traversability criteria, although they contribute implicitly to the effective geometry through the construction of the metric.
	}
	
	At the limit $r \rightarrow r_0$, we have
	\begin{eqnarray*}
		(\rho_{\text{tot}} + p_{r,\text{tot}}) &\;\longrightarrow\;& -\frac{1}{\kappa^{2} r_0^{\,2}}, \\
		(\rho_{\text{tot}} + p_{t,\text{tot}}) &\;\longrightarrow\;& 
		\frac{1 + r_0\,\nu'(r_0)}{2 \kappa^{2} r_0^{\,2}},
	\end{eqnarray*}
	which explicitly shows that the NEC is violated in the vicinity of the wormhole throat, and that our wormhole is traversable.
	
	An important remark from Ref.~\cite{Nojiri2024} should be emphasized: 
	the violation of the energy conditions usually induces instabilities in the configuration. For instance, the sound speed may exceed the speed of 
	light. In the wormhole spacetime constructed from the two-scalar model used in this work, however, such instabilities do not arise. This is because the two scalar fields are nondynamical; they neither propagate nor fluctuate.
	Consequently, no sound mode is generated from oscillations of the effective fluid produced by the scalar fields, even though this effective fluid violates the energy conditions. The absence of the sound mode is really important for the formulation of radial perturbation that is discussed in Sec.~\ref{stabilityanalysissss}.

\section{Non-magnetized configuration of CSSTW}
	\label{nonmagnetized}
	\textcolor{black}{Before discussing the CSSTW, we briefly clarify how the corresponding pure NS solutions are obtained. 
		The NS configurations considered in this work are constructed using the standard TOV equations for a spherically symmetric relativistic fluid, without a wormhole throat.
		These solutions serve solely as reference configurations, allowing for a direct comparison between ordinary NSs and the CSSTW discussed below.
		No modification of the standard TOV formalism is assumed for the pure NS case.}
	
	From Eq.~(\ref{fieldscalar}), for the CSSTW model considered here, one obtains
	\begin{eqnarray}
		0&=& \frac{1}{2}A_\chi e^{-2\nu}+\frac{1}{2}C_\chi e^{-2\lambda}\nonumber\\
		&&+Ce^{-2\lambda}\left(\frac{2}{r}+\nu'-\lambda'\right)-V_\chi. \label{fieldscalar2}
	\end{eqnarray}
	Recalling that $\chi=r$, Eq.~(\ref{fieldscalar2}) can be written as
	\begin{eqnarray}
		0&=&\frac{1}{2}\frac{dA}{dr} e^{-2\nu}+\frac{1}{2}\frac{dC}{dr} e^{-2\lambda}\nonumber\\
		&&+Ce^{-2\lambda}\left(\frac{2}{r}+\nu'-\lambda'\right)-\frac{dV}{dr}.
	\end{eqnarray}
	By using the expression of $A$, $C$, and $V$ shown by Eqs. (\ref{A})-(\ref{V}), we are left with
	\begin{eqnarray}
		\nu'(P_r+\varrho)+P_r'+\frac{2\sigma}{r}=0,\label{conservation law}
	\end{eqnarray}
	which is the conservation law for anisotropic fluid.
	
	\textcolor{black}{It is worth emphasizing that the conservation equation (\ref{conservation law}) is not imposed as an independent assumption for the neutron fluid alone.
		Instead, it emerges naturally from the equations of motion obtained by varying the action with respect to the auxiliary scalar fields.
		This procedure guarantees the internal consistency of the constrained scalar-gravity system that underlies the CSSTW construction.}
	
	\textcolor{black}{Physically, this approach ensures that the presence of auxiliary scalar fields, introduced to generate the effective exotic matter sector required by the wormhole geometry, does not spoil the standard conservation properties of the ordinary matter.
		Although the scalar fields contribute implicitly through the geometry and the effective EMT, the resulting conservation law involves only the ordinary matter and magnetic field variables.
		In this sense, the conservation equation should be interpreted as a consistency condition of the full system.}
	
	For non-magnetized configuration, we have
	\begin{eqnarray}
		\nu'(p_r+\rho)+p_r'+\frac{2\sigma}{r}=0.\label{conservationlawnonmag}
	\end{eqnarray}
	By using the anisotropy profile shown by Eq.~(\ref{CHEW}), Eq.~(\ref{conservationlawnonmag}) can be integrated to give
	\begin{eqnarray}
		\nu=-\left(2-\frac{1}{h}\right)\int\frac{dp_r}{\rho+p_r}.
	\end{eqnarray}
	
	Note that in this work we use the polytropic equation of state (EOS), which reads \cite{Dzhunushaliev2012,Nojiri2024,daSilva2021,Godani2024,Ray:2007}
	\begin{eqnarray}
		p_r=K \rho^{1+\frac{1}{n}}.\label{polytripiceosss}
	\end{eqnarray}
	In this paper, we adopt the same parameters as those used by Ref.~\cite{Nojiri2024}, namely 
	$K = 100~\mathrm{km}^2$ and $n=1$. Therefore, the EOS reduces to $p_r = K \rho^2$.
	For the matter-energy density profile, we also follow Ref.~\cite{Nojiri2024} and employ a 
	Tolman--VII-like profile, i.e.
	\begin{eqnarray}
		\rho=\rho_c\left[1-\left(\frac{r-r_0}{R_s-r_0}\right)^2\right],
	\end{eqnarray}
	where $\rho_c$ denotes the energy density of the NS matter at $r=r_0$. \textcolor{black}{The present analysis is intentionally restricted to the adopted density profile and matter model. Consequently, the obtained quantitative results should be interpreted as solutions of this particular model rather than universal features of wormhole embedded in compact star-like systems. A systematic investigation of the dependence on different EOS and density profiles is left for future work.}
	
	With straightforward integration of Eq.~(\ref{conservationlawnonmag}), one finds
	\begin{eqnarray}
		\nu=\nu_c-\left(4-\frac{2}{h}\right)\ln{\left(1+K\rho\right)},\label{nunonmagnetized}
	\end{eqnarray}
	where $\nu_c$ is the integration constant. Consequently, the temporal metric function reads
	\begin{eqnarray}
		e^{2\nu}=\frac{e^{2\nu_c}}{(1+K \rho)^{8-\frac{4}{h}}}.\label{enunonmag}
	\end{eqnarray}
	
	At $r = R_s$, the energy density vanishes, $\rho(R_s)=0$, and the temporal metric
	function satisfies $e^{2\nu(R_s)} = 1 - \dfrac{r_0}{R_s}$. Substituting this condition into Eq.~(\ref{enunonmag}) yields
	\begin{eqnarray}
		e^{2\nu_c}=1-\frac{r_0}{R_s}.\label{constantnonmag}
	\end{eqnarray}
	
	We can also compute the radial derivative of Eq.~(\ref{enunonmag}) at $r=R_s$.
	Using $\rho(R_s)=0$ and $\rho'(R_s) = -2\rho_c/(R_s - r_0)$ for the Tolman--VII-like
	profile, one obtains.
	\begin{eqnarray}
		(e^{2\nu})'|_{r=R_s}=\left(16-\frac{8}{h}\right)\frac{K\rho_c}{R_s}.
	\end{eqnarray}
	As $(e^{2\nu})'|_{r=R_s}=\dfrac{r_0}{R_s^2}$, we obtain the relation
	\begin{eqnarray}
		r_0=\left(16-\frac{8}{h}\right)K\rho_c R_s.\label{throatradiusnonnnn}
	\end{eqnarray}
	The metric given by Eq. (\ref{metric}) implies that the ADM mass $M$ satisfies
	\begin{eqnarray}
		M=\frac{r_0}{2}=\left(8-\frac{4}{h}\right)K\rho_c R_s.\label{ADM}
	\end{eqnarray}
	Generally, the total mass of the CSSTW is given by~\cite{Nojiri2024}
	\begin{eqnarray}
		M(r)=\frac{r_0}{2}+\int_{r_0}^{r}\rho_{\text{tot}}(r') r'^2 dr'.
	\end{eqnarray}
	Since $\rho_{\text{tot}}=0$, the total mass is only taken from the ADM mass contribution, i.e. Eq.~(\ref{ADM}). It is also important to highlight that for the isotropic case where $h=1$, the calculations recover to the one presented by Ref.~\cite{Nojiri2024}.
	
	To facilitate a comparison with pure polytropic NS configurations, we adopt the following two approaches. Approach~1: for a given central energy density $\rho_c$, the radius of the surface, $R_s$, is assumed to coincide with the radius of an ordinary NS obtained at the same $\rho_c$. Approach~2: for a given central energy density $\rho_c$, the total mass of the system, $M$, is taken to be equal to that of an ordinary NS computed at the same $\rho_c$.
	
	Both approaches have also been employed by Refs.~\cite{Nojiri2024,Pattersons2026mpla}, which serve as methodological predecessors to the procedure adopted in this work. \textcolor{black}{However, it is worth noting that Approaches 1 and 2 are ad hoc methods introduced to force the mass or radius of the CSSTW to match that of a pure NS at a given central density. Both approaches are purely mathematical construction schemes and do not represent any physical formation scenario.} In the present analysis, we consider $h = 1,;1.5,;2$.
	
	\textcolor{black}{At this stage, it is important to consider junction conditions for our model. The method related to junction conditions can be referred to Ref.~\cite{Sakti:2025dkf}. However, it is worth noting that junction conditions are not required for our model. Eqs.~(\ref{rhotot})-(\ref{pttot}) are valid throughout the entire spacetime, both inside and outside the neutron fluid. These equations show that the total pressures are smoothly continuous across the fluid boundary. This behavior arises from the presence of the nondynamical scalar fields throughout the spacetime, whose contributions of total pressures remain nonzero outside the neutron fluid. Consequently, the exterior spacetime is not vacuum, even though its metric takes a Schwarzschild-like form. In addition to the continuity of the pressures, the metric is also continuous across the fluid boundary. This continuity is ensured by the conditions given in Eqs.~(\ref{enunonmag})-(\ref{constantnonmag}), where the exterior metric is given by Eq.~(\ref{exterior}). Note that the first derivative of the temporal metric function is also continuous.}
	
	\textcolor{black}{We also verify that the configurations are horizonless. Since the systems are static and spherically symmetric, the temporal component of the metric, $g_{tt}=-e^{2\nu(r)}$, provides a criterion for identifying a potential horizon. In particular, a horizon may occur where the temporal metric function vanishes, $e^{2\nu(r)}=0$. We therefore examine the behavior of the interior temporal metric function given by Eq.~(\ref{enunonmag}) by plotting its values. As shown in Fig.~\ref{figinteriormetricnon}, the temporal metric function does not vanish anywhere in the interior of the systems. For the exterior spacetime, where $e^{2\nu(r)}=1-\frac{r_0}{r}$, the temporal metric function vanishes only at $r=r_0$. However, this point does not lie within the exterior region considered in our model. Therefore, the temporal metric function remains nonzero throughout the whole spacetime, indicating that no horizon is present in our systems.}

\textcolor{black}{In order to ensure that the geometry is regular, we calculate Kretschmann scalar, which is given by~\cite{Henry:1999rm}}
\begin{eqnarray}
	\textcolor{black}{\mathcal{K}=R^{\alpha\beta\gamma\delta}R_{\alpha\beta\gamma\delta}.}
\end{eqnarray}

\textcolor{black}{In this calculation, we transform, $r\rightarrow l$, where $l$ is shown by Eq.~(\ref{coordinatelllll}). Thus, the interior metric is given by}

\textcolor{white}{.}
\begin{eqnarray}
	\textcolor{black}{ds^2}&\textcolor{black}{=}&\textcolor{black}{-e^{2\nu(l)}dt^2+dl^2+\left(r_0+\frac{l}{2r_0}\right)^2}
    \textcolor{black}{(d\theta^2 +\sin^2\theta d\varphi).}
\end{eqnarray}
\textcolor{black}{For such a metric, the throat is at $l=0$. As mentioned earlier in Subsec.~\ref{energyconditions}, $l<0$ corresponds to another universe connected by the throat. The Kretschmann scalar is given by}
\begin{equation}
		\begin{aligned}
			\textcolor{black}{\mathcal{K}} \;\textcolor{black}{=}\; \textcolor{black}{4\Biggl\{}\; 
			&\textcolor{black}{64\left(1-2h\right)^{2}h^{2}K^{2}\,r_0\left(R_s-r_0\right)\mathcal{U}^{4}\rho_c^{2}\,\mathcal{I}^{2}\mathcal{H}^{18}+64\left(1-2h\right)^{2}h^{2}K^{2}\,r_0\left(R_s-r_0\right)\mathcal{U}^{4}\rho_c^{2}\,\mathcal{H}^{20}} \\[2pt]
			\textcolor{black}{+}\;&\textcolor{black}{h^{4}\left(1-4r_0^{2}\right)^{2}\mathcal{U}^{2}\mathcal{H}^{22}+8\left(1-2h\right)^{2}K^{2}\mathcal{U}^{6}\rho_c^{2}\,\mathcal{I}^{18}\mathcal{J}^{2}+8\left(1-2h\right)^{2}K^{2}\mathcal{U}^{6}\rho_c^{2}\,\mathcal{H}^{18}\mathcal{J}^{2} \;
				\Biggr\}} \\[2pt]
			\textcolor{black}{\times}\;&\textcolor{black}{\Bigl[\,h^{4}\mathcal{U}^{6}\mathcal{I}^{2}\mathcal{H}^{20}\,\Bigr]^{-1},}
		\end{aligned}
		\label{kretschmannnon}
	\end{equation}
	\textcolor{black}{where}
	\begin{align}
		\textcolor{black}{\mathcal{Y}} &\textcolor{black}{= \sqrt{r_0\left(R_s-r_0\right)}} \\[4pt]
		\textcolor{black}{\mathcal{U}} &\textcolor{black}{= 2r_0^{2}+2\mathcal{Y}} \\[4pt]
		\textcolor{black}{\mathcal{I}} &\textcolor{black}{= -4r_0^{2}\left(R_s-r_0\right)^{2}
			+K\rho_c\Bigl[-4r_0^{2}\left(R_s-r_0\right)^{2}+4r_0\left(R_s-r_0\right)\Bigr]} \\[4pt]
		\textcolor{black}{\mathcal{H}} &\textcolor{black}{= 4Kr_0\left(R_s-r_0\right)\rho_c-4r_0^{4}\left(1+K\rho_c\right)}
		\textcolor{black}{-4r_0^{2}R_s^{2}\left(1+K\rho_c\right)+8r_0^{3}\left(R_s+KR_s\rho_c\right)} \\[4pt]
		\textcolor{black}{\mathcal{J}} &\textcolor{black}{= -16Kr_0\left(R_s-r_0\right)\rho_c
			+h\Bigl[36Kr_0\left(R_s-r_0\right)\rho_c+4r_0^{4}\left(1+K\rho_c\right)}\nonumber\\
		&\qquad\textcolor{black}{+4r_0^{2}R_s^{2}\left(1+K\rho_c\right)-8r_0^{3}\left(R_s+KR_s\rho_c\right)\Bigr]}
	\end{align}

\textcolor{black}{In Fig.~\ref{figkretschmannnon}, we plot the Kretschmann scalar for all numerical configurations considered in the interior spacetime. The values of $\rho_c$, $r_0$, and $R_s$ correspond to those used in all of our numerical configurations. We observe that the Kretschmann scalar remains finite throughout the considered parameter space, indicating the absence of curvature singularities in the interior spacetime of the CSSTW.}

\textcolor{black}{The Kretschmann scalar for the exterior of CSSTW, whose metric is given by Eq.~(\ref{exterior}) is given by $\mathcal{K}_{ext}=\frac{12 r_0^2}{r^6}$. Since $r\geq R_s>r_0$ in the exterior region, $\mathcal{K}_{ext}$ remains finite throughout the entire exterior spacetime. This indicates that the exterior spacetime is free of curvature singularities.}

\textcolor{black}{Another point worth noting in Fig.~\ref{figkretschmannnon} concerns the behavior of the Kretschmann scalar as $\rho_c\rightarrow 0$ and $l\rightarrow 0$. Although the contour appears to exhibit singular-like behavior in this region, the Kretschmann scalar itself remains finite. If we take $\rho_c\rightarrow0$, the Kretschmann scalar becomes}
\begin{eqnarray}
	\textcolor{black}{ \mathcal{K}\approx\frac{(8r_0^3 R_s-4r_0^2 R_s^2-4r_0^4)^2(1-4 r_0^2)^2}{4r_0^4(l+2r_0^2)^4(R_a-r_0)^4}.}\label{rhocmendekati0}
\end{eqnarray}
\textcolor{black}{We can see that this singular-like behavior arises from the denominator of Eq.~(\ref{rhocmendekati0}), since, according to Eq.~(\ref{throatradiusnonnnn}), $r_0\propto\rho_c$. However, it is important to emphasize that neither $\rho_c$ and $r_0$ vanishes for our CSSTW configurations. Therefore, the entire spacetime of all non-magnetized CSSTW considered in this section remains regular.}

\textcolor{black}{Furthermore, to ensure that the wormhole is traversable, we do not rely solely on the violation of the NEC. A traversable wormhole should have a redshift function $\nu(r)$ that is free of singularities and horizons, together with the throat condition $b(r_0)=r_0$ and flare-out confition $b'(r_0)<1$~\cite{Dzhunushaliev2011}, where $b(r)$ is the shape function defined through $e^{-2\lambda(r)}=1-\frac{b(r)}{r}$. The absence of curvature singularities and horizons is demonstrated in Figs.~\ref{figkretschmannnon} and~\ref{figinteriormetricnon}, respectively. In our system, $b(r)=r_0$, as can be seen from Eqs.~(\ref{elambdaansatz})--(\ref{exterior}). Therefore, we obtain $b(r_0)=r_0$ and $b'(r_0)=0<1$, satisfying the throat and flare-out conditions, respectively. Thus, in addition to the NEC violation, all the required geometric conditions for traversability are satisfied.}

\begin{figure*}[htbp]
	\centering
	\begin{tabular}{cc}
		\includegraphics[width=0.4\textwidth]{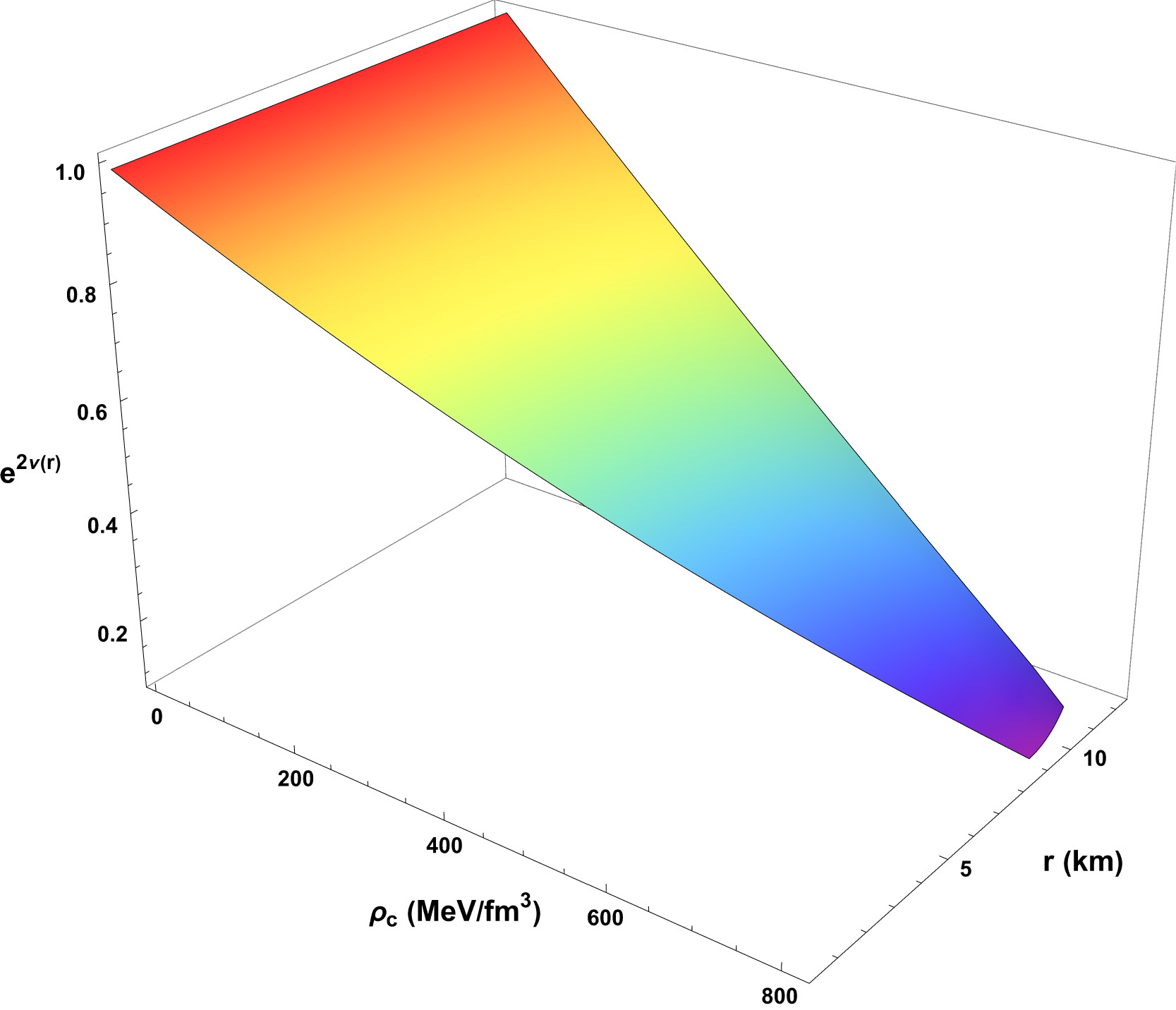} &
		\includegraphics[width=0.4\textwidth]{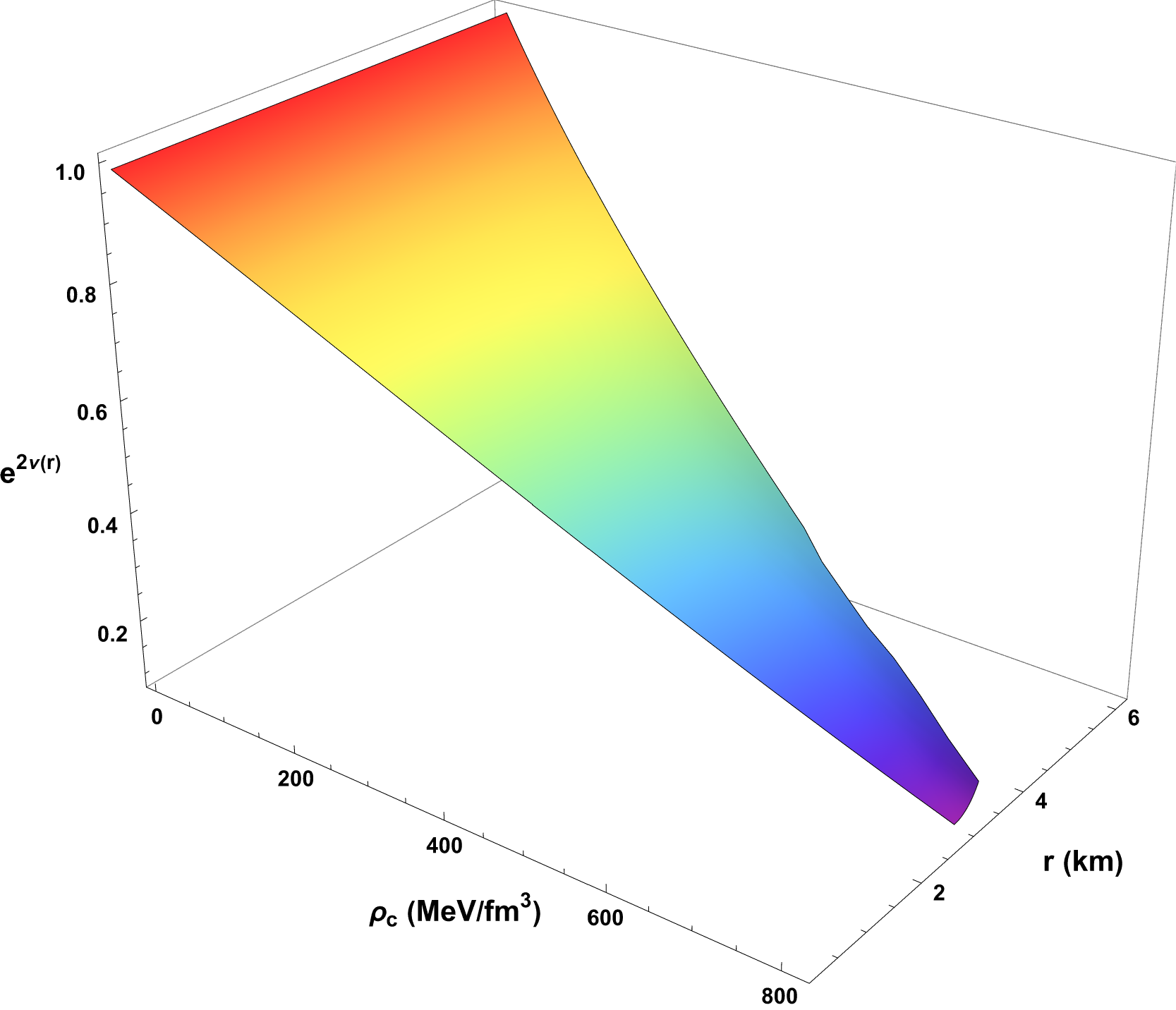} \\
		(a) & (b) \\[1em]
		\includegraphics[width=0.4\textwidth]{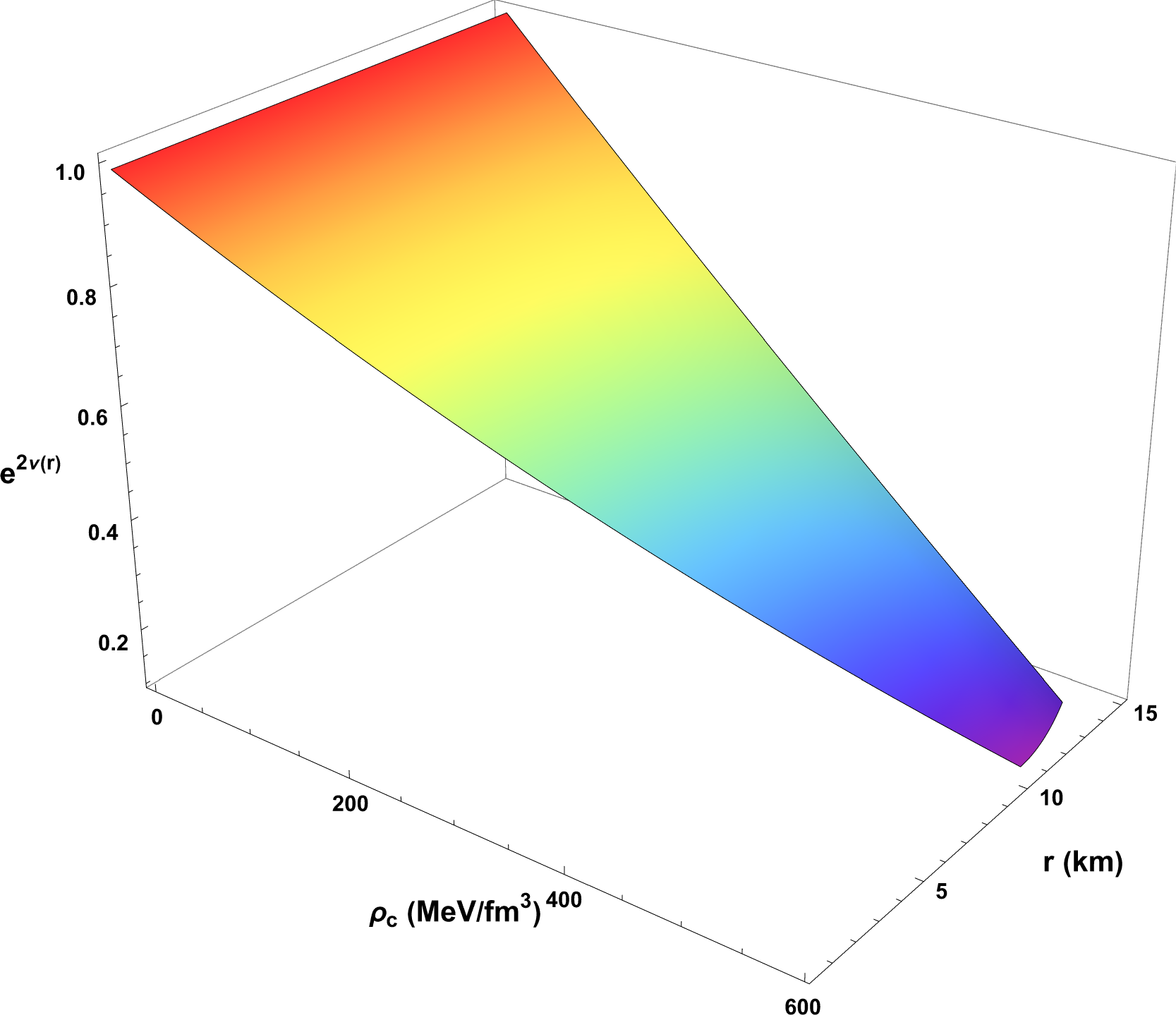} &
		\includegraphics[width=0.4\textwidth]{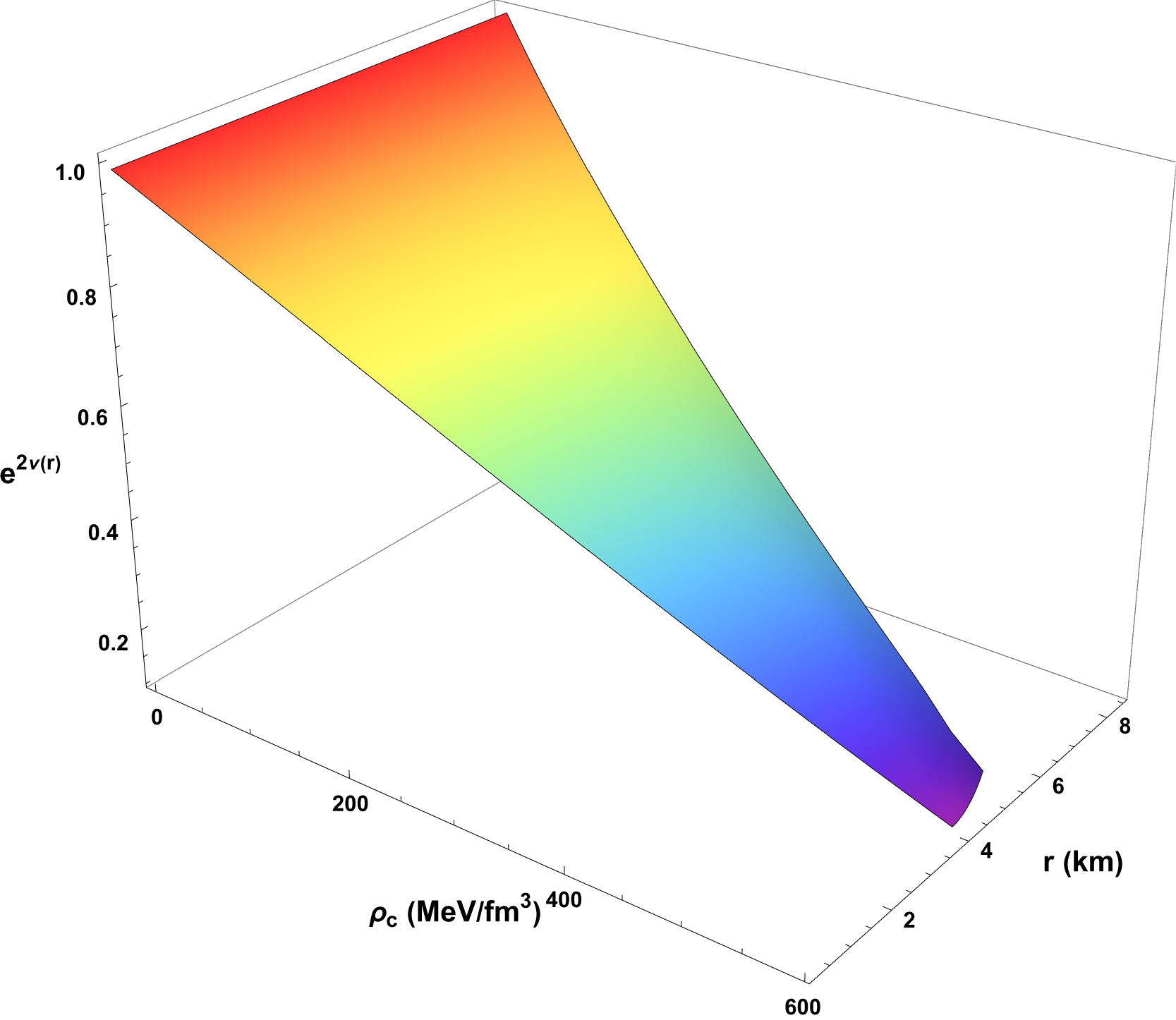} \\
		(c) & (d) \\[1em]
		\includegraphics[width=0.4\textwidth]{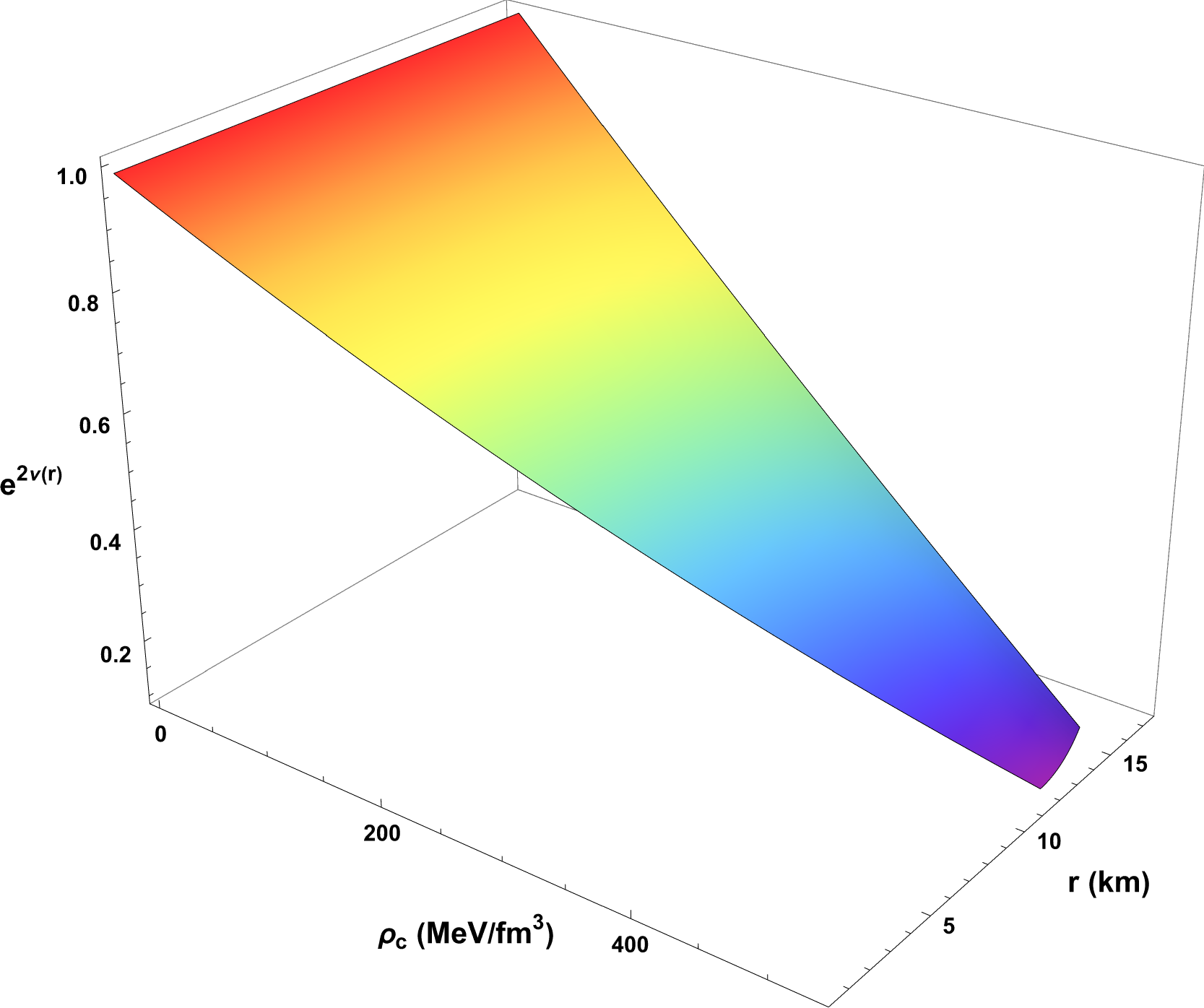} &
		\includegraphics[width=0.4\textwidth]{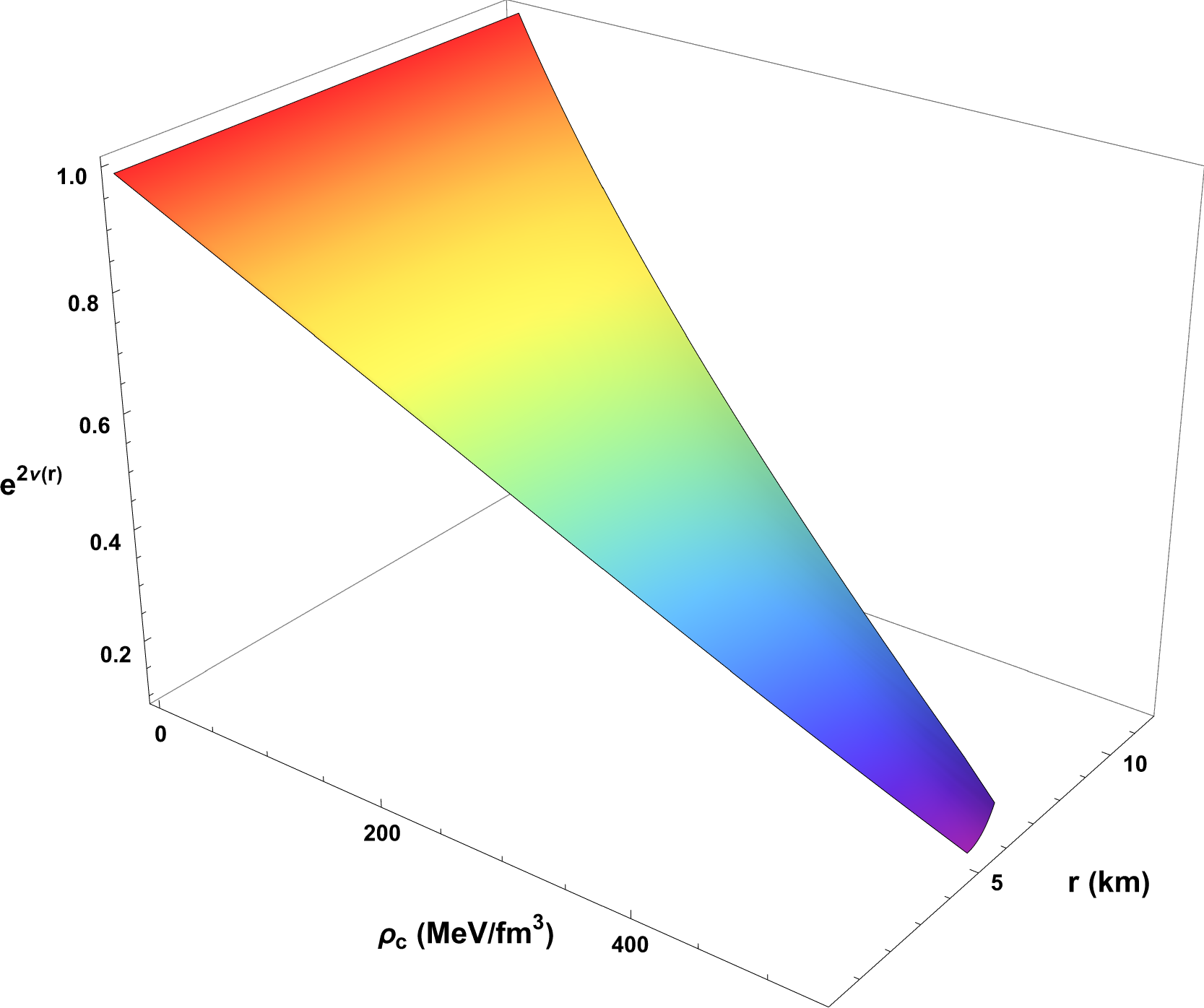} \\
		(e) & (f)
	\end{tabular}
	\caption{\textcolor{black}{Three-dimensional plot of interior temporal metric function as a function of $\rho_c$ and $r$ for (a) $h=1$ with Approach 1, (b) $h=1$ with Approach 2, (c) $h=1.5$ with Approach 1, (d) $h=1.5$ with Approach 2, (e) $h=2$ with Approach 1, and (f) $h=2$ with Approach 2.}}
	\label{figinteriormetricnon}
\end{figure*}

\begin{figure*}[htbp]
	\centering
	\begin{tabular}{cc}
		\includegraphics[width=0.48\textwidth]{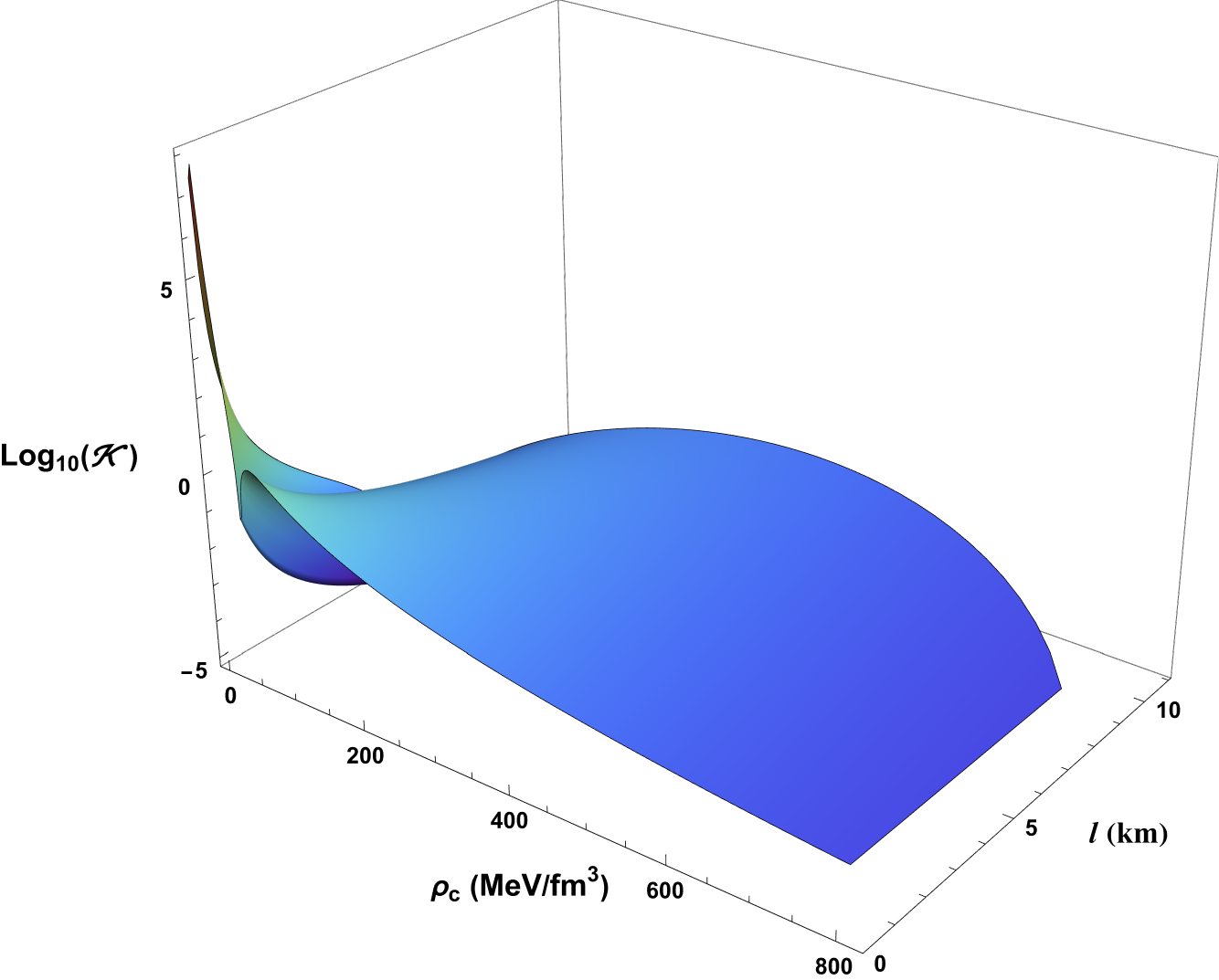} &
		\includegraphics[width=0.48\textwidth]{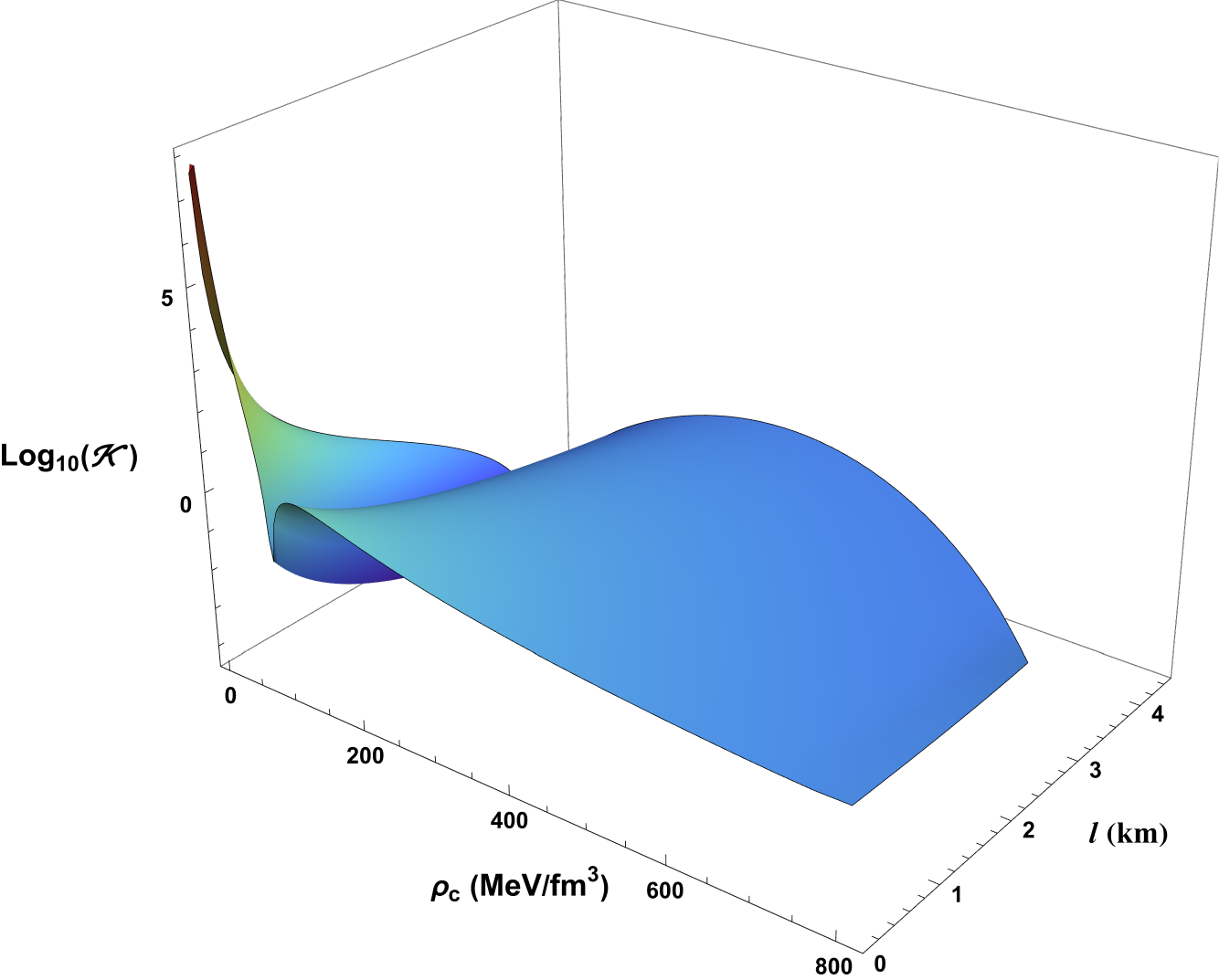} \\
		(a) & (b) \\[1em]
		\includegraphics[width=0.48\textwidth]{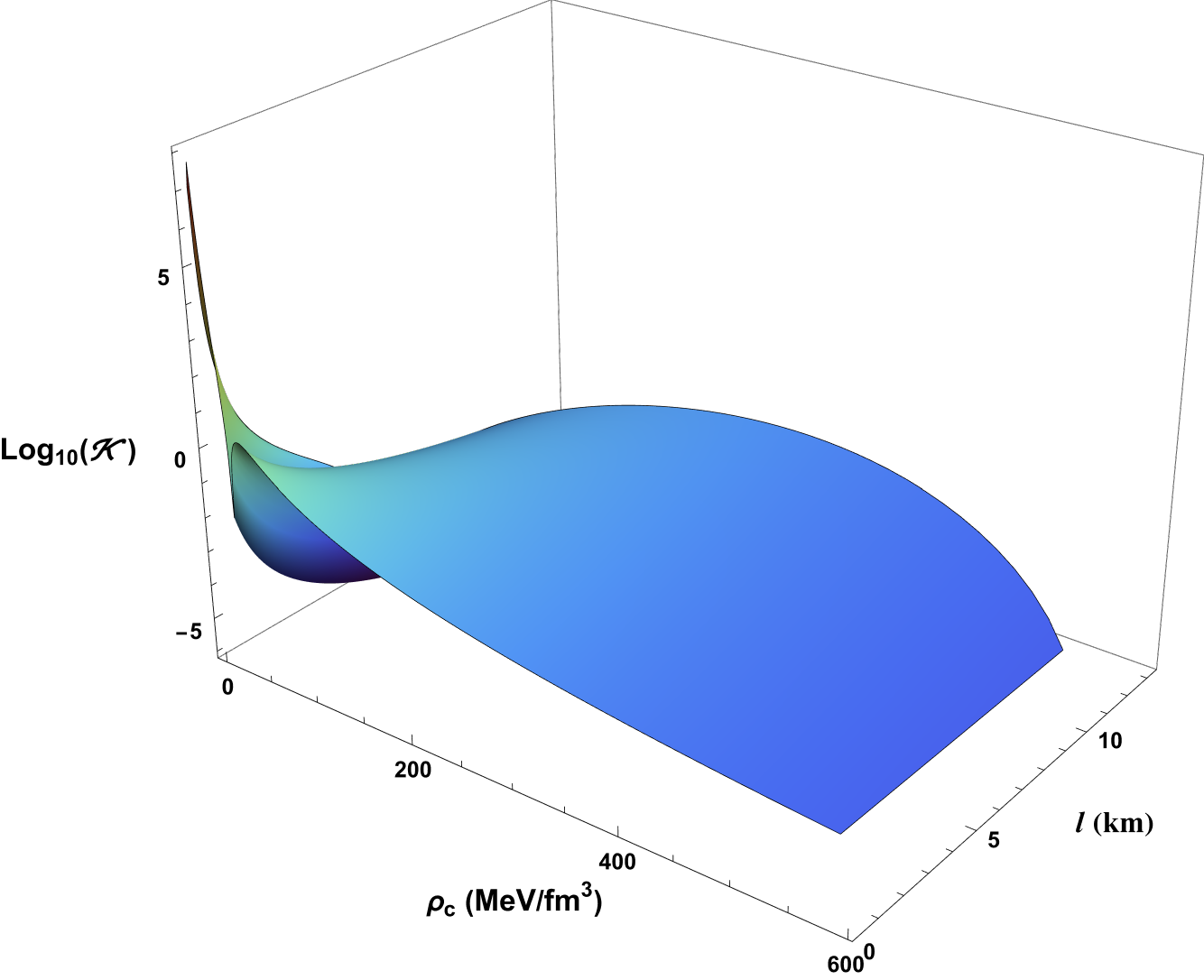} &
		\includegraphics[width=0.48\textwidth]{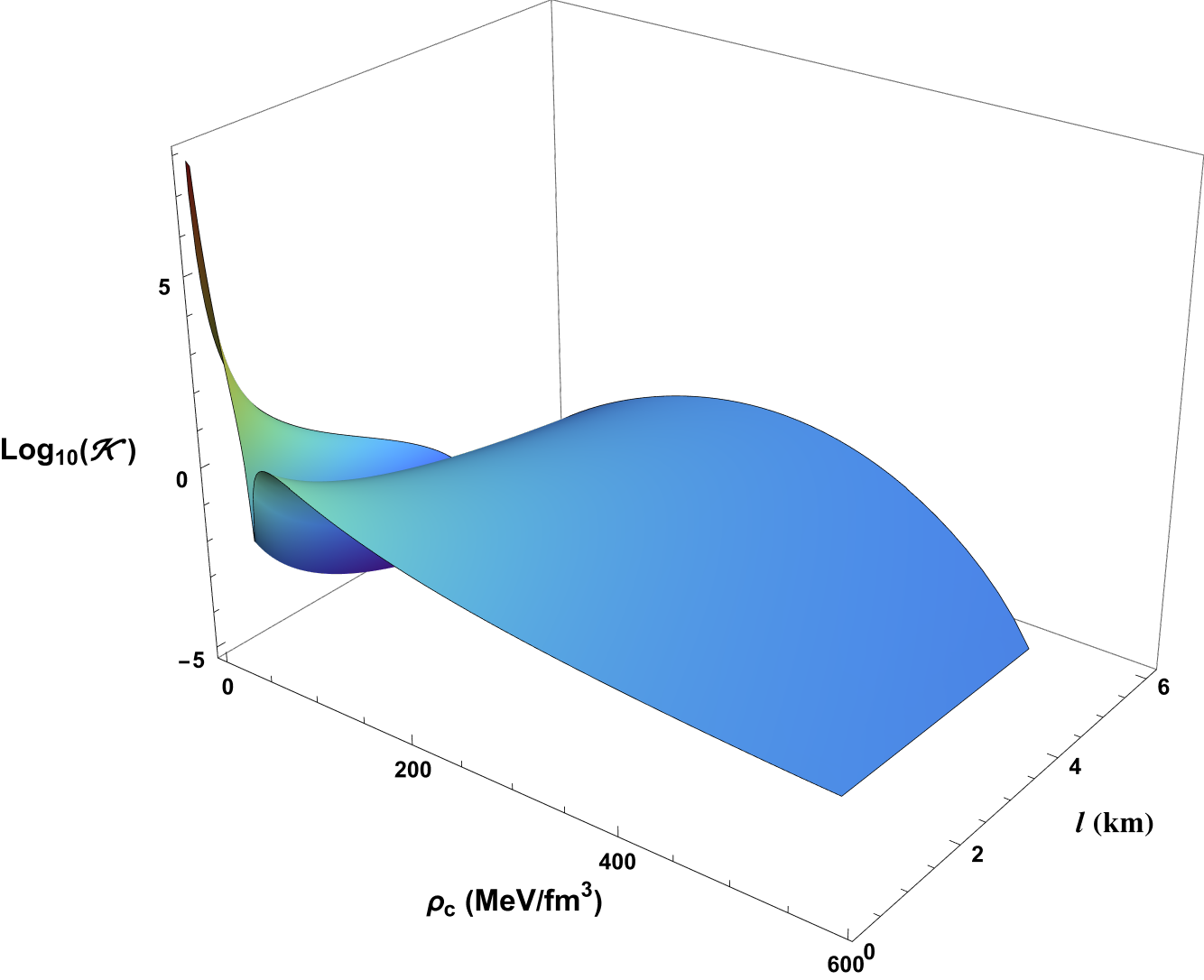} \\
		(c) & (d) \\[1em]
		\includegraphics[width=0.48\textwidth]{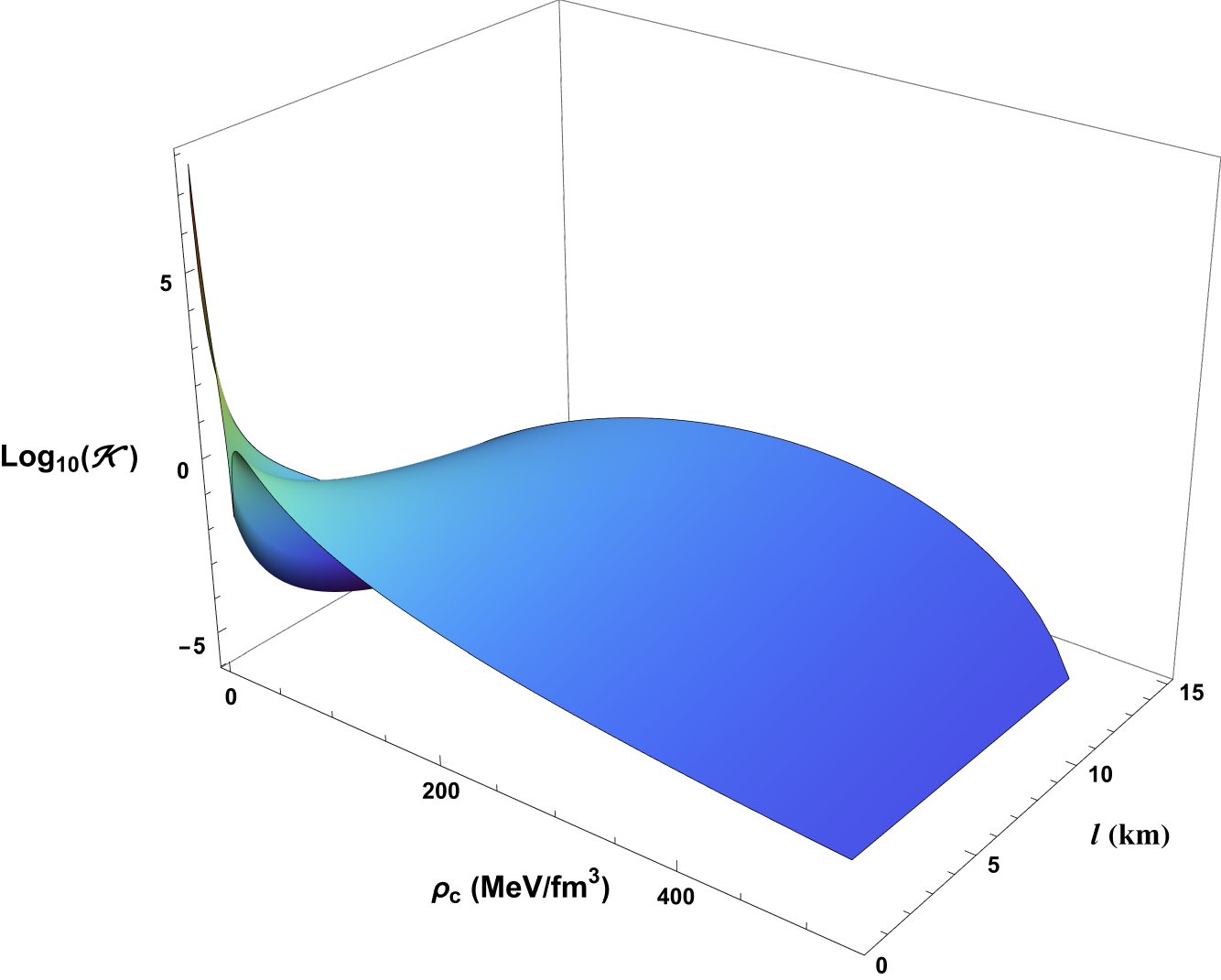} &
		\includegraphics[width=0.48\textwidth]{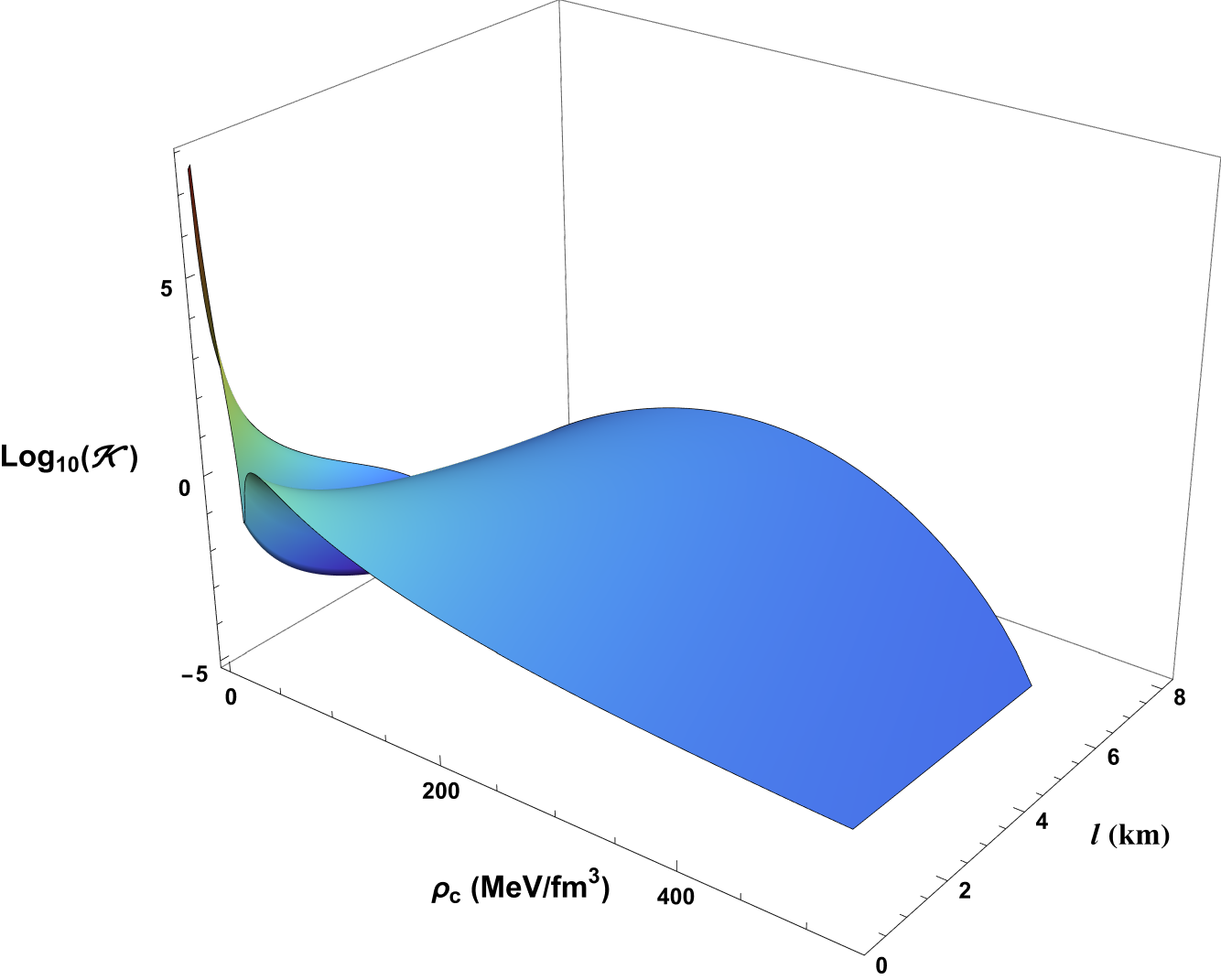} \\
		(e) & (f)
	\end{tabular}
	\caption{\textcolor{black}{Three-dimensional plot of $\log_{10}{(\mathcal{K})}$ as a function of $\rho_c$ and $l$ for (a) $h=1$ with Approach 1, (b) $h=1$ with Approach 2, (c) $h=1.5$ with Approach 1, (d) $h=1.5$ with Approach 2, (e) $h=2$ with Approach 1, and (f) $h=2$ with Approach 2.}}
	\label{figkretschmannnon}
\end{figure*}


\begin{figure*}[tb]
	\centering
	
	\begin{minipage}{0.48\textwidth}
		\centering
		\includegraphics[width=\textwidth]{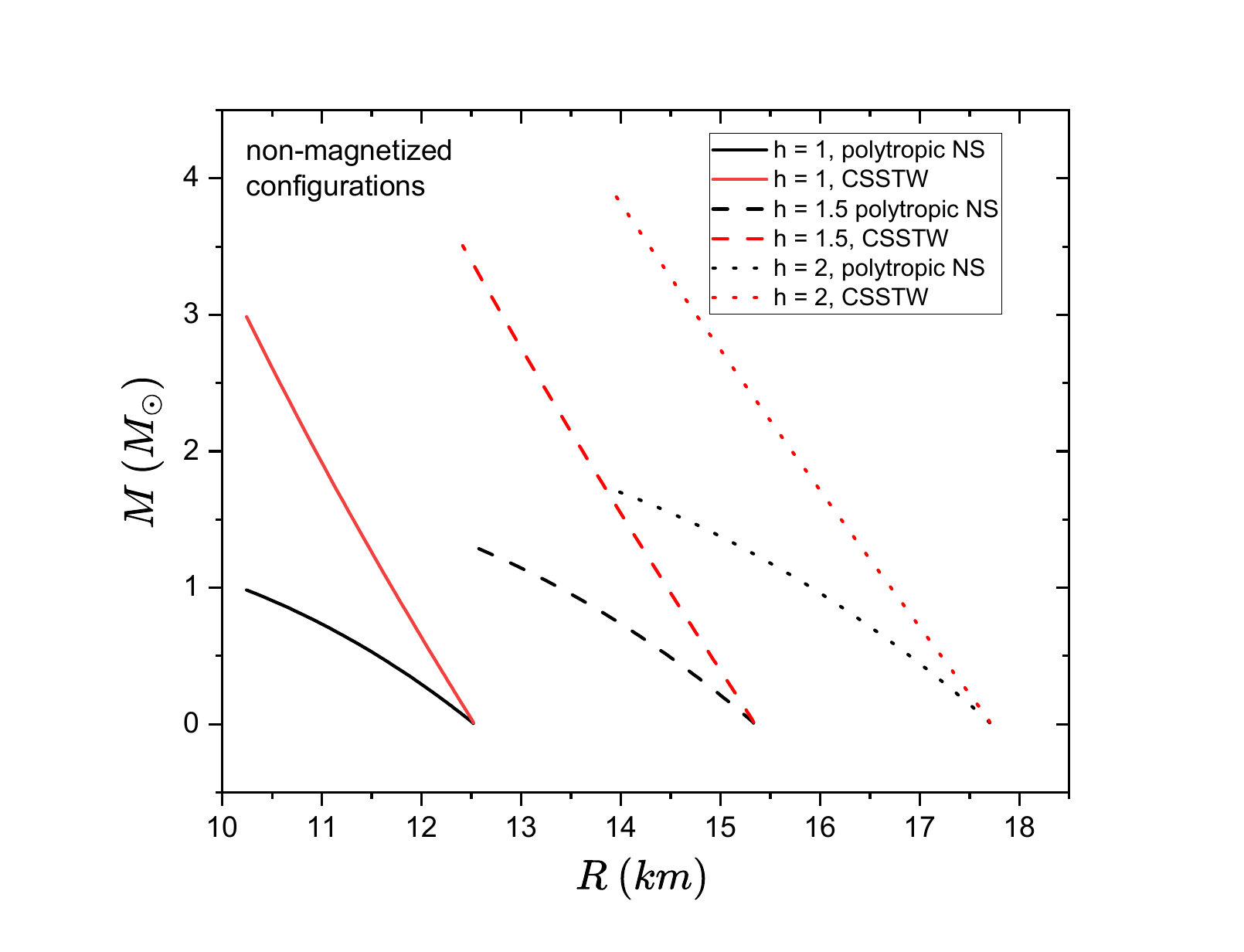}
		\textbf{(a)} Approach 1
	\end{minipage}
	\hfill
	\begin{minipage}{0.48\textwidth}
		\centering
		\includegraphics[width=\textwidth]{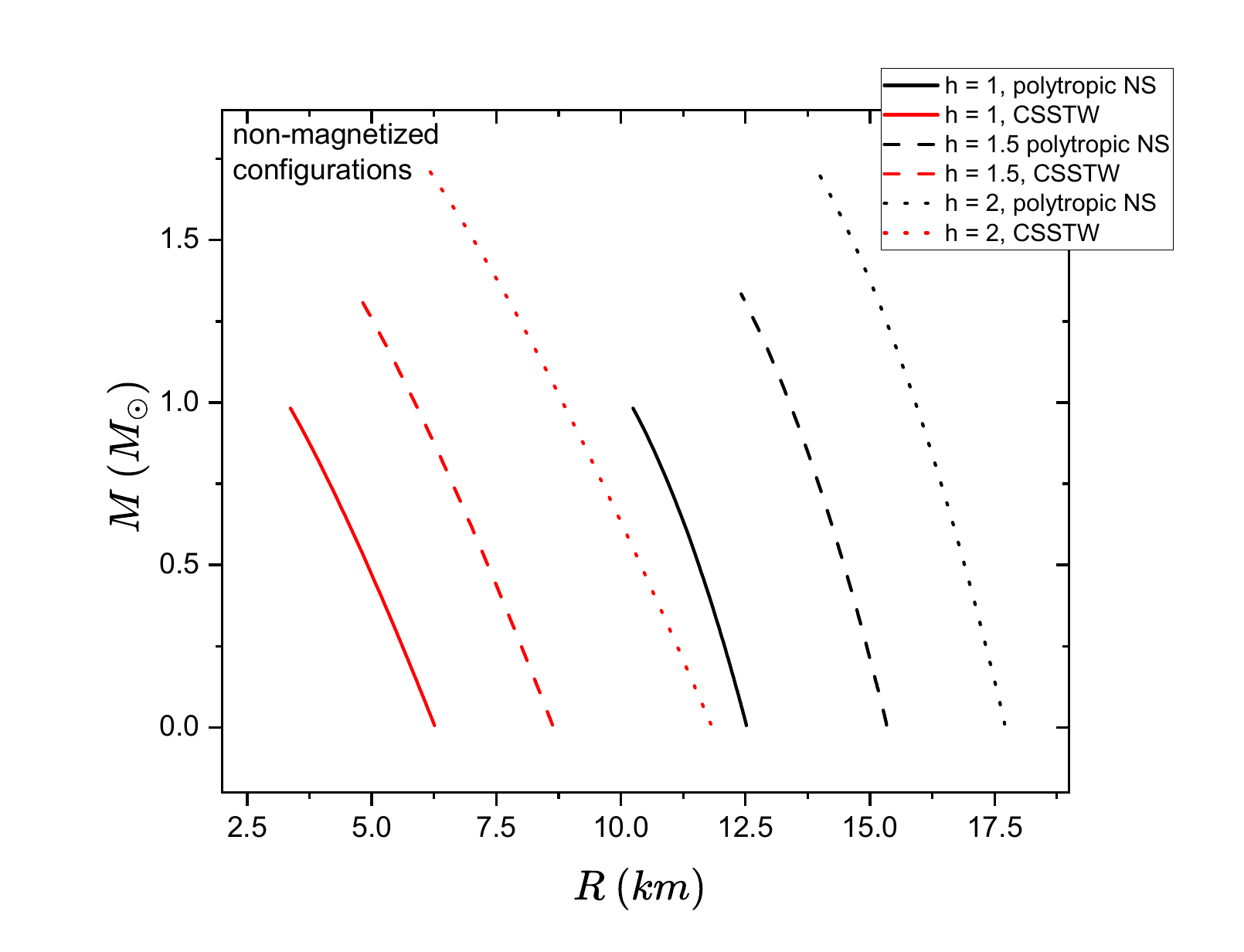}
		\textbf{(b)} Approach 2
	\end{minipage}
	
	\caption{Mass–radius relation of the pure NSs and the CSSTW for different values of $h$. Panels (a) and (b) correspond to Approach~1 and Approach~2, respectively.}
	\label{figMRnonmagnetized}
\end{figure*}


\begin{figure*}[tb]
	\centering
	
	\begin{minipage}{0.48\textwidth}
		\centering
		\includegraphics[width=\textwidth]{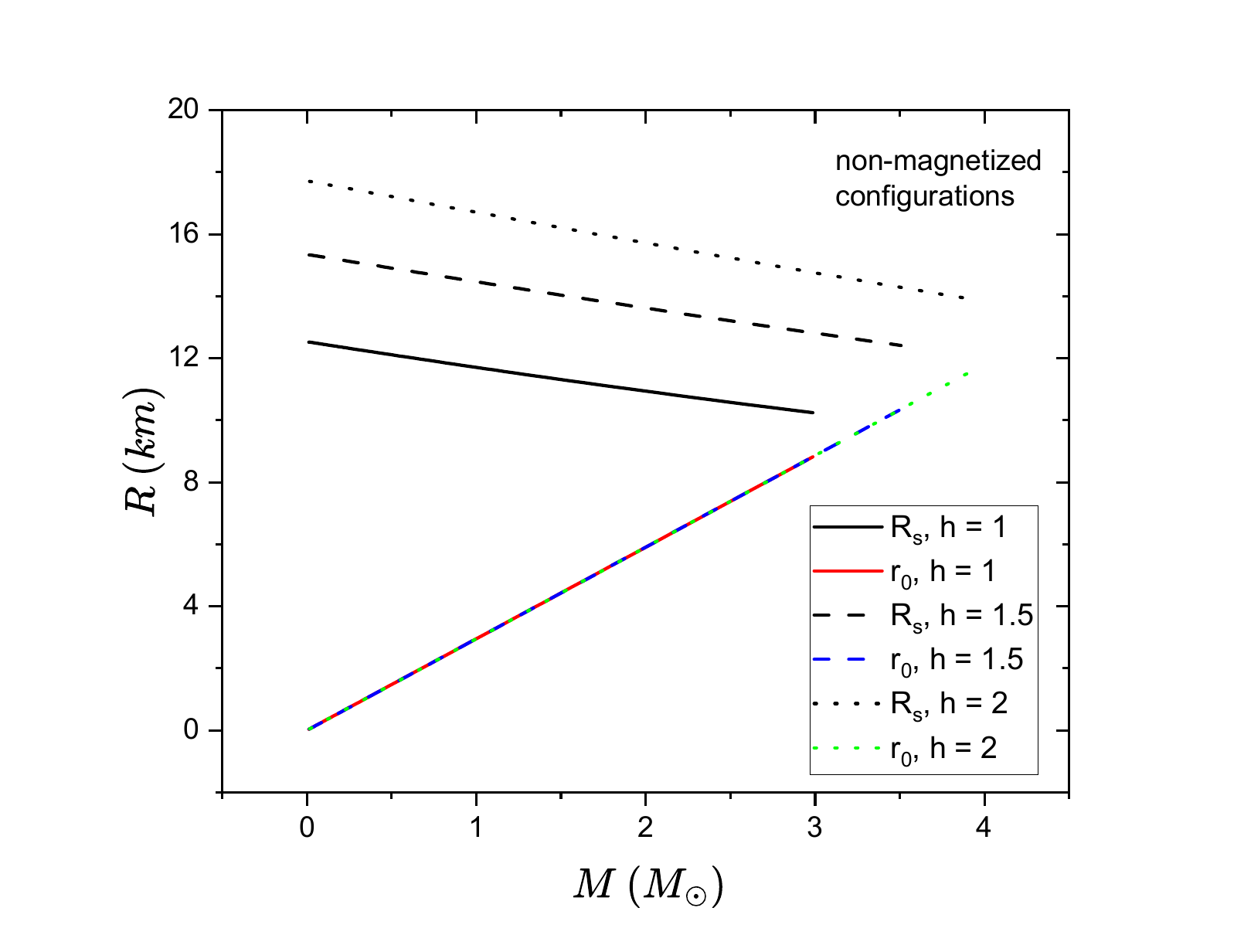}
		\textbf{(a)} Approach 1
	\end{minipage}
	\hfill
	\begin{minipage}{0.48\textwidth}
		\centering
		\includegraphics[width=\textwidth]{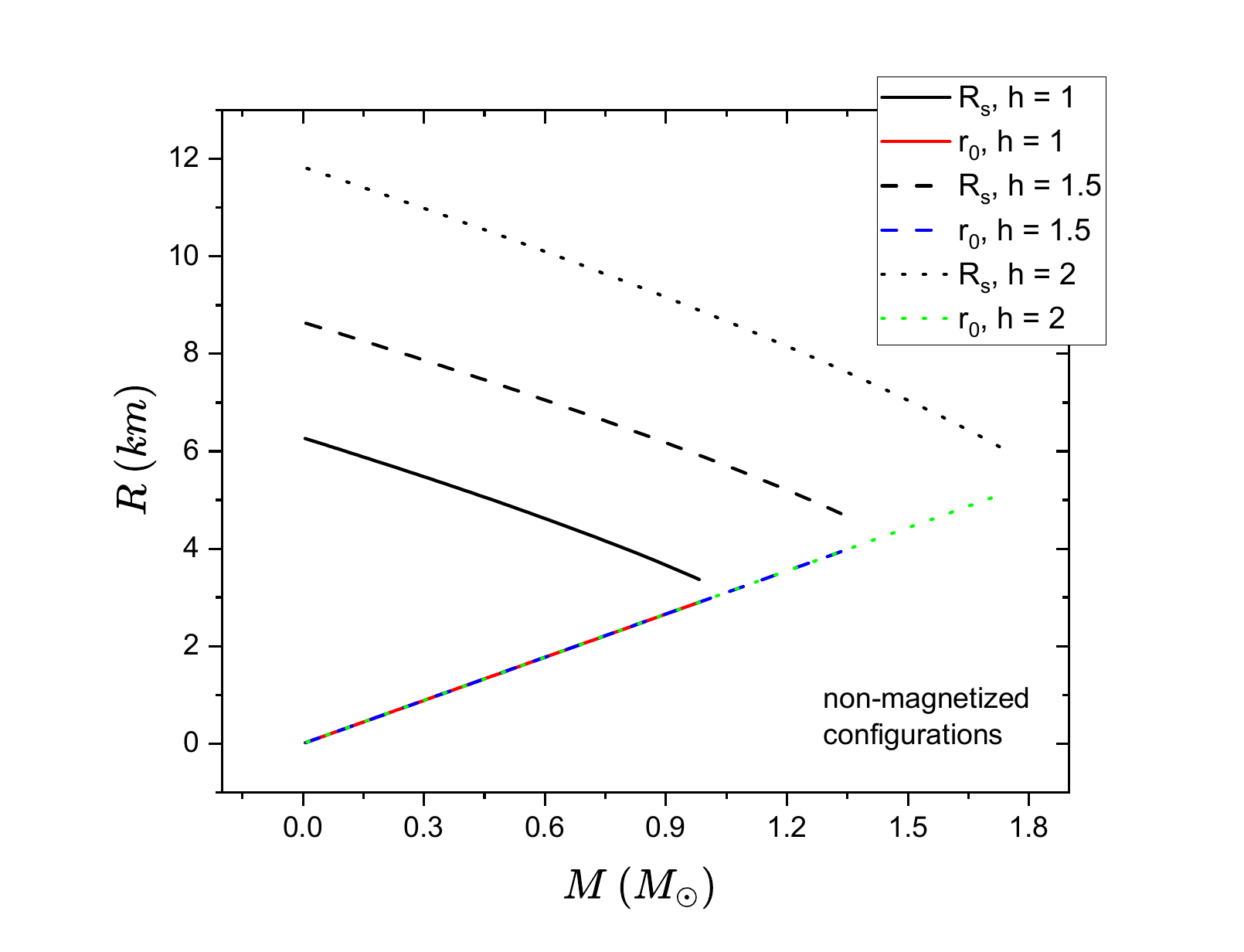}
		\textbf{(b)} Approach 2
	\end{minipage}
	
	\caption{Radii of the stellar surface and the wormhole throat in the CSSTW as functions of the ADM mass for different values of $h$. Panels (a) and (b) correspond to Approach~1 and Approach~2, respectively.}
	\label{figRsr0nonmagnetized}
\end{figure*}


\begin{figure*}[tb]
	\centering
	
	\begin{minipage}{0.48\textwidth}
		\centering
		\includegraphics[width=\textwidth]{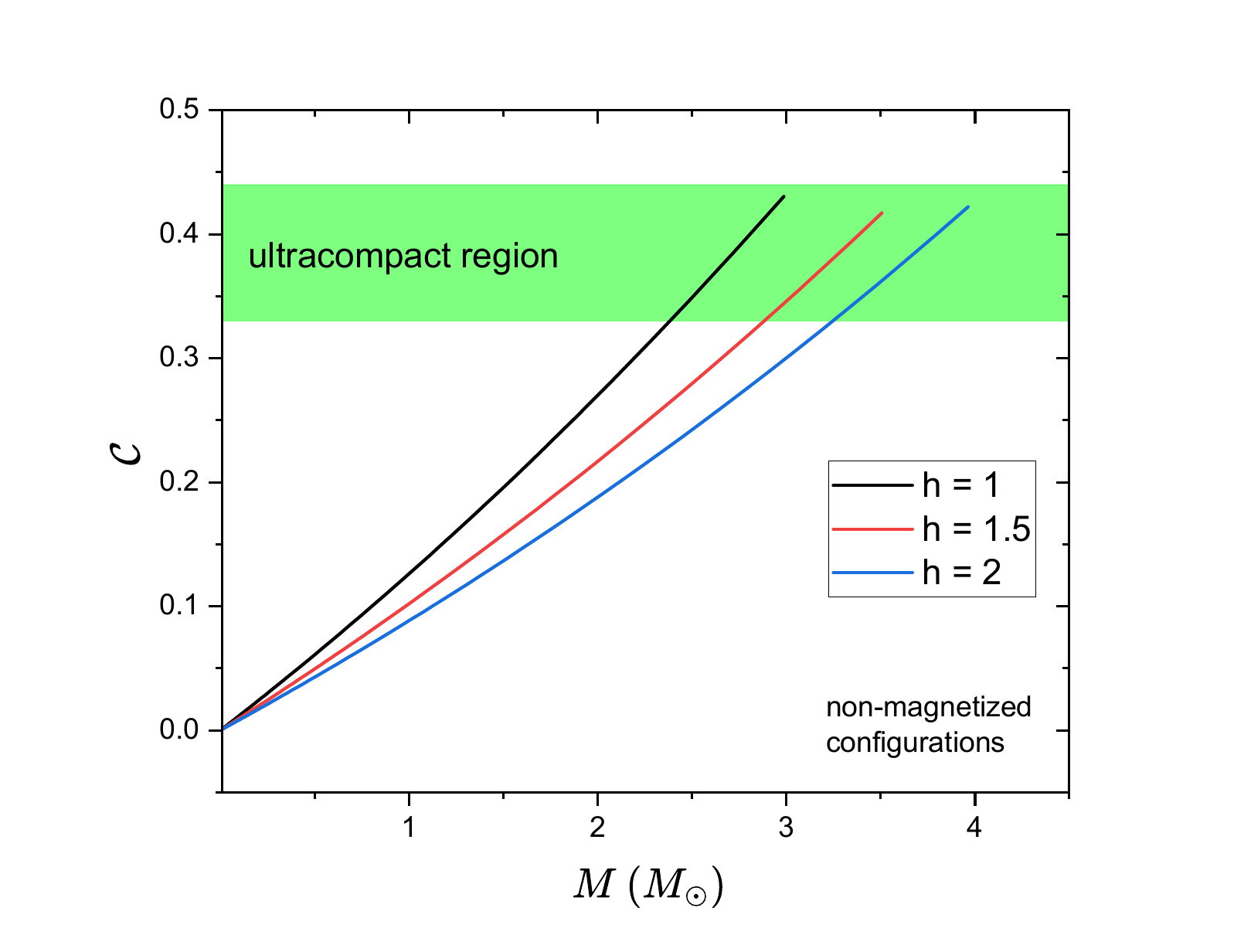}
		\textbf{(a)} Approach 1
	\end{minipage}
	\hfill
	\begin{minipage}{0.48\textwidth}
		\centering
		\includegraphics[width=\textwidth]{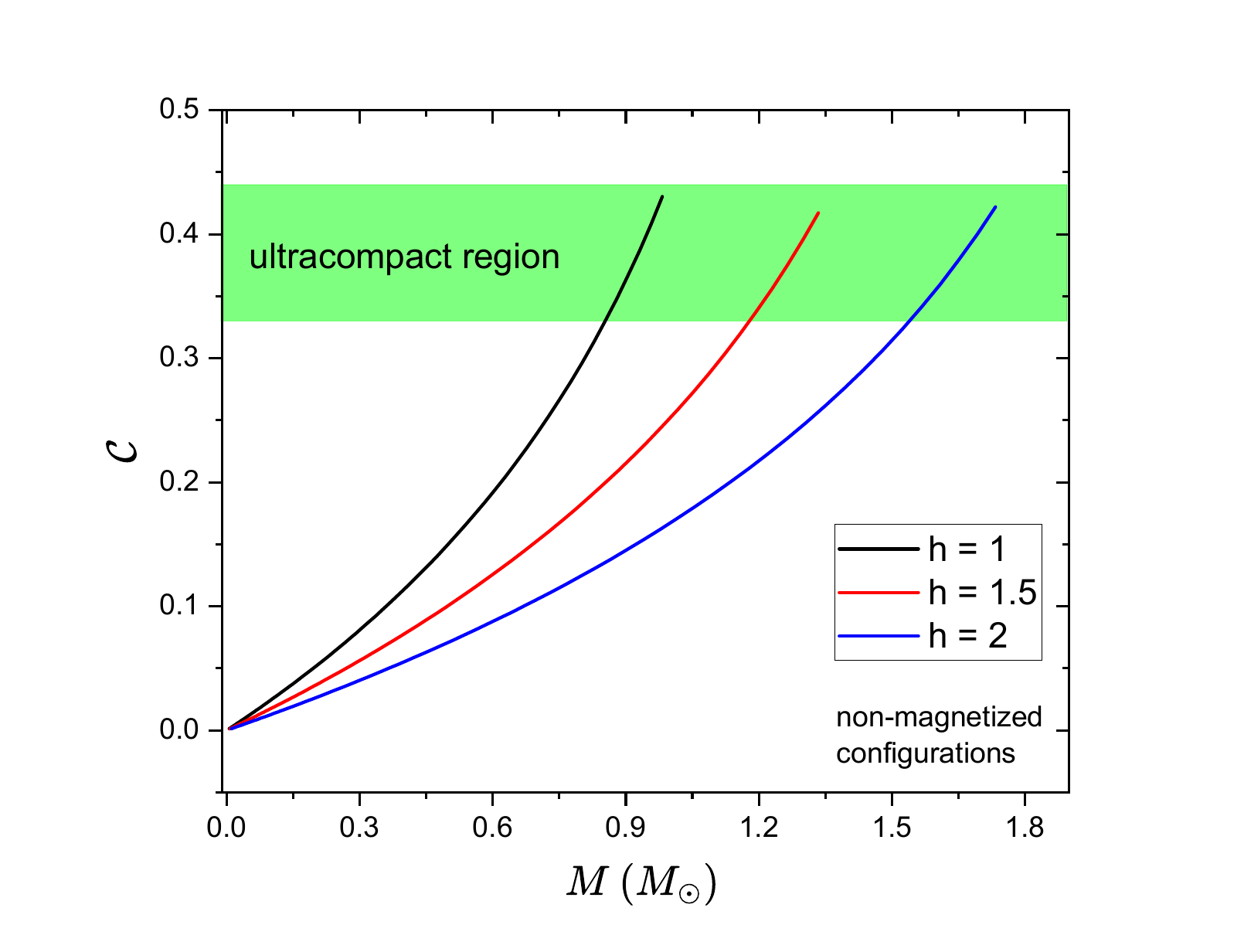}
		\textbf{(b)} Approach 2
	\end{minipage}
	
	\caption{Compactness of the CSSTW as a function of the ADM mass for different values of $h$. Panels (a) and (b) correspond to Approach~1 and Approach~2, respectively.}
	\label{figcompactnessnonmagnetized}
\end{figure*}

Fig.~\ref{figMRnonmagnetized} presents the mass-radius (MR) relations of the CSSTW for several values of the parameter $h$. For comparison, we also display the MR curve of pure neutron stars (NSs), allowing the impact of the wormhole component to be directly identified.

For all considered values of $h$ within Approach~1, as we can see in Fig.~\ref{figMRnonmagnetized}(a), the mass of the CSSTW can exceed \textcolor{black}{3~$M_\odot$}, indicating that the resulting configurations are exceptionally massive. Similar mass enhancement has already been reported for isotropic systems in Ref.~\cite{Nojiri2024}. However, our inclusion of anisotropic matter allows the existance of even more massive configurations, accompanied by a larger surface radius.

Fig.~\ref{figMRnonmagnetized}(b) presents the MR relations obtained using Approach~2. Since pure polytropic NSs do not reach the large masses found in the configurations constructed under Approach~1, the CSSTW obtained via Approach~2 naturally cannot achieve comparable masses. As a consequence, the radii of the CSSTW become significantly smaller and may even fall below 5~km.

Fig.~\ref{figRsr0nonmagnetized}(a) shows the comparison between the radius of the CSSTW system and the corresponding wormhole throat radius obtained using Approach~1. \textcolor{black}{We emphasize that the coincidence of the wormhole throat radii observed in Figs.~\ref{figRsr0nonmagnetized}(a) and (b) is exact.
	This is a direct consequence of the relation $r_0 = 2M$.
	Since the ADM mass $M$ is used as the horizontal axis in the figures, the functional dependence $r_0(M)$ is fixed and identical for all configurations.}

\textcolor{black}{The role of the anisotropy parameter $h$ and other matter parameters is not to modify the relation between $r_0$ and $M$, but to determine the values of the ADM mass that are realized by admissible solutions.
	Different values of $h$ therefore lead to different mass ranges and, consequently, to different accessible throat radii through the relation $M = r_0/2$, while preserving the exact coincidence of the $r_0(M)$ curves.
}

\textcolor{black}{The wormhole throat radii obtained using Approach~2, shown in Fig.~\ref{figRsr0nonmagnetized}(b), exhibit a trend similar to that found in Approach~1. The primary difference between the two approaches appears in the overall mass and radius scales: Approach~1 yields systematically larger masses and radii, consistent with the corresponding MR relations obtained under that approach.}

As mentioned earlier, the CSSTW system can reach masses that may even exceed \textcolor{black}{3~$M_\odot$}. This fact implies that the system can become very massive and, consequently, highly compact, as illustrated in Fig.~\ref{figcompactnessnonmagnetized}. The compactness $\mathcal{C}=M/R_s$ can even enter the region typically associated with ultracompact objects. The range of $\mathcal{C}$ for ultracompact object is $0.33<\mathcal{C}<0.44$ \cite{ByonBurikham,ByonAnto}. Recently, research on ultracompact objects has attracted significant interest (see, e.g., Refs.~\cite{ByonBurikham,ByonAnto,Prasetyo2021,Fauzi2025,Benetti2025,Urbano2019,Prasetyo2023,Hidalgo2021,Tasinato2022,Kleihaus2020,Zhu2020}). Those objects possess radii smaller than $3M$, i.e., inside the photon-sphere radius. In such regimes, gravitational-wave echoes may occur, and the corresponding echo time can be computed~\cite{Sakti2021}.

\begin{figure*}[tb]
	\centering
	
	\begin{minipage}{0.48\textwidth}
		\centering
		\includegraphics[width=\textwidth]{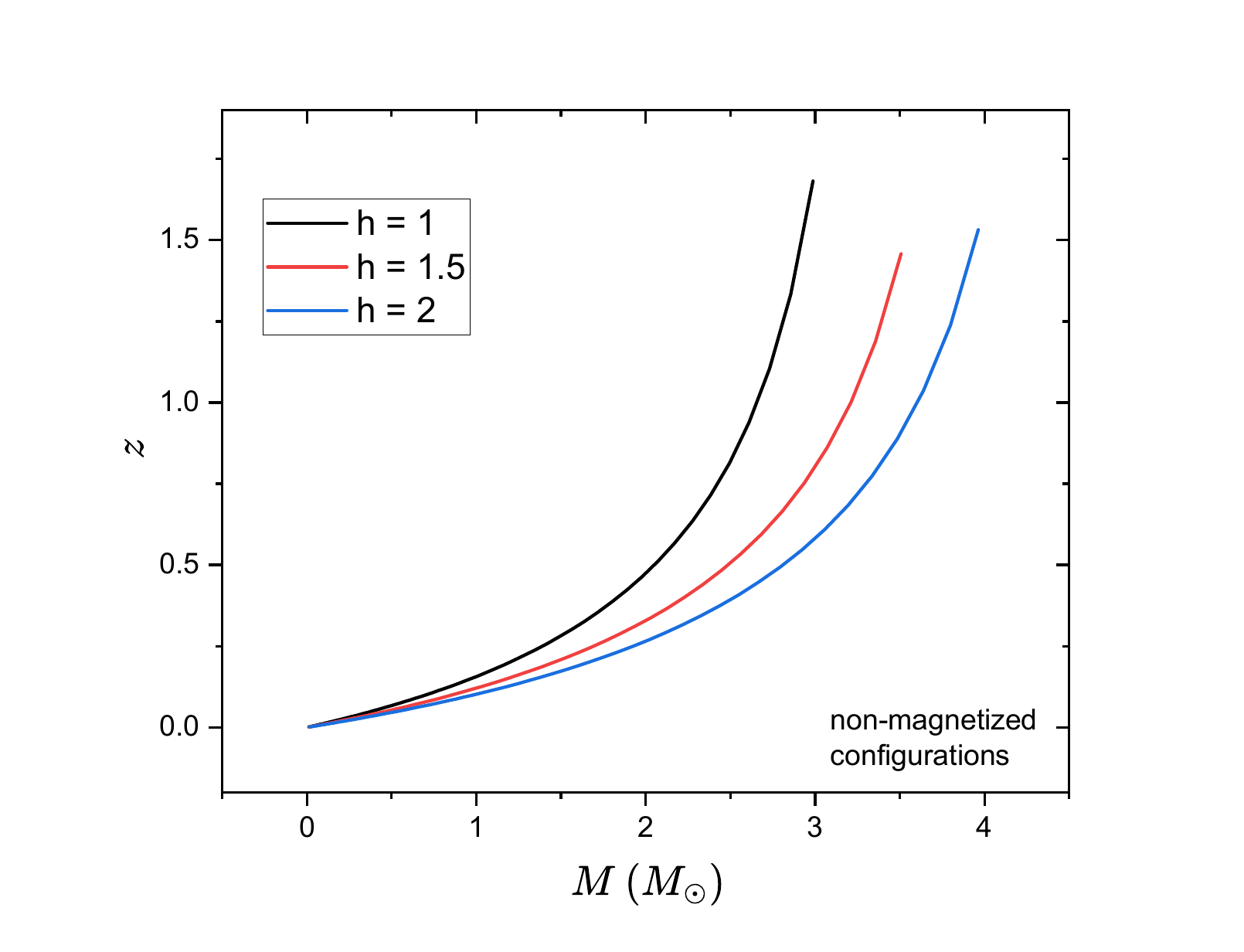}
		\textbf{(a)} Approach 1
	\end{minipage}
	\hfill
	\begin{minipage}{0.48\textwidth}
		\centering
		\includegraphics[width=\textwidth]{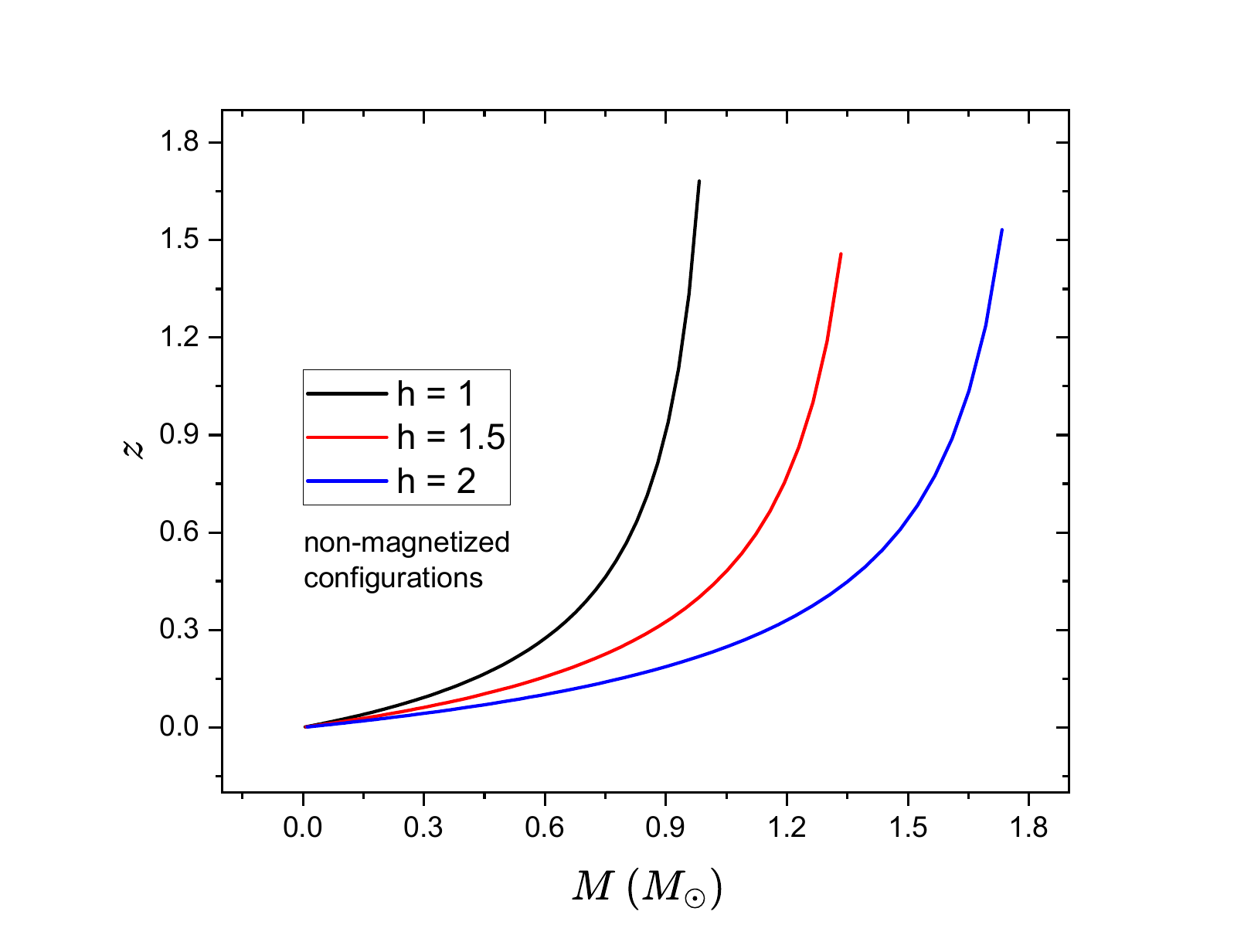}
		\textbf{(b)} Approach 2
	\end{minipage}
	
	\caption{Surface redshift of the CSSTW as a function of the ADM mass for different values of $h$. Panels (a) and (b) correspond to Approach~1 and Approach~2, respectively.}
	\label{figredshiftnonmagnetized}
\end{figure*}

\begin{figure*}[tb]
	\centering
	
	\begin{minipage}{0.48\textwidth}
		\centering
		\includegraphics[width=\textwidth]{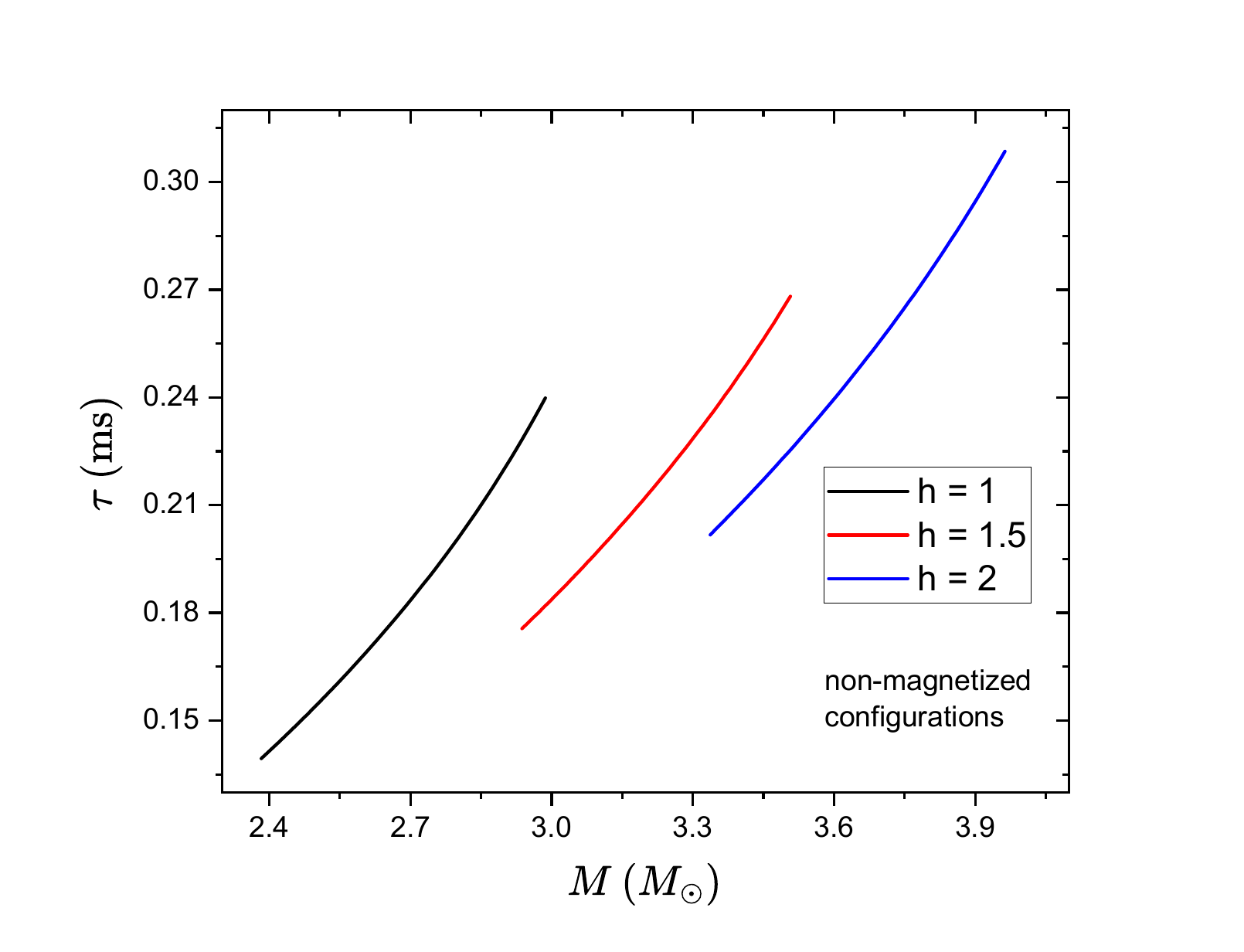}
		\textbf{(a)} Approach 1
	\end{minipage}
	\hfill
	\begin{minipage}{0.48\textwidth}
		\centering
		\includegraphics[width=\textwidth]{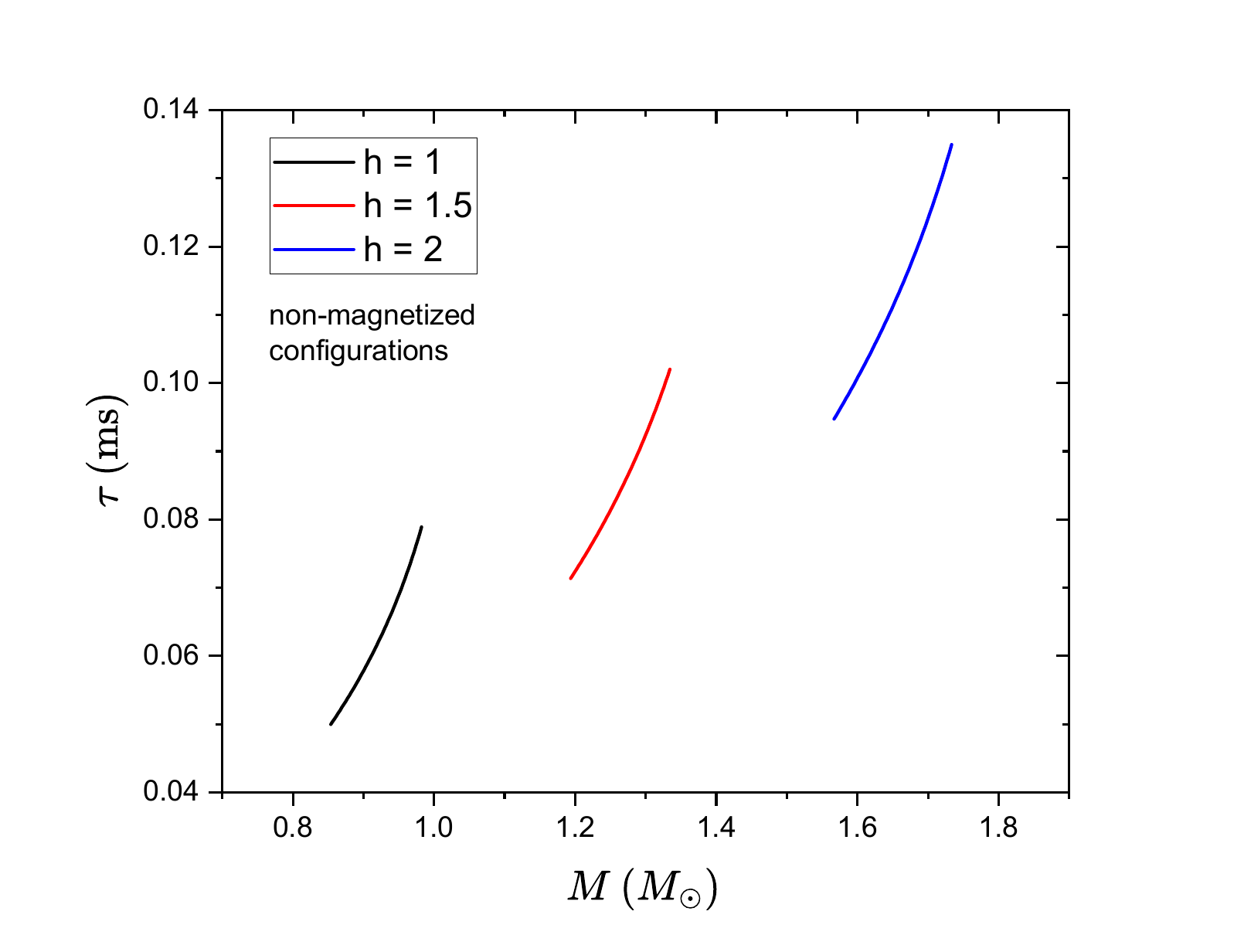}
		\textbf{(b)} Approach 2
	\end{minipage}
	
	\caption{Echo time of the CSSTW as a function of the ADM mass for different values of $h$. Panels (a) and (b) correspond to Approach~1 and Approach~2, respectively.}
	\label{figechononmagnetized}
\end{figure*}

One of the physical properties of CSSTW that can be calculated is the surface redshift $\mathcal{Z}$. Mathematically, it is defined as~\cite{Sakti2021}
\begin{eqnarray}
	\mathcal{Z} = \frac{1}{e^{\nu(R_s)}} - 1.\label{redshiftformula}
\end{eqnarray}
For ordinary NSs with typical masses $M = 1.2\text{--}2.0\,M_\odot$ and radii $R_s = 10\text{--}14$ km, the expected surface redshift falls within the range $\mathcal{Z} \simeq 0.1\text{--}0.5$. Ref.~\cite{Cottam2002}~reported a gravitational redshift of $\mathcal{Z} = 0.35$ from 28 bursts of the low-mass X-ray binary EXO~0748$-$676, consistent with theoretical expectations for regular NSs.

Fig.~\ref{figredshiftnonmagnetized} shows the surface redshift of the non-magnetized CSSTW as a function of the ADM mass for various values of $h$. The results indicate that the redshift can exceed $\mathcal{Z}=1$, meaning that the predicted $\mathcal{Z}$ lies well above typical values for ordinary NSs. This suggests that the configurations considered here behave as exotic compact objects. We also find that both Approach~1 and Approach~2 yield comparable values of the parameter $\mathcal{Z}$. However, these values appear in different mass ranges, reflecting the fact that the two approaches lead to systematically different mass distributions. This indicates that the parameter $\mathcal{Z}$ is relatively insensitive to the choice of approach, while the associated mass scale depends strongly on the adopted construction scheme. Furthermore, in both cases, increasing anisotropy tends to reduce the resulting value of $\mathcal{Z}$. Hence, the disappearance of anisotropy with positive $h$ values prolongs the wavelength that is measured by the observer at $r=\infty$.

Another physical property of the CSSTW that have reach the ultracompact regime is the echo time. By assuming that the wormhole is connected to an identical object residing in another spacetime—following the setup considered by Ref.~\cite{Lai2025}—gravitational waves can propagate through the wormhole mouths and become trapped between the two photon spheres. This process gives rise to gravitational echoes (see Fig.~1 of Ref.~\cite{Lai2025} for a schematic illustration). A distinctive feature of such systems is that the resulting signal is expected to have an extremely short duration and to lack a clearly identifiable inspiral phase, which is commonly observed in standard compact binary mergers. In this scenario, the detected gravitational-wave signal essentially consists of the echo component alone, originating from waves repeatedly traversing the wormhole throat.

Note that in this work, we limit the echo time calculation as a simplified geometric-optics-inspired prescription. \textcolor{black}{Therefore, the echo times estimated in this work are intended only as exploratory estimates based on the adopted background geometry. They should not be interpreted as predictions of an observable gravitational-wave echo signal.} It is important to emphasize that in this work, we do not derive the perturbation equations related to gravitational waves; present the effective potential for axial or polar perturbations; and
calculate reflection/transmission coeffcients, mode spectra, or the actual waveform. \textcolor{black}{These aspects are left as future works. Consequently, the echo times obtained here alone are not sufficient to establish an observational signature of the CSSTW.}

Within this framework, the echo time $\tau$ can be expressed as~\cite{Lai2025}
\begin{eqnarray}
	\tau=2\int_{r_0}^{R_p} e^{\lambda-\nu}dr.\label{echotime}
\end{eqnarray}
Here, $R_p$ is the radius of the photon sphere. At the photon sphere, the photon must move at constant values of $r$ and $\theta$. This condition gives
\begin{eqnarray}
	\left(\frac{d\varphi}{dt}\right)^2=\frac{1-\frac{r_0}{r}}{r^2\sin^2\theta}.\label{dphidt}
\end{eqnarray}
On the other hand, from geodesic equation, one can obtain
\begin{eqnarray}
	\left(\frac{d\varphi}{dt}\right)^2=\frac{r_0}{2r^3\sin^2\theta}.\label{geodesic}
\end{eqnarray}
By combining Eqs.~(\ref{dphidt}) and (\ref{geodesic}), one obtains $r=3r_0/2=3M$, which corresponds to the location of the photon sphere. It is important to emphasize that this result is model dependent. For instance, in alternative compact object configurations such as dark energy stars with phantom fields, the photon sphere radius $R_p$ can deviate from the value $3M$. In such scenarios, the upper limit of the integral in Eq.~(\ref{echotime}) is consequently modified, as demonstrated by Ref.~\cite{Sakti2021}.

Since the functional form of $e^{2\nu(r)}$ differs between the regions $r_0 \leq r < R_s$ and $r \geq R_s$, the integration in Eq.~(\ref{echotime}) must be performed piecewise. Accordingly, we split the integration domain into two intervals: from $r_0$ to $R_s$, and from $R_s$ to $R_p$. In this work, the integral over the inner region, $r_0 \leq r < R_s$, is evaluated numerically due to the complicated form of $e^{2\nu(r)}$. On the other hand, the integral over the outer region, $R_s \leq r \leq R_p$, admits a closed-form analytical expression, given by
\begin{eqnarray}
	\int_{R_s}^{R_p}e^{\lambda-\nu}dr=(R_p-R_s)+r_0\ln\left(\frac{R_p-r_0}{R_s-r_0}\right).\label{integralecho2}
\end{eqnarray}
\textcolor{black}{Details regarding the numerical calculation of echo time are given in Subsec.~\ref{appechonum} of Appendix~\ref{appnumerical}.}

Fig.~\ref{figechononmagnetized} displays the resulting echo time as a function of the system's total mass. The echo time is found to lie within a very short interval, namely of order $10^{-1}\,\mathrm{ms}$ for Approach~1 and between $10^{-2}$ and $10^{-1}\,\mathrm{ms}$ for Approach~2. This behavior is expected: the presence of a wormhole at the center effectively creates a ``hollow'' region in the geometry, such that the echo signal is abruptly transported to a different branch of the spacetime. As a consequence, the optical path length decreases. This effect becomes even more pronounced when the wormhole throat radius approaches the stellar surface radius.

From Fig.~\ref{figechononmagnetized}, it can also be seen that in Approach~1, configurations with the same mass but larger values of $h$ exhibit shorter echo times. In Approach~2, although configurations with identical masses for different $h$ cannot be directly compared on the same plot, one may imagine extending each curve following its visible trend; under this qualitative continuation, larger values of $h$ would again correspond to shorter echo times. This situation--at least for Approach~1--is consistent with the behavior of the surface redshift in Fig.~\ref{figredshiftnonmagnetized}, where smaller $h$ at fixed mass yields a larger redshift $\mathcal{Z}$. Physically, a larger $\mathcal{Z}$ indicates stronger gravitational trapping, making it more difficult not only for light to escape but also for the echo signal to climb out of the gravitational well.

\section{Magnetized configuration of CSSTW}
\label{magnetized}
\textcolor{black}{Before discussing the CSSTW, we clarify the role of the parameter
	$\rho_0$ used throughout this work.
	Here, $\rho_0$ denotes the central energy density of the corresponding non-magnetized
	pure NS configurations and is treated as a scanning parameter.} \textcolor{black}{The range $10^{-4} \le \rho_0 \le 10^{4}\,\mathrm{MeV\,fm^{-3}}$ is chosen to encompass the
	physically relevant domain of central energy densities typically considered in NS
	structure calculations, from low-density solutions to very dense cores.
	Thus, $\rho_0$ must exist within this range.}

\textcolor{black}{For each value of the anisotropy parameter $h$, the value of $\rho_0$ is determined
	numerically from the corresponding pure NS solutions.
	Specifically, we solve the standard TOV equations for non-magnetized NSs
	by varying the central energy density $\rho_c$ and identify the value of $\rho_c$
	that yields the maximum-mass configuration.}

\textcolor{black}{For the specific values of the anisotropy parameter considered in this work,
	the maximum-mass pure NS configurations are found at
	$\rho_0 = 3240.46$ MeV fm$^{-3}$ for $h=1$,
	$\rho_0 = 2008.27$ MeV fm$^{-3}$ for $h=1.5$,
	and $\rho_0 = 1459.81$ MeV fm$^{-3}$ for $h=2$,
	respectively.
	All these values lie well within the range
	$10^{-4} \le \rho_0 \le 10^{4}\,\mathrm{MeV\,fm^{-3}}$
	adopted in our analysis.
}

The EMT for magnetized configuration is shown by Eq.~(\ref{EMTmagnetized}). The conservation law gives
\begin{eqnarray}
	\nu=-\left(2-\frac{1}{h}\right)\int\frac{dP_r}{\varrho+P_r}.
\end{eqnarray}
With a little calculation, we obtain
	\begin{eqnarray}
		\nu &=& -2K\left( 2 - \frac{1}{h} \right)
		\int \frac{d\rho}{\rho + p_r + \frac{1}{6\pi}\left( B_s + B_0 \frac{\rho}{\rho_0} \right)^2}\nonumber\\
		&&- \frac{1}{12\pi} \int \frac{B\, dB}{
			\rho_0 \left( \frac{B - B_s}{B_0} \right)
			+ K \left[ \rho_0 \left( \frac{B - B_s}{B_0} \right) \right]^2
			+ \frac{B^2}{6\pi} } .\label{longnu}
	\end{eqnarray}
	
	Now the first and second terms of the right-hand side in Eq.~(\ref{longnu}) can be defined as $I_1$ and $I_2$, respectively. We have
	\begin{eqnarray}
		I_1=-2K\left( 2 - \frac{1}{h} \right)\int\frac{\rho d\rho}{\alpha\rho^2+\beta\rho+\gamma},\label{I1}
	\end{eqnarray}
	\begin{eqnarray}
		I_2=-\frac{z}{12}\int\frac{(z\rho+B_s)d\rho}{\alpha\rho^2+\beta\rho+\gamma},\label{I2}
	\end{eqnarray}
	where
	\begin{equation}
		\alpha=K+\frac{B_0^2}{6\pi\rho_0^2},\qquad\beta=1+\frac{B_0 B_s}{3\pi\rho_0},\qquad\gamma=\frac{B_s^2}{6\pi},\qquad z=\frac{B_0}{\rho_0}.\label{alphabetagamma}
	\end{equation}
	
	By defining $D = \beta^{2} - 4 \alpha \gamma$, the integral in Eq.~(\ref{I1}) will give
	\begin{eqnarray}
		\int \frac{\rho\, d\rho}{\alpha \rho^{2} + \beta \rho + \gamma}
		&=& \frac{1}{2\alpha} \ln\left( \alpha \rho^{2} + \beta \rho + \gamma \right)
		\left\{
		\begin{array}{ll}
			\displaystyle
			-\, \frac{\beta}{\alpha} \frac{1}{\sqrt{|D|}}
			\tan^{-1}\!\left( \frac{2\alpha \rho + \beta}{\sqrt{|D|}} \right),
			& \text{for } D < 0,
			\\[3ex]
			\displaystyle
			-\,\frac{\beta}{2\alpha}\,\frac{1}{\sqrt{D}}
			\ln\!\left(
			\frac{2\alpha \rho + \beta - \sqrt{\beta^{2}-4\alpha\gamma}}
			{2\alpha \rho + \beta + \sqrt{\beta^{2}-4\alpha\gamma}}
			\right),
			& \text{for } D > 0 .
		\end{array}
		\right.\label{integrallongI1}
	\end{eqnarray}
	
	Considering that we face two possibilities for the second term on the right-hand side of Eq.~(\ref{integrallongI1}), we decide to perform the calculation using four ranges of surface magnetic field strengths, while keeping the range of $\rho_0$ fixed between $1\times 10^{-4}$ and $1\times10^{4}$ MeV~fm$^{-3}$. The four ranges are summarized in Table~\ref{tab:BsB0}.\\
	
	The choice of these values is motivated by the fact that the typical surface magnetic field of NSs lies in the range $10^{12}$--$10^{15}$~G. Furthermore, according to the simulations reported in Ref.~\cite{Lopes2020}, a surface field of $B_s = 10^{15}$~G corresponds to a core magnetic field of $B_0 = 3 \times 10^{18}$~G. Therefore, in our analysis we restrict the order of magnitude of $B_0$ to be at most 
	three orders of magnitude larger than $B_s$. For the values of $\rho_0$ used in this work, 
	our non-magnetized pure NS calculations give $\rho_0 = 3240.46~\text{MeV fm}^{-3}$ for 
	$h=1$, $\rho_0 = 2008.27~\text{MeV fm}^{-3}$ for $h=1.5$, and 
	$\rho_0 = 1459.81~\text{MeV fm}^{-3}$ for $h=2$. 
	We see that all these values remain inside the range 
	$1 \times 10^{-4}$--$1 \times 10^{4}~\text{MeV fm}^{-3}$. 
	Thus, all values of $\rho_0$, $B_s$, and $B_0$ used throughout this study lie within 
	the parameter ranges described above. The small value of $\rho_0$ is intentionally taken so that the chosen second term of the right-hand side of Eq.~(\ref{integrallongI1}) remains applicable even in the small-order regime, without requiring concern over how small $\rho_0$ is. \\
	
	In Fig.~\ref{figdiscriminant}, the gray region corresponds to the range \(0 \leq D < 10^{-40}\), while the black region represents negative values of \(D\). Note that the threshold $10^{-40}$ is adopted as a classification cutoff for visualization purposes only, and is set far below the range of discriminant values relevant to the parameter space considered here. We observe that no gray or black regions appear in the figure, indicating that the discriminant remains positive and always above \(1 \times 10^{-40}\) throughout all computed regimes. This shows that we can use the logarithmic function for the second term on the right-hand side of Eq.~(\ref{integrallongI1}), instead of the inverse tangent function. Now Eq.~(\ref{integrallongI1}) reads
	\begin{eqnarray}
		\int\frac{\rho d\rho}{\alpha\rho^2+\beta\rho+\gamma}&=&\frac{1}{2\alpha} \ln\left(\alpha\rho^2+\beta\rho+\gamma\right)-\frac{\beta}{2\alpha}\frac{1}{\sqrt{D}}\ln\left(\frac{2\alpha\rho+\beta-\sqrt{\beta^2-4\alpha\gamma}}{2\alpha\rho+\beta+\sqrt{\beta^2-4\alpha\gamma}}\right).\label{choicework}
	\end{eqnarray}
	So we have
	\begin{align}
		I_1=\ln\left[(\alpha\rho^2+\beta\rho+\gamma)^{1/2\alpha}\bigg(\frac{2\alpha\rho+\beta-\sqrt{\beta^2-4\alpha\gamma}}{2\alpha\rho+\beta+\sqrt{\beta^2-4\alpha\gamma}}\bigg)^{-\frac{\beta}{2\alpha\sqrt{\beta^2-4\alpha\gamma}}}\right]^{-2K\left(2-\frac{1}{h}\right)}+C_1.
	\end{align}
	With the same calculation method, we obtain
	\begin{align}
		I_2=\ln\left[(\alpha\rho^2+\beta\rho+\gamma)^{z^2/2\alpha}\left(\frac{2\alpha\rho+\beta-\sqrt{\beta^2-4\alpha\gamma}}{2\alpha\rho+\beta+\sqrt{\beta^2-4\alpha\gamma}}\right)^{\frac{z\left(B_s-\frac{z\beta}{2\alpha}\right)}{\sqrt{\beta^2-4\alpha\gamma}}}\right]^{-\frac{1}{12\pi}}+C_2.
	\end{align}
	
	The metric function $\nu(r)$ is now given by
	\begin{align}
		\nu
		=&\:\nu_c
		+\ln\left\{
		\left[
		(\alpha\rho^2+\beta\rho+\gamma)^{\frac{1}{2\alpha}}
		\left(
		\frac{2\alpha\rho+\beta-\sqrt{\beta^2-4\alpha\gamma}}
		{2\alpha\rho+\beta+\sqrt{\beta^2-4\alpha\gamma}}
		\right)^{-\frac{\beta}{2\alpha\sqrt{\beta^2-4\alpha\gamma}}}
		\right]^{-2K\left(2-\frac{1}{h}\right)}
		\right.
		\nonumber\\[1ex]
		&\qquad\qquad\left.
		\times
		\left[
		(\alpha\rho^2+\beta\rho+\gamma)^{\frac{z^2}{2\alpha}}
		\left(
		\frac{2\alpha\rho+\beta-\sqrt{\beta^2-4\alpha\gamma}}
		{2\alpha\rho+\beta+\sqrt{\beta^2-4\alpha\gamma}}
		\right)^{\frac{z\left(B_s-\frac{z\beta}{2\alpha}\right)}{\sqrt{\beta^2-4\alpha\gamma}}}
		\right]^{-\frac{1}{12\pi}}
		\right\},\label{numagnetized}
	\end{align}
	
	\textcolor{white}{.}
	
	\textcolor{white}{.}
	
	where $\nu_c=C_1+C_2$ is the integration constant. The function $e^{2\nu}$ can be easily obtain, i.e.
	\begin{align}
		e^{2\nu}=&\:e^{2\nu_c}\left\{
		\left[
		(\alpha\rho^2+\beta\rho+\gamma)^{\frac{1}{2\alpha}}
		\left(
		\frac{2\alpha\rho+\beta-\sqrt{\beta^2-4\alpha\gamma}}
		{2\alpha\rho+\beta+\sqrt{\beta^2-4\alpha\gamma}}
		\right)^{-\frac{\beta}{2\alpha\sqrt{\beta^2-4\alpha\gamma}}}
		\right]^{-4K\left(2-\frac{1}{h}\right)}
		\right.
		\nonumber\\[1ex]
		&\qquad\qquad\left.
		\times
		\left[
		(\alpha\rho^2+\beta\rho+\gamma)^{\frac{z^2}{2\alpha}}
		\left(
		\frac{2\alpha\rho+\beta-\sqrt{\beta^2-4\alpha\gamma}}
		{2\alpha\rho+\beta+\sqrt{\beta^2-4\alpha\gamma}}
		\right)^{\frac{z\left(B_s-\frac{z\beta}{2\alpha}\right)}{\sqrt{\beta^2-4\alpha\gamma}}}
		\right]^{-\frac{1}{6\pi}}
		\right\}.\label{enumagggg}
	\end{align}
	At $r=R_s$, $\rho=0$ and $e^{2\nu(R_s)}=1-\frac{r_0}{R_s}$, so we have
	\begin{align}
		\textcolor{black}{e^{2\nu_c}=\left(1-\frac{r_0}{R_s}\right)\left(\gamma^{1/2\alpha}\right)^{\left[4K\left(2-\frac{1}{h}\right)+\frac{z^2}{6\pi}\right]}\left[\left(\frac{\beta-\sqrt{\beta^2-4\alpha\gamma}}{\beta+\sqrt{\beta^2-4\alpha\gamma}}\right)^{\frac{1}{\sqrt{\beta^2-4\alpha\gamma}}}\right]^{\left[\frac{B_s z}{6\pi}-\frac{\beta}{2\alpha}\left\{\frac{z^2}{6\pi}+4K\left(2-\frac{1}{h}\right)\right\}\right]}}.\label{integrationconstantmagnetized}
	\end{align}
	In Appendix~\ref{metricnojiri}, we show that for the isotropic configuration with the limit $\mathcal{B}\rightarrow 0$, $e^{2\nu}$ is reduces to the expression of $e^{2\nu}$ shown by Eq.~(24) in Ref.~\cite{Nojiri2024}.
	
	The next step is calculating the differential of $e^{\nu}$ with respect to $r$, i.e.
	\begin{align}
		\left(e^{2\nu}\right)'=e^{2\nu}\frac{\rho_c(r-r_0)}{(R_s-r_0)^2}\left[\frac{4K\left(2-\frac{1}{h}\right)}{\alpha}\left(\frac{2\alpha\rho+\beta-1}{\alpha\rho^2+\beta\rho+\gamma}\right)+\frac{1}{3\pi}\left\{\frac{z}{2\alpha}\frac{(2\alpha\rho+\beta)z+\left(B_s-\frac{z\beta}{2\alpha}\right)}{\alpha\rho^2+\beta\rho+\gamma}\right\}\right].\label{derivativenu}
	\end{align}

\begin{figure*}[htbp]
	\centering
	\begin{tabular}{cc}
		\includegraphics[width=0.48\textwidth]{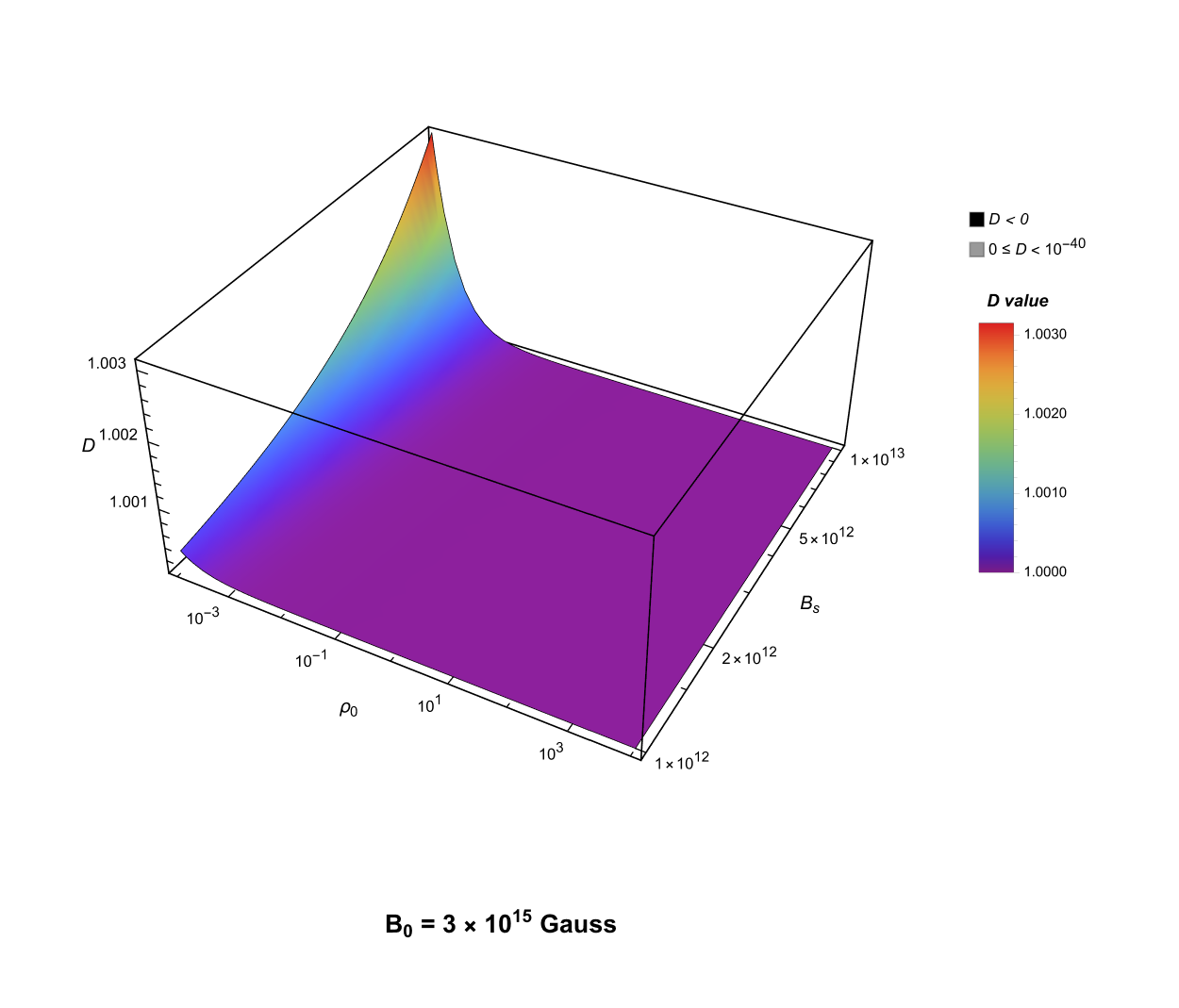} &
		\includegraphics[width=0.48\textwidth]{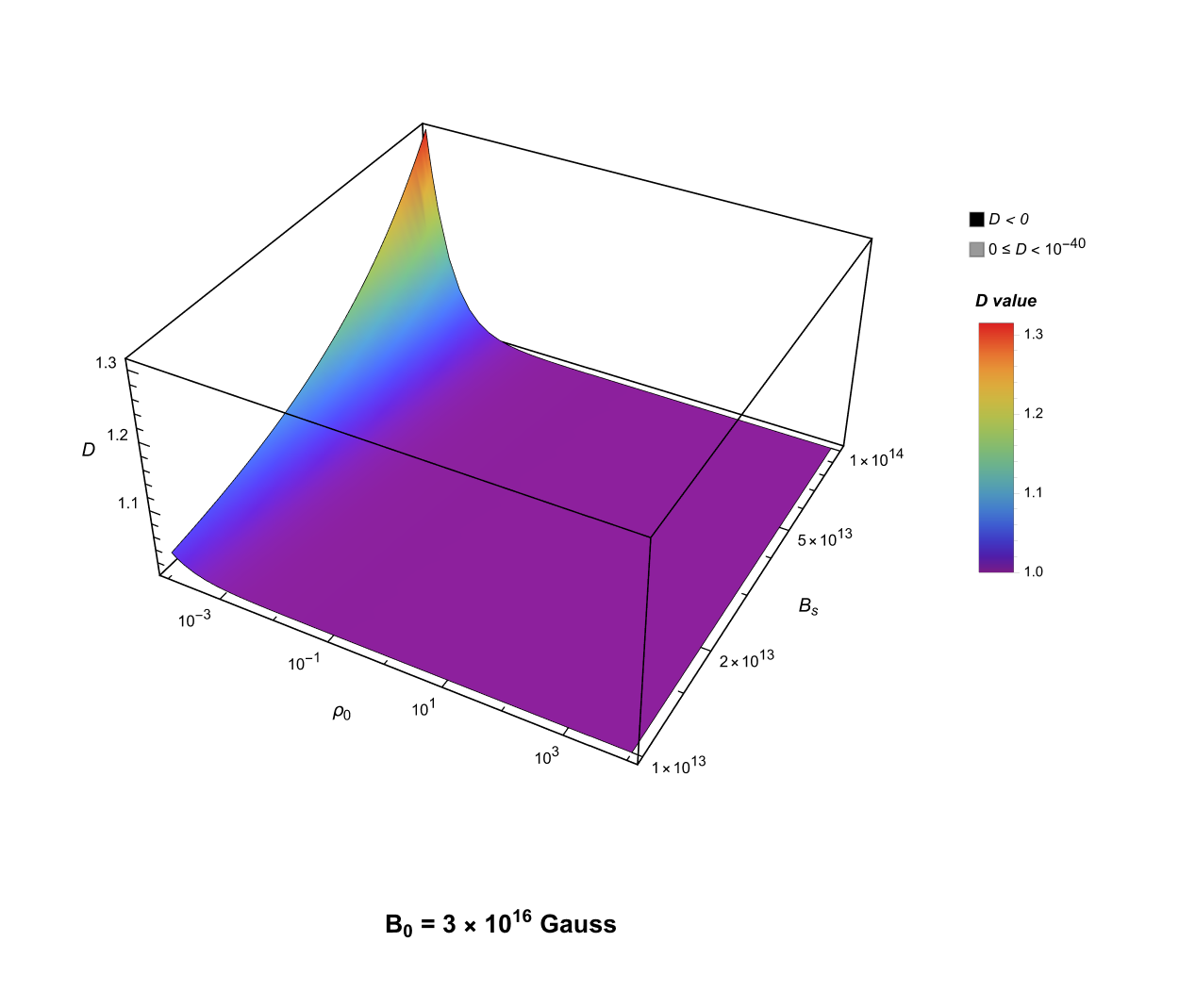} \\
		(a) & (b) \\[1em]
		\includegraphics[width=0.48\textwidth]{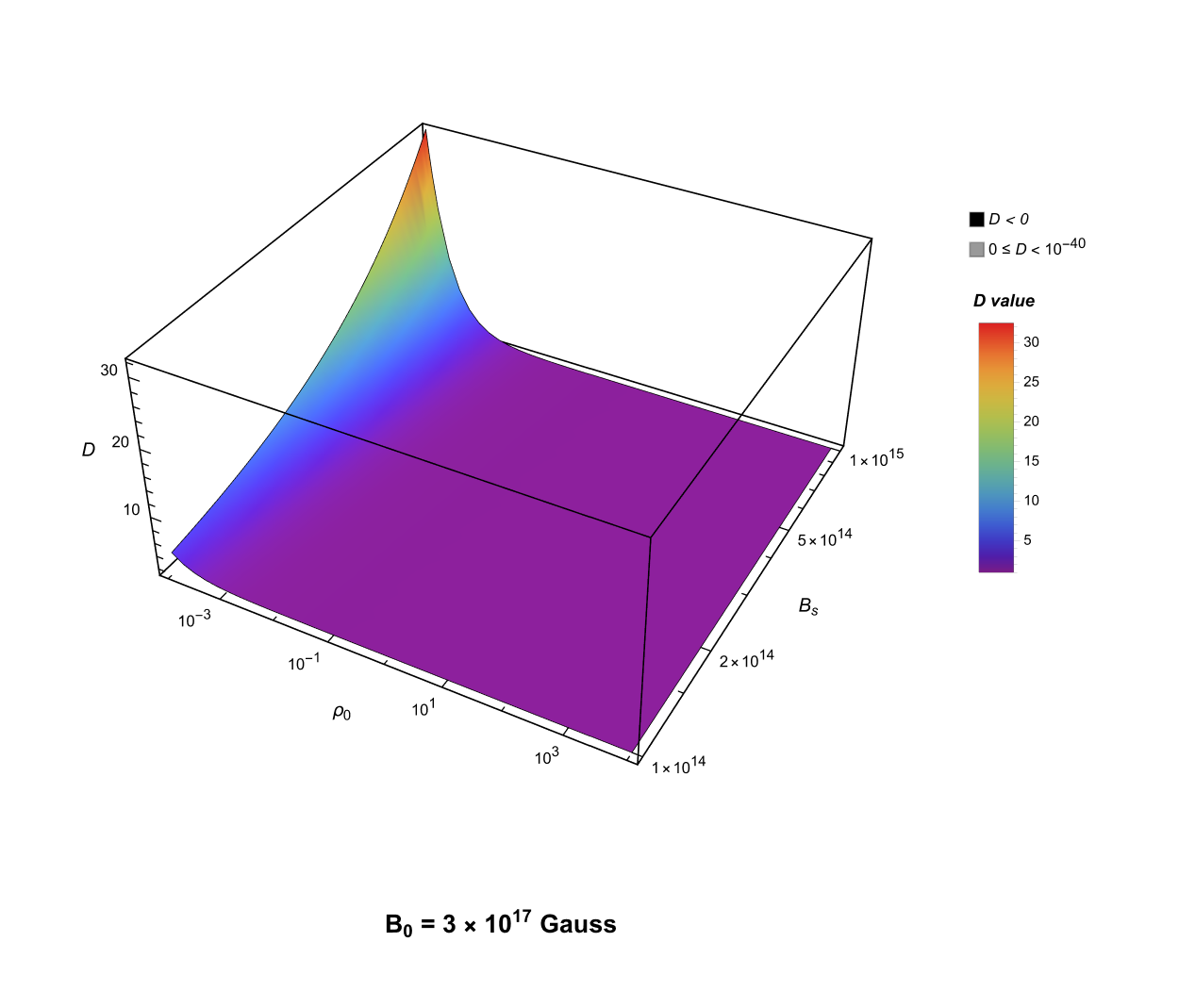} &
		\includegraphics[width=0.48\textwidth]{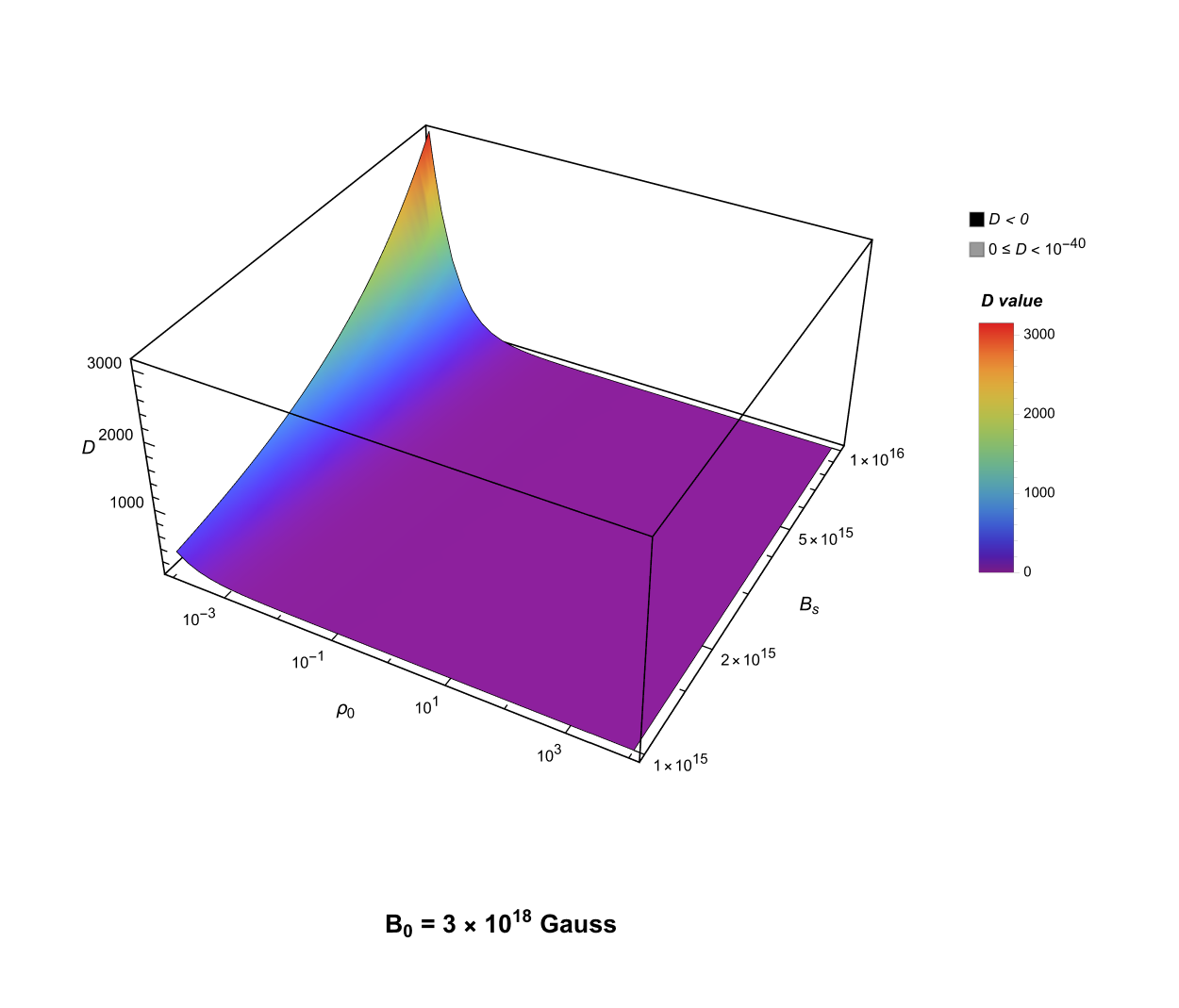} \\
		(c) & (d)
	\end{tabular}
	\caption{Three-dimensional plot of the discriminant $D = \beta^2 - 4\alpha\gamma$ as a function of $\rho_0$ and $B_s$ for (a) $B_0=3\times10^{15}$ G, (b) $B_0=3\times10^{16}$ G, (c) $B_0=3\times10^{17}$ G, and (d) $B_0=3\times10^{18}$ G.}
	\label{figdiscriminant}
\end{figure*}

\begin{figure*}[htbp]
	\centering
	\begin{tabular}{cc}
		\includegraphics[width=0.4\textwidth]{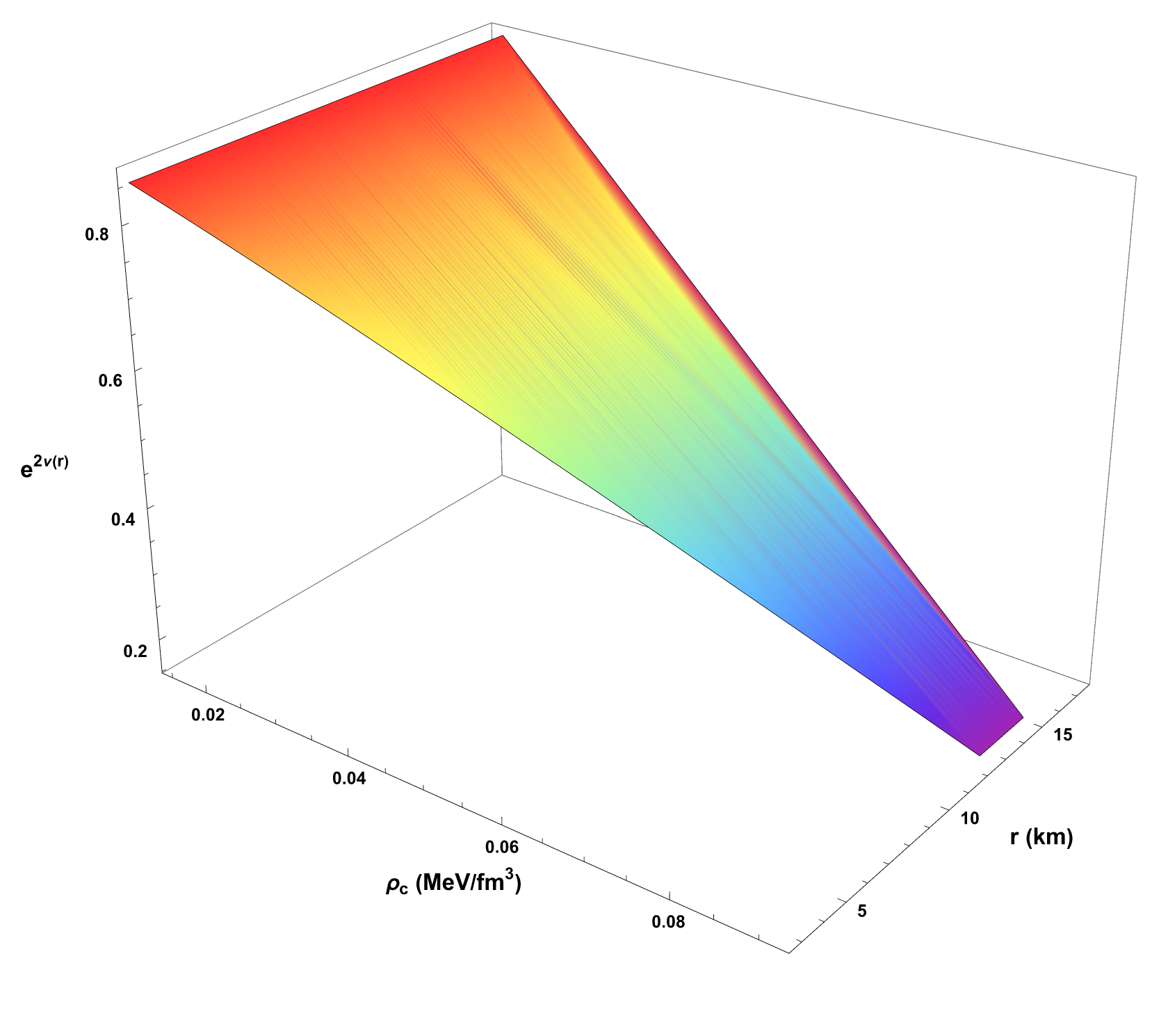} &
		\includegraphics[width=0.4\textwidth]{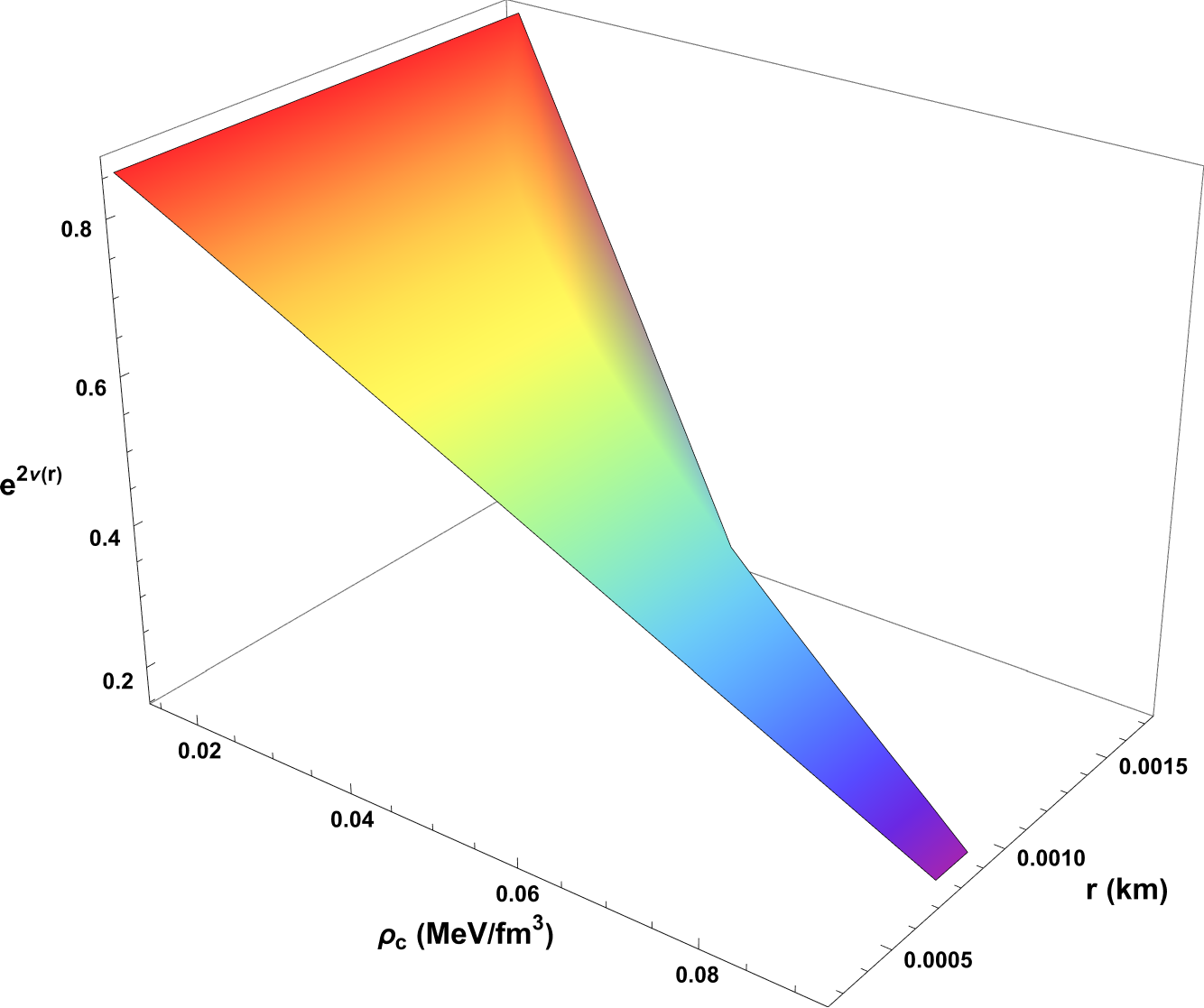} \\
		(a) & (b) \\[1em]
		\includegraphics[width=0.4\textwidth]{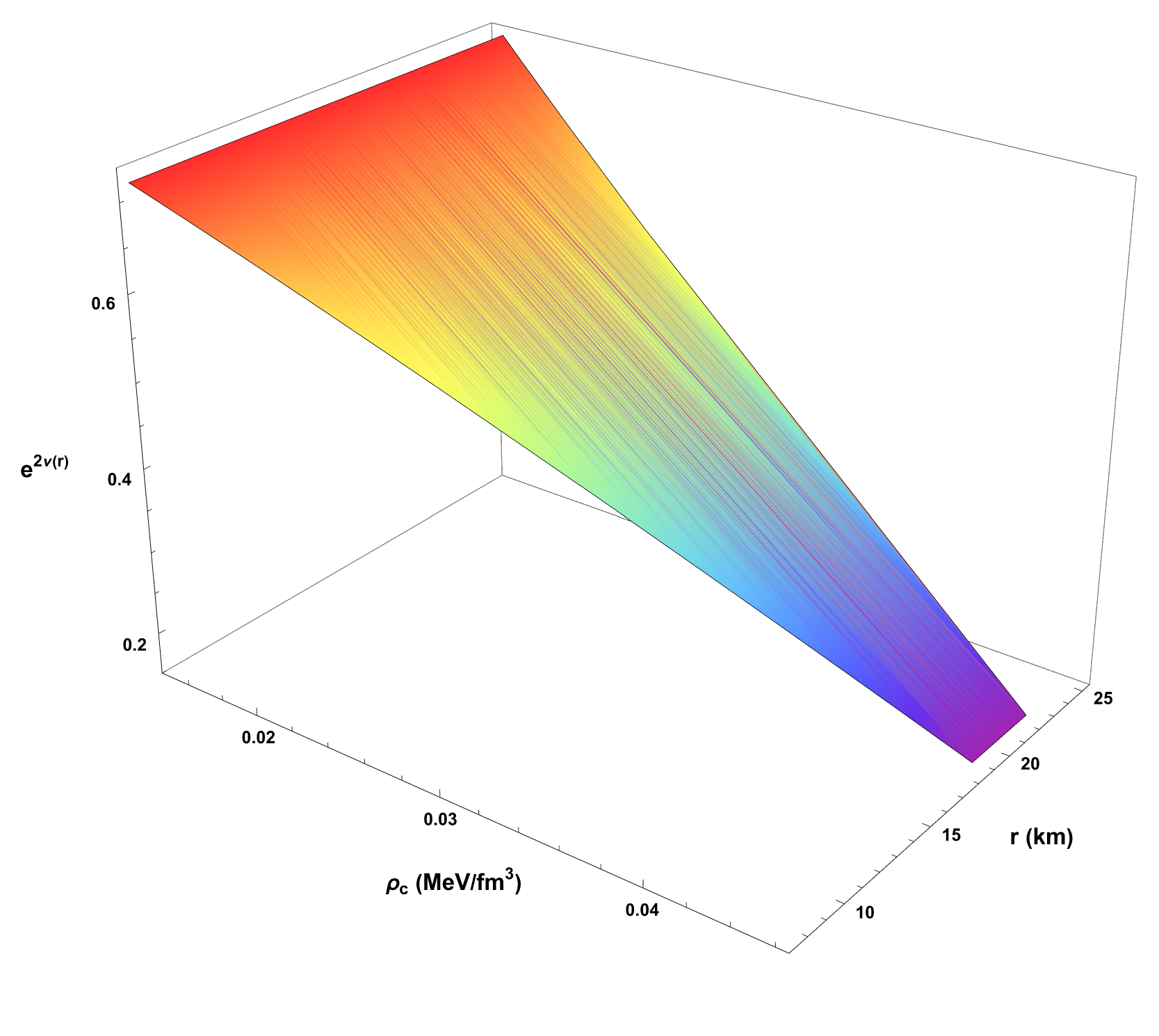} &
		\includegraphics[width=0.4\textwidth]{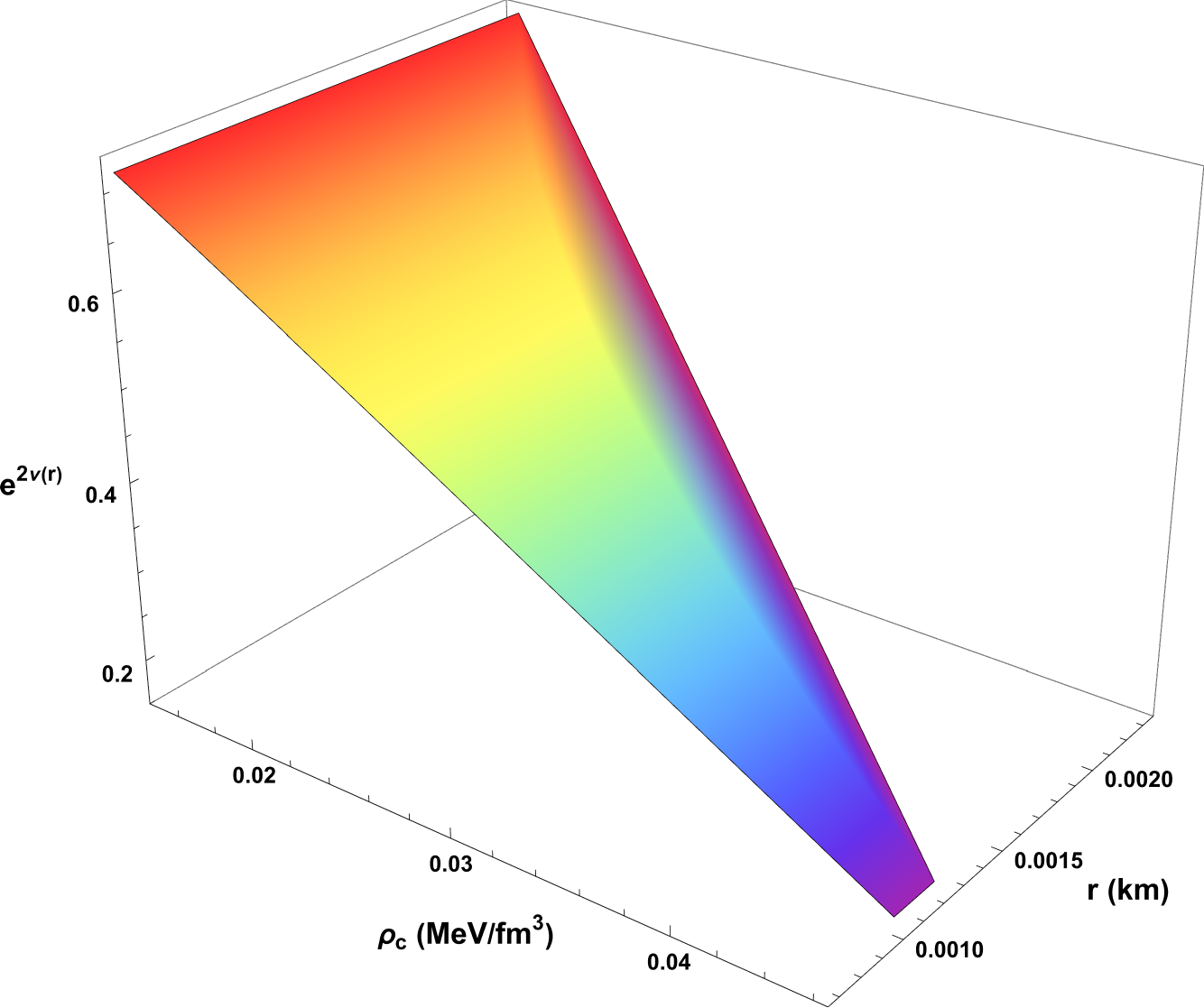} \\
		(c) & (d) \\[1em]
		\includegraphics[width=0.4\textwidth]{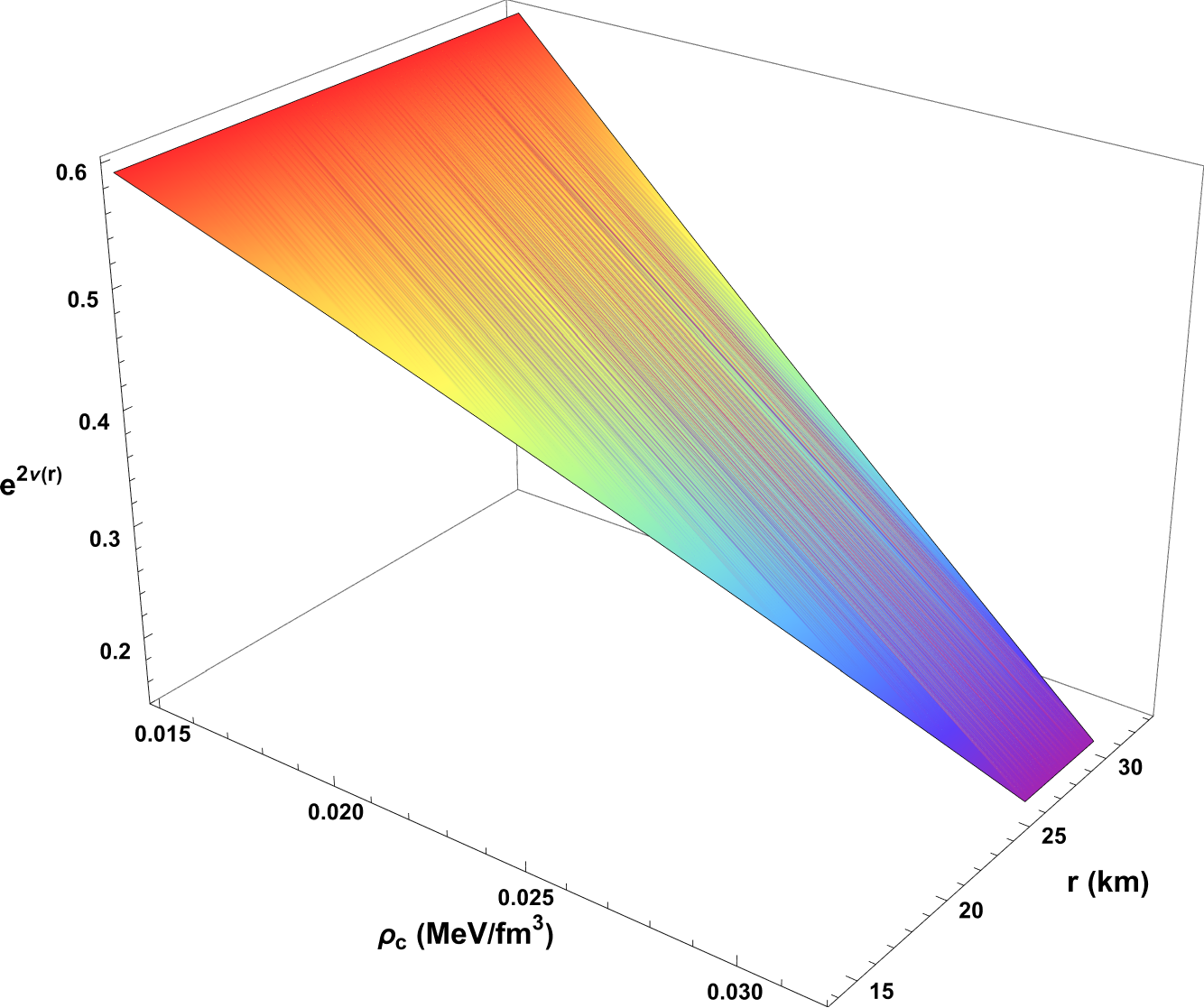} &
		\includegraphics[width=0.4\textwidth]{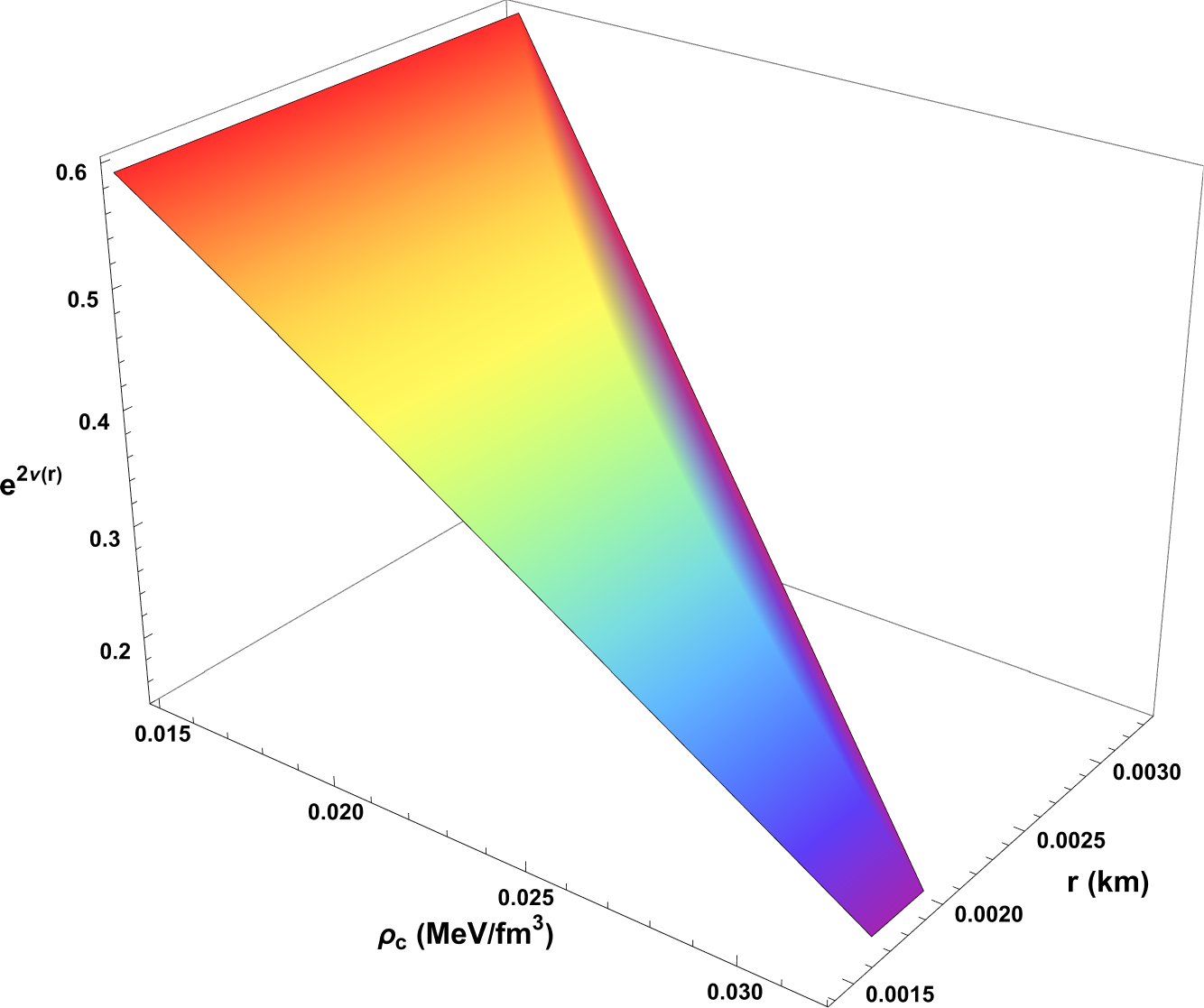} \\
		(e) & (f)
	\end{tabular}
	\caption{\textcolor{black}{Three-dimensional plot of interior temporal metric function as a function of $\rho_c$ and $r$ of configurations with $B_s=1\times10^{15}$ G for (a) $h=1$ with Approach 1, (b) $h=1$ with Approach 2, (c) $h=1.5$ with Approach 1, (d) $h=1.5$ with Approach 2, (e) $h=2$ with Approach 1, and (f) $h=2$ with Approach 2.}}
	\label{figinteriormetric1015}
\end{figure*}

\begin{figure*}[htbp]
	\centering
	\begin{tabular}{cc}
		\includegraphics[width=0.4\textwidth]{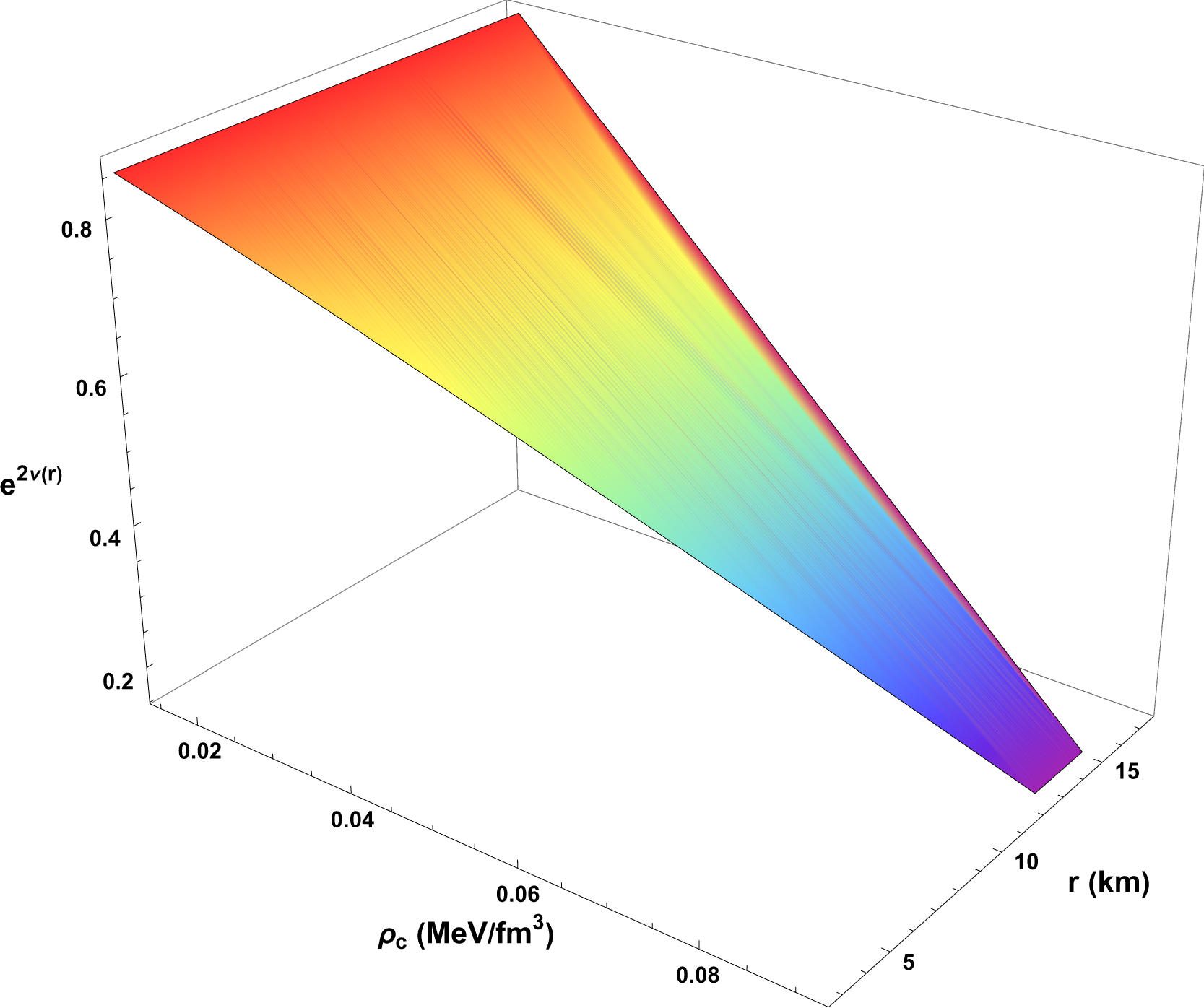} &
		\includegraphics[width=0.4\textwidth]{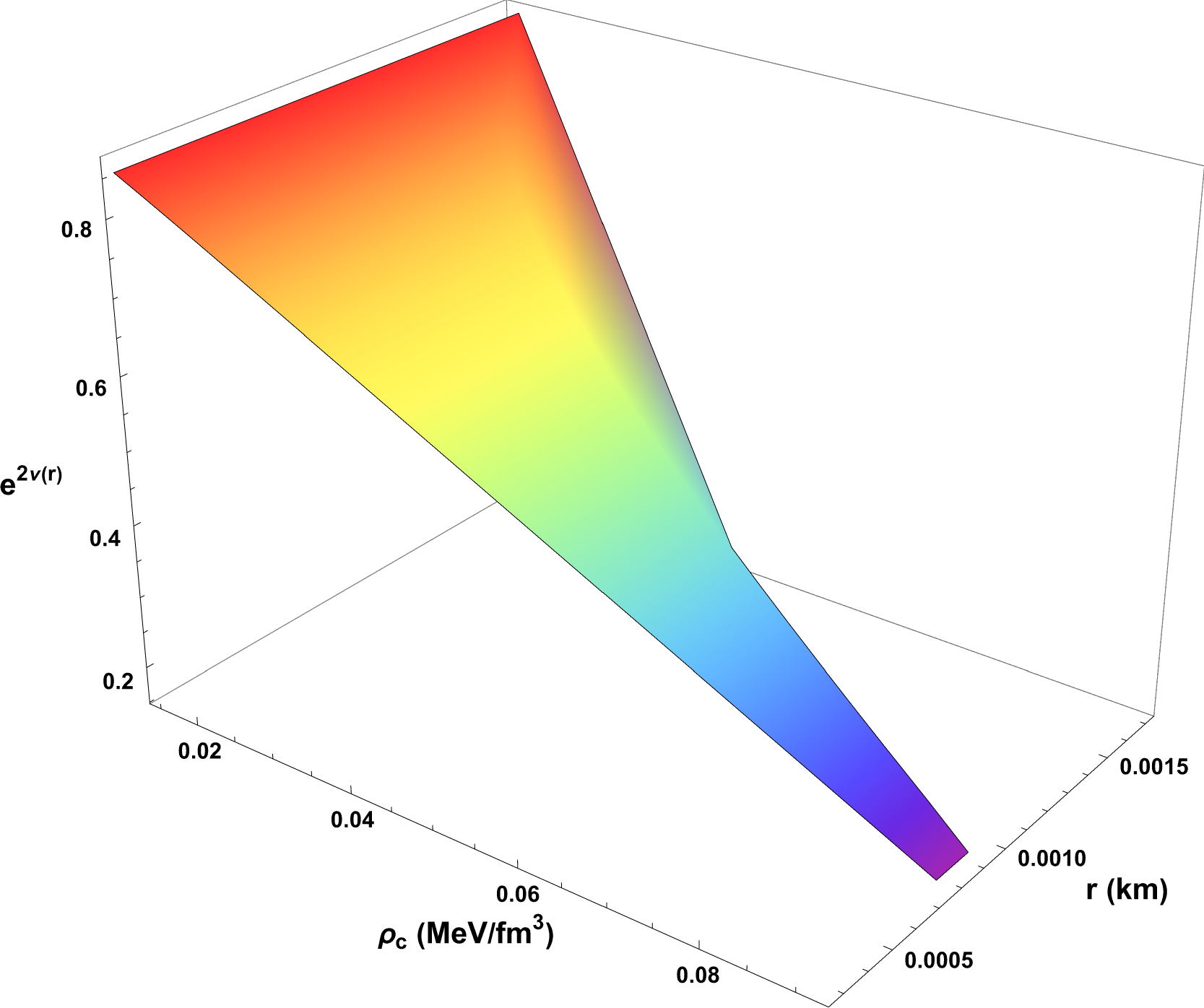} \\
		(a) & (b) \\[1em]
		\includegraphics[width=0.4\textwidth]{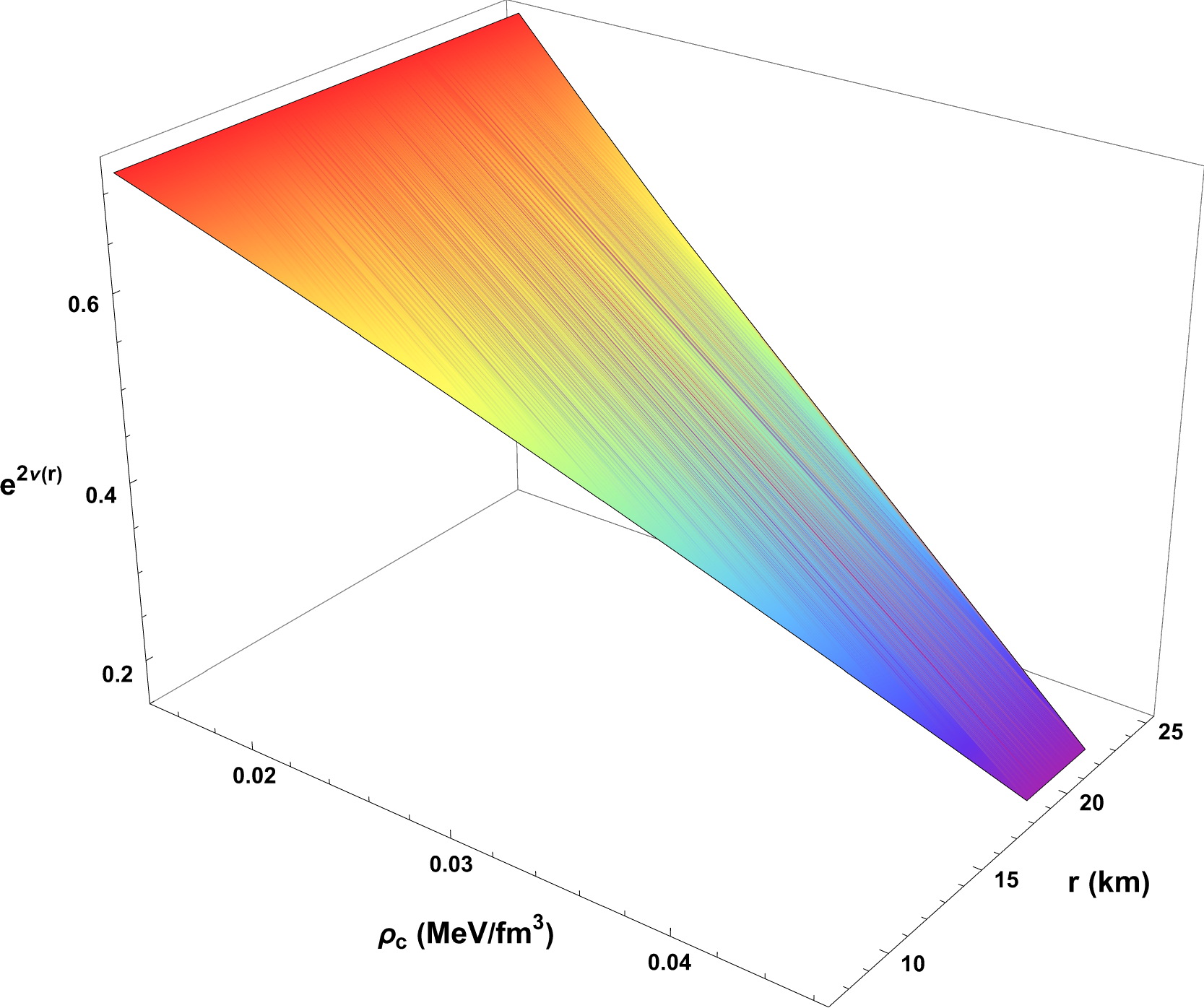} &
		\includegraphics[width=0.4\textwidth]{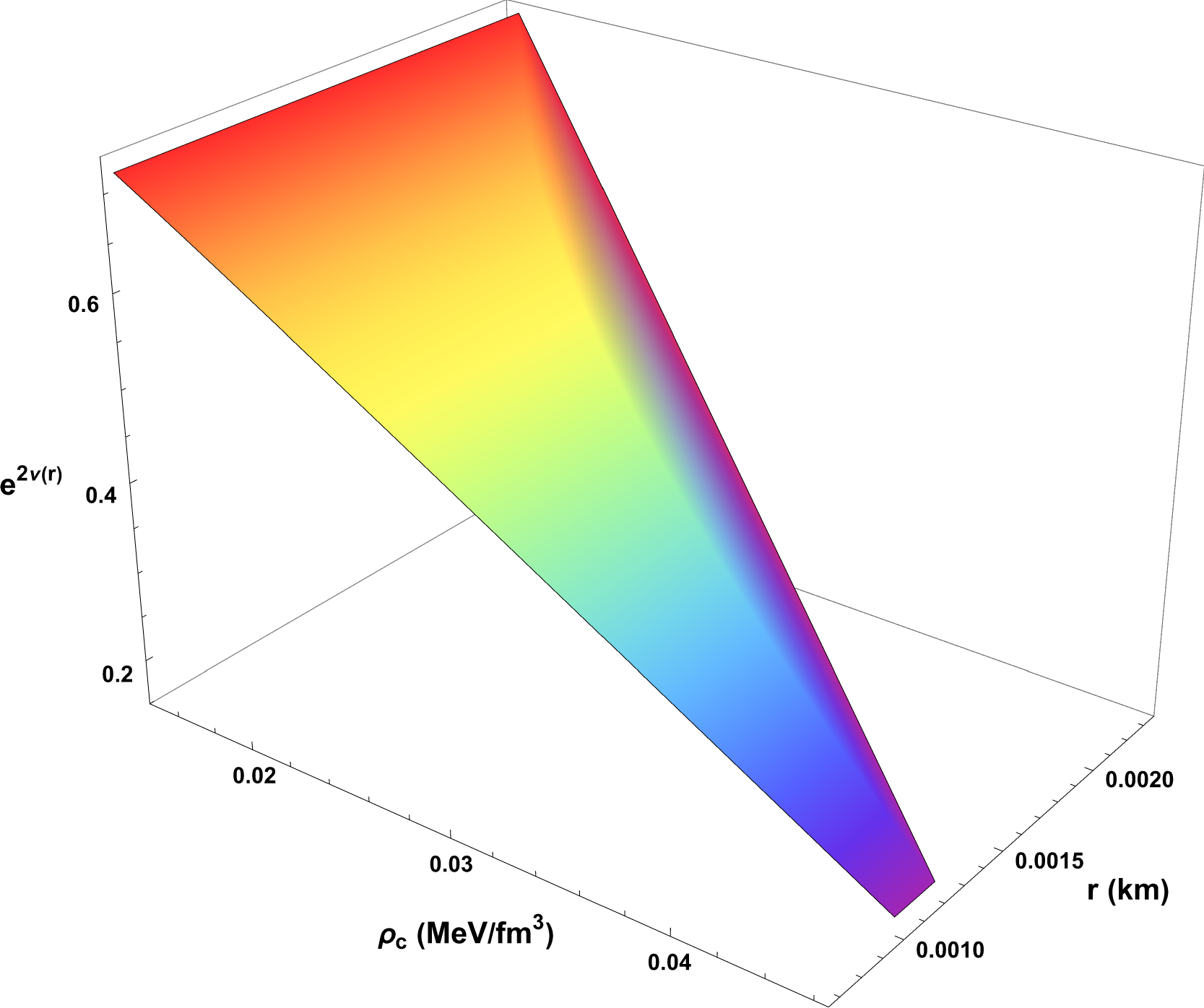} \\
		(c) & (d) \\[1em]
		\includegraphics[width=0.4\textwidth]{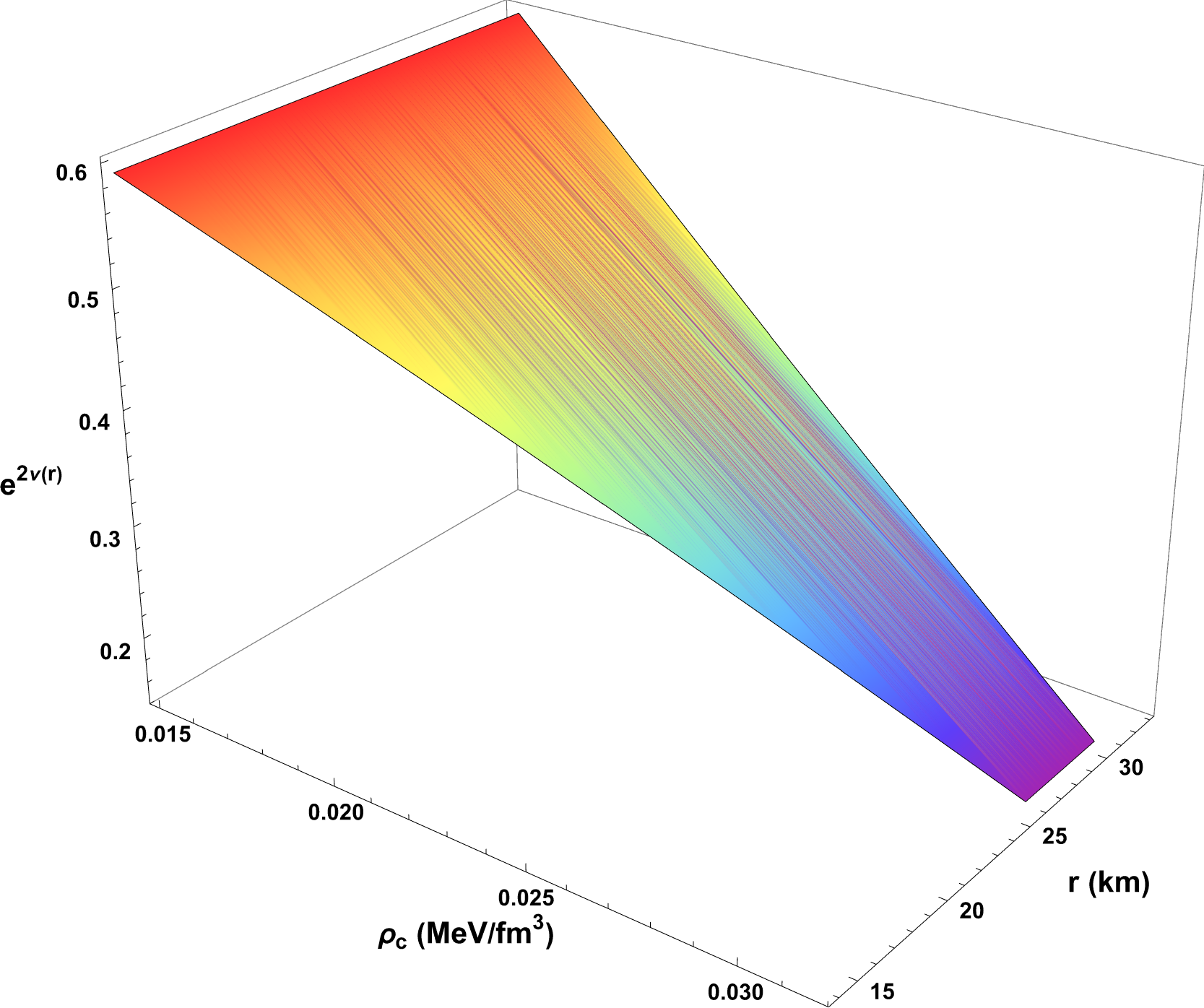} &
		\includegraphics[width=0.4\textwidth]{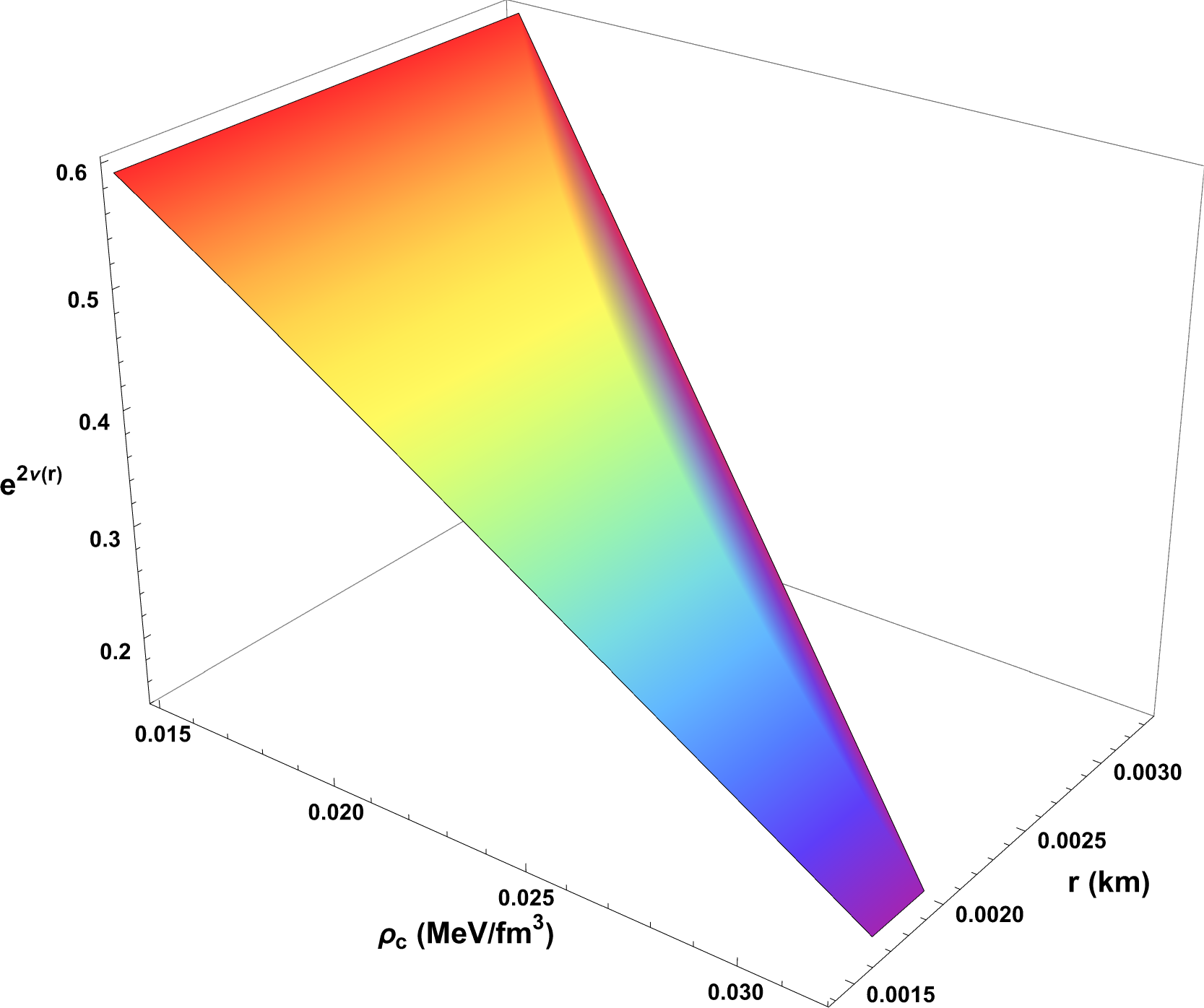} \\
		(e) & (f)
	\end{tabular}
	\caption{\textcolor{black}{Three-dimensional plot of interior temporal metric function as a function of $\rho_c$ and $r$ of configurations with $B_s=1\times10^{16}$ G for (a) $h=1$ with Approach 1, (b) $h=1$ with Approach 2, (c) $h=1.5$ with Approach 1, (d) $h=1.5$ with Approach 2, (e) $h=2$ with Approach 1, and (f) $h=2$ with Approach 2.}}
	\label{figinteriormetric1016}
\end{figure*}

\begin{figure*}[!t] 
	\centering
	
	\begin{minipage}{0.48\textwidth}
		\centering
		\includegraphics[width=\linewidth]{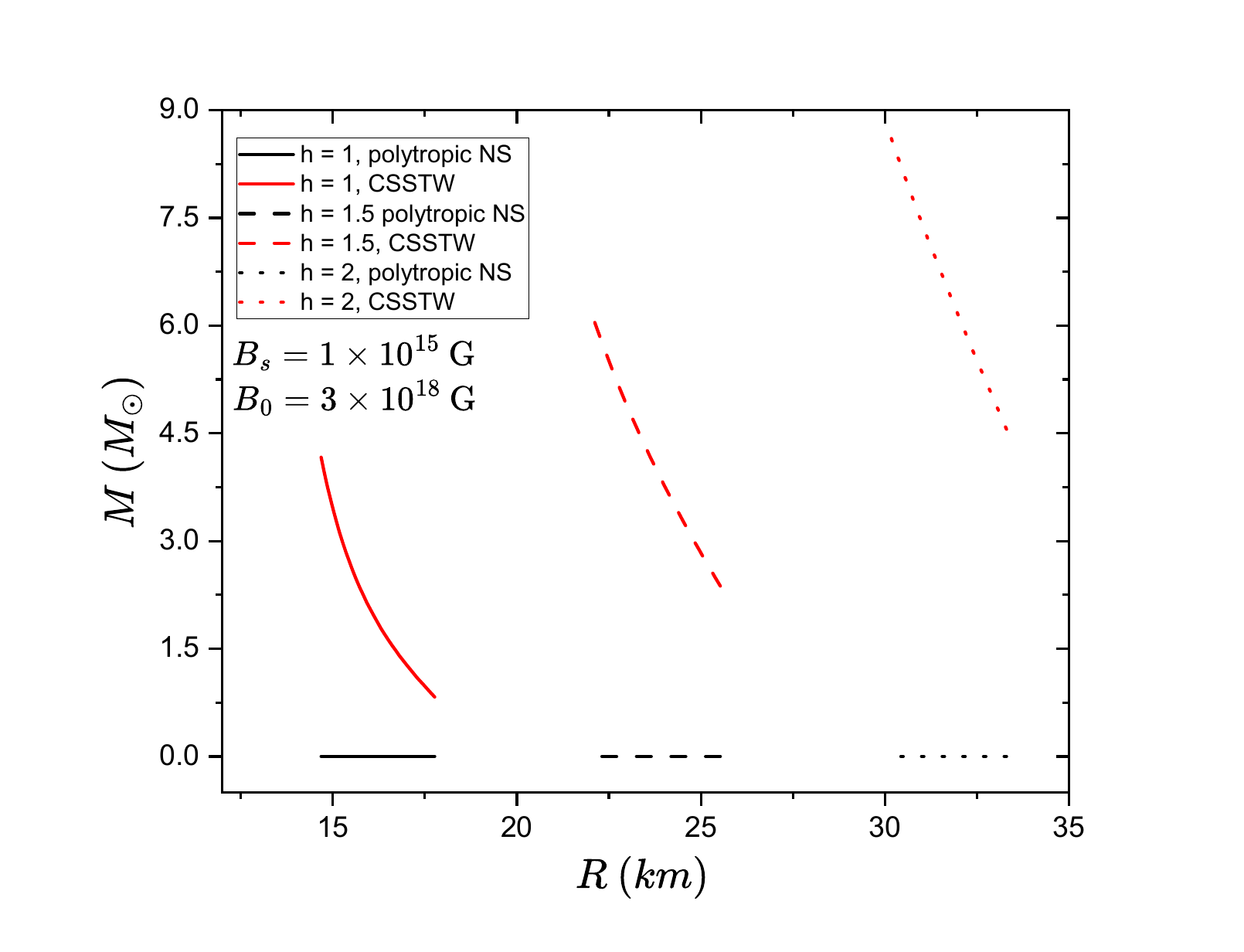}
		\vspace{-0.3cm}
		{\small (a) Approach~1}
	\end{minipage}
	\hfill
	\begin{minipage}{0.48\textwidth}
		\centering
		\includegraphics[width=\linewidth]{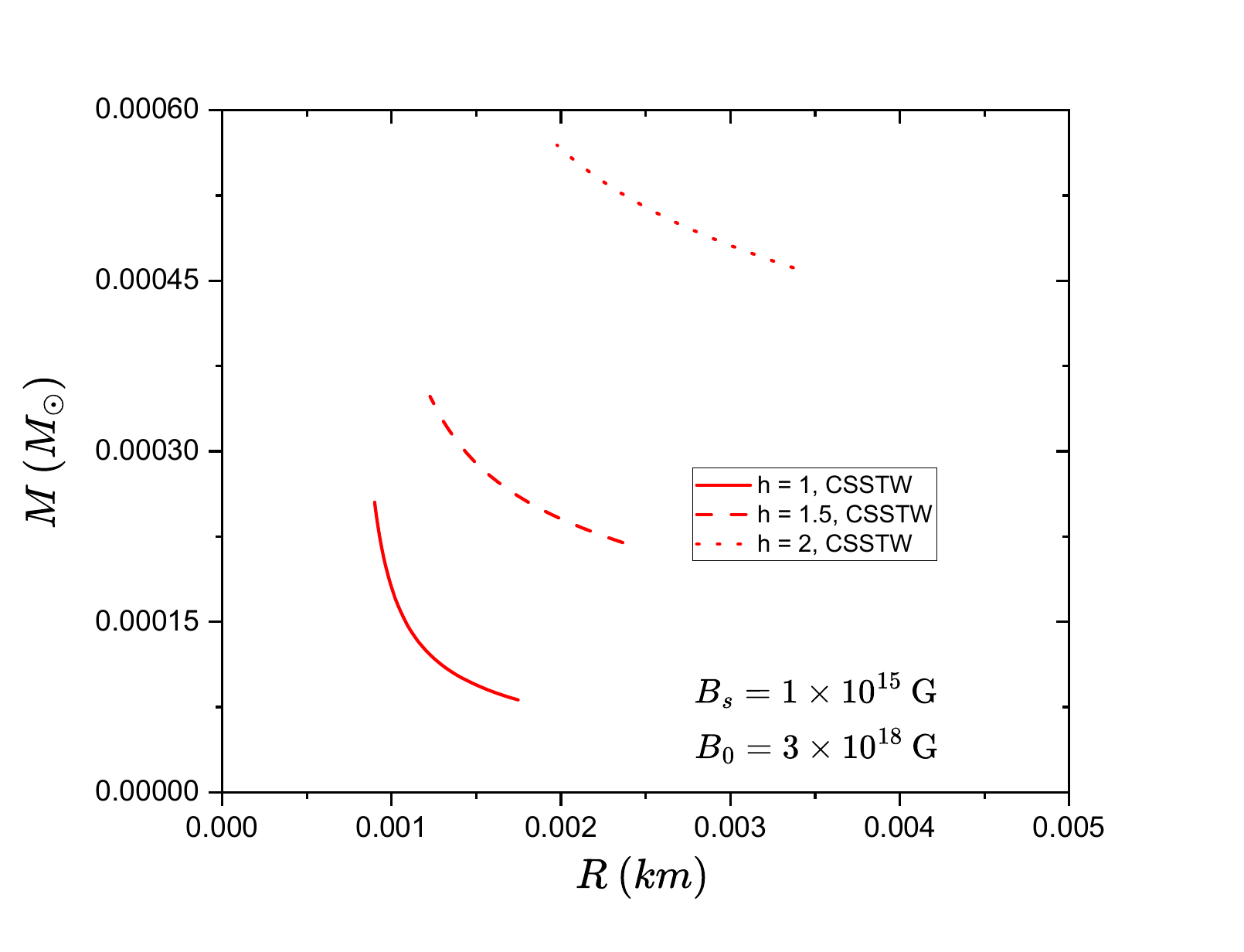}
		\vspace{-0.3cm}
		{\small (b) Approach~2}
	\end{minipage}
	
	\vspace{0.4cm}
	
	\begin{minipage}{0.48\textwidth}
		\centering
		\includegraphics[width=\linewidth]{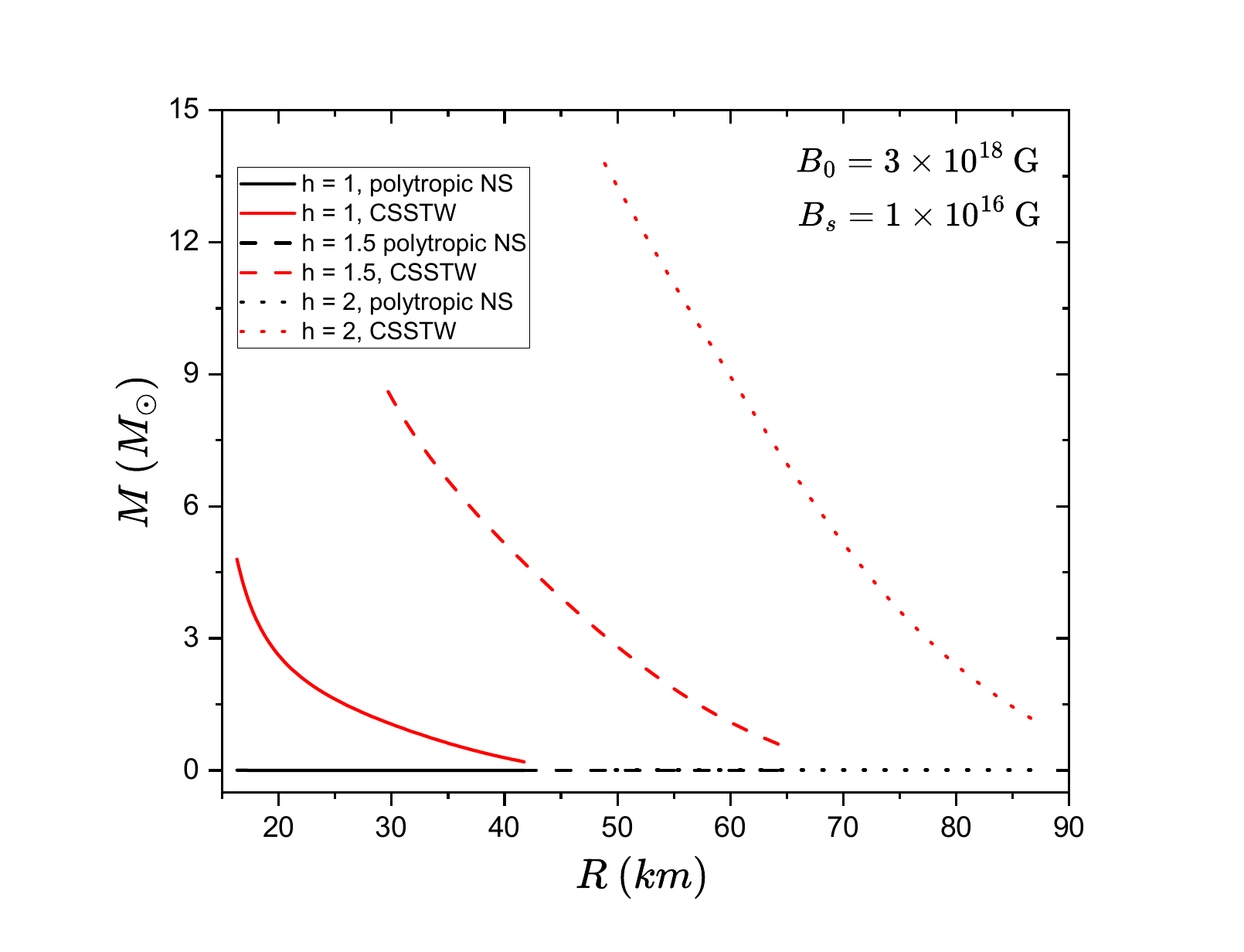}
		\vspace{-0.3cm}
		{\small (c) Approach~1}
	\end{minipage}
	\hfill
	\begin{minipage}{0.48\textwidth}
		\centering
		\includegraphics[width=\linewidth]{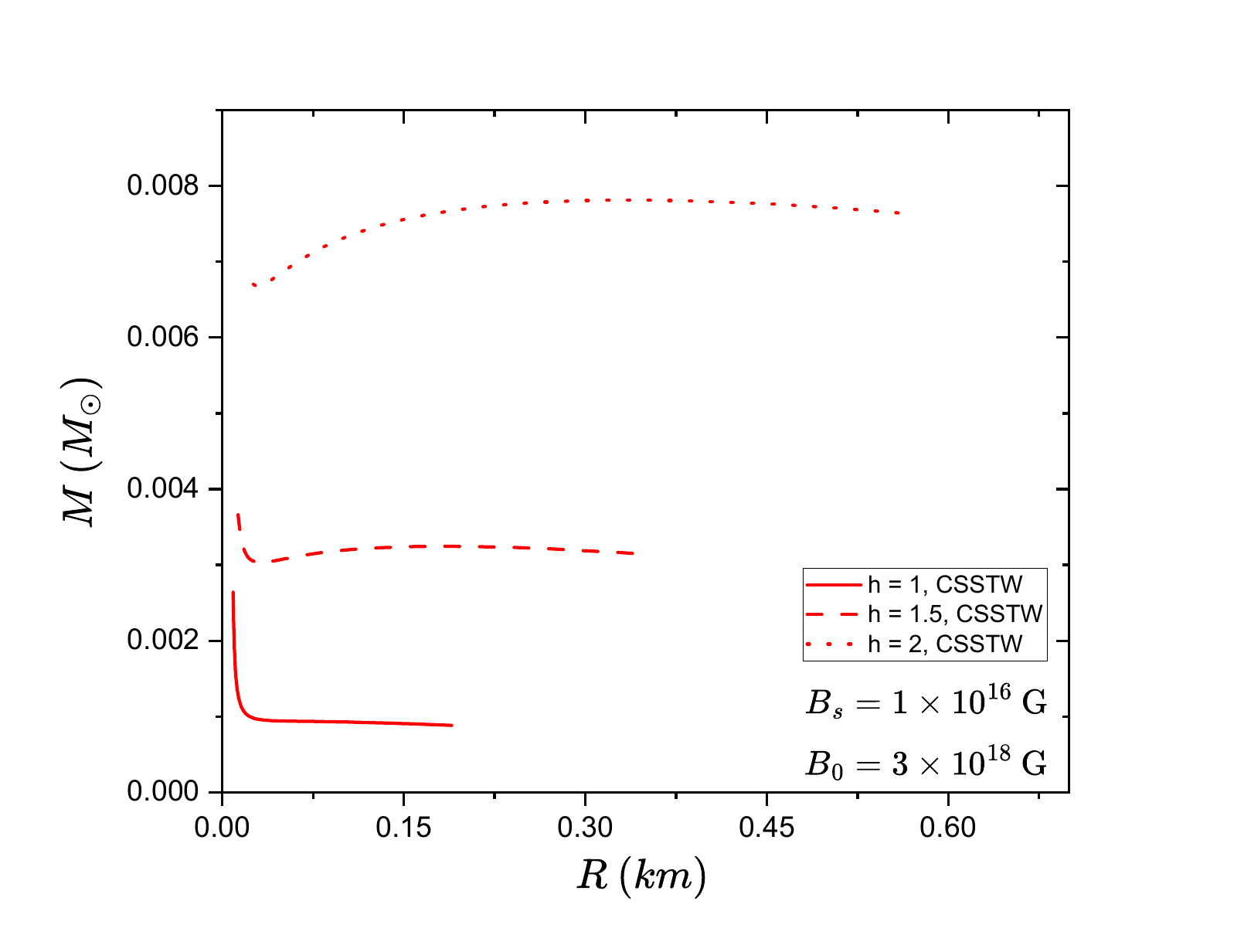}
		\vspace{-0.3cm}
		{\small (d) Approach~2}
	\end{minipage}
	
	\caption{
		MR relations of the pure NSs and the CSSTW for different values of the anisotropy parameter $h$.
		Panels (a) correspond to $B_s = 10^{15}\,\mathrm{G}$ and $B_0 = 10^{18}\,\mathrm{G}$ with Approach~1,
		(b) $B_s = 10^{15}\,\mathrm{G}$ and $B_0 = 10^{18}\,\mathrm{G}$ with Approach~2,
		(c) $B_s = 10^{16}\,\mathrm{G}$ and $B_0 = 10^{18}\,\mathrm{G}$ with Approach~1, and
		(d) $B_s = 10^{16}\,\mathrm{G}$ and $B_0 = 10^{18}\,\mathrm{G}$ with Approach~2.
	}
	\label{fig:mr_magnetized}
\end{figure*}

\begin{figure*}[!t] 
	\centering
	
	\begin{minipage}{0.48\textwidth}
		\centering
		\includegraphics[width=\linewidth]{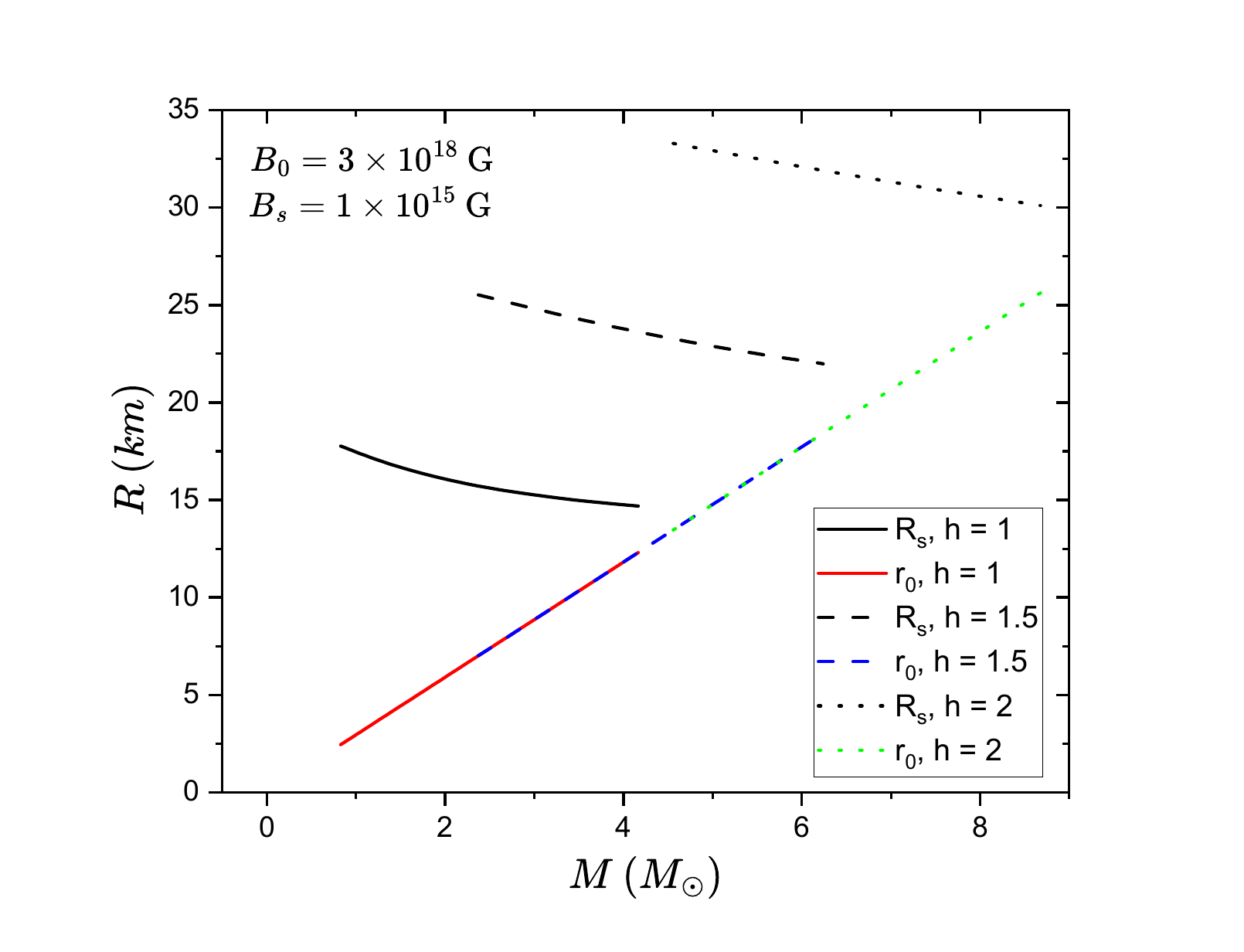}
		\vspace{-0.3cm}
		{\small (a) Approach~1}
	\end{minipage}
	\hfill
	\begin{minipage}{0.48\textwidth}
		\centering
		\includegraphics[width=\linewidth]{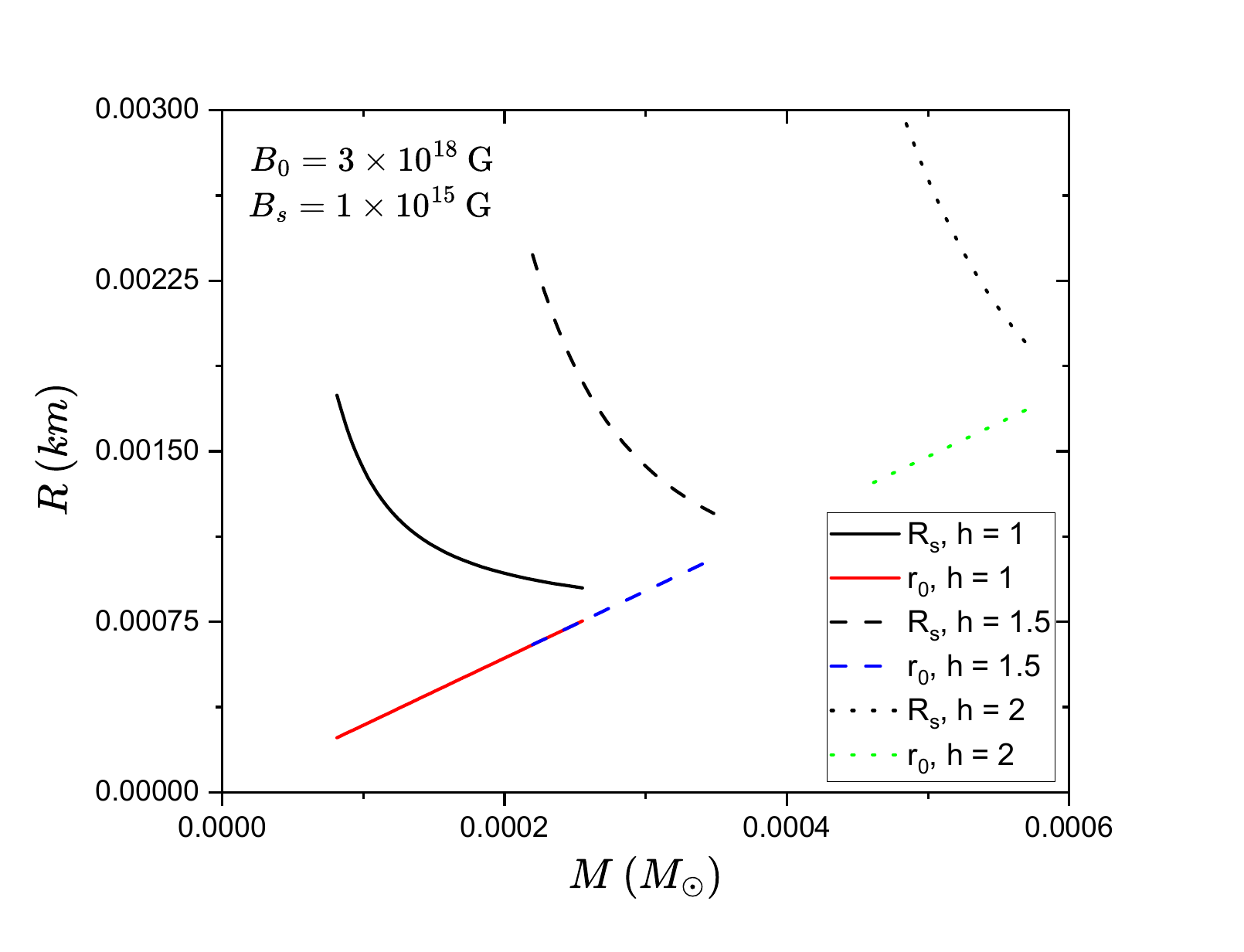}
		\vspace{-0.3cm}
		{\small (b) Approach~2}
	\end{minipage}
	
	\vspace{0.4cm}
	
	\begin{minipage}{0.48\textwidth}
		\centering
		\includegraphics[width=\linewidth]{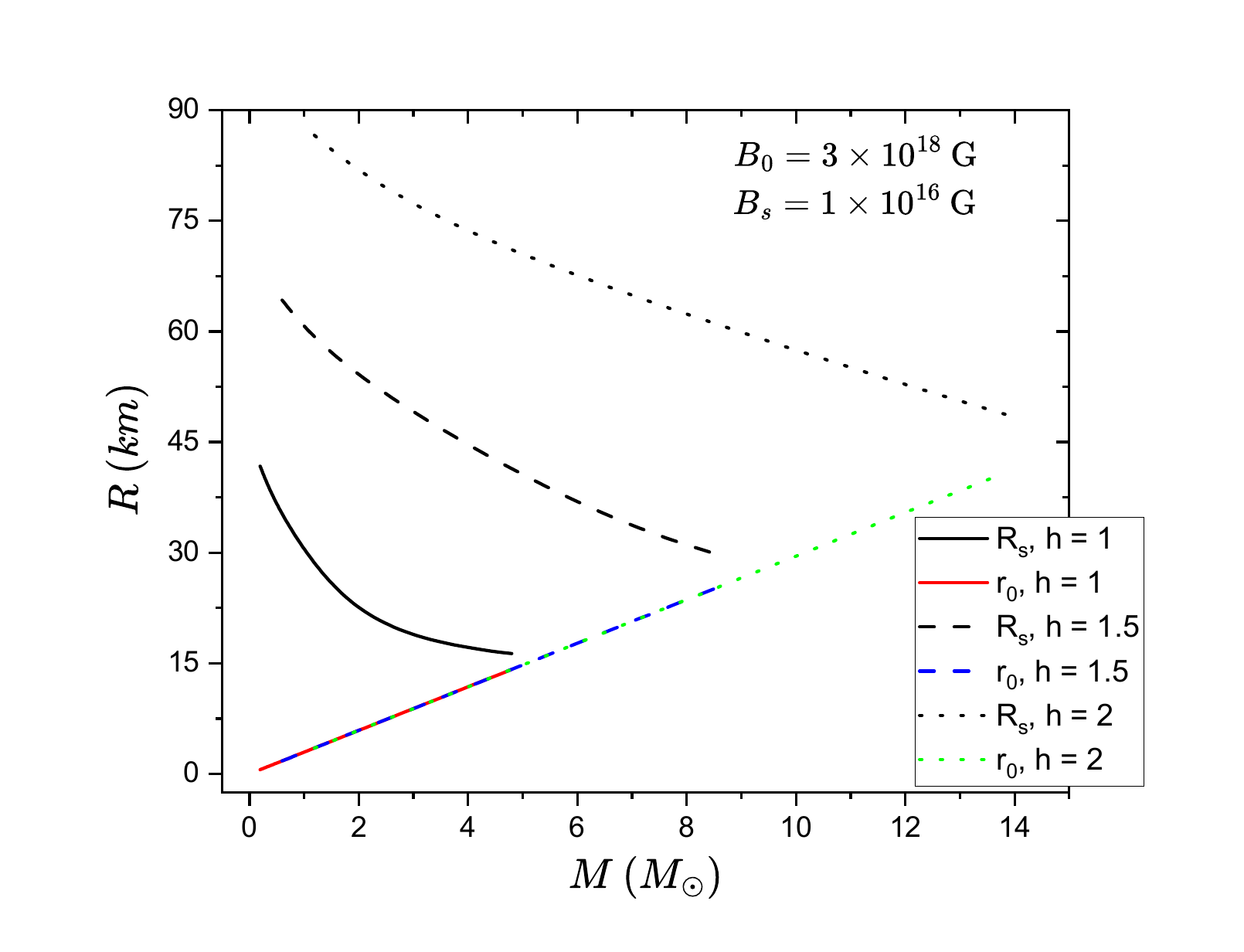}
		\vspace{-0.3cm}
		{\small (c) Approach~1}
	\end{minipage}
	\hfill
	\begin{minipage}{0.48\textwidth}
		\centering
		\includegraphics[width=\linewidth]{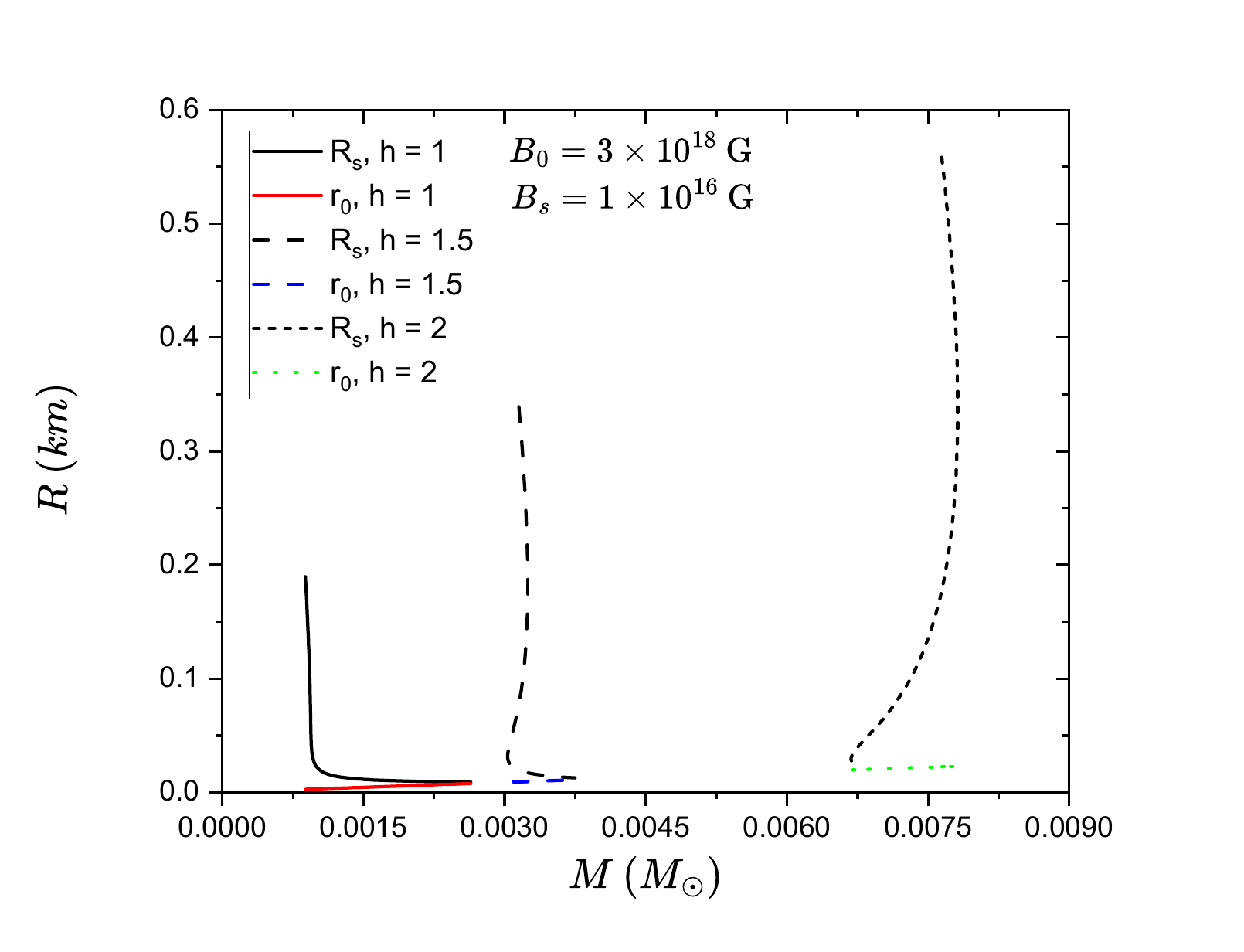}
		\vspace{-0.3cm}
		{\small (d) Approach~2}
	\end{minipage}
	
	\caption{
		Radii of the stellar surface and the wormhole throat in the CSSTW with $B_s=10^{15}$ G and $B_0=10^{18}$ G as functions of the ADM mass for different values of $h$.
		Panels (a) correspond to $B_s = 10^{15}\,\mathrm{G}$ and $B_0 = 10^{18}\,\mathrm{G}$ with Approach~1,
		(b) $B_s = 10^{15}\,\mathrm{G}$ and $B_0 = 10^{18}\,\mathrm{G}$ with Approach~2,
		(c) $B_s = 10^{16}\,\mathrm{G}$ and $B_0 = 10^{18}\,\mathrm{G}$ with Approach~1, and
		(d) $B_s = 10^{16}\,\mathrm{G}$ and $B_0 = 10^{18}\,\mathrm{G}$ with Approach~2.}
	\label{figRsrmagnetized}
\end{figure*}

\begin{figure*}[!t] 
	\centering
	
	\begin{minipage}{0.48\textwidth}
		\centering
		\includegraphics[width=\linewidth]{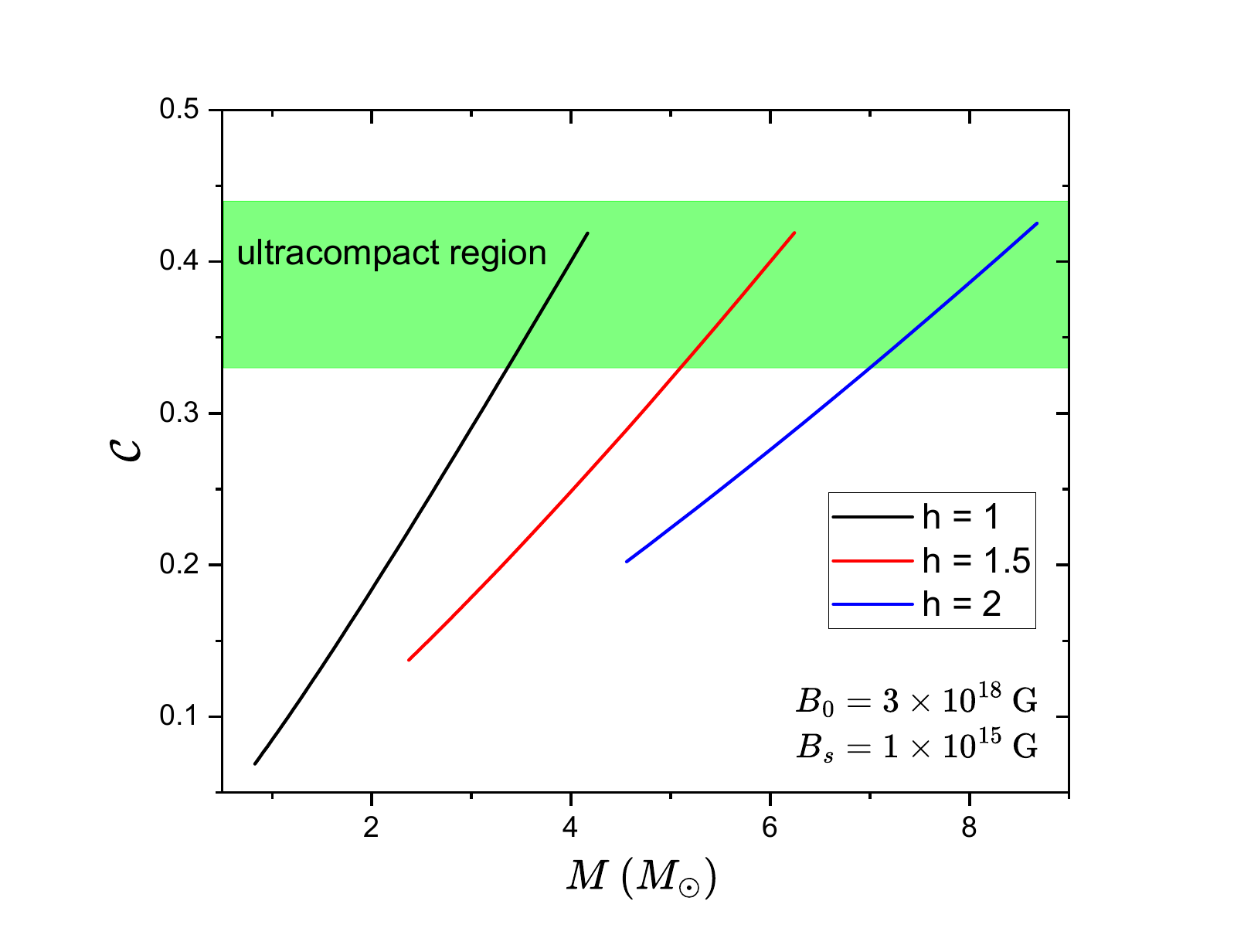}
		\vspace{-0.3cm}
		{\small (a) Approach~1}
	\end{minipage}
	\hfill
	\begin{minipage}{0.48\textwidth}
		\centering
		\includegraphics[width=\linewidth]{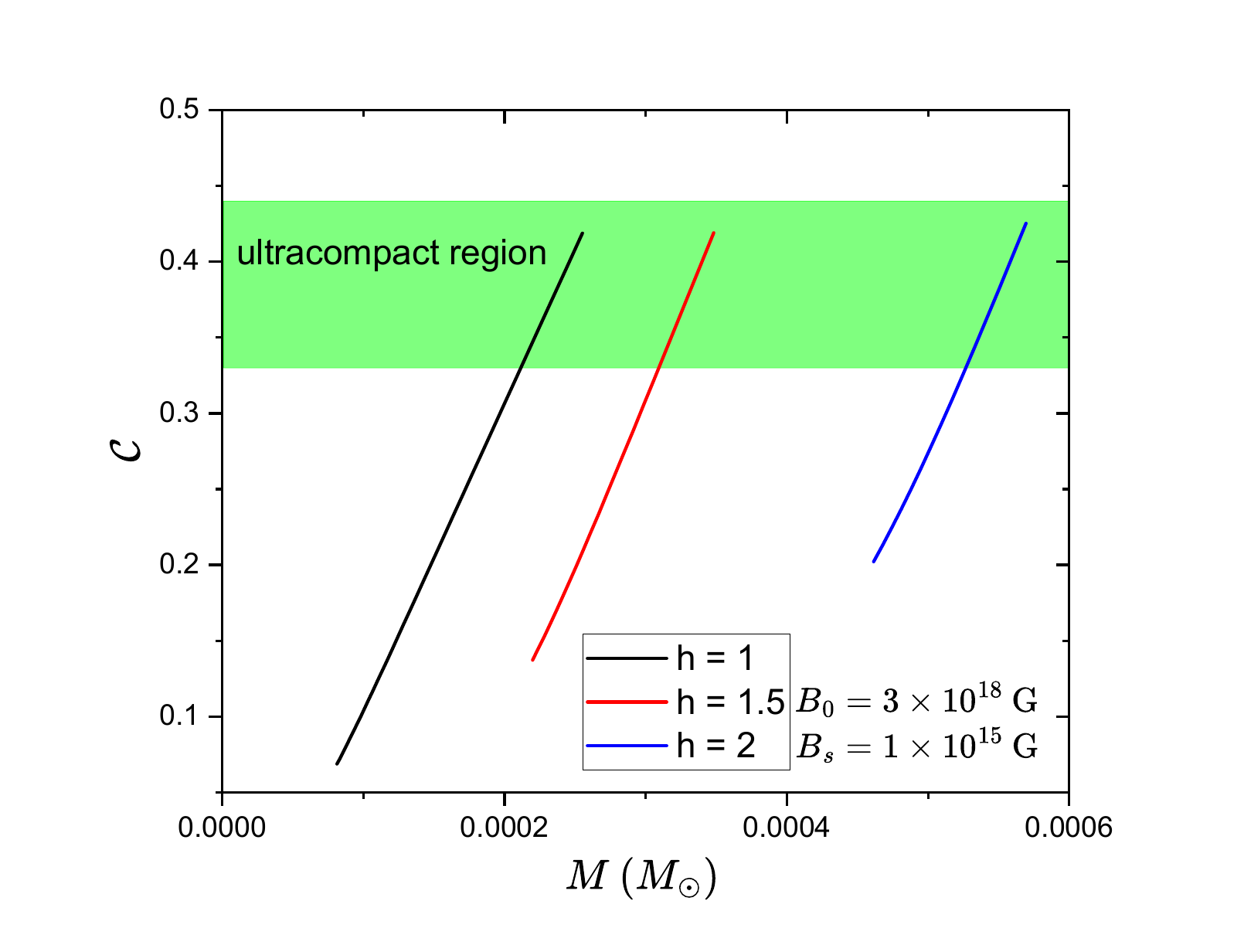}
		\vspace{-0.3cm}
		{\small (b) Approach~2}
	\end{minipage}
	
	\vspace{0.4cm}
	
	\begin{minipage}{0.48\textwidth}
		\centering
		\includegraphics[width=\linewidth]{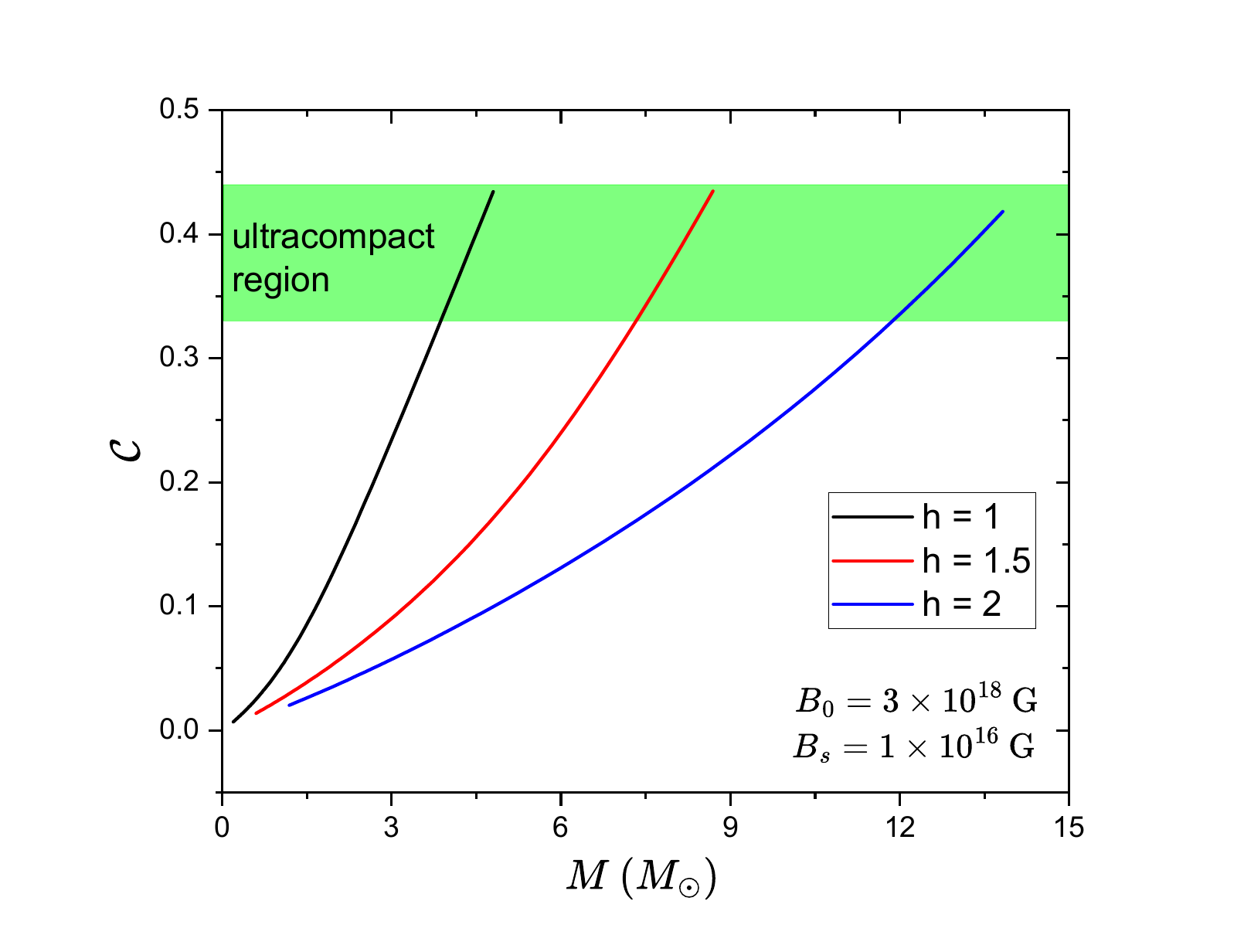}
		\vspace{-0.3cm}
		{\small (c) Approach~1}
	\end{minipage}
	\hfill
	\begin{minipage}{0.48\textwidth}
		\centering
		\includegraphics[width=\linewidth]{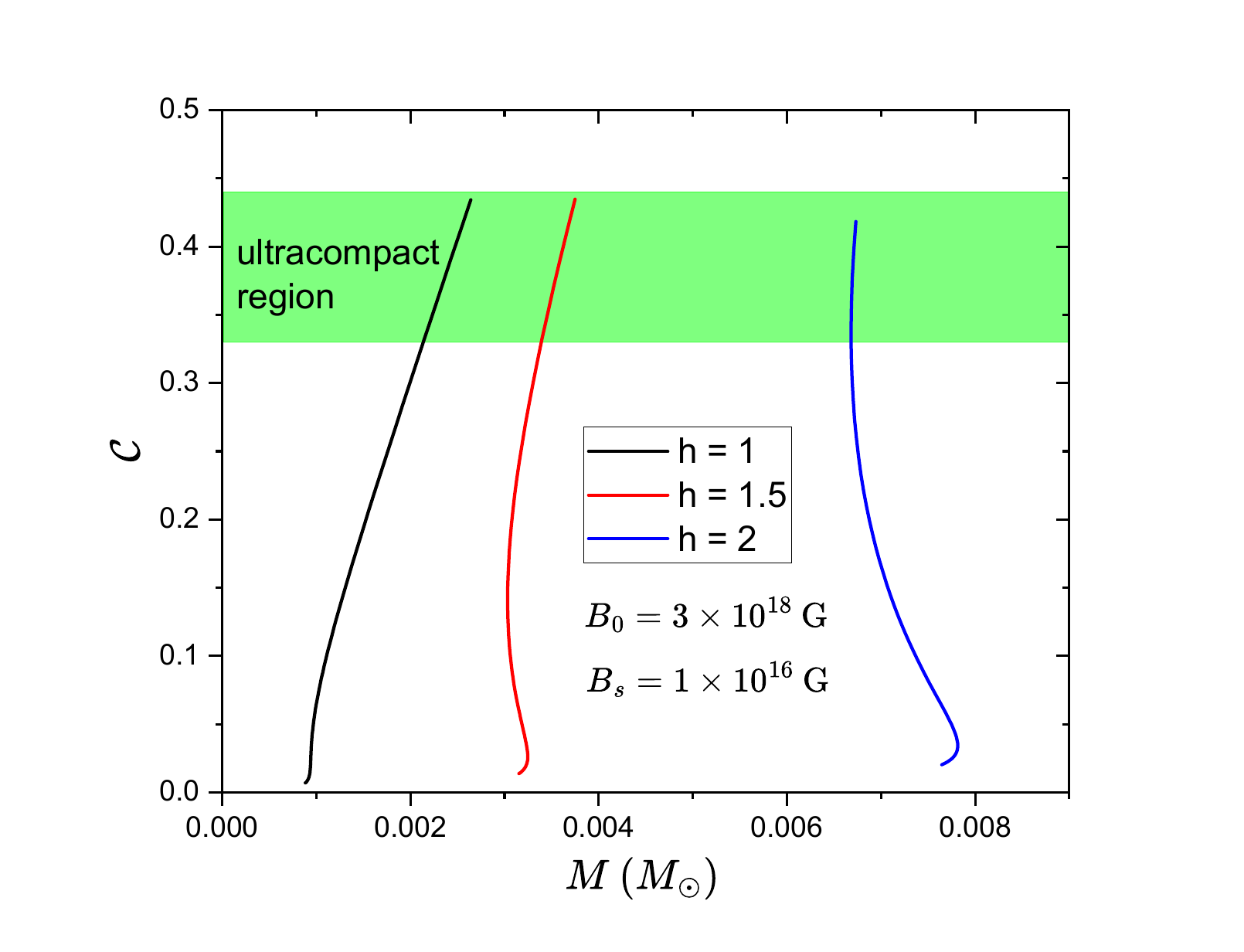}
		\vspace{-0.3cm}
		{\small (d) Approach~2}
	\end{minipage}
	
	\caption{
		Compactness of the CSSTW with $B_s=10^{15}$ G and $B_0=10^{18}$ G as functions of the ADM mass for different values of $h$.
		Panels (a) correspond to $B_s = 10^{15}\,\mathrm{G}$ and $B_0 = 10^{18}\,\mathrm{G}$ with Approach~1,
		(b) $B_s = 10^{15}\,\mathrm{G}$ and $B_0 = 10^{18}\,\mathrm{G}$ with Approach~2,
		(c) $B_s = 10^{16}\,\mathrm{G}$ and $B_0 = 10^{18}\,\mathrm{G}$ with Approach~1, and
		(d) $B_s = 10^{16}\,\mathrm{G}$ and $B_0 = 10^{18}\,\mathrm{G}$ with Approach~2.}
	\label{figcompactnessmagnetized}
\end{figure*}

\begin{figure*}[!t] 
	\centering
	
	\begin{minipage}{0.48\textwidth}
		\centering
		\includegraphics[width=\linewidth]{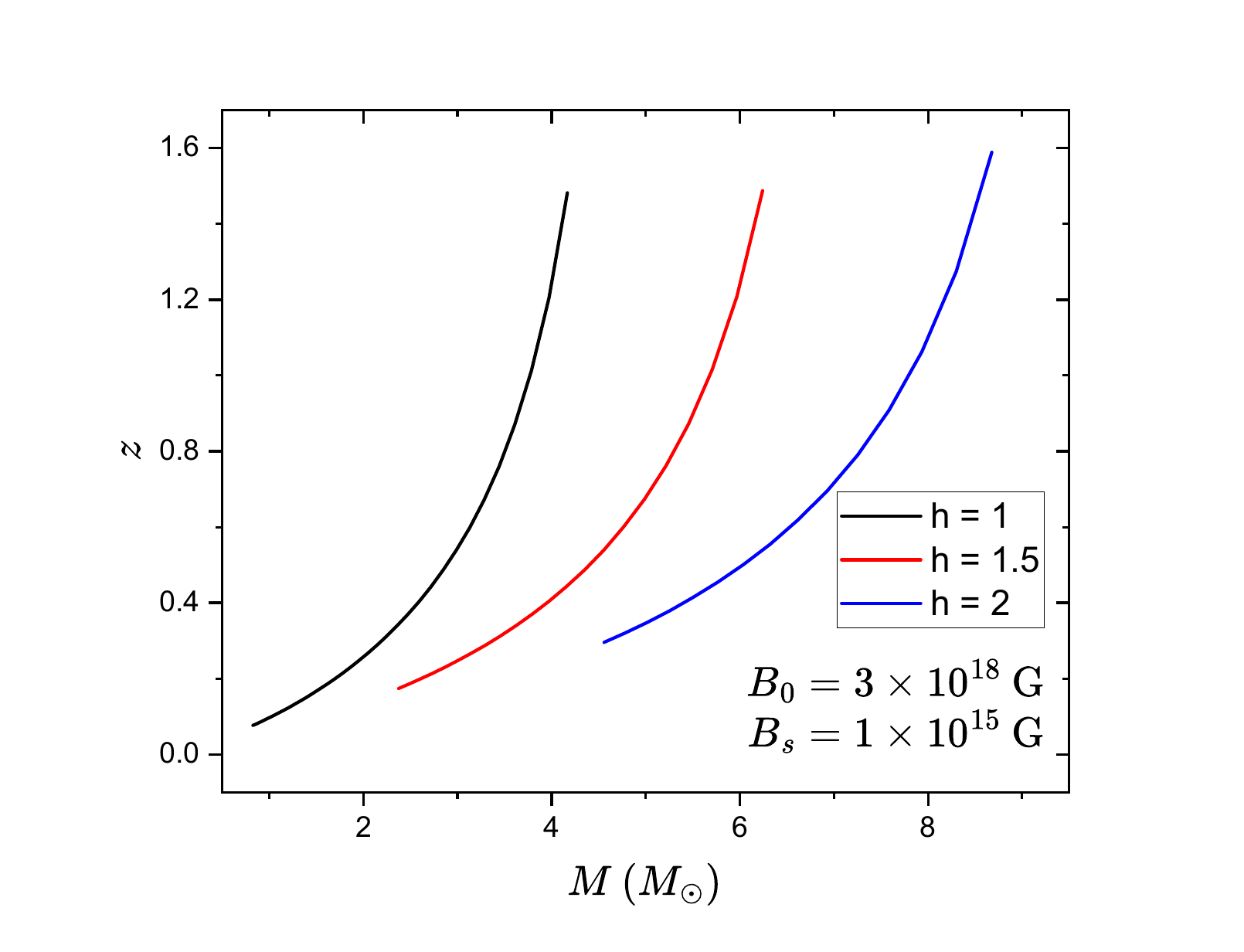}
		\vspace{-0.3cm}
		{\small (a) Approach~1}
	\end{minipage}
	\hfill
	\begin{minipage}{0.48\textwidth}
		\centering
		\includegraphics[width=\linewidth]{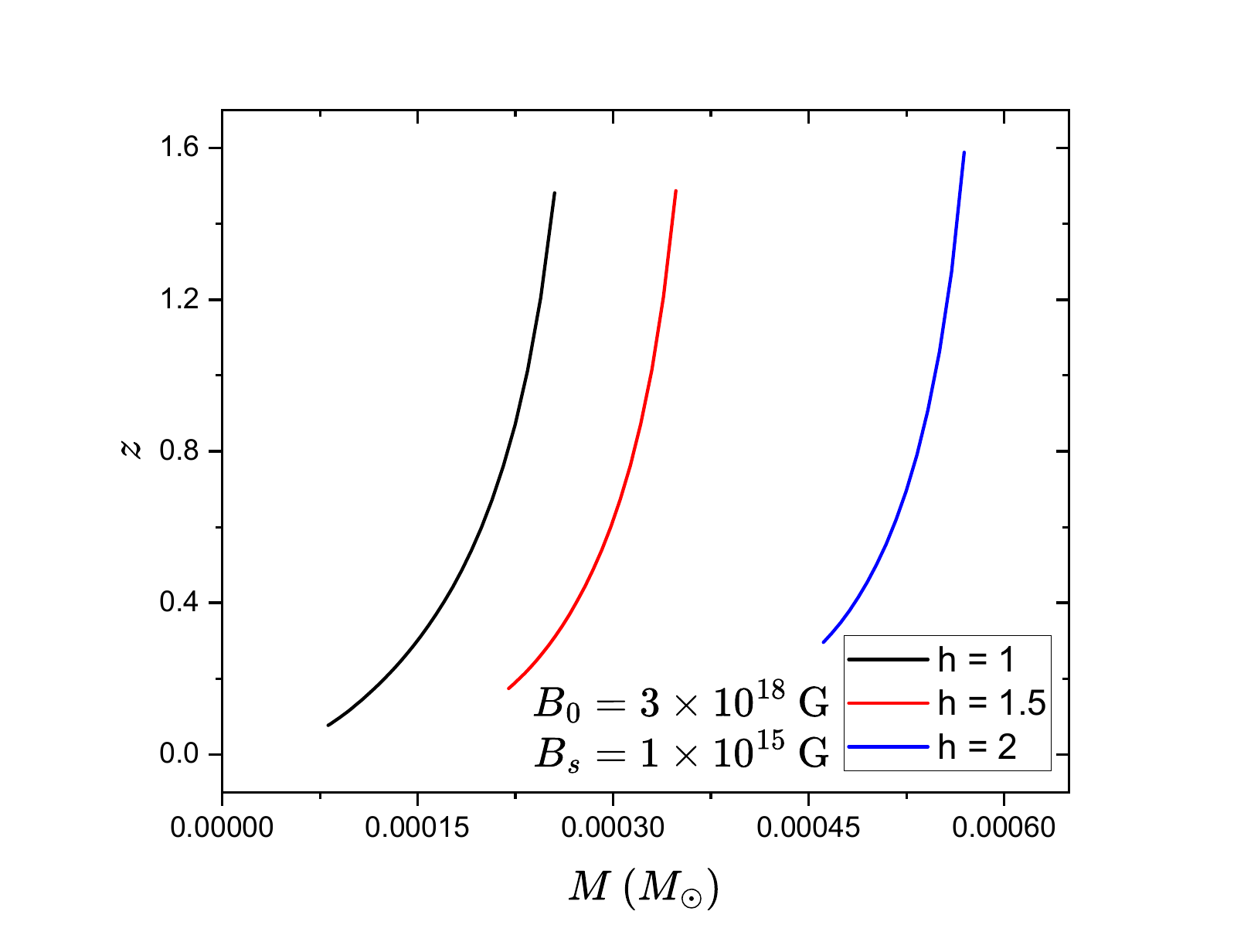}
		\vspace{-0.3cm}
		{\small (b) Approach~2}
	\end{minipage}
	
	\vspace{0.4cm}
	
	\begin{minipage}{0.48\textwidth}
		\centering
		\includegraphics[width=\linewidth]{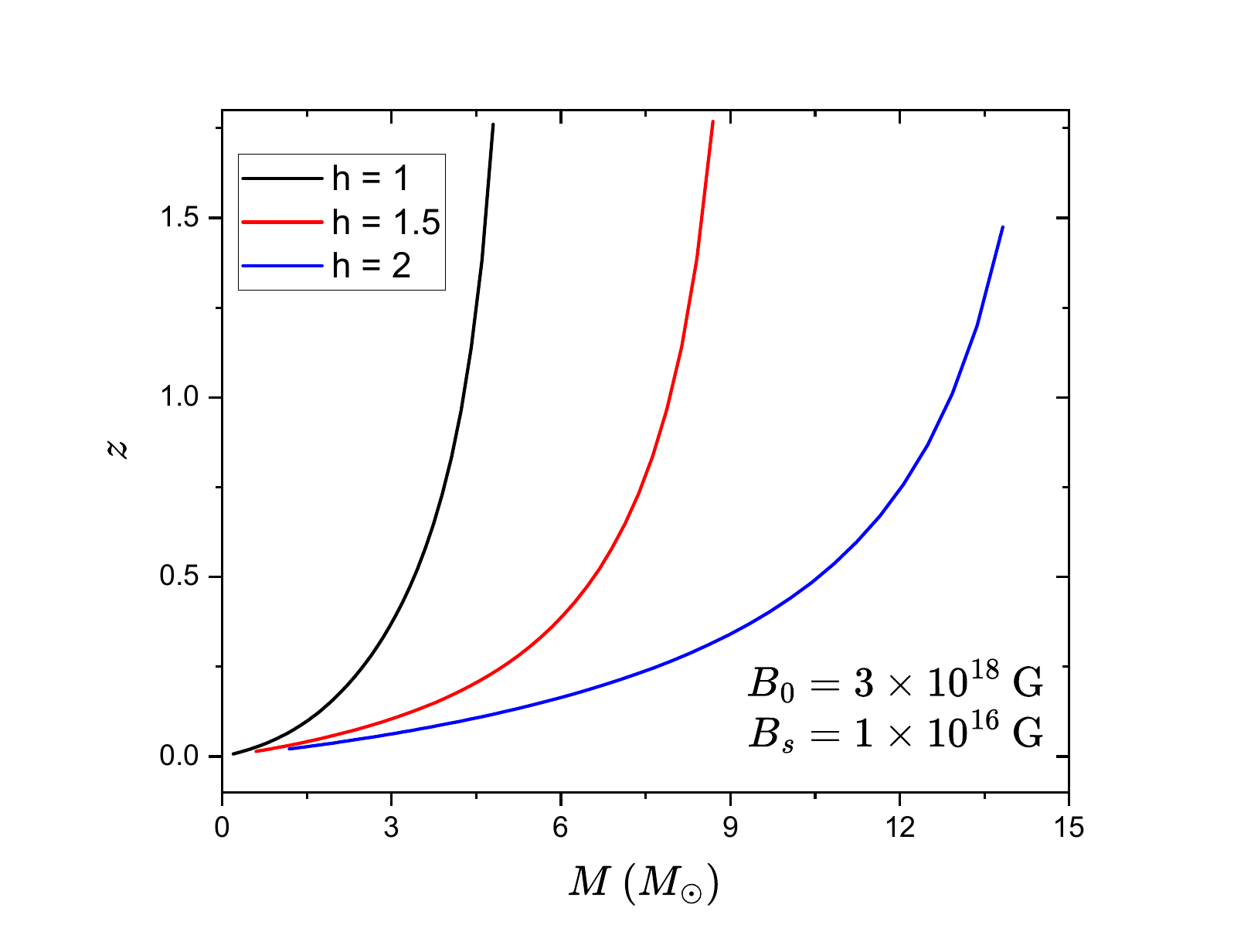}
		\vspace{-0.3cm}
		{\small (c) Approach~1}
	\end{minipage}
	\hfill
	\begin{minipage}{0.48\textwidth}
		\centering
		\includegraphics[width=\linewidth]{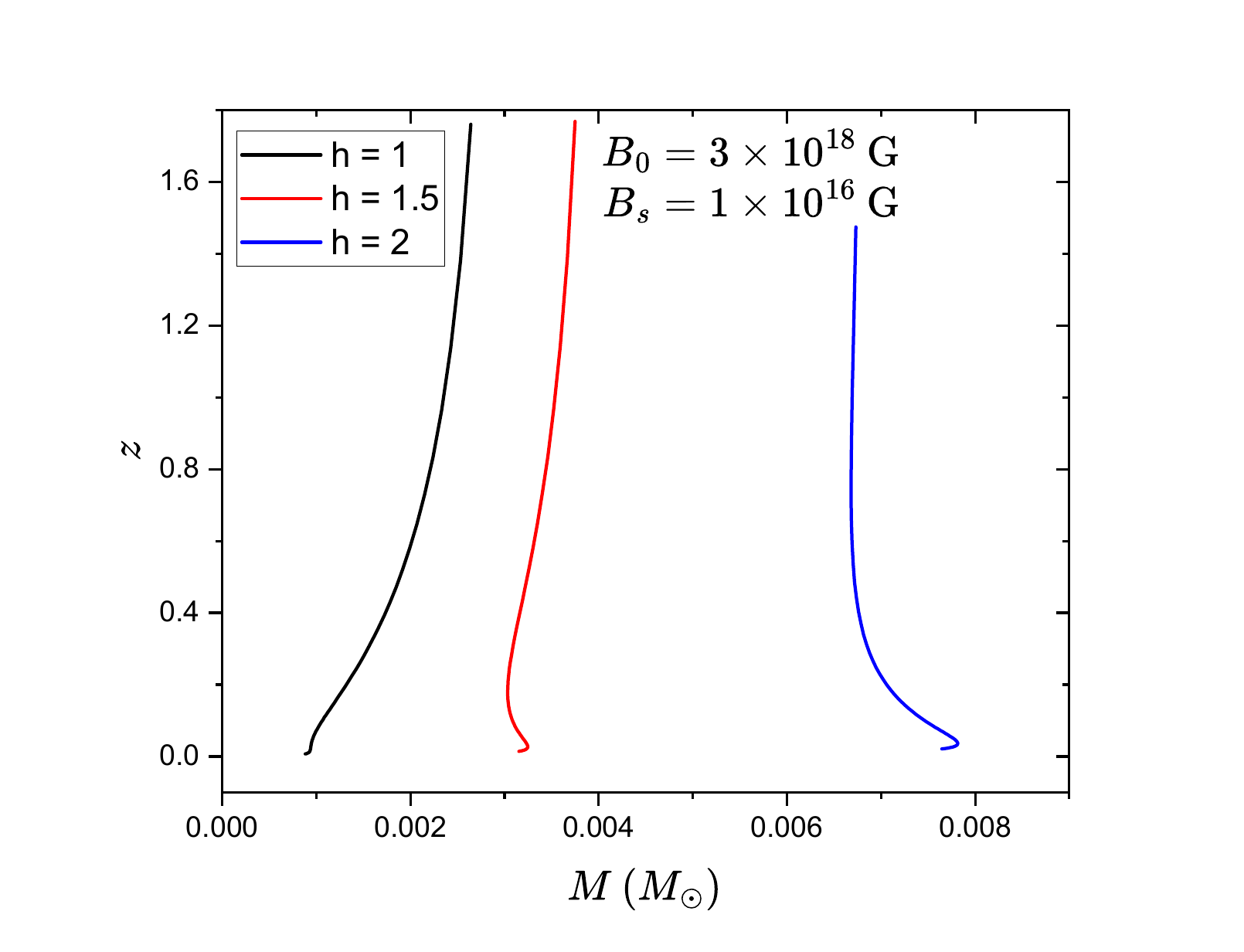}
		\vspace{-0.3cm}
		{\small (d) Approach~2}
	\end{minipage}
	
	\caption{
		Surface redshift of the CSSTW with $B_s=10^{15}$ G and $B_0=10^{18}$ G for different values of $h$.
		Panels (a) correspond to $B_s = 10^{15}\,\mathrm{G}$ and $B_0 = 10^{18}\,\mathrm{G}$ with Approach~1,
		(b) $B_s = 10^{15}\,\mathrm{G}$ and $B_0 = 10^{18}\,\mathrm{G}$ with Approach~2,
		(c) $B_s = 10^{16}\,\mathrm{G}$ and $B_0 = 10^{18}\,\mathrm{G}$ with Approach~1, and
		(d) $B_s = 10^{16}\,\mathrm{G}$ and $B_0 = 10^{18}\,\mathrm{G}$ with Approach~2.}
	\label{figredshiftmagnetized}
\end{figure*}

\begin{figure*}[!t] 
	\centering
	
	\begin{minipage}{0.48\textwidth}
		\centering
		\includegraphics[width=\linewidth]{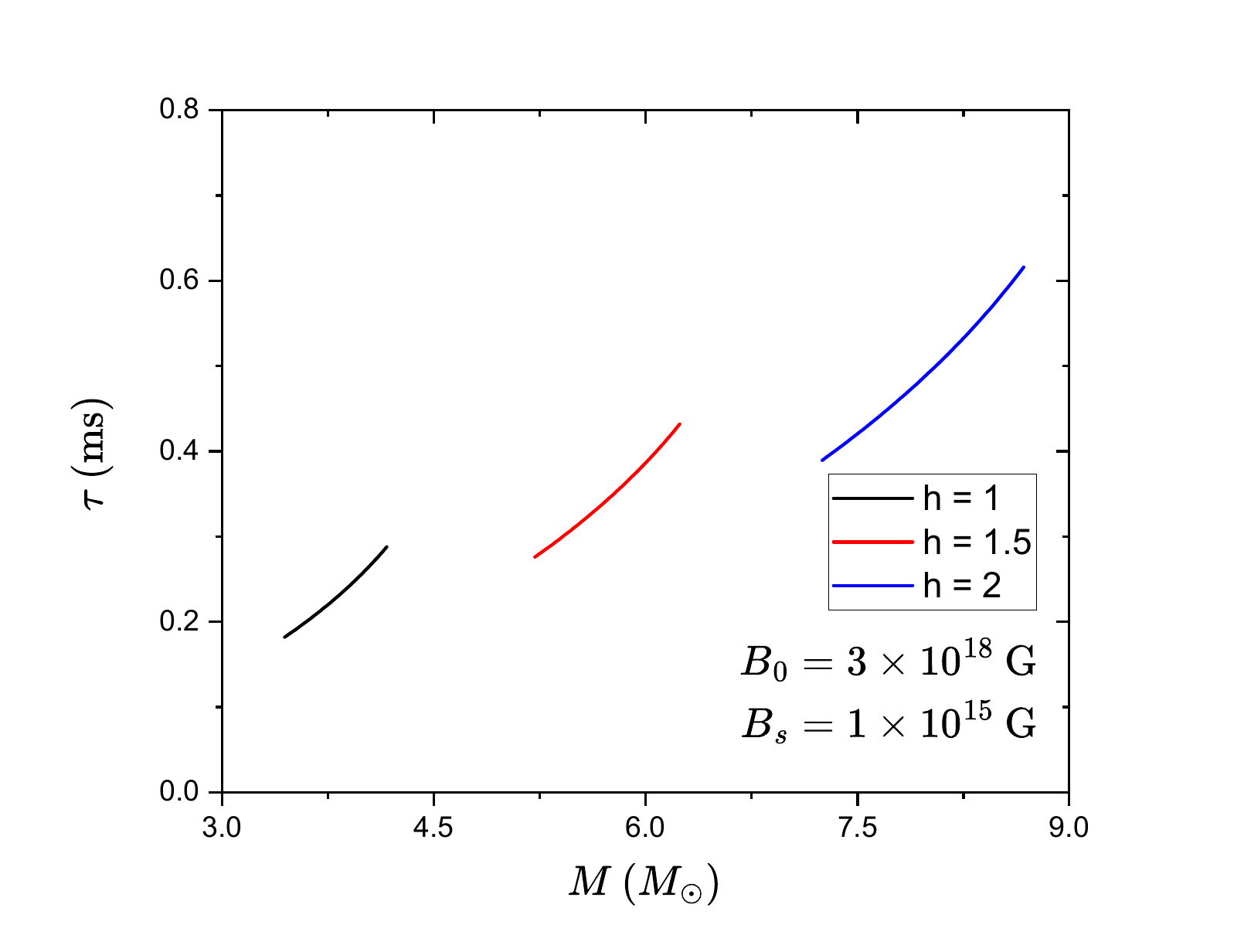}
		\vspace{-0.3cm}
		{\small \textcolor{black}{(a) Approach~1}}
	\end{minipage}
	\hfill
	\begin{minipage}{0.48\textwidth}
		\centering
		\includegraphics[width=\linewidth]{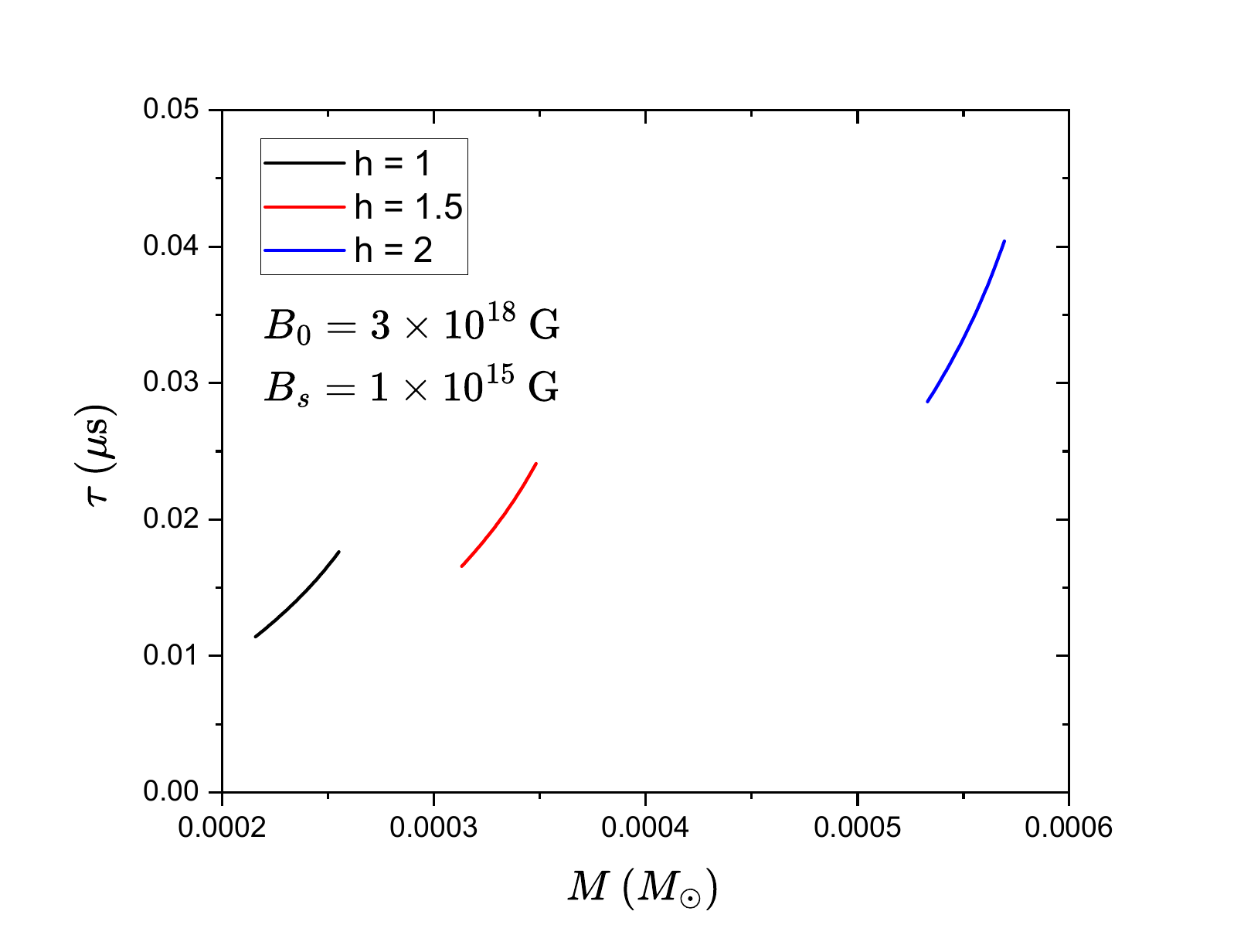}
		\vspace{-0.3cm}
		{\small \textcolor{black}{(b) Approach~2}}
	\end{minipage}
	
	\vspace{0.4cm}
	
	\begin{minipage}{0.48\textwidth}
		\centering
		\includegraphics[width=\linewidth]{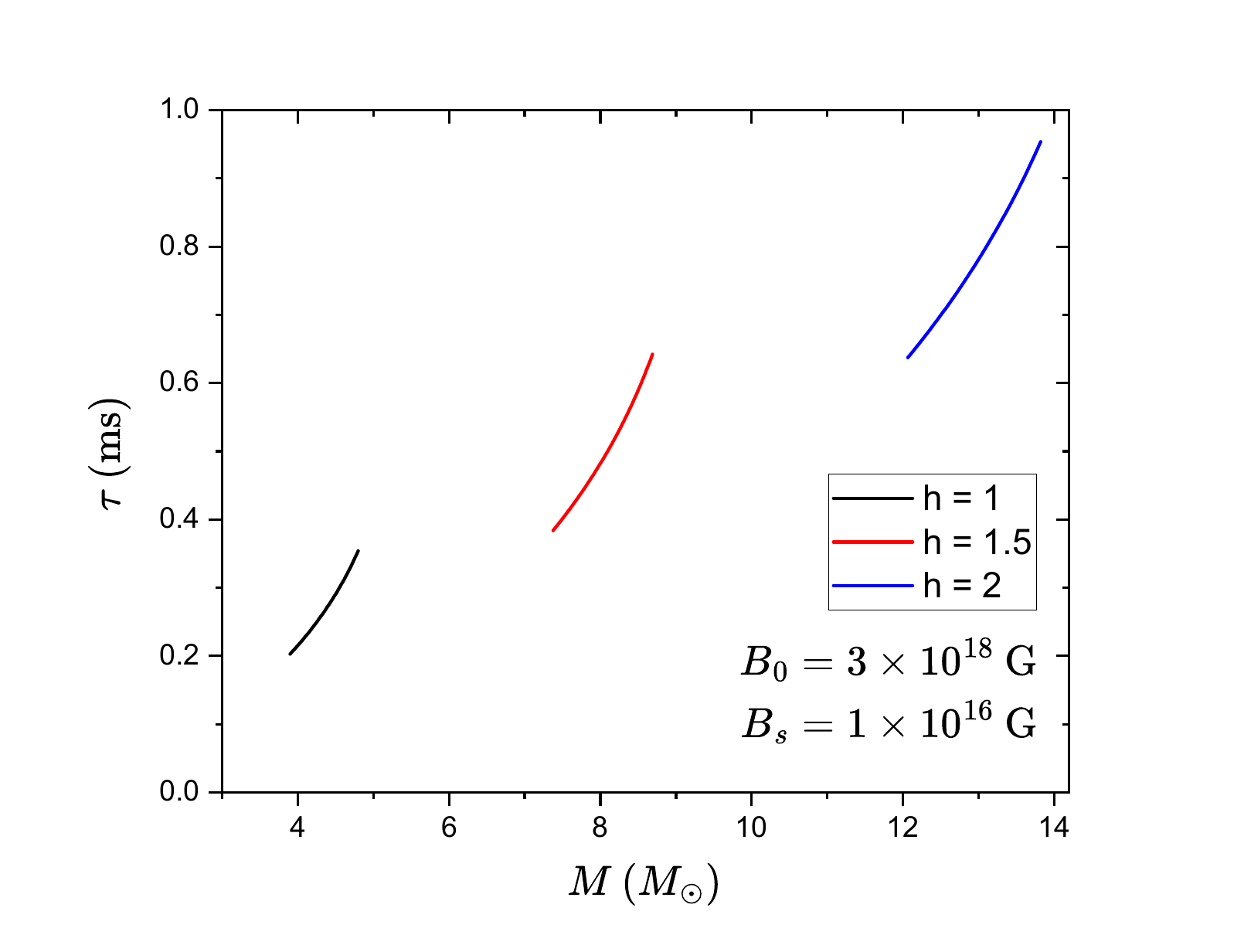}
		\vspace{-0.3cm}
		{\small \textcolor{black}{(c) Approach~1}}
	\end{minipage}
	\hfill
	\begin{minipage}{0.48\textwidth}
		\centering
		\includegraphics[width=\linewidth]{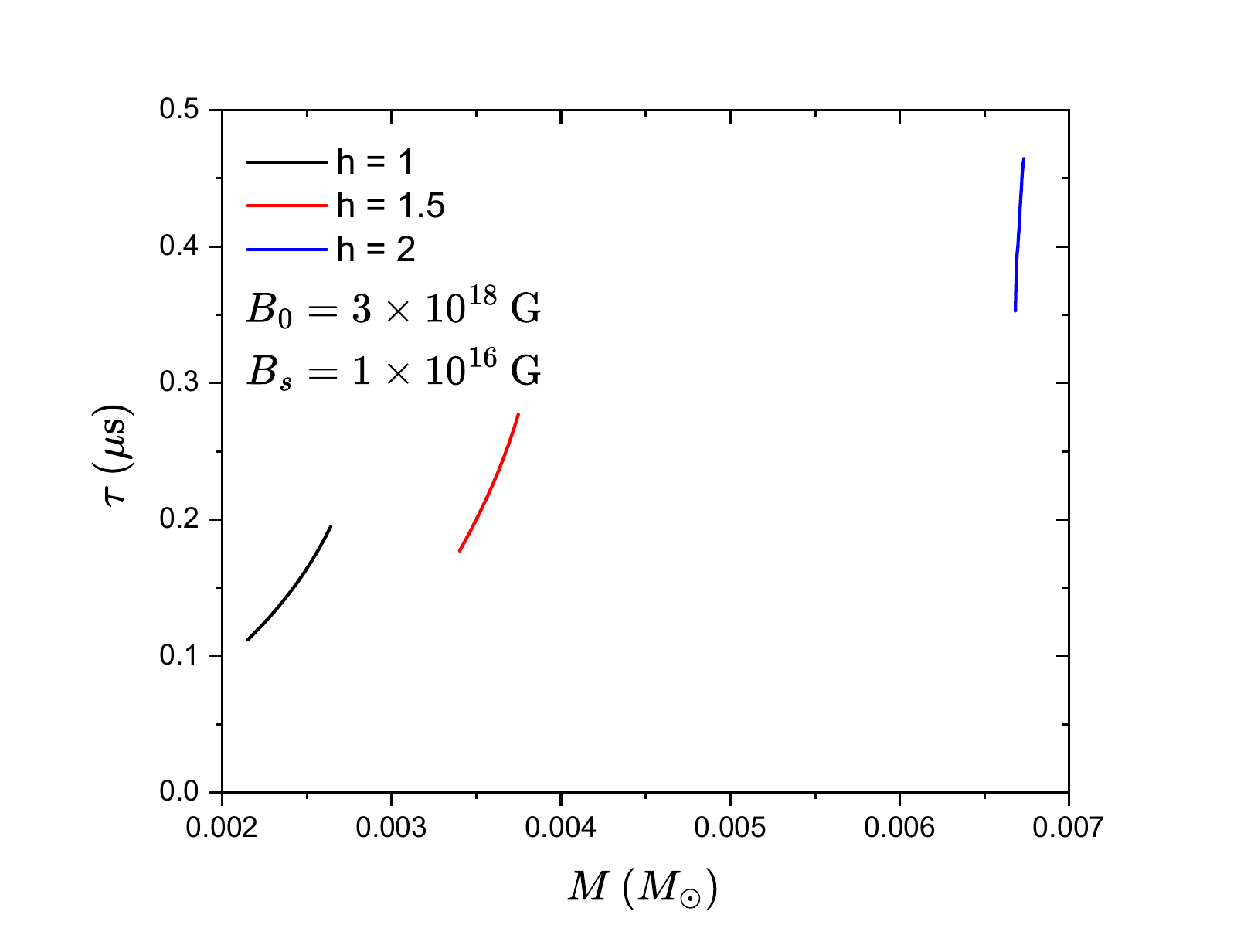}
		\vspace{-0.3cm}
		{\small \textcolor{black}{(d) Approach~2}}
	\end{minipage}
	
	\caption{
		\textcolor{black}{Echo time of the CSSTW with $B_s=10^{15}$ G and $B_0=10^{18}$ G as functions of the ADM mass for different values of $h$.
			Panels (a) correspond to $B_s = 10^{15}\,\mathrm{G}$ and $B_0 = 10^{18}\,\mathrm{G}$ with Approach~1,
			(b) $B_s = 10^{15}\,\mathrm{G}$ and $B_0 = 10^{18}\,\mathrm{G}$ with Approach~2,
			(c) $B_s = 10^{16}\,\mathrm{G}$ and $B_0 = 10^{18}\,\mathrm{G}$ with Approach~1, and
			(d) $B_s = 10^{16}\,\mathrm{G}$ and $B_0 = 10^{18}\,\mathrm{G}$ with Approach~2.}}
	\label{figechomagnetized}
\end{figure*}

\begin{table*}[h!]
	\centering
	\caption{Four considered ranges of the surface magnetic field $B_s$ and the corresponding core magnetic field $B_0$ considered in this work.}
	\label{tab:BsB0}
	\begin{tabular}{c @{\hspace{1.5cm}} c @{\hspace{1.5cm}} c}
		\hline
		Fig.~\ref{figdiscriminant} & $B_s$ (G) & $B_0$ (G) \\
		\hline
		(a)   & $10^{12} - 10^{13}$ & $3 \times 10^{15}$ \\
		(b)  & $10^{13} - 10^{14}$ & $3 \times 10^{16}$ \\
		(c) & $10^{14} - 10^{15}$ & $3 \times 10^{17}$ \\
		(d)  & $10^{15} - 10^{16}$ & $3 \times 10^{18}$ \\
		\hline
	\end{tabular}
\end{table*}

In Appendix~\ref{derivativemetricnojiri}, we show that for the isotropic configuration with the limit $\mathcal{B}\rightarrow 0$, $\left(e^{2\nu}\right)'$ is reduces to the expression of $\left(e^{2\nu}\right)'$ shown by Eq.~(24) in Ref.~\cite{Nojiri2024}.

At the surface, $\left(e^{2\nu}\right)'\big|_{r=R_s}=\frac{r_0}{R_s}$, so we have
\begin{eqnarray}
	r_0=\left[\frac{4K\left(2-\frac{1}{h}\right)\rho_c}{\alpha\gamma}(\beta-1)+\frac{\rho_c z B_s}{3\pi\gamma}\right]R_s.\label{r0alphabeta}
\end{eqnarray}
Furthermore, by substituting the expressions of $\alpha$, $\beta$, $\gamma$, and $z$ in Eq.~(\ref{alphabetagamma}) into Eq.~(\ref{r0alphabeta}), we are left with
\begin{eqnarray}
	r_0=2\left[\frac{4K\left(2-\frac{1}{h}\right)}{K+\frac{B_0^2}{6\pi\rho_0^2}}+1\right]\frac{\rho_c B_0}{\rho_0 B_s}R_s.\label{r0fix}
\end{eqnarray}
From Eq.~(\ref{r0fix}), we can easily obtain the ADM mass
\begin{eqnarray}
	M=\left[\frac{4K\left(2-\frac{1}{h}\right)}{K+\frac{B_0^2}{6\pi\rho_0^2}}+1\right]\frac{\rho_c B_0}{\rho_0 B_s}R_s.\label{ADMmagnetized}
\end{eqnarray}

\textcolor{black}{Similar to the non-magnetized configurations, junction conditions are not required for the magnetized configurations. The same reasoning applies here: Eqs.~(\ref{rhotot})--(\ref{pttot}) are valid throughout the entire spacetime, both inside and outside the neutron fluid, regardless of whether the magnetic field is present. Consequently, the total pressures remain smoothly continuous across the fluid boundary. This continuity arises because the contributions from the nondynamical scalar fields remain nonzero outside the neutron fluid. For the magnetized configurations, the metric is also continuous across the fluid boundary. This continuity is ensured by the conditions given in Eqs.~(\ref{enumagggg})--(\ref{integrationconstantmagnetized}), while the exterior metric is again described by Eq.~(\ref{exterior}).}

\textcolor{black}{A remark is stressed to be emphasized in order concerning the limit of vanishing magnetic field. In the magnetized configurations, the magnetic field modifies the geometric structure of the solution. As a result, the limit $\mathcal{B}\to 0$ is non-uniform: setting $\mathcal{B}=0$ directly at the level of intermediate quantities does not, in general, reproduce the non-magnetized configuration obtained by solving the full system.}

\textcolor{black}{As shown explicitly in~\ref{metricnojiri}~and~\ref{derivativemetricnojiri}, the metric function $e^{2\nu}$ and its radial derivative $\left(e^{2\nu}\right)'$ smoothly reduce to their non-magnetized counterparts when the limit $\mathcal{B}\to 0$ is taken at the level of the full metric functions.
	This point is crucial because the wormhole throat radius $r_0$ is determined from $\left(e^{2\nu}\right)'$ evaluated at the surface $r=R_s$, where the energy density vanishes.
	At this stage, $\left(e^{2\nu}\right)'$ contains the coefficients $\alpha$, $\beta$, and $\gamma$, which depend on the magnetic field and appear both in the numerator and the denominator, leading to the finite expression given in Eq.~(\ref{r0fix}).}

\textcolor{black}{If the limit $\mathcal{B}\to 0$ is taken prematurely at the level of individual quantities, unphysical results may arise.
	For instance, the coefficient $e^{2\nu_c}$ appearing in the magnetized solution, shown by Eq. (\ref{integrationconstantmagnetized}), explicitly depends on the magnetic-field parameter $\gamma$ and would vanish identically if the limit were applied naively, causing the metric function $e^{2\nu}$ itself to collapse.
	This pathology is avoided once the limit is taken at the level of the full metric function, where the cancellation of magnetic field dependent terms occurs consistently, and the system is reduced to non-magnetized configuration. In other words, this behavior shows that the magnetic field genuinely modifies the geometric structure of the system. Consequently, the magnetized and non-magnetized configurations are treated separately throughout this work.}

Note that in this work, regarding the influence of the magnetic field on the EOS, we follow the approach used by Refs.\cite{Mallick2014,Konno1999,Konno2000,Rizaldy2019}, where the magnetic field contribution to the EOS is taken to be negligible. In particular, Refs.~\cite{Konno1999,Konno2000} also employed a polytropic EOS in their analysis. Ref.~\cite{Mallick2014} argued that the dominant magnetic effects arise from the additional stress and pressure terms, while the contribution from magnetization remains subdominant even for extremely strong magnetic fields, especially when such high central fields tend to drive the star toward instability. They used magnetic fields of the order of $10^{18}$~G in their analysis. The neglect of Landau quantization effects on the EOS has also been discussed by Ref.~\cite{Sinha2013}.

\textcolor{black}{In the numerical calculations of magnetized CSSTW we consider two representative combinations of magnetic fields: (i) $B_0 = 3\times10^{18}\,\mathrm{G}$ and $B_s = 10^{15}\,\mathrm{G}$, and (ii) $B_0 = 3\times10^{18}\,\mathrm{G}$ and $B_s = 10^{16}\,\mathrm{G}$. 
	Although the typical surface magnetic field of NSs lies in the range $10^{12}$--$10^{15}\,\mathrm{G}$ as mentioned in Sec.~\ref{intro}, Ref.~\cite{Gomes2019} reports that the limiting magnetic moment of NSs corresponds to a surface field of order $B_s\sim 10^{16}\,\mathrm{G}$. For this reason we also include $B_s=10^{16}\,\mathrm{G}$ among the cases examined in our numerical calculations. Configurations (i) and (ii) lie within the parameter ranges shown in Fig.~\ref{figdiscriminant}. The anisotropy parameter is chosen to be the same as in the non-magnetized case, namely $h = 1,\ 1.5,$ and $2$.}

\textcolor{black}{We next verify the absence of horizons by examining the temporal metric function in the interior, given by Eq.~(\ref{enumagggg}). Figs.~\ref{figinteriormetric1015} and \ref{figinteriormetric1016} show that $e^{2\nu(r)}$ remains nonzero over the entire interior region of configurations with $B_s=1\times10^{15}$ G and $B_s=1\times10^{16}$ G, respectively. In the exterior, the temporal metric function is given by $e^{2\nu(r)}=1-\frac{r_0}{r}$, which becomes zero at $r=r_0$. Since the exterior region in our model starts outside this point, $r=r_0$ is not part of the exterior spacetime. Thus, $e^{2\nu(r)}$ does not vanish anywhere in the spacetime considered here, confirming the absence of an event horizon.}

\textcolor{black}{We also calculate the Kretschmann scalar for the magnetized configurations, i.e. $\mathcal{K}_{mag}$. The general expression for arbitrary $l$ follows the same derivation as that of the non-magnetized configurations, and we obtain}
\begin{eqnarray}
	\textcolor{black}{\mathcal{K}_{mag}}&\textcolor{black}{=}&\textcolor{black}{\frac{4(1-4r_0^2)^2}{(l+2r_0^2)^4}.}\label{kretschmannmagggggg}
\end{eqnarray}

\textcolor{black}{From Eq.~(\ref{kretschmannmagggggg}); together with the fact that in our considered model, $r_0$ never becomes zero; we can obviously see that the Kretschmann scalar of the interior spacetime of the magnetized configurations is always finite, and therefore the spacetime is always regular. The exterior metric given by Eq.~(\ref{exterior}) yields the same Kretschmann scalar as that of the corresponding non-magnetized configuration, i.e. $K_{ext}=\frac{12r_0^2}{r^6}$. This quantity remains finite throughout the exterior region, $r\geq R_s$, and approaches zero as $r\rightarrow\infty$. Therefore, the exterior spacetime contains no curvature singularity.}

\textcolor{black}{Similar to the non-magnetized configurations, we also examine the traversability of the wormholes in the magnetized configurations by verifying that the redshift function $\nu(r)$ remains free of singularities and horizons, together with the throat condition $b(r_0)=r_0$ and the flare-out condition $b'(r_0)<1$~\cite{Dzhunushaliev2011}. The absence of horizons is shown in Figs.~\ref{figinteriormetric1015} and~\ref{figinteriormetric1016} for the same configurations. Recall that, in our system, $b(r)=r_0$, as follows from Eqs.~(\ref{elambdaansatz})--(\ref{exterior}). Therefore, we obtain $b(r_0)=r_0$ and $b'(r_0)=0<1$, satisfying the throat and flare-out conditions, respectively. Thus, in addition to the violation of the NEC, all the required geometric conditions for traversability are satisfied.
}

The MR relations obtained using Approach~1 are shown in Fig.~\ref{fig:mr_magnetized}(a), while those corresponding to Approach~2 are presented in Fig.~\ref{fig:mr_magnetized}(b). The two approaches lead to markedly different mass and radius ranges. Approach~1 yields configurations with large masses and radii, whereas Approach~2 produces systems with extremely small masses and radii. For this reason, in the case of Approach~2, a direct comparison with ordinary NSs is not meaningful, as the mass–radius ranges differ by several orders of magnitude. Including the ordinary NS curves would render the  MR relations of CSSTW visually indistinguishable in the plot.

Another important point is that observational constraints indicate that the typical radius of neutron stars lies in the range of approximately $10$-$11.5$ km~\cite{ozel2016}, while the canonical neutron-star mass is $1.4\,M_\odot$~\cite{Pattersons2025}. Therefore, the very large radii in the range $20$–$90$ km obtained in Fig.~\ref{fig:mr_magnetized} are incompatible with current observational evidence.

However, it should be emphasized that these large-radius configurations arise from the use of a polytropic EOS in the low density range, for which such extended radii are mathematically allowed. The absence of observational counterparts for these configurations is further supported by the fact that they correspond to extremely small ADM masses, approaching zero and thus lying far below the canonical mass of $1.4\,M_\odot$.

In contrast, within Approach~1, CSSTW with similarly large radii can be realized only at the cost of extremely large masses. In particular, for $h=2$, the total mass obtained in Approach~1 can exceed $8\,M_\odot$. Moreover, increasing the surface magnetic field further enhances both the mass and radius of the system: for $B_s = 10^{16}\,\mathrm{G}$, the resulting configurations can attain total masses exceeding $14\:M_\odot$.

\textcolor{black}{\textcolor{black}{It is worth noting that most of the explored parameter space is incompatible with current astrophysical constraints. However, the CSSTW considered in this work is intended as a toy model for investigating the properties of a compact star-like system harboring a traversable wormhole, rather than as a realistic model of an astrophysical compact object.}}

Approach~2 yields configurations with radii between $0$-$0.7$~km, as shown in Fig.~\ref{fig:mr_magnetized}(b)~and~(d). The corresponding MR curves exhibit significantly steeper trends compared to those obtained with Approach~1.
For $B_s = 10^{16}\,\mathrm{G}$, systems with low anisotropy parameter $h$ display the usual steep MR behavior, with only a mild flattening forming a shallow hump. In contrast, for anisotropic cases, a pronounced hump gradually develops as $h$ increases, while the steep branch becomes progressively shorter.
In particular, for $h=2$, the maximum mass of the system no longer occurs along the steep branch of the MR curve, but instead is reached at the peak of the hump structure, although this peak remains relatively moderate and does not form an extreme maximum.

\textcolor{black}{The behavior observed in Fig.~\ref{fig:mr_magnetized} can be understood by recalling how the reference
	scales are fixed in the two approaches.
	In both cases, the calibration is performed with respect to the corresponding pure
	NS solutions.
	In Approach~1, the surface radius of the CSSTW system is fixed to coincide with the
	radius of the pure NS, while in Approach~2 the total ADM mass is fixed to the mass of
	the pure NS at the same central density.}

\textcolor{black}{In the non-magnetized configuration, the masses and radii of pure NSs lie within a
	moderate and physically familiar range.
	The obtain masses and radii of the CSSTW also lie within moderate range, and comparable to MR relation of the pure NS, as illustrated in Fig.~\ref{figMRnonmagnetized}.}

\textcolor{black}{The inclusion of a magnetic field qualitatively alters this picture.
	In magnetized configurations, the wormhole throat radius $r_0$, and therefore the ADM
	mass $M=r_0/2$, explicitly depend on the magnetic field parameters.
	At large radii, where the mass of a pure magnetized NS becomes very small, a CSSTW
	configuration constructed using Approach~1 necessarily develops an extremely large
	mass as the consequence of the Eq.~(\ref{r0fix}).
	This leads to the appearance of very massive CSSTW, as shown in
	Figs.~\ref{fig:mr_magnetized}(a)~and~(c).}

On the other hand, when using Approach~2, the ADM mass of the CSSTW system is constrained to remain equal to that of the pure magnetized NS. Under this condition, the relation $M=r_0/2$ forces the throat radius and the surface radius to shrink to very small values. Overall, the magnetic field gives rise to CSSTW with MR properties that differ qualitatively from those of the non-magnetized case: Approach~1 naturally leads to extremely massive systems, whereas Approach~2 produces configurations with extremely small radii.

\textcolor{black}{Fig.~\ref{figRsrmagnetized} shows the wormhole throat radius $r_0$ and surface radius $R_s$ as the functions of the ADM mass $M$ for different values of the anisotropy parameter $h$ and the different surface magnetic field $B_s$, within two construction approaches considered in this work.} \textcolor{black}{In panels (a) and (c), corresponding to Approach~1, we can obvioulsy see that the throat-radius curves coincide.
	In panel (b), corresponding to Approach~2, the solutions exist only within distinct
	mass intervals.
	As a result, the throat radius curves for $h=1.5$ and $h=2$ appear separated.
	This behavior does not indicate a modification of the functional relation $r_0(M)$,
	but rather reflects the restricted mass ranges over which the corresponding solutions
	are physically admissible.} \textcolor{black}{A different behavior is seen in panel (d), which also corresponds to Approach~2.
	Here, we can see that the curves of the throat radius are separated to several disconnected mass intervals.
	Within each interval the relation $r_0 = 2M$ is still satisfied, but the allowed ranges
	of the throat radius differ, which explains the clearly separated curves shown in the
	figure.} \textcolor{black}{It is important to note that anisotropy does not modify the geometric relation
	between $r_0$ and $M$, but it does affect the domain of existence of CSSTW solutions,
	with a particularly strong impact in Approach~2.
}

Fig.~\ref{figcompactnessmagnetized} shows the compactness of the CSSTW for the magnetized configurations considered in this work. We find that all curves can enter the ultracompact regime, indicating the existence of parameter ranges in which the echo time can be evaluated. Systems with a higher surface magnetic field generally populate higher mass ranges; consequently, the ultracompact region is reached at larger masses compared to configurations with lower surface magnetic fields. Notably, the curves corresponding to $B_s = 1\times10^{16}\,\mathrm{G}$ within Approach~2 exhibit trends that are markedly different from the other cases. This behavior can be understood as a direct consequence of the distinct MR relations obtained in this configuration.

Another important point to emphasize is that, in the magnetized configurations, \textcolor{black}{as mentioned before}, Approach~2 yields systems with extremely small masses, even though their compactness values are comparable to those obtained using Approach~1. 
\textcolor{black}{Approach~2 is not relevant for obtaining physically detectable objects. Nevertheless, they remain of theoretical interest, and for this reason we still include them in our calculations of the surface redshift~$\mathcal{Z}$ and the echo time~$\tau$.}

The mathematical expression used to calculate the surface redshift is the same as that given in Eq.~(\ref{redshiftformula}). 
From Fig.~\ref{figredshiftmagnetized}, we observe that the overall trend of the surface redshift closely resembles that found in the non-magnetized configurations, where larger values of the anisotropy parameter $h$ lead to smaller values of the surface redshift $\mathcal{z}$.  
An exception occurs for the configuration with $B_s = 1\times10^{16}\,\mathrm{G}$ within Approach~2, in which each curve spans a different mass range. In this case, a direct point-by-point comparison between the curves is not meaningful; nevertheless, it can still be seen that smaller values of $h$ correspond to narrower mass ranges.  

Although the qualitative behavior of $\mathcal{Z}$ remains similar to that of the non-magnetized case, it is important to note that the numerical range of the surface redshift values is also comparable to those obtained without magnetic fields. 
As expected, Approach~1 produces a much broader and higher mass range than the non-magnetized configurations, whereas Approach~2 yields a significantly smaller mass range in comparison.

Fig.~\ref{figechomagnetized} presents the echo time for each magnetized configuration obtained using Approach~1 and Approach~2. \textcolor{black}{For Approach~1, both configurations with $B_s = 1\times10^{15}\,\mathrm{G}$ and
	$B_s = 1\times10^{16}\,\mathrm{G}$ yield echo times of order $10^{-1}$ ms,
	which is the same order of magnitude as in the non-magnetized case.
	However, the echo times in the magnetized configurations are larger
	than those obtained without magnetic fields. In the non-magnetized configuration, the longest echo time considered corresponds to
	$h=2$ and remains below $0.4$ ms.
	By contrast, for $B_s = 1\times10^{15}\,\mathrm{G}$ the maximum echo time at $h=2$
	exceeds $0.6$ ms, while for $B_s = 1\times10^{16}\,\mathrm{G}$ it can exceed $0.9$ ms.
	This behavior can be attributed to the larger surface radii induced by the magnetic
	field, which increase the characteristic propagation distance of the echo signals.
}

\textcolor{black}{On the other hand, Approach~2, shown in Figs.~\ref{figechomagnetized}(b) and (d),
	yields extremely short echo times, of order $10^{-2}\,\mu\mathrm{s}$ for
	$B_s = 1\times10^{15}\,\mathrm{G}$ and $10^{-1}\,\mu\mathrm{s}$ for
	$B_s = 1\times10^{16}\,\mathrm{G}$.
	This behavior can be understood as a consequence of the very small separation
	between the wormhole throat radius $r_0$ and the stellar surface $R_s$, which
	confines the echo propagation to a very short radial distance. We note that such short echo times would be extremely challenging to detect. \textcolor{black}{It is worth mentioning once again that the echo time considered in this work is exploratory and approximate in nature. This does not imply that our results represent the actual detectability of gravitational-wave echoes. A realistic formulation of the echo time and any assessment of detectability would require a more rigorous treatment of gravitational perturbations and their waveforms, which is beyond the scope of the present work.
	}
	Nevertheless, as emphasized earlier, Approach~2 is primarily of theoretical
	interest and is included to illustrate the range of possible behaviors allowed
	by the model.
}

\textcolor{black}{In general, a direct comparison between the systems with
	$B_s = 1\times10^{15}\,\mathrm{G}$ and $B_s = 1\times10^{16}\,\mathrm{G}$ is not
	straightforward, since the mass ranges over which echo signals exist differ
	between the two cases.
	An exception occurs for the $h=1$ configuration in Approach~1, where the
	admissible mass intervals overlap in the range
	$M \simeq 3.90$--$4.17\,M_\odot$, allowing a meaningful comparison. The comparison is shown by Table~\ref{table:echo_time_magnetized}.}

\begin{table}[t]
	\centering
	\caption{\textcolor{black}{Echo time for magnetized CSSTW with $h=1$ for different masses.
			The comparison is shown for surface magnetic fields
			$B_s = 1\times10^{15}\,\mathrm{G}$ and
			$B_s = 1\times10^{16}\,\mathrm{G}$.}}
	\label{table:echo_time_magnetized}
	\begin{tabular}{c cc}
		\hline
		\textcolor{black}{$M\,(M_\odot)$} & \multicolumn{2}{c}{\textcolor{black}{$\tau$ (ms)}} \\
		\cline{2-3}
		& \textcolor{black}{$B_s = 1\times10^{15}\,\mathrm{G}$} 
		& \textcolor{black}{$B_s = 1\times10^{16}\,\mathrm{G}$} \\
		\hline
		\textcolor{black}{3.90} & \textcolor{black}{0.24} & \textcolor{black}{0.20} \\
		\textcolor{black}{3.97} & \textcolor{black}{0.25} & \textcolor{black}{0.21} \\
		\textcolor{black}{4.03} & \textcolor{black}{0.26} & \textcolor{black}{0.22} \\
		\textcolor{black}{4.12} & \textcolor{black}{0.27} & \textcolor{black}{0.23} \\
		\textcolor{black}{4.17} & \textcolor{black}{0.29} & \textcolor{black}{0.24} \\
		\hline
	\end{tabular}
\end{table}

\begin{figure*}[htbp]
	\centering
	\begin{tabular}{cc}
		\includegraphics[width=0.48\textwidth]{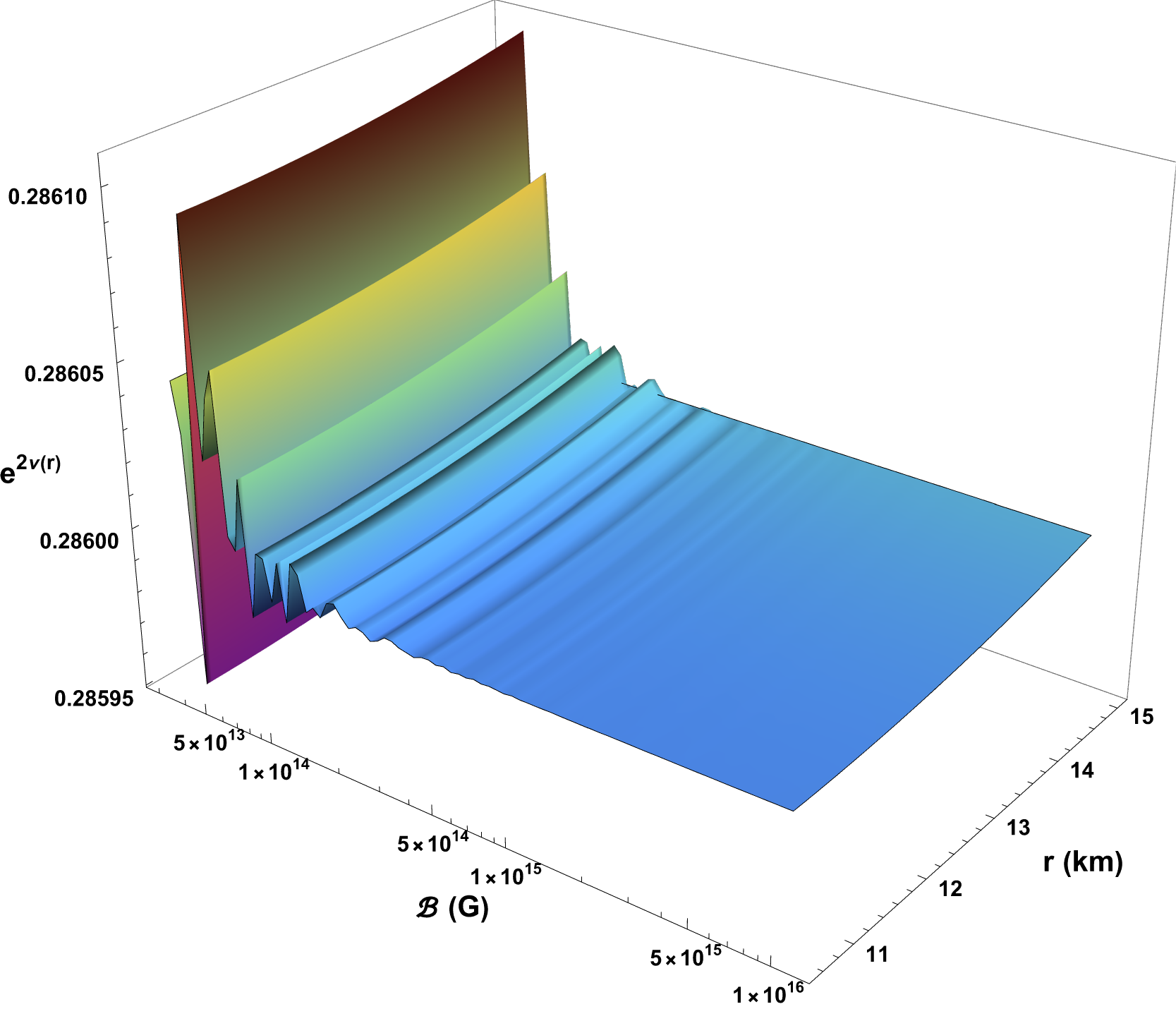} &
		\includegraphics[width=0.48\textwidth]{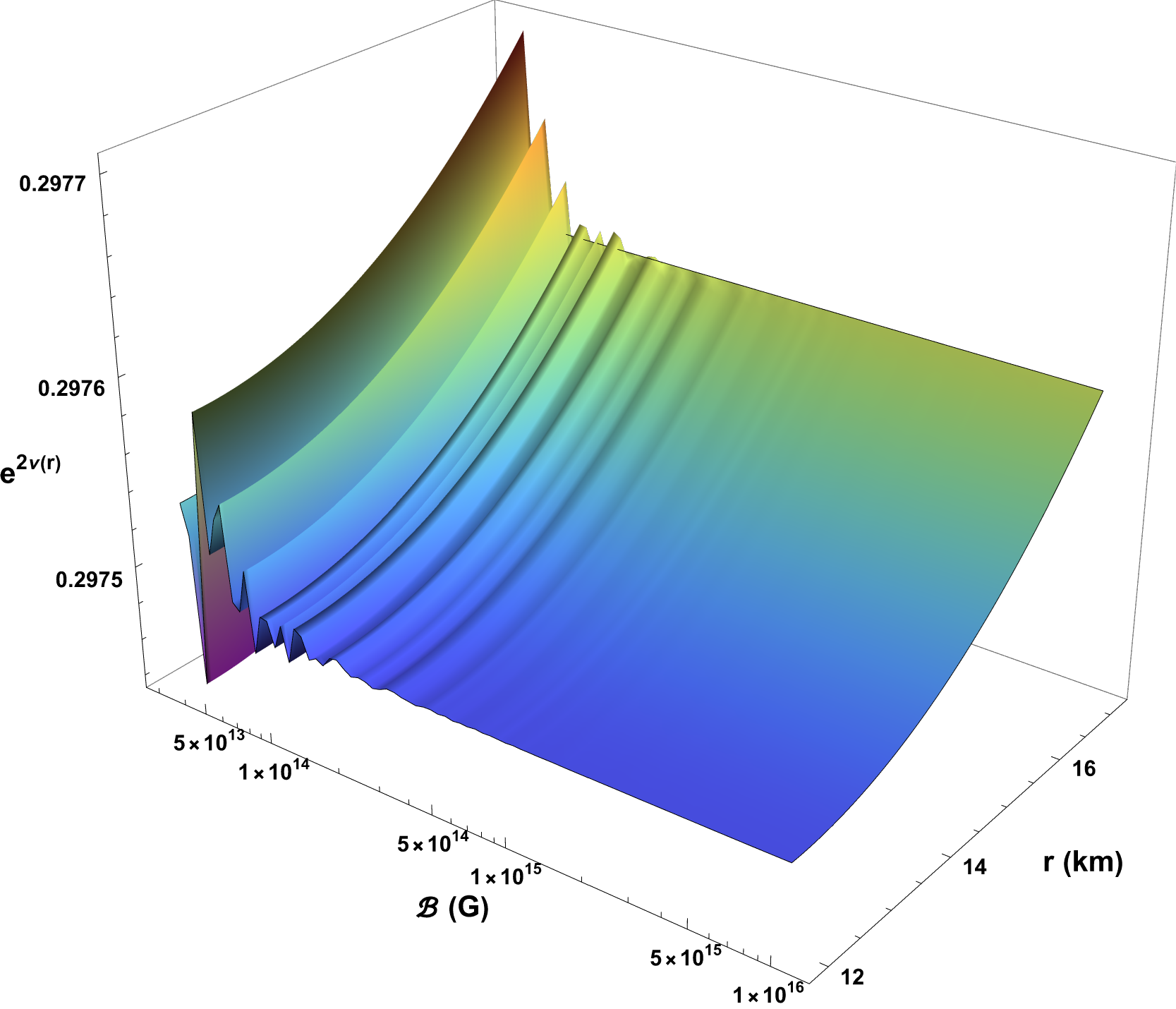} \\
		(a) & (b)
	\end{tabular}
	\caption{\textcolor{black}{Three-dimensional plot of interior temporal metric of the simplified toy model as a function of $\mathcal{B}$ and $r$ for (a) Set~1, and (b) Set~2.}}
	\label{figmetrictoy}
\end{figure*}

\begin{figure*}[htbp]
	\centering
	\begin{tabular}{cc}
		\includegraphics[width=0.48\textwidth]{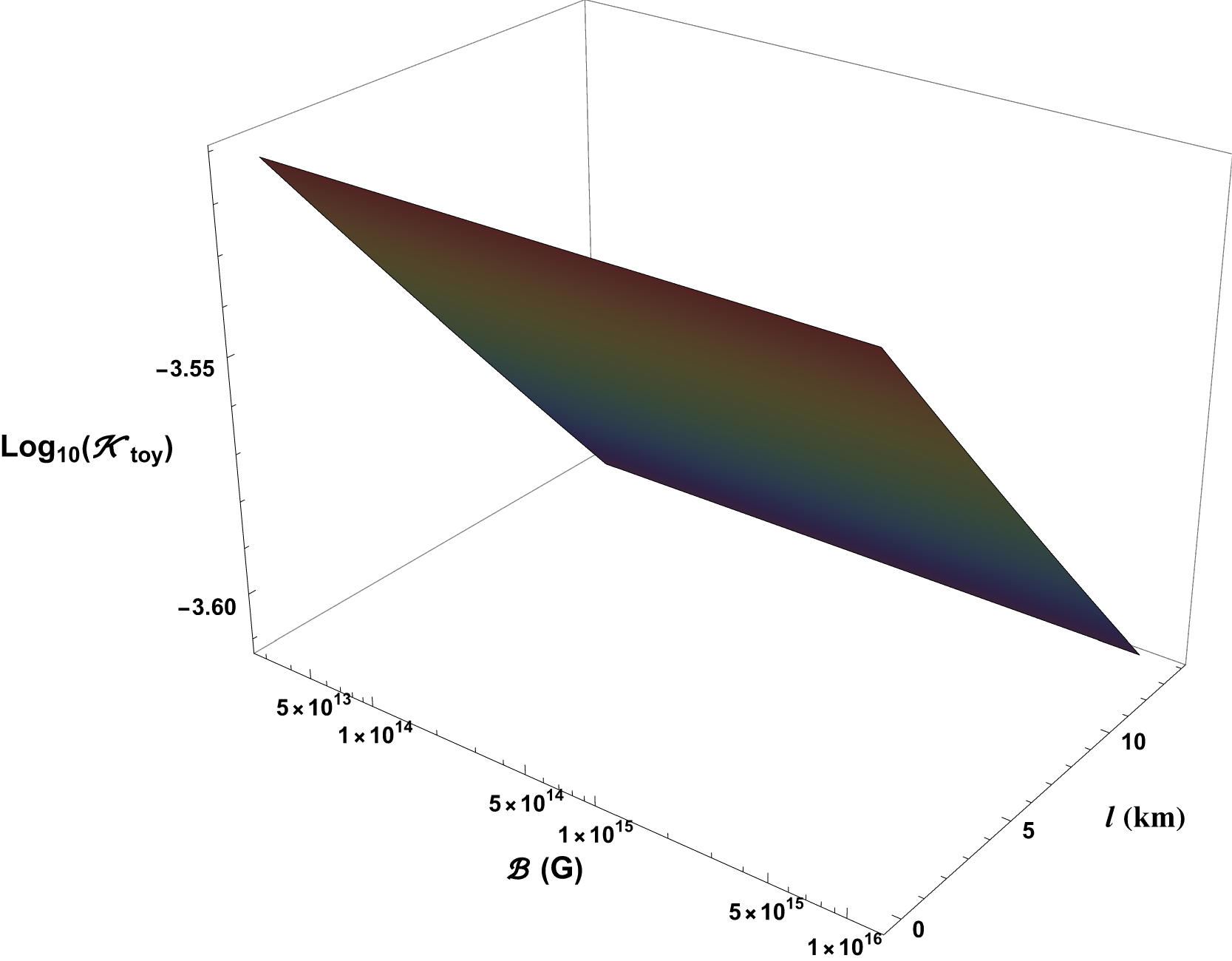} &
		\includegraphics[width=0.48\textwidth]{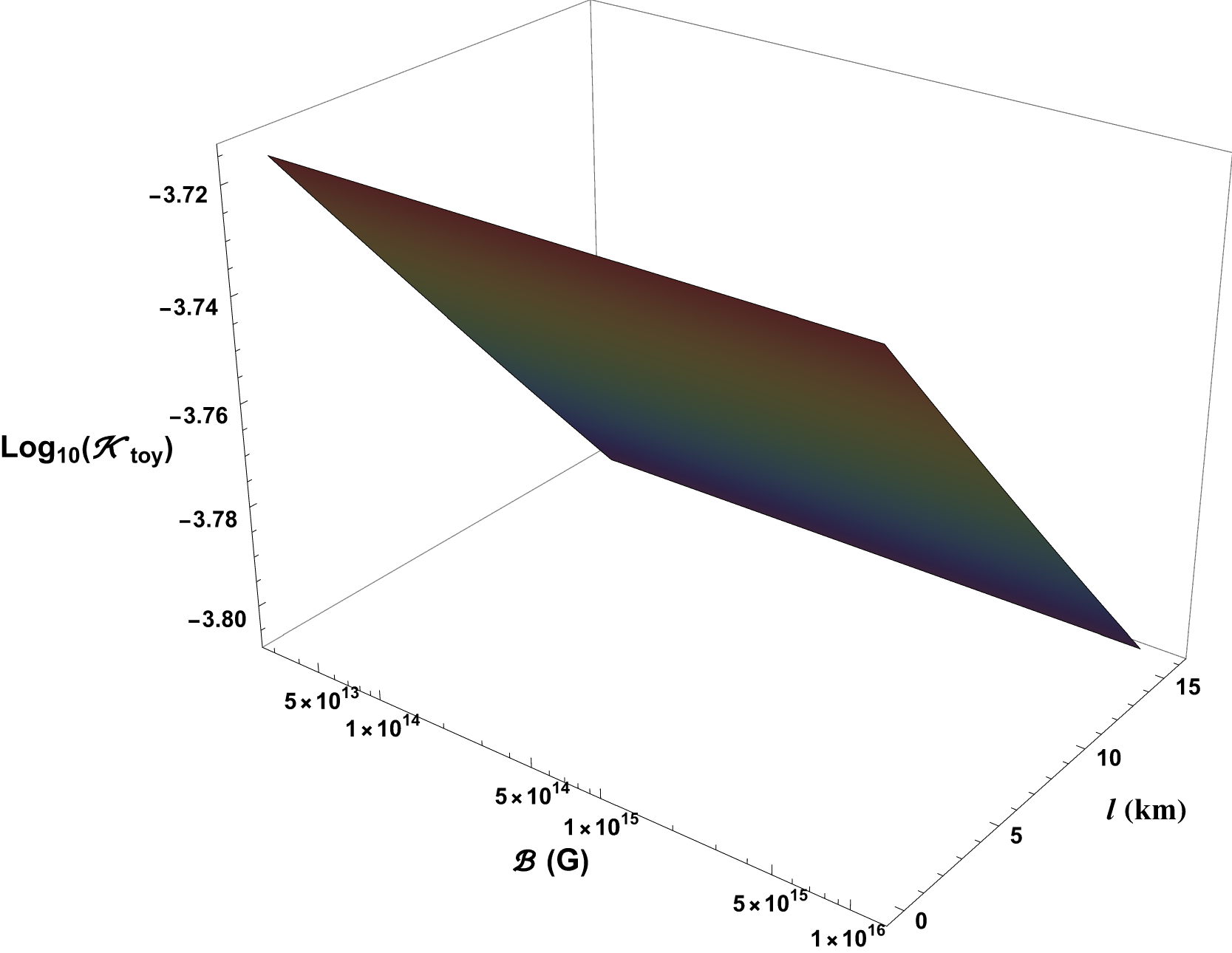} \\
		(a) & (b)
	\end{tabular}
	\caption{\textcolor{black}{Three-dimensional plot of $\log_{10}{(\mathcal{K}_{toy})}$ of the simplified toy model as a function of $\mathcal{B}$ and $l$ for (a) Set~1, and (b) Set~2.}}
	\label{figkretschmanntoy}
\end{figure*}

\begin{figure*}
	\centering
	\includegraphics[width=0.7\linewidth]{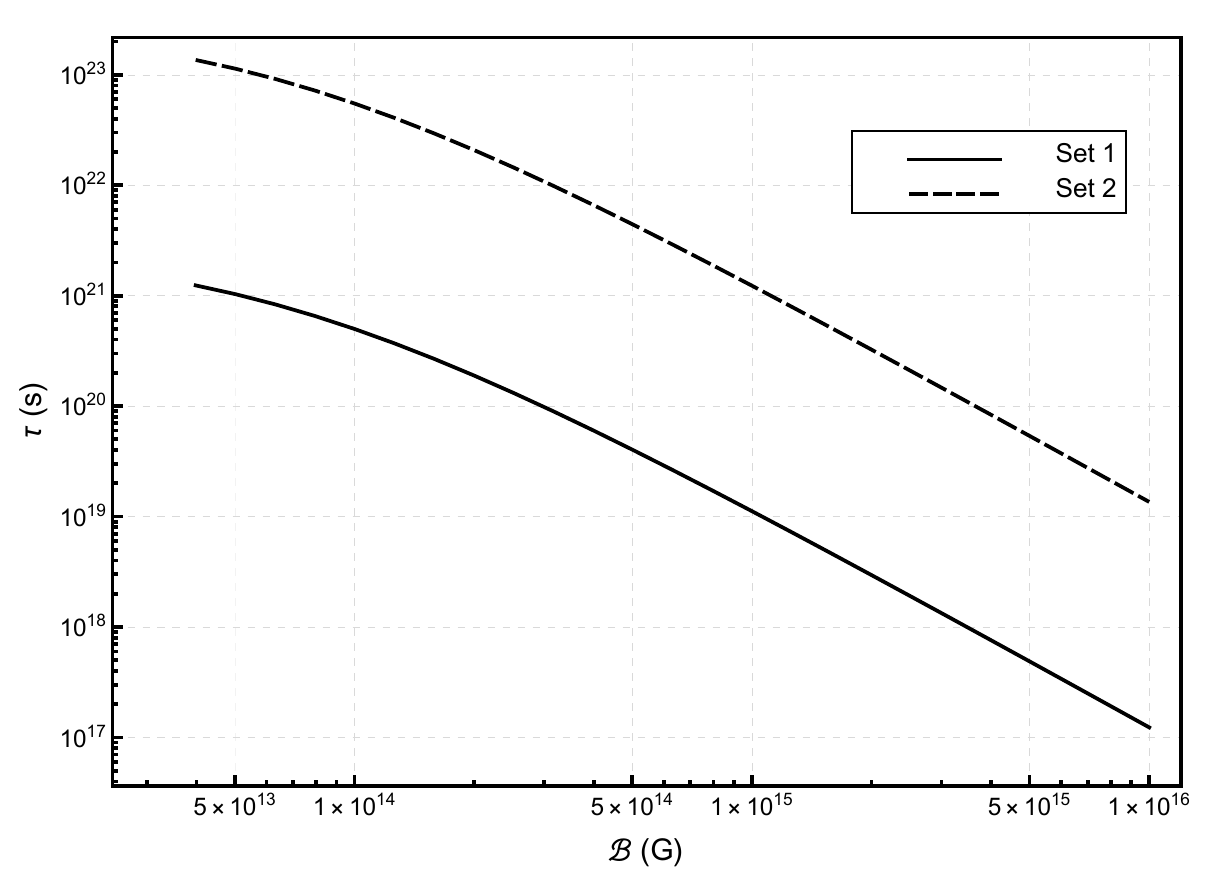}
	\caption{Echo time as a function of $\mathcal{B}$.}
	\label{figechoB}
\end{figure*}

\textcolor{black}{Since we obtain the echo times in magnetized configurations that are larger than those in the
	non-magnetized case, one might expect the echo time to increase monotonically with
	the magnetic-field strength.
	However, Table~\ref{table:echo_time_magnetized} shows that, within the mass range
	where a direct comparison is possible, the systems with
	$B_s = 1\times10^{16}\,\mathrm{G}$ exhibit shorter echo times than those with
	$B_s = 1\times10^{15}\,\mathrm{G}$.
} 

\textcolor{black}{In this work, we also analyze the dependence of the echo time on the magnetic field strength $\mathcal{B}$ only. To address this, we consider a simplified toy model in which the magnetic field $\mathcal{B}$ is taken to be uniform from $r_0$ to $R_s$, and the pressure is assumed to be isotropic. The purpose of this simplified toy model is to reconcile the results shown in Table~\ref{table:echo_time_magnetized} with those obtained from the simplified model. The conservation law yields}
\begin{eqnarray}
	\nu&=&-2K\int\frac{\rho d\rho}{\rho+p_r+\frac{\mathcal{B}^2}{6\pi}}\nonumber\\
	\nu&=&\nu_c+\frac{2\sqrt{\pi}\arctan{\left[\frac{\sqrt{\pi}(1+2K\rho)}{\sqrt{\frac{2K\mathcal{B}^2}{3}-\pi}}\right]}}{\sqrt{\frac{2K\mathcal{B}^2}{3}-\pi}}\nonumber\\
	&&-\ln[\mathcal{B}^2+6\pi\rho(1+K\rho)].\label{nutoymodel}
\end{eqnarray}
The entity $e^{2\nu}$ will be
\begin{eqnarray}
	e^{2\nu}&=&e^{2\nu_c}\exp{\left[\frac{4\sqrt{\pi}\arctan{\left[\frac{\sqrt{\pi}(1+2K\rho)}{\sqrt{\frac{2K\mathcal{B}^2}{3}-\pi}}\right]}}{\sqrt{\frac{2K\mathcal{B}^2}{3}-\pi}}\right]}\nonumber\\
	&&\times\:[\mathcal{B}^2+6\pi\rho(1+K\rho)]^{-2}.\label{enutoy}
\end{eqnarray}
By recalling $e^{2\nu(R_s)}=1-\frac{r_0}{R_s}$, we can easily find
\begin{eqnarray}
	e^{2\nu_c}&=&\exp{\left[-\frac{4\sqrt{\pi}\arctan{\left[\frac{\sqrt{\pi}}{\sqrt{\frac{2K\mathcal{B}^2}{3}-\pi}}\right]}}{\sqrt{\frac{2K\mathcal{B}^2}{3}-\pi}}\right]}\nonumber\\
	&&\times\:\mathcal{B}^4\left(1-\frac{r_0}{R_s}\right).\label{enuctoyyyy}
\end{eqnarray}

With the $e^{2\nu}$ of the \textcolor{black}{simplified} toy model in our hands, the echo time can be calculated using the same expression as in Eq.~(\ref{echotime}).
In this stage of calculation, as in the previous cases, the integral over the inner region, $r_0 \leq r < R_s$, is evaluated numerically.
The integral over the outer region, $R_s \leq r \leq R_p$, on the other hand, also admits a closed-form analytical expression, given by the second term on the right-hand side of Eq.~(\ref{integralecho2}).

Here, we consider two sets of calculation inputs, as summarized in Table~\ref{tabinputcalcul}.
The compactness of the system in Set~1 is $0.36$, while that of Set~2 is $0.35$.
We emphasize that the mass and radii of the systems generally depend strongly on the magnetic field. However, since the present analysis is intended as \textcolor{black}{simplified} a toy model, these dependencies are neglected. Our aim is solely to investigate the influence of the magnetic field on the echo time.

\begin{table*}[h!]
	\centering
	\caption{Sets of input parameters used for calculating the echo time as a function of $\mathcal{B}$ in the \textcolor{black}{simplified} toy model}
	\label{tabinputcalcul}
	\begin{tabular}{c @{\hspace{0.5cm}} c @{\hspace{0.5cm}} c @{\hspace{0.5cm}} c @{\hspace{0.5cm}} c}
		\hline
		Set & $R_s$ (km) & $r_0$ (km) & $R_p$ (km) & $\rho_c$ (MeV fm$^{-3}$) \\
		\hline
		1   & 14.93 & 10.66 & 15.99 & 0.078\\
		2  & 17.07 & 11.99 & 17.98 & 0.769\\
		\hline
	\end{tabular}
\end{table*}

\textcolor{black}{For the simplified toy model, we also check the temporal metric function to determine whether a horizon is present. In the interior, the function is given by Eq.~(\ref{enutoy}), and its behavior is shown in Fig.~\ref{figmetrictoy}. It remains nonzero throughout the interior region. Outside the neutron fluid, we have $e^{2\nu(r)}=1-\frac{r_0}{r}$, which vanishes at $r=r_0$. However, this radius is not included in the exterior region of the model. Consequently, the temporal metric function remains nonzero over the entire spacetime under consideration, indicating that the simplified toy model is horizonless.}

\textcolor{black}{Similar to the previous cases, to ensure that the spacetime of the simplified toy model is regular, we calculate the corresponding Kretschmann scalar, which is given by}

\begin{align}
		\textcolor{black}{\mathcal{K}_{toy}} \;\textcolor{black}\; \textcolor{black}{\frac{1}{\Big[(l+2r_{0}^{2})^{4}\,\mathcal{D}^{4}\Big]}
			\Bigg\{}
		& \textcolor{black}{4\Big[
			576\,K^{2}l^{4}\pi^{2}(l+2r_{0}^{2})^{4}\rho_{c}^{4}\,N_{1}^{2}
			+1152\,K^{2}l^{6}\pi^{2}(l+2r_{0}^{2})^{2}}\nonumber \\[4pt]
		& \times \big(l^{4}-4r_{0}^{2}(R_{s}-r_{0})^{2}\big)^{2}\rho_{c}^{4}\,N_{2}^{2}\;\textcolor{black}{+}\;\textcolor{black}{(1-4r_{0}^{2})^{2}\,\mathcal{D}^{4}
			\Big]
			\Bigg\},}
	\end{align}
	
	\vspace{4pt}
	\noindent \textcolor{black}{where}
	
	\begin{align}
		\textcolor{black}{\mathcal{D}} &\textcolor{black}{=} \textcolor{black}{8\mathcal{B}^{2}r_{0}^{4}(R_{s}-r_{0})^{4}
			+3\pi\big[l^{4}-4r_{0}^{2}(R_{s}-r_{0})^{2}\big]\rho_{c}
			\Big[Kl^{4}\rho_{c}-4r_{0}^{4}(1+K\rho_{c}) -4r_{0}^{2}R_{s}^{2}(1+K\rho_{c})}\nonumber\\
		&\quad\textcolor{black}{+8r_{0}^{3}(R_{s}+KR_{s}\rho_{c})\Big],} \\[8pt]
		\textcolor{black}{N_{1}} &\textcolor{black}{= 8\mathcal{B}^{2}r_{0}^{4}\big[7l^{4}-12r_{0}^{2}(R_{s}-r_{0})^{2}\big](R_{s}-r_{0})^{4} -9\pi\big[l^{4}-4r_{0}^{2}(R_{s}-r_{0})^{2}\big]^{2}\rho_{c}
			\Big[3Kl^{4}\rho_{c}} \nonumber\\
		&\quad \textcolor{black}{+4r_{0}^{4}(1+K\rho_{c})+4r_{0}^{2}R_{s}^{2}(1+K\rho_{c})-8r_{0}^{3}(R_{s}+KR_{s}\rho_{c})\Big],} \\[8pt]
		\textcolor{black}{N_{2}} &\textcolor{black}{= 8\mathcal{B}^{2}r_{0}^{4}(R_{s}-r_{0})^{4}
			+3\pi\big[l^{4}-4r_{0}^{2}(R_{s}-r_{0})^{2}\big]\rho_{c}\Big[Kl^{4}\rho_{c}-4r_{0}^{4}(1+K\rho_{c})}\nonumber \\
		&\textcolor{black}{\quad -4r_{0}^{2}R_{s}^{2}(1+K\rho_{c})+8r_{0}^{3}(R_{s}+KR_{s}\rho_{c})\Big].}
	\end{align}

\textcolor{black}{Fig.~\ref{figkretschmanntoy} shows the logarithm of the Kretschmann scalar, $\log_{10}(K_{\rm toy})$, for the interior spacetime of the simplified toy model as a function of $\mathcal{B}$ and $l$. As shown in Figs.~\ref{figkretschmanntoy}(a) and (b), the Kretschmann scalar remains finite throughout the parameter ranges considered for both Set~1 and Set~2, indicating that the corresponding interior spacetimes are free from curvature singularities within these parameter ranges. Similarly, the Kretschmann scalar for the exterior spacetime of the simplified toy model is given by $K^{toy}_{ext}=\frac{12r_0^2}{r^6}.$ Thus, for $r>R_s$, the exterior Kretschmann scalar remains finite, indicating that the exterior spacetime is free from curvature singularities.}

\textcolor{black}{We further assess the traversability of the simplified toy model, following the same geometric criteria used for the non-magnetized and magnetized configurations with $B_s=1\times10^{15}$ G and $B_s=1\times10^{16}$ G. In particular, the redshift function $\nu(r)$ is required to be free of singularities and horizons, while the shape function must satisfy the throat and flare-out conditions, $b(r_0)=r_0$ and $b'(r_0)<1$~\cite{Dzhunushaliev2011}, respectively. For the simplified toy model, the absence of curvature singularities and horizons is shown in Figs.~\ref{figkretschmanntoy} and \ref{figmetrictoy}, respectively. Moreover, since $b(r)=r_0$ in our system, as follows from Eqs.~(\ref{elambdaansatz})--(\ref{exterior}), we have $b(r_0)=r_0$ and $b'(r_0)=0<1$. Hence, the throat and flare-out conditions are both satisfied. Together with the NEC violation, this confirms that the simplified toy model satisfies the geometric requirements for a traversable wormhole.
}

\textcolor{black}{In Fig.~\ref{figechoB}, we find that the echo time decreases with increasing magnetic field strength $\mathcal{B}$. The results shown in Fig.~\ref{figechoB} reconcile the trend reported in Table~\ref{table:echo_time_magnetized}, where configurations with a surface magnetic field of $B_s=10^{15}$ G exhibit longer echo times than those with $B_s=10^{16}$ G.}

\textcolor{black}{Finally, it is worth noting that quantitative values of physical properties reported in this work should be interpreted as properties of the adopted phenomenological construction and ansatzes, which is including a specific equation of state, density profile, anisotropy prescription, magnetic field model, and matching procedure, and not as robust predictions.}

\section{\textcolor{black}{Stability analysis of CSSTW}}
\label{stabilityanalysissss}

\begin{figure*}[htbp]
	\centering
	\begin{tabular}{cc}
		\includegraphics[width=0.43\textwidth]{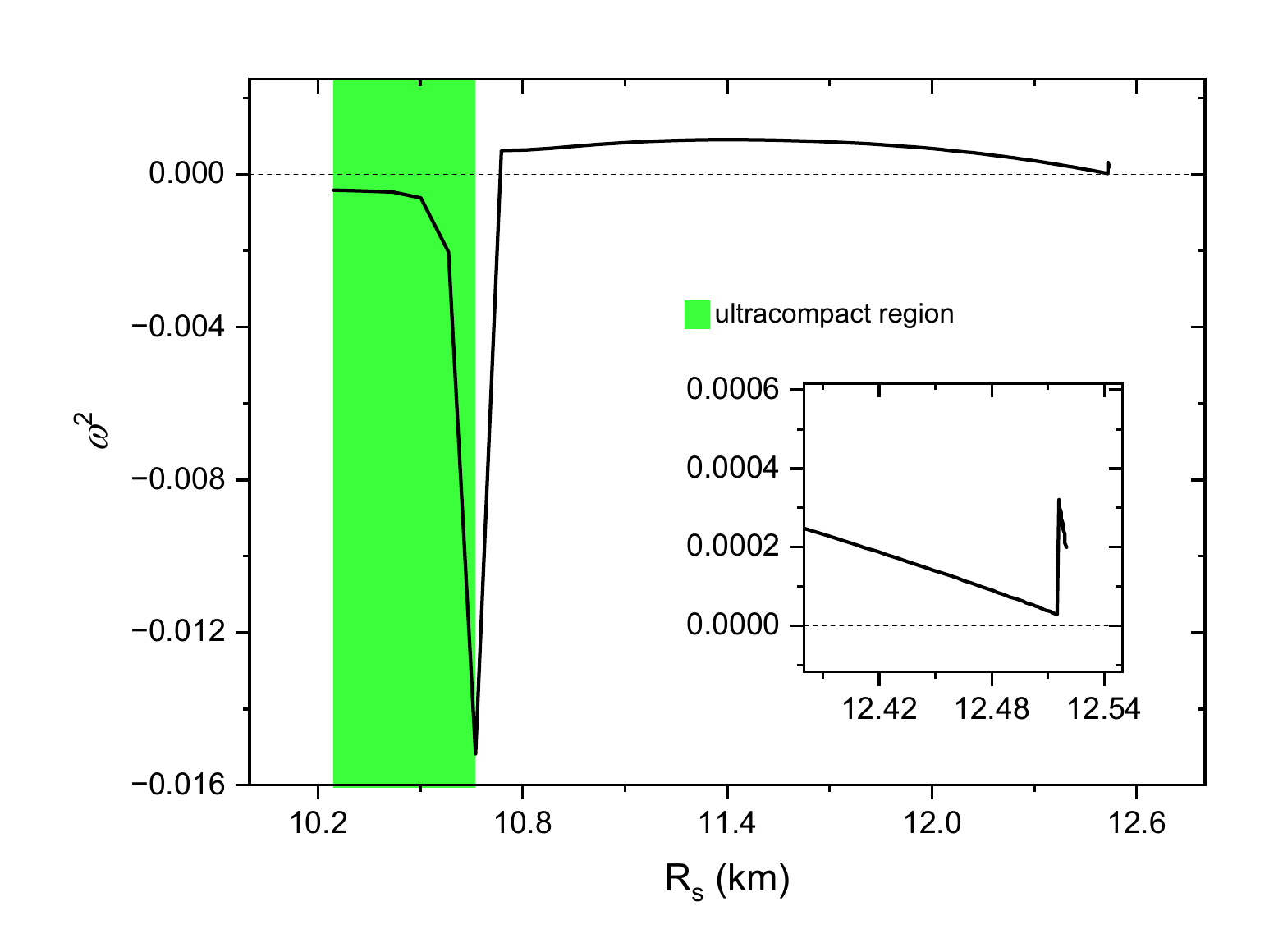} &
		\includegraphics[width=0.43\textwidth]{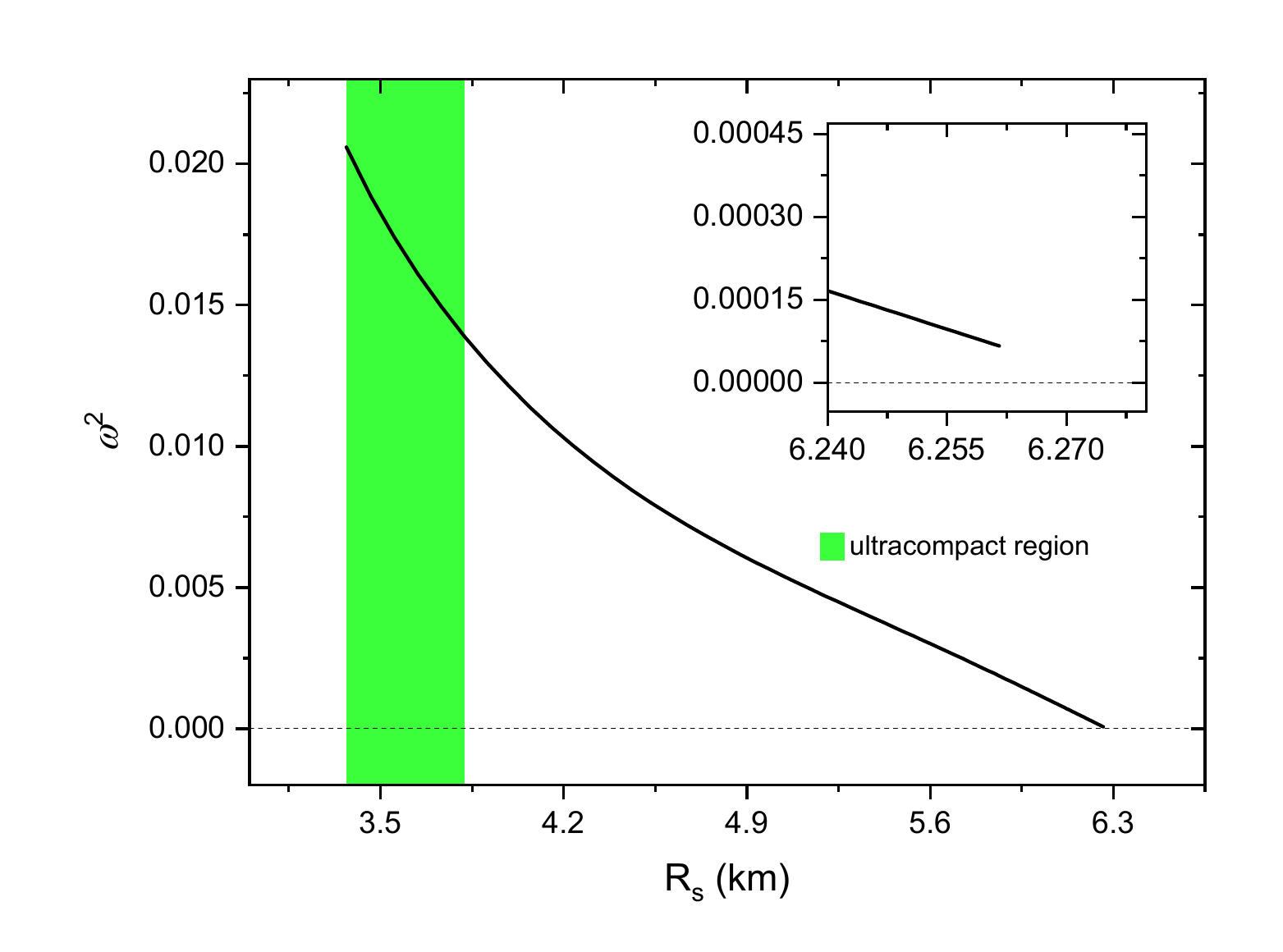} \\
		(a) & (b) \\[1em]
		\includegraphics[width=0.43\textwidth]{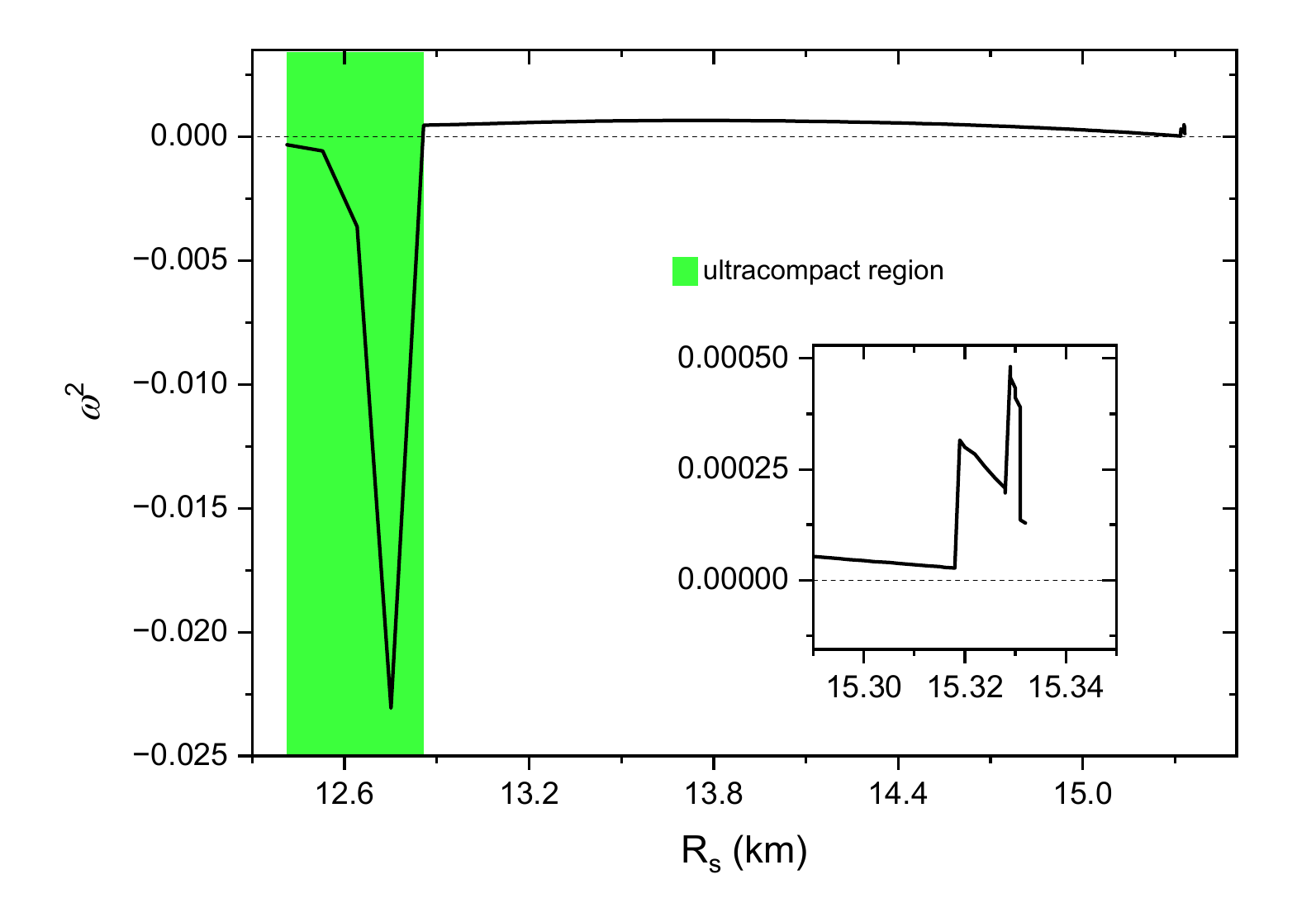} &
		\includegraphics[width=0.43\textwidth]{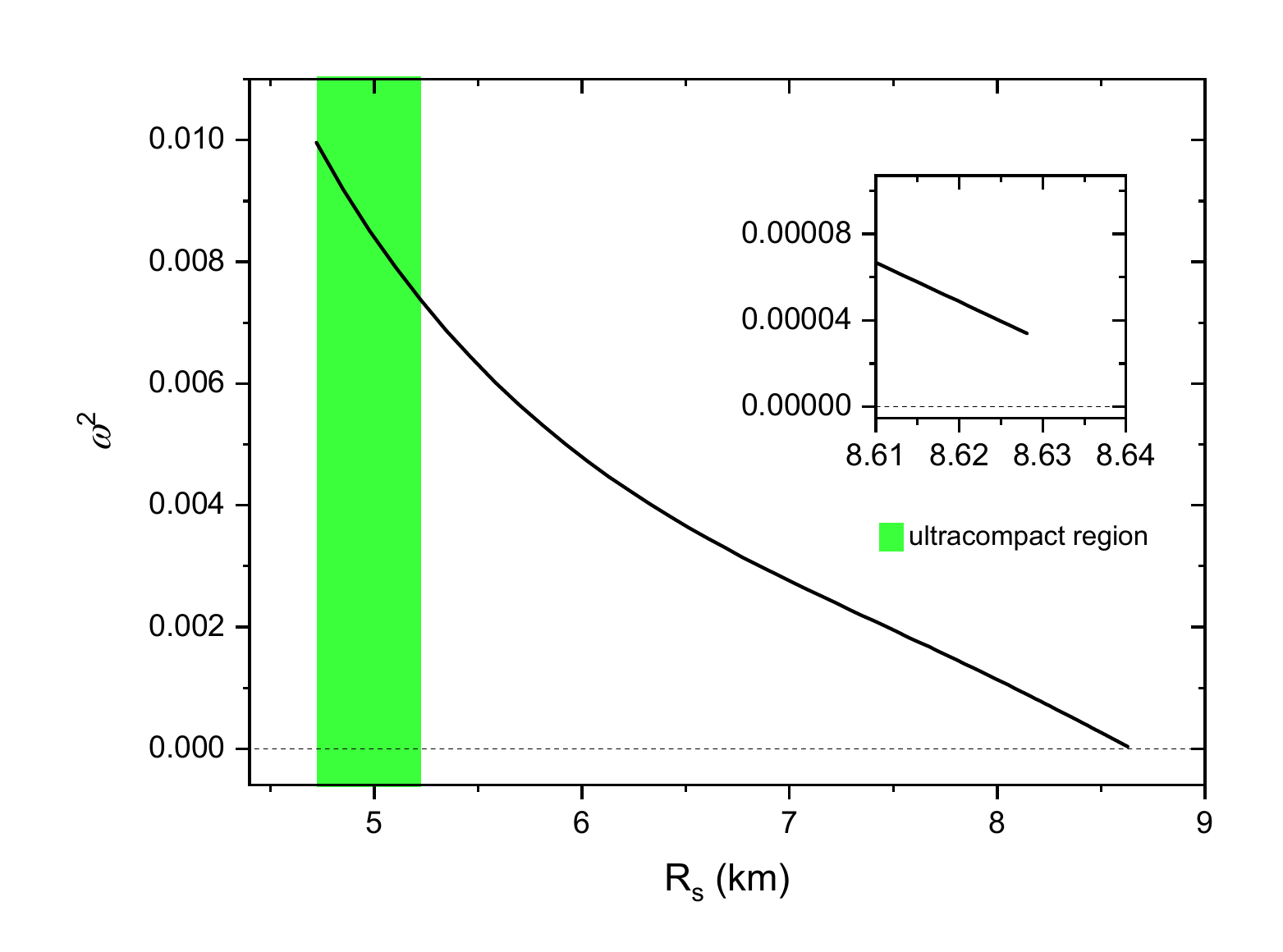} \\
		(c) & (d) \\[1em]
		\includegraphics[width=0.43\textwidth]{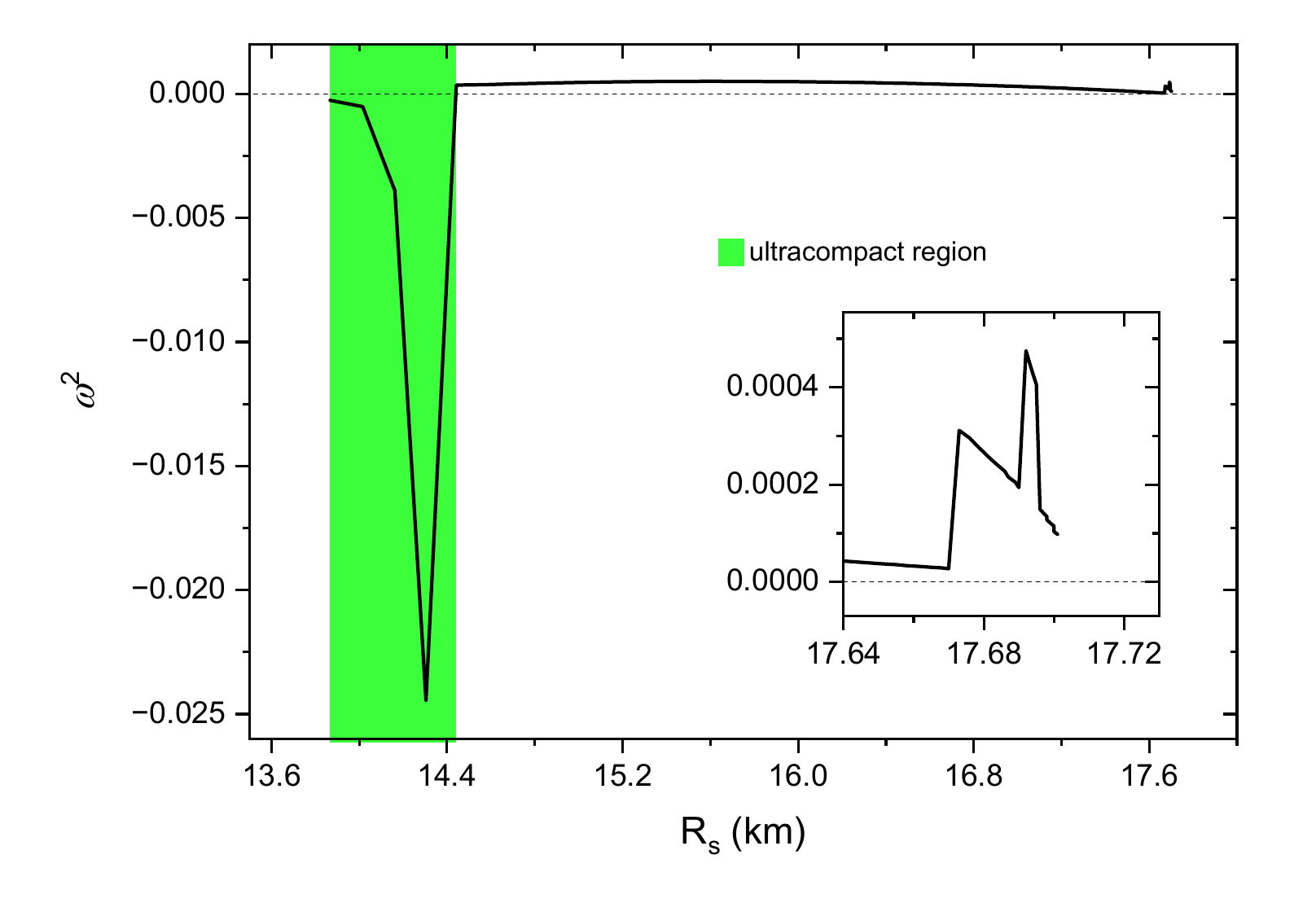} &
		\includegraphics[width=0.43\textwidth]{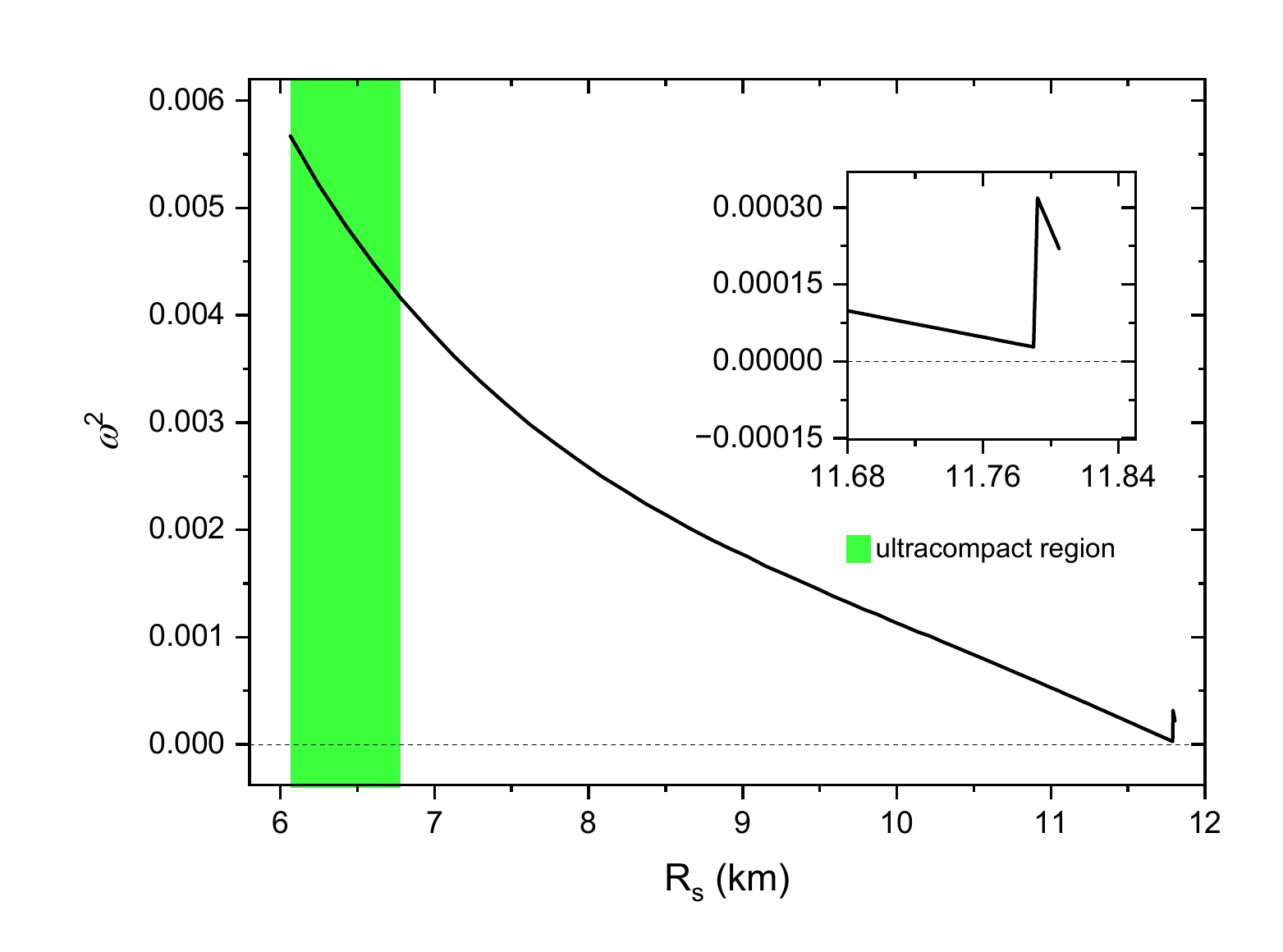} \\
		(e) & (f)
	\end{tabular}
	\caption{\textcolor{black}{Radial perturbation's $\omega^2-R_s$ relations of non-magnetized CSSTW configurations for (a)~$h=1$ with Approach 1, (b)~$h=1$ with Approach 2, (c)~$h=1.5$ with Approach 1, (d)~$h=1.5$ with Approach 2, (e)~$h=2$ with Approach 1, and (f)~$h=2$ with Approach 2.}}
	\label{figstabilityanalysisnon}
\end{figure*}

\begin{figure*}[htbp]
	\centering
	\begin{tabular}{cc}
		\includegraphics[width=0.43\textwidth]{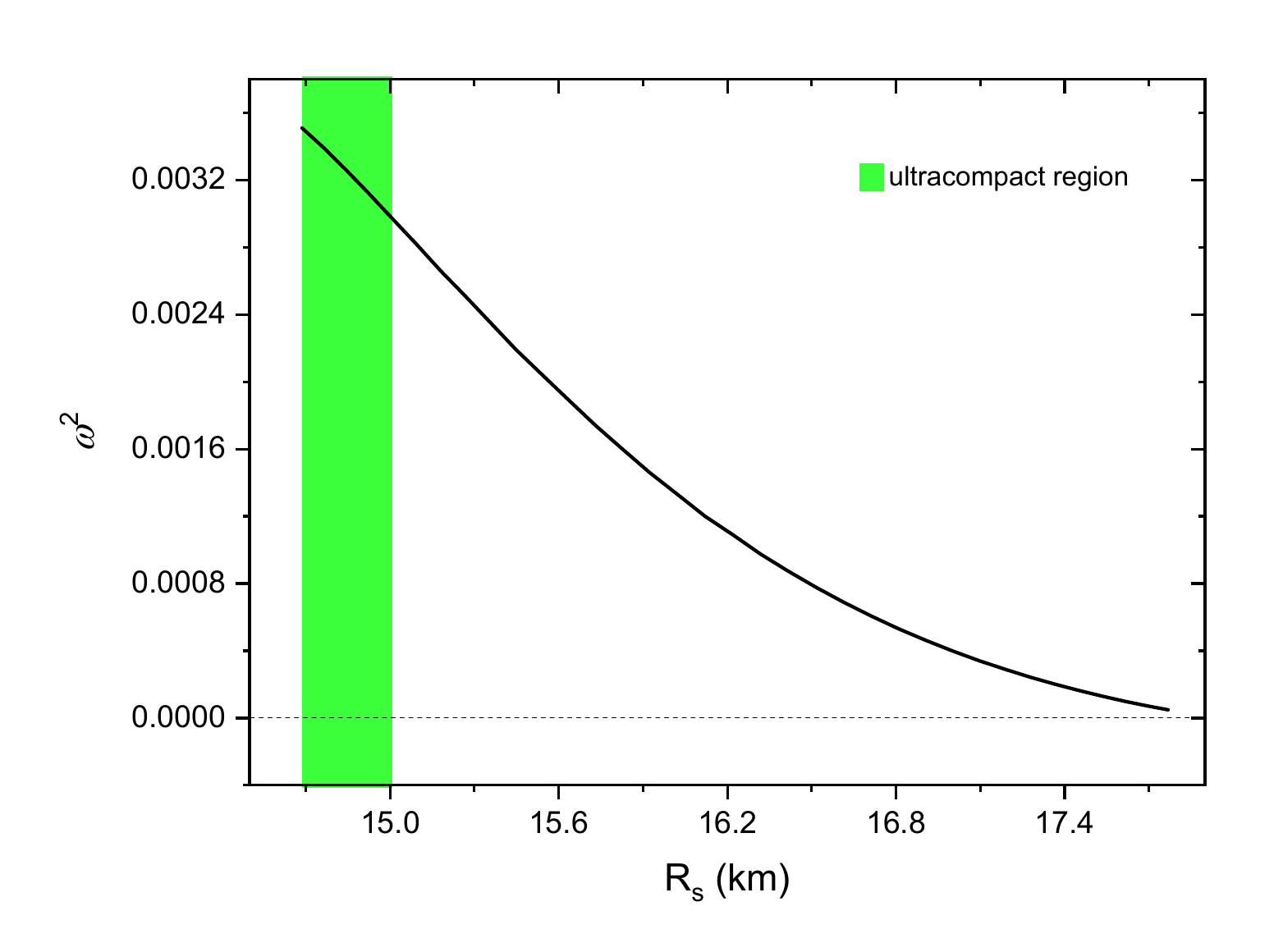} &
		\includegraphics[width=0.43\textwidth]{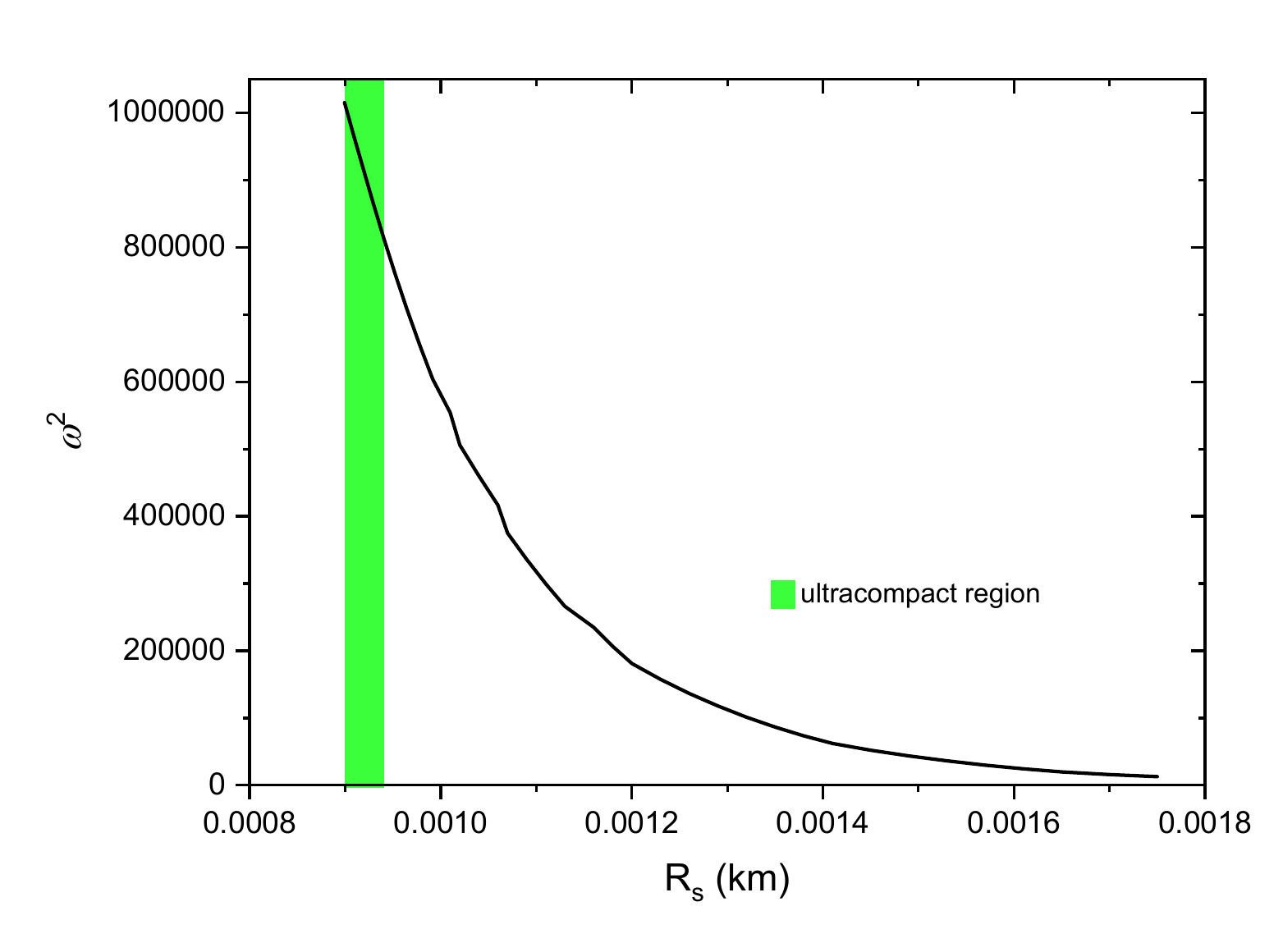} \\
		(a) & (b) \\[1em]
		\includegraphics[width=0.43\textwidth]{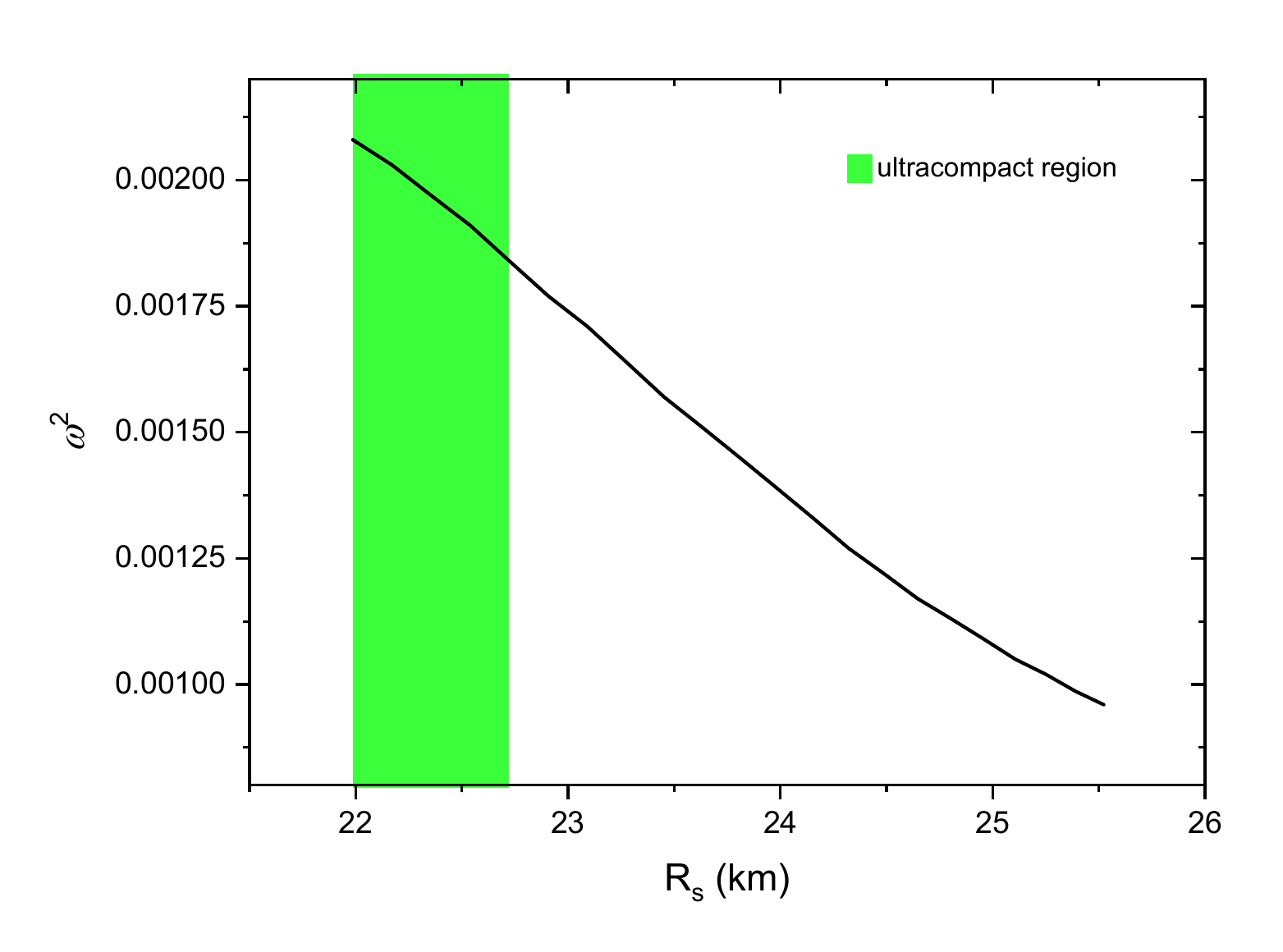} &
		\includegraphics[width=0.43\textwidth]{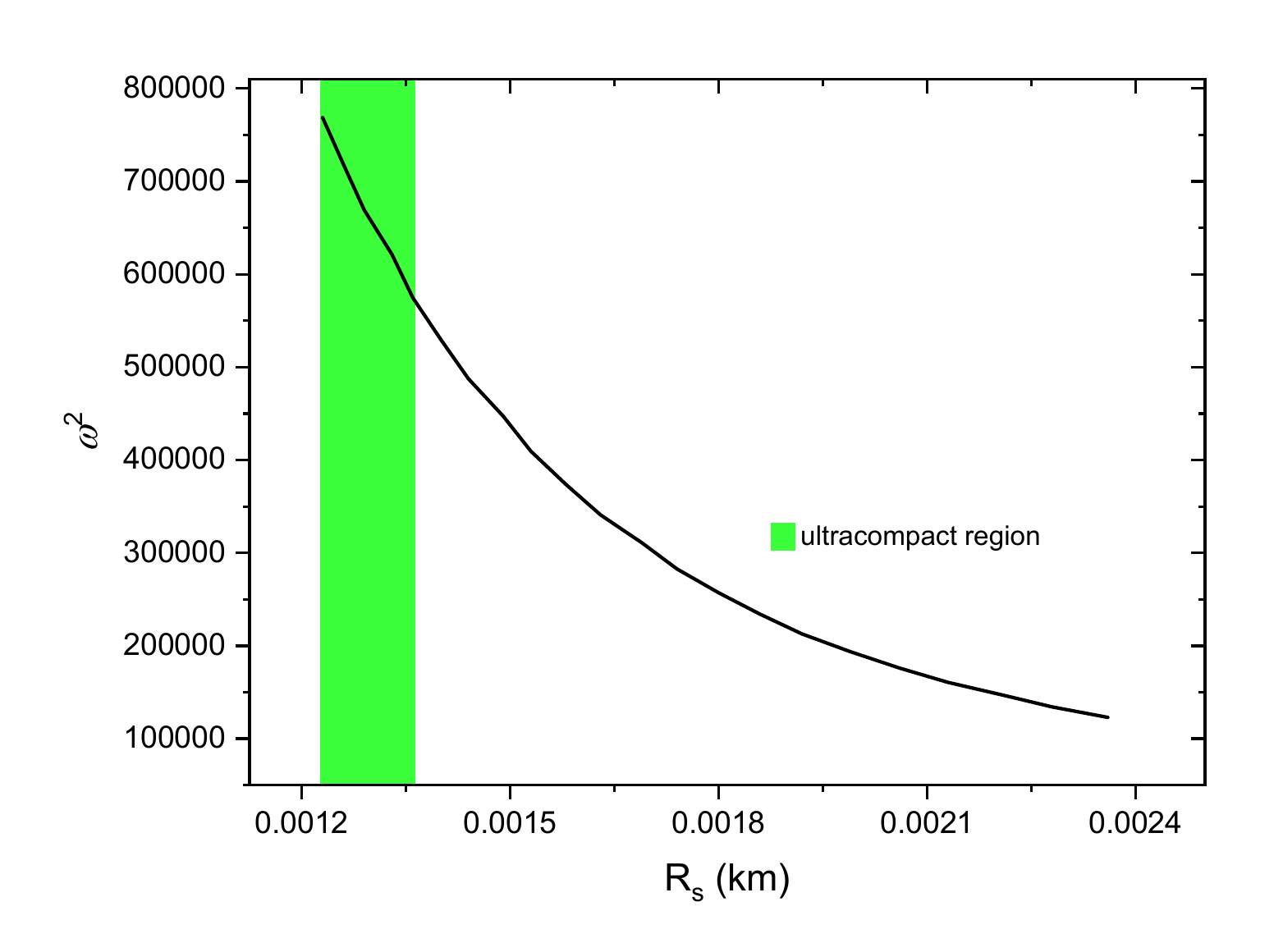} \\
		(c) & (d) \\[1em]
		\includegraphics[width=0.43\textwidth]{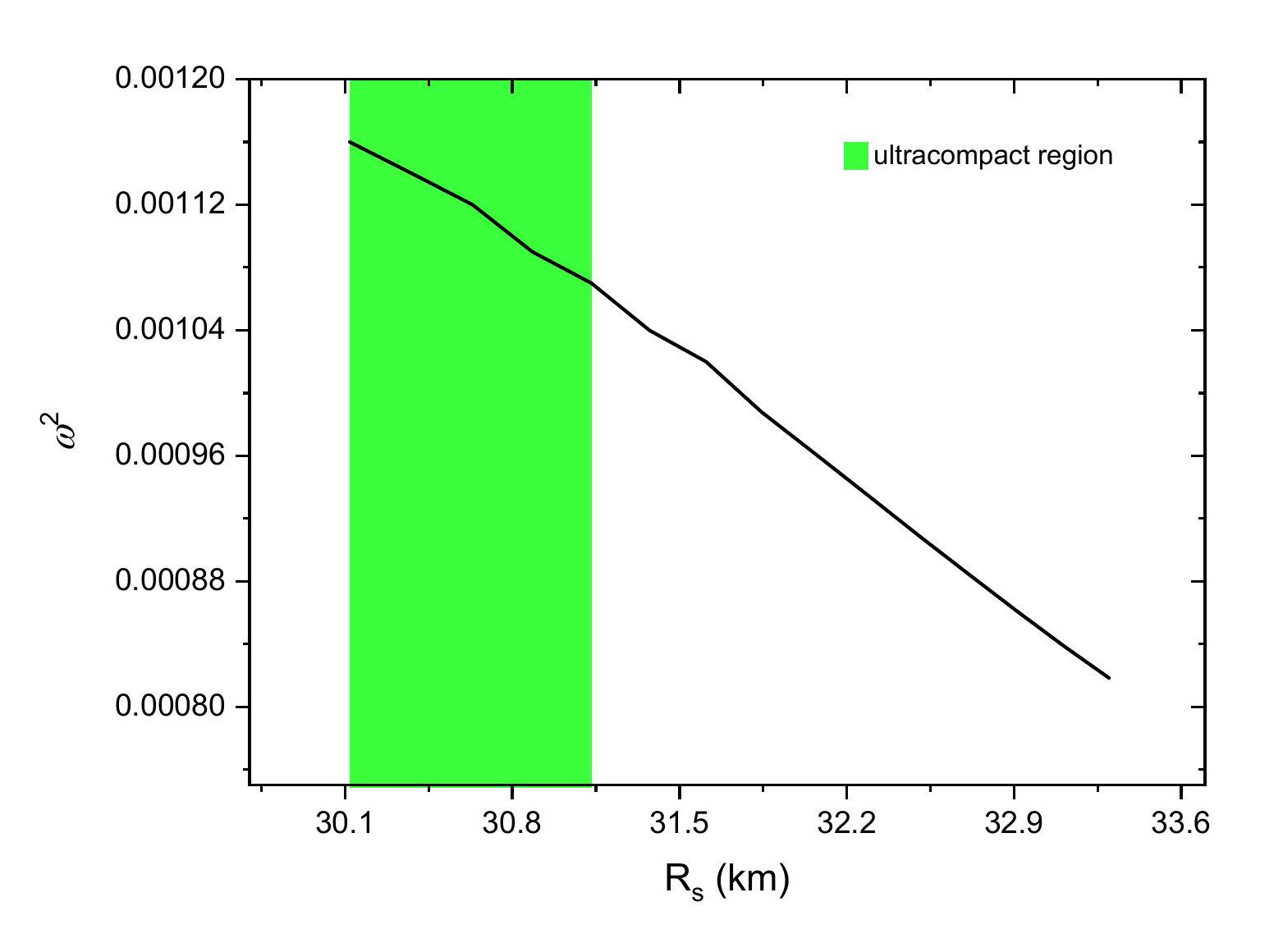} &
		\includegraphics[width=0.43\textwidth]{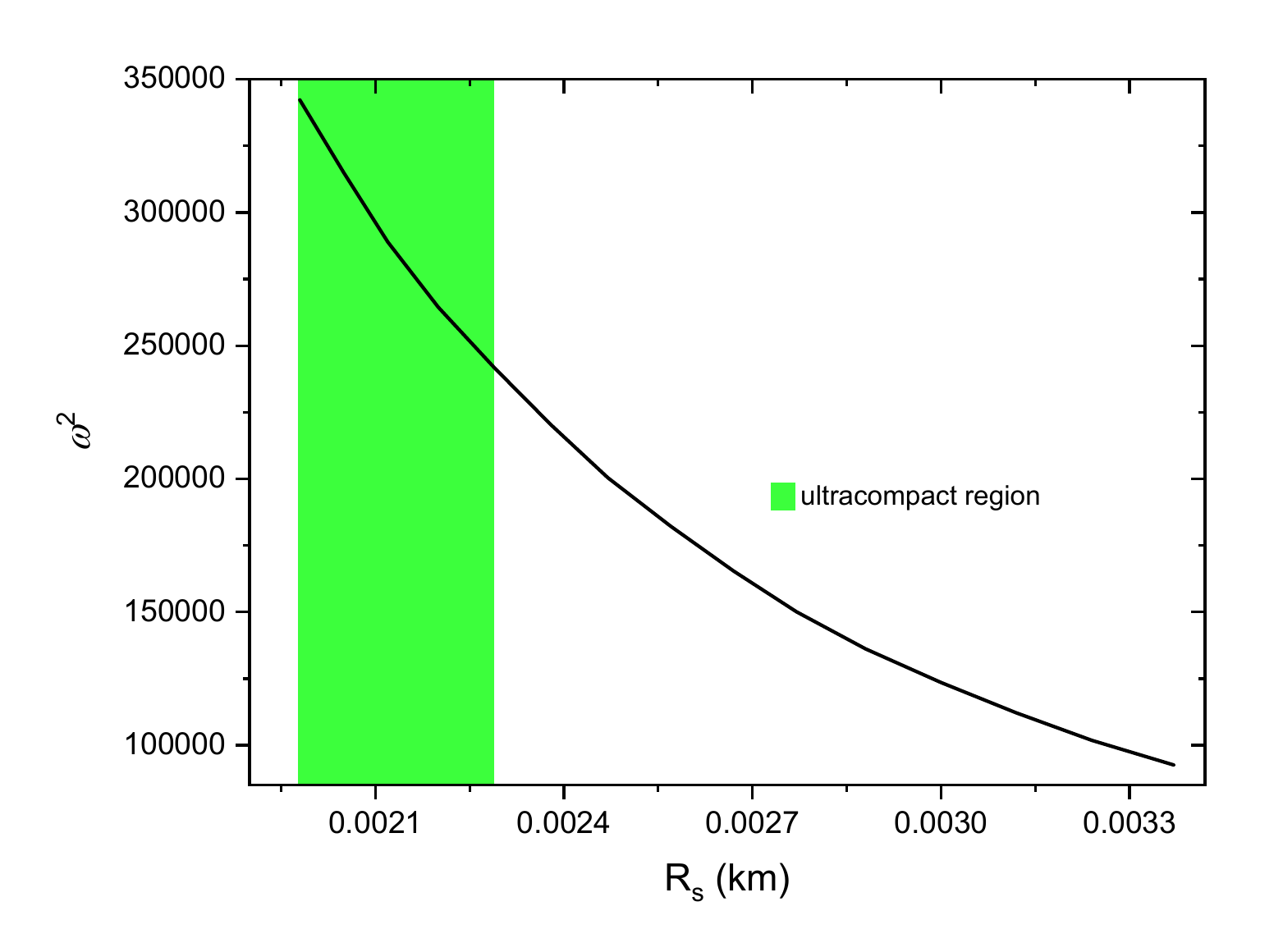} \\
		(e) & (f)
	\end{tabular}
	\caption{\textcolor{black}{Radial perturbation's $\omega^2-R_s$ relations of magnetized CSSTW configurations with $B_s=1\times10^{15}$~G for (a)~$h=1$ with Approach 1, (b)~$h=1$ with Approach 2, (c)~$h=1.5$ with Approach 1, (d)~$h=1.5$ with Approach 2, (e)~$h=2$ with Approach 1, and (f)~$h=2$ with Approach 2.}}
	\label{figstabilityanalysisBs1015}
\end{figure*}

\begin{figure*}[htbp]
	\centering
	\begin{tabular}{cc}
		\includegraphics[width=0.43\textwidth]{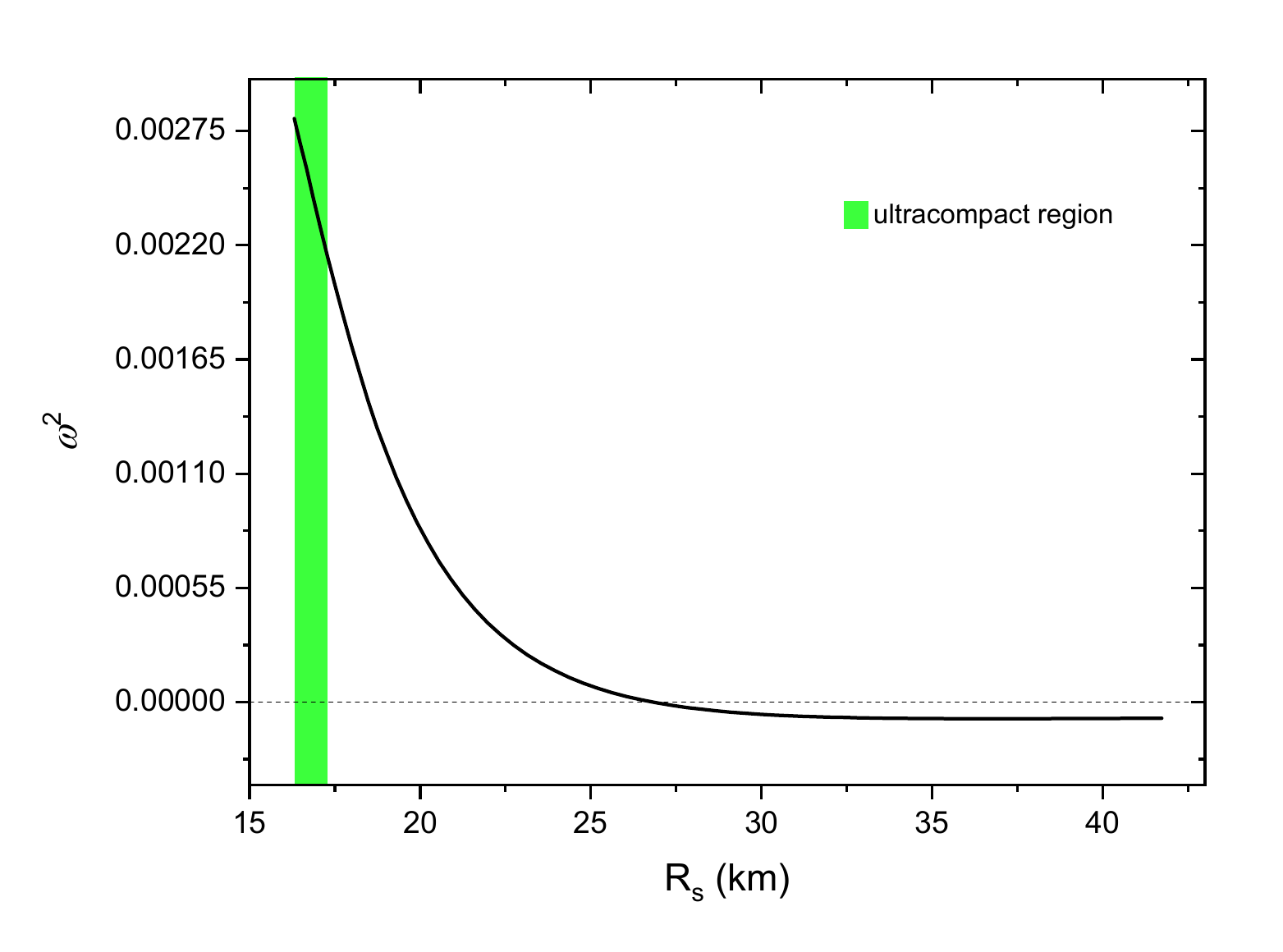} &
		\includegraphics[width=0.43\textwidth]{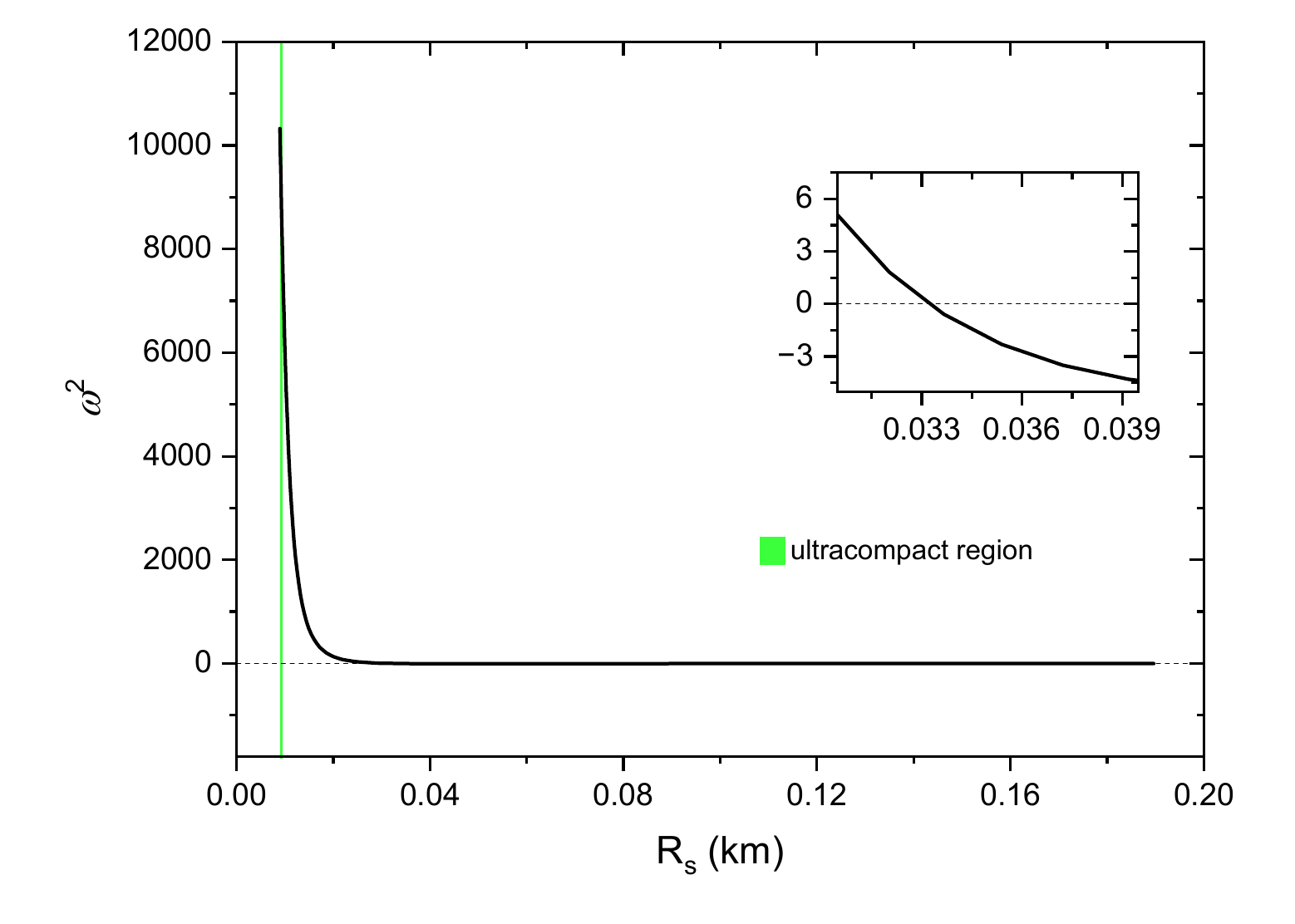} \\
		(a) & (b) \\[1em]
		\includegraphics[width=0.43\textwidth]{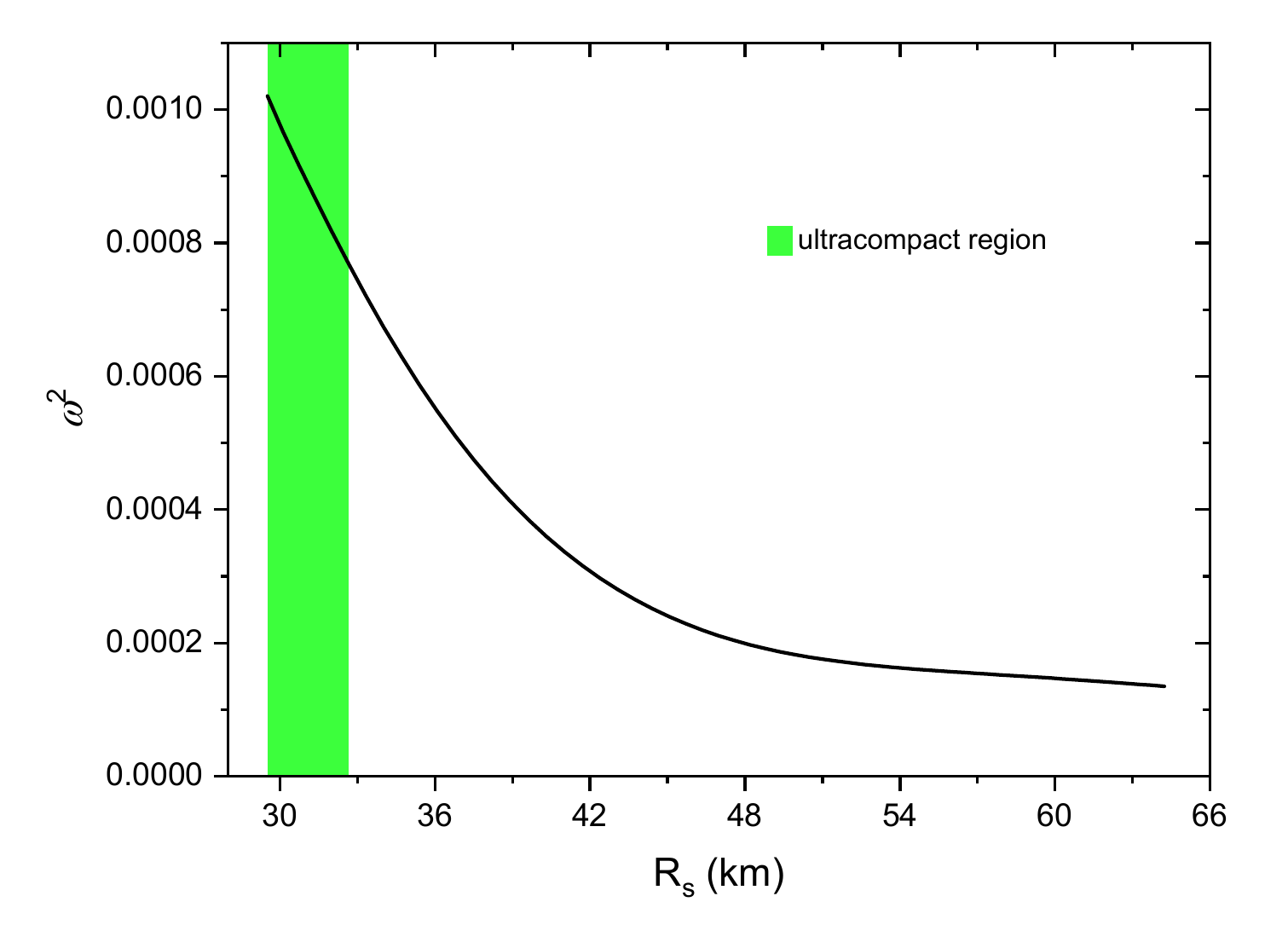} &
		\includegraphics[width=0.43\textwidth]{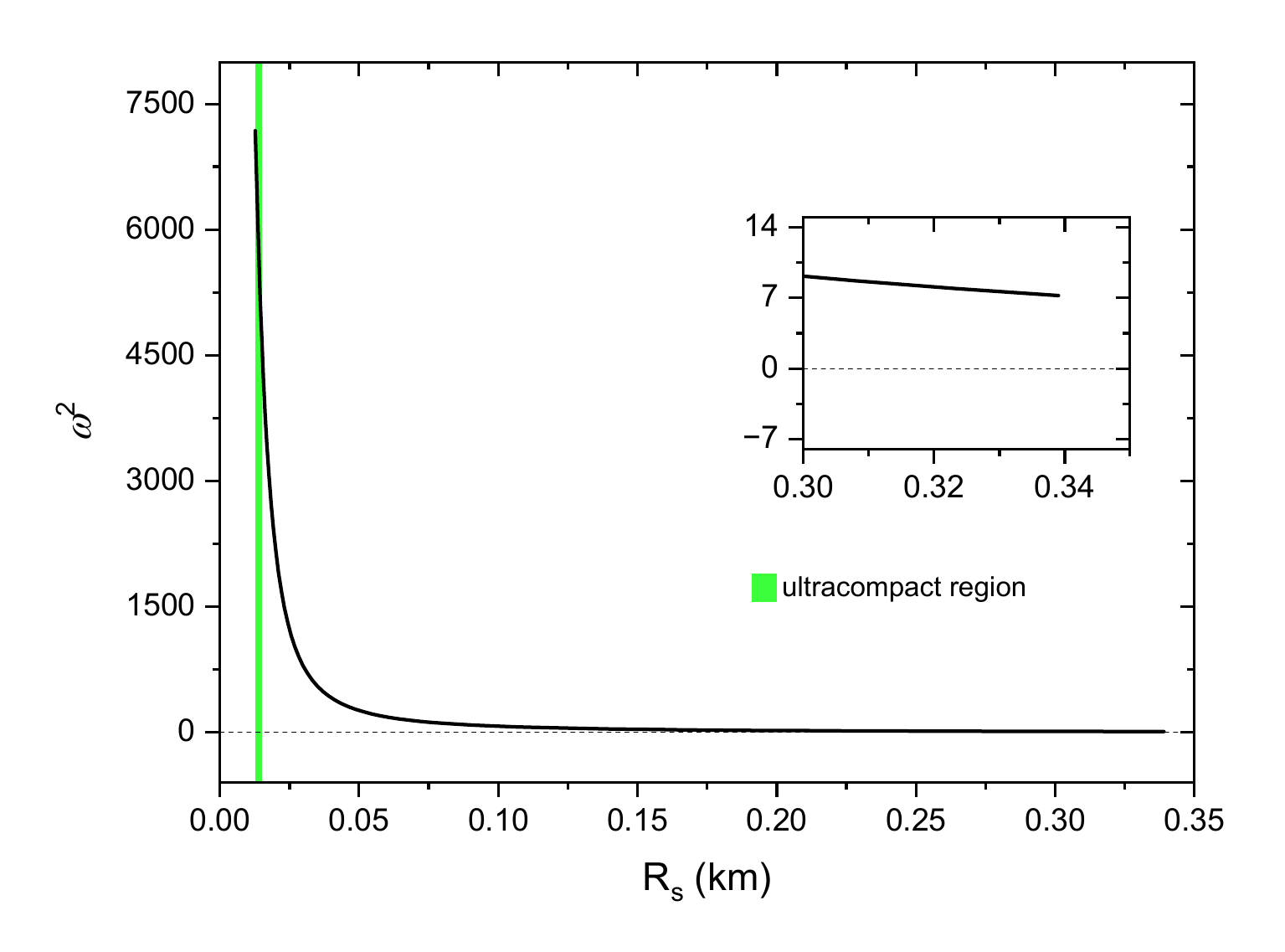} \\
		(c) & (d) \\[1em]
		\includegraphics[width=0.43\textwidth]{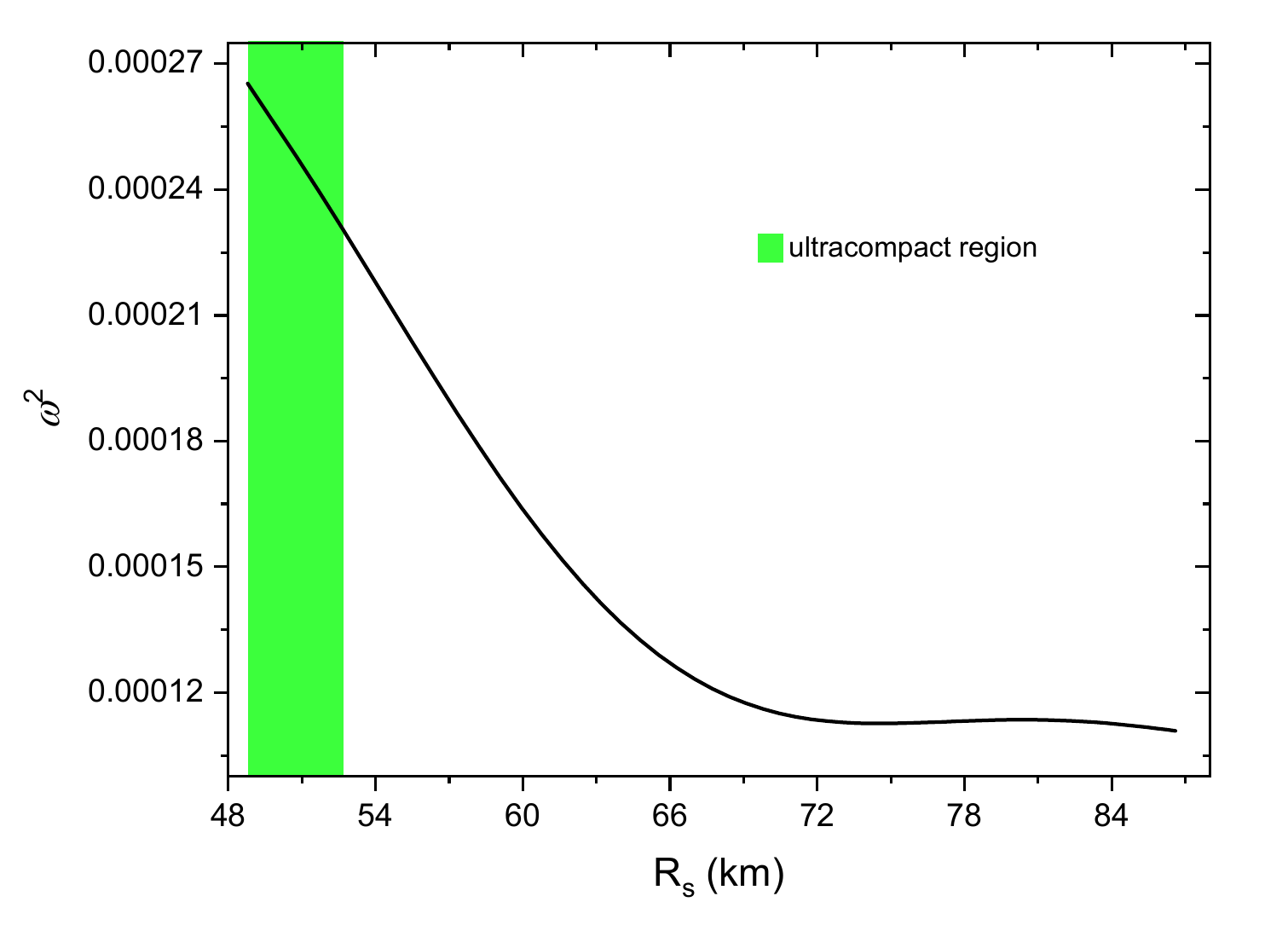} &
		\includegraphics[width=0.43\textwidth]{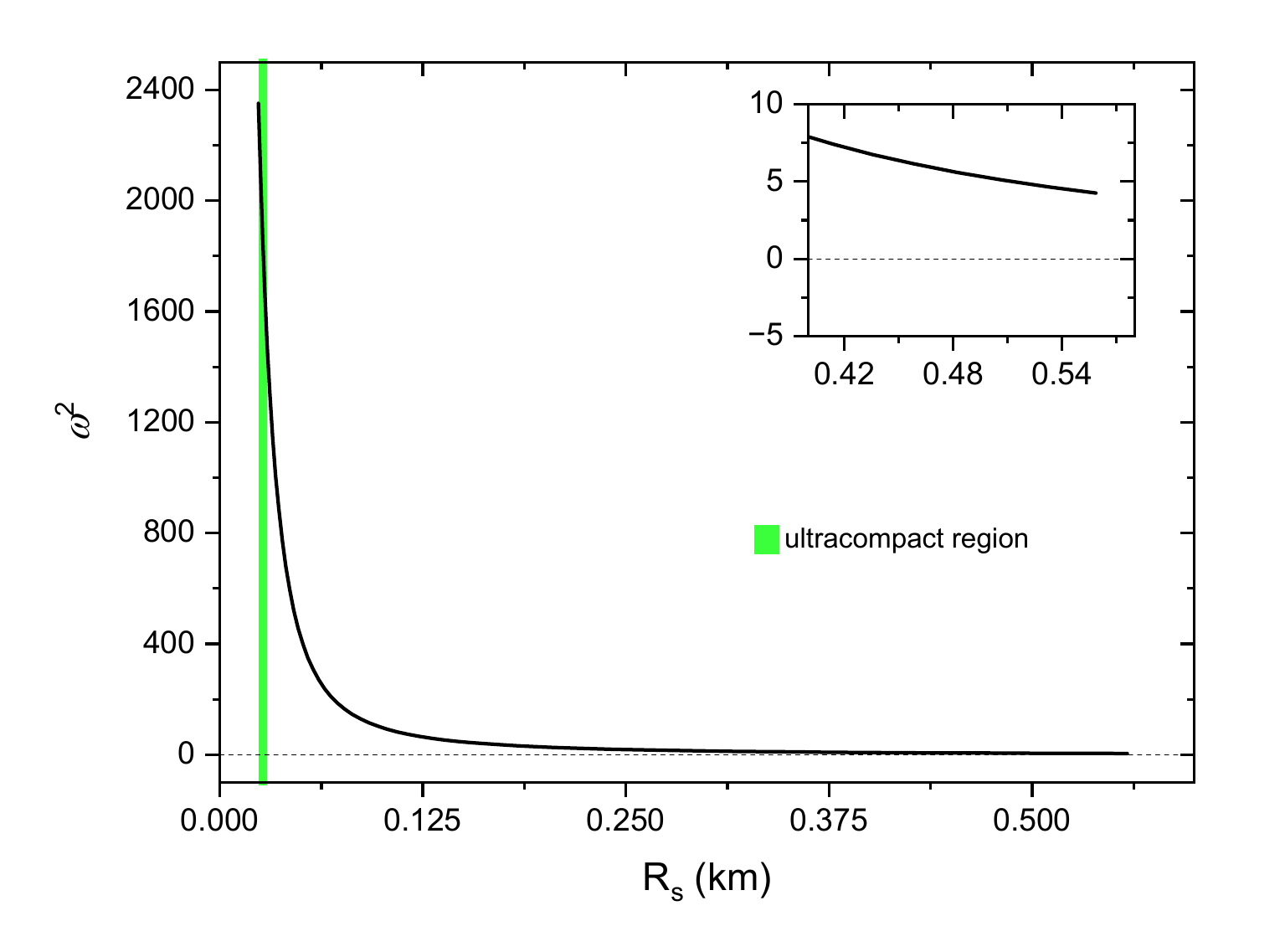} \\
		(e) & (f)
	\end{tabular}
	\caption{\textcolor{black}{Radial perturbation's $\omega^2-R_s$ relations of magnetized CSSTW configurations with $B_s=1\times10^{16}$~G for (a)~$h=1$ with Approach 1, (b)~$h=1$ with Approach 2, (c)~$h=1.5$ with Approach 1, (d)~$h=1.5$ with Approach 2, (e)~$h=2$ with Approach 1, and (f)~$h=2$ with Approach 2.}}
	\label{figstabilityanalysisBs1016}
\end{figure*}

\textcolor{black}{Chandrasekhar~\cite{Chandrasekhar:1964zz} proposed a formalism for analyzing the stability of relativistic bodies composed of a perfect fluid. However, this formalism assumes isotropic pressures. For anisotropic pressure, a stability analysis formalism was developed in Ref.~\cite{Mohanty:2023kin}.
}

In this work, we employ radial perturbations to investigate the stability analysis of the system. Under radial perturbations, both the spacetime and the neutron fluid are perturbed. However, as discussed in Ref.~\cite{Nojiri2024}, the scalar fields are frozen and therefore nondynamical, so their oscillations do not generate sound in the effective fluid generated by those scalar fields. Consequently, the radial velocity of this effective fluid vanishes, i.e., $u_r^{\phi\chi}=0$. Thus, the perturbations affect only the spacetime and neutron-fluid sectors, leaving the scalar-field contribution to the total EMT unperturbed.

The unperturbed quantities $\varrho$, $P_r$, and $P_t$ enter terms of the perturbation equations through the radial fluid velocity $u_r$, as the component $\mathcal{T}_{tr}$ depends on both $u_t$ and $u_r$ and does not vanish (see Chandrasekhar's paper~\cite{Chandrasekhar:1964zz} for the detailed derivation). Since $u_r^{\phi\chi}=0$, the scalar fields' sector makes no contribution to $\mathcal{T}_{tr}$ and hence does not contribute to the perturbation equations through this term.

\textcolor{black}{As we work on an anisotropic neutron fluid, we follow Ref.~\cite{Mohanty:2023kin} for the procedure of the radial perturbation. We first assume a harmonic time dependence for the radial displacement of a fluid element that is initially located at $r$ in the unperturbed configuration. This displacement, also known as the Lagrangian displacement, is given by
	$\delta r = \xi(r,t) = \xi(r)e^{i\omega t}$.}

\textcolor{black}{By solving the perturbed field equations, i.e. $\delta G_{\mu\nu}=\kappa^2 \delta T_{\text{matter}\:\mu\nu}$, linearized radial perturbation equations can be obtained, which writes}
\begin{align}
	\textcolor{black}{\omega^2 (\varrho+P_r)e^{2(\lambda-\nu)}\xi
		=}{}&
	\textcolor{black}{\Delta P_r \frac{d}{dr}(2\nu+\lambda)+\frac{d}{dr}(\Delta P_r)}
	\nonumber\\
	&\textcolor{black}{+8\pi P_t e^{2\lambda}(\varrho+P_r)\xi-\xi(\varrho+P_r)}
	\nonumber\\
	&\textcolor{black}{\times \left(\frac{d\nu}{dr}\right)^2+\frac{4\xi}{r}\frac{dP_r}{dr}+\frac{2\sigma\xi}{r}}
	\nonumber\\
	&\textcolor{black}{\times\left[\frac{d}{dr}(2\nu+\lambda)
		+\frac{4}{r}\right]}\nonumber\\
	&\textcolor{black}{+\frac{d}{dr}\left[\frac{2\sigma\xi}{r}\right]+\frac{2}{r}\delta\sigma,}
	\label{perturbationfromfield}
\end{align}
\textcolor{black}{where perturbations in $\varrho$, $P_r$, $\lambda$, and $\sigma$ are given by}
\begin{align}
	\textcolor{black}{\delta\varrho
		=}& \textcolor{black}{-\frac{1}{r^2}
		\frac{\partial}{\partial r}
		\left[r^2(\varrho+P_r)\xi\right],}\\
	\textcolor{black}{\delta P_r
		=}& \textcolor{black}{-\xi \frac{dP_r}{dr}
		-\Gamma P_r \frac{e^\nu}{r^2}
		\frac{\partial}{\partial r}
		\left(r^2\xi e^{-\nu}\right)}\nonumber\\
	&\textcolor{black}{-\frac{2}{r}\sigma\xi
		\frac{\partial P_r}{\partial\varrho},}\label{enambelas}\\
	\textcolor{black}{\delta\lambda
		=}& \textcolor{black}{-\xi \frac{d}{dr}(\nu+\lambda)
		= -4\pi r(\varrho+P_r)e^{2\lambda}\xi,}\\
	\textcolor{black}{\delta\sigma=}&\textcolor{black}{-\frac{r}{2}\frac{1-h}{h}\frac{d\delta P_r}{dr}.}
\end{align}
\textcolor{black}{Here, $\Gamma$ denotes adiabatic index defined by}
\begin{eqnarray}
	\textcolor{black}{\Gamma=\left(1+\frac{\varrho}{P_r}\right)\frac{dP_r}{d\varrho}.}
\end{eqnarray}

\textcolor{black}{It is convenient to redefine the Lagrngian displacement as $\zeta=\xi/r$. Therefore, Eqs.~(\ref{perturbationfromfield}) and (\ref{enambelas}) can be rewritten~as}
\begin{align}
	\textcolor{black}{\frac{d\zeta}{dr}}
	&\textcolor{black}{=
		-\frac{1}{r}
		\left(
		3\zeta
		+\frac{\Delta P_r}{\Gamma P_r}
		-\frac{2\sigma\zeta}{\varrho+P_r}
		\right)
		+\frac{d\nu}{dr}\zeta,}
	\label{eq:dzetadr}
\end{align}
\begin{align}
	\textcolor{black}{\frac{d(\Delta P_r)}{dr}
		=}{}&
	\textcolor{black}{\zeta\left\{
		\omega^2 e^{2(\lambda-\nu)}
		(\varrho+P_r)r
		-4\frac{dP_r}{dr}
		\right.}
	\nonumber\\
	&\left.
	\quad \textcolor{black}{-8\pi(\varrho+P_r)e^{2\lambda}rP_r+r(\varrho+P_r)
		\left(\frac{d\nu}{dr}\right)^2}\right.
	\nonumber\\
	&\textcolor{black}{\left.
		\quad
		-2\sigma
		\left(
		\frac{4}{r}+\frac{d\nu}{dr}
		\right)
		-2\frac{d\sigma}{dr}
		\right\}+2\sigma\frac{d\zeta}{dr}}\nonumber\\
	&
	\quad
	\textcolor{black}{-\Delta P_r
		\left[
		\frac{d\nu}{dr}
		+4\pi(\varrho+P_r)re^{2\lambda}
		\right]}\nonumber\\
	&
	\quad
	\textcolor{black}{-\frac{2}{r}\delta\sigma.}
	\label{eq:dDeltaPrdr}
\end{align}

\textcolor{black}{
	The differential equations~(\ref{eq:dzetadr}) and~(\ref{eq:dDeltaPrdr}), subject to the aforementioned boundary conditions, constitute a Sturm--Liouville eigenvalue problem that can be used to determine the radial oscillation modes and their corresponding eigenvalues,
	\[
	\omega_0^2 < \omega_1^2 < \cdots < \omega_n^2 < \cdots,
	\]
	where $n$ denotes the number of nodes within the fluid. According to Ref.~\cite{Mohanty:2023kin}, $\Delta P(R_s)=0$ and, in the case of NSs, $\xi(0)=0$. Since there is no $r=0$ in our case, we instead assume $\xi(r_0)\to 0$. Since $\zeta=\xi/r$, it follows that $\zeta(r_0)\to 0$. Details of the numerical calculation of Eqs.~(\ref{eq:dzetadr})--(\ref{eq:dDeltaPrdr}) to obtain $\omega^2$ are given in Subsec.~\ref{sec:shooting} of Appendix~\ref{appnumerical}.
}

\textcolor{black}{Fig.~\ref{figstabilityanalysisnon} shows the numerical results of the radial perturbation analysis for non-magnetized configurations. We observe that the construction of Approach~1 results in both stable and unstable systems. Non-magnetized systems in the ultracompact region are more prone to instability. For all non-magnetized configurations obtained with Approach~1, the unstable systems are found to be ultracompact. However, we can see that increasing $h$ can reduce the susceptibility of ultracompact systems to instability. In particular, for $h=1.5$ and $h=2$, there exist ultracompact systems that remain stable, as shown in Figs.~\ref{figstabilityanalysisnon}(b) and~(c). This provides the physical insight that anisotropy can reduce the susceptibility of ultracompact non-magnetized CSSTW to instability. On the other hand, all considered non-magnetized systems obtained using Approach~2 are stable.}

\textcolor{black}{Moreover, many of the considered magnetized CSSTW configurations are stable. Fig.~\ref{figstabilityanalysisBs1015} shows that all considered magnetized CSSTW configurations with $B_s=10^{15}$~G and $B_0=3\times10^{18}$~G are stable. Fig.~\ref{figstabilityanalysisBs1016} shows that all considered magnetized CSSTW configurations with $B_s=10^{16}$~G and $B_0=3\times10^{18}$~G are stable, except for the cases with $h=1$ in both Approaches~1 and~2. For these exceptional cases, unstable systems exist, but they are not located in the ultracompact region. This result once again indicates that anisotropy can prevent the appearance of unstable CSSTW configurations.
}

\section{Conclusion}
\label{conclusion}
In this study, we have constructed the formalism of anisotropic CSSTW, both in the absence and in the presence of magnetic fields. In general, the underlying model admits ghosts; however, these ghosts can be eliminated by imposing appropriate constraints through the introduction of Lagrange multiplier fields in the extended action. \textcolor{black}{We find that the wormhole remains traversable irrespective of whether fluid anisotropy and/or magnetic fields are present, in the sense that the NEC is violated in the vicinity of the wormhole throat both with and without these effects.
}

\textcolor{black}{The Kretschmann scalar for all considered configurations show that the whole spacetime are regular. Furthermore, the temporal metric functions of all considered configurations show that there are no horizon in the considered cases. The conditions for traversable wormhole where $b(r_0)=r_0$ and $b'(r_0)<1$ are also satisfied.}

We investigate the physical properties of the resulting configurations, including the ADM mass, stellar radius, wormhole throat radius, and compactness. In addition, we analyze other physical properties, namely the surface redshift and the gravitational-wave echo time. Throughout our analysis, we adopt two complementary approaches, following the methodology introduced by Ref.~\cite{Nojiri2024}, and also applied in Ref.~\cite{Pattersons2026mpla}. In Approach~1, for a given central energy density $\rho_c$, the stellar radius is assumed to coincide with that of an ordinary NS obtained at the same $\rho_c$. In Approach~2, for the same $\rho_c$, the total mass of the CSSTW system is taken to be equal to that of an ordinary NS computed at the same central energy density.

We find that, within Approach~1, both non-magnetized and magnetized CSSTW can become extremely massive. In particular, the total mass of the system can surpass $8\,M_\odot$. In contrast, Approach~2 yields significantly lower masses by construction, since the total mass of the CSSTW system is fixed to be equal to that of an ordinary polytropic neutron star with the same central energy density. In the magnetized configurations of Approach~2, the resulting masses can even approach values close to zero.

Despite these differences in mass, the compactness of all configurations considered in this work can reach the ultracompact regime. Regarding the role of the magnetic field, we find that a larger surface magnetic field generally leads to larger masses and stellar radii. However, magnetized configurations obtained within Approach~2 are expected to be extremely difficult to observe, due to their very small masses and radii. Nevertheless, such configurations remain of theoretical interest.

%
In the context of the surface redshift, we find that all configurations considered in this work can exceed values of $\mathcal{Z}\simeq1.5$, which are significantly larger than those typically expected for ordinary neutron stars. Furthermore, for a fixed configuration and at a given mass, larger values of the anisotropy parameter $h$ lead to smaller surface redshift values.

The echo time for non-magnetized configurations is of the order of $10^{-2}$--$10^{-1}$~ms. \textcolor{black}{For magnetized configurations, the echo time in Approach~1 is of order
	$10^{-1}\,\mathrm{ms}$ and is larger than that of the corresponding non-magnetized
	configurations.
	In contrast, Approach~2 yields much shorter echo times for magnetized systems, of
	order $10^{-2}$--$10^{-1}\,\mu\mathrm{s}$.
}

In addition, we construct a formulation of the echo time as an explicit function of the magnetic field, assuming a constant magnetic field throughout the system. it \textcolor{black}{reconciles} the trend reported in Table~\ref{table:echo_time_magnetized}, where configurations with a surface magnetic field of $B_s=10^{15}$~G exhibit longer echo times than those with $B_s=10^{16}$ G. Therefore, the curves of echo time as a function of $\mathcal{B}$ offer a plausible interpretation of the behavior observed in the magnetized configurations listed in Table~\ref{table:echo_time_magnetized}.

\textcolor{black}{The quantitative values of the physical properties reported in this work has to be interpreted as properties of the adopted phenomenological construction and ansatzes, not as robust predictions.}

\textcolor{black}{The stability analysis of radial perturbations shows that the construction of Approach~1 results in both stable and unstable systems. Non-magnetized systems in the ultracompact region are more prone to instability. For all non-magnetized configurations obtained with Approach~1, the unstable systems are found to be ultracompact. For $h=1.5$ and $h=2$, there exist ultracompact systems that remain stable. On the other hand, all non-magnetized systems obtained using Approach~2 are stable. All considered configurations with a surface magnetic field of $B_s=10^{15}$~G are stable; while all considered magnetized CSSTW configurations with $B_s=10^{16}$~G are stable, except for the cases with $h=1$ in both Approaches~1 and~2. For these exceptional cases, unstable systems are not located in the ultracompact region.}

\section*{Declaration of generative AI and AI-assisted technologies in the writing process}
During the preparation of this work, Muhammad Lawrence Pattersons used ChatGPT (OpenAI) and Claude in order to support code development and computation; and to improve the language and readability of the manuscript. After using these tools, the authors reviewed and edited the content as needed and take full responsibility for the content of the paper.

\section*{Acknowledgments}
MLP gratefully acknowledges Byon Nugraha Jayawiguna for precious discussions on the echo time, and Anna Campoy-Ordaz for publicly providing code in Ref.~\cite{ordaz} for the calculation of TOV equation. MLP is also financially supported by Indonesia Endowment Fund for Education Agency~(LPDP). FPZ would like to thank PPMI ITB for financial support. HLP was funded by the postdoctoral program in BRIN (National Research and Innovation Agency), Indonesia. MFARS is supported by the Second Century Fund (C2F) and C2F research abroad scholarship. We would like to thank the members of the Theoretical High Energy Physics Group at Institut Teknologi Bandung for the discussions, insights, and hospitality.

\appendix

\section{The expression of $e^{2\nu}$ for the isotropic configuration with the limit $\mathcal{B}\rightarrow 0$}
\label{metricnojiri}
For the isotropic non-magnetized configuration,
\begin{eqnarray}
	\alpha = K,\:\:\:\: \beta=1,\:\:\:\: \gamma=0.\label{conditionalphaetc}
\end{eqnarray}
In this case, Eq.~(\ref{numagnetized}) gives
\begin{eqnarray}
	\nu&=&\mathscr{C}-(2-1)\bigg[\ln\{K\rho^2+\rho\}\nonumber\\
	&&-\ln\bigg\{\frac{2K\rho+1-1}{2K\rho+1+1}\bigg\}\bigg].\label{a1}
\end{eqnarray}
Here, $\mathscr{C}$ is a constant. Further calculation gives
\begin{eqnarray}
	\nu=\mathscr{C}-\ln[\rho(K\rho+1)]+\ln\bigg[\frac{K\rho}{K\rho+1}\bigg].
\end{eqnarray}
\begin{eqnarray}
	\nu=\mathscr{C}+\ln K-\ln(K\rho+1)-\ln(K\rho+1).
\end{eqnarray}
We can define $\nu_c\equiv\mathscr{C}+\ln K$, so that
\begin{eqnarray}
	\nu=\nu_c-2\ln(K\rho+1).
\end{eqnarray}
The metric function $e^{2\nu(r)}$ now reads
\begin{eqnarray}
	e^{2\nu}=\frac{e^{2\nu_c}}{\bigg[1+K\rho_c\left\{1-\frac{(r-r_0)^2}{(R_s-r_0)^2}\right\}\bigg]^4}.\label{nurecovered}
\end{eqnarray}
Eq.~(\ref{nurecovered}) is nothing but Eq.~(24) in Ref.~\cite{Nojiri2024}.

\section{The expression of $(e^{2\nu})'$ for the isotropic configuration with the limit $\mathcal{B}\rightarrow 0$}
\label{derivativemetricnojiri}
Note that the condition shown by Eq.~(\ref{conditionalphaetc}) is also satisfied in this case. Eq.~(\ref{derivativenu}) is now reduced and becomes
\begin{eqnarray}
	\left(e^{2\nu}\right)' &=&
	e^{2\nu}\,
	4(2-1)\,
	\frac{\rho_c (r-r_0)}{(R_s-r_0)^2}
	\nonumber\\
	&&\times
	\left(
	\frac{2K\rho + 1}{K\rho^2 + \rho}
	-
	\frac{1}{K\rho^2 + \rho}
	\right)\nonumber\\
	&=&\frac{8K\rho_c(r-r_0)(R_s-r_0)^{-2}}{K\rho+1}\nonumber\\
	&&\times\frac{e^{2\nu_c}}{(1+K\rho)^4}.\label{b1}
\end{eqnarray}
Finally, Eq.~(\ref{b1}) can be written as
\begin{eqnarray}
	(e^{2\nu})'=\frac{8K\rho_c e^{2\nu_c}(r-r_0)(R_s-r_0)^{-2}}{\bigg[1+K\rho_c\left\{1-\frac{(r-r_0)^2}{(R_s-r_0)^2}\right\}\bigg]^5},
\end{eqnarray}
which is actually Eq.~(24) in Ref.~\cite{Nojiri2024}.

\section{\textcolor{black}{Details on numerical methods and precision}}\label{appnumerical}

\textcolor{black}{The summary of the main assumptions and model parameters adopted in this work that play role in the numerical calculation are given in Table~\ref{tab:model_assumptions}.} \textcolor{black}{Moreover, the explanation regarding the generation of EOS table used in this work is given in Subsec.~\ref{appnumericaleosss}, the numerical method for the TOV equation is given in Subsec.~\ref{tovnumerical}, 
	the echo time integration is discussed in Subsec.~\ref{appechonum}, 
	and the radial perturbation eigenvalue calculation is explained in Subsec.~\ref{sec:shooting}.}
\begin{table*}
	\centering
	\caption{\textcolor{black}{Summary of the main assumptions and model parameters adopted in this work.}}
	\label{tab:model_assumptions}
	\begin{tabular}{ll}
		\hline
		\textbf{\textcolor{black}{Component}} & \textbf{\textcolor{black}{Assumption / Parameterization}} \\
		\hline
		\\
		\textcolor{black}{Equation of state (EOS)} 
		& \textcolor{black}{$p=K\rho^2$; $K$=100 km$^2$.} \\
		\\
		\textcolor{black}{Density profile }
		& \textcolor{black}{$\rho=\rho_c\left[1-\left(\frac{r-r_0}{R_s-r_0}\right)^2\right]$.} 
		\\
		\\
		\textcolor{black}{Range of $\rho_c$}
		& \textcolor{black}{$1\times10^{-4}-1\times10^4$ MeV fm$^{-3}$ for discriminant plots; $1\times10^{-3}-8\times10^3$ MeV fm$^{-3}$ for EOS table.}
		\\
		\\
		\textcolor{black}{Anisotropy prescription} 
		& \textcolor{black}{$\sigma=-\frac{r}{2}\frac{1-h}{h}\frac{dp_r}{dr}$.} \\
		\\
		\textcolor{black}{Values of $h$}
		& \textcolor{black}{$1,\:1.5,\:2$.}
		\\
		\\
		\textcolor{black}{Magnetic-field ansatz} 
		& \textcolor{black}{$\mathcal{B}=B_s+B_0(\rho/\rho_0)^\eta$; $\eta=1$; $B_0=3\times10^{18}$ G.} \\
		\\
		\textcolor{black}{Values of $B_s$}
		& \textcolor{black}{$10^{15}$ G, $10^{16}$ G.}
		\\
		\\
		\textcolor{black}{Values of $\rho_0$} 
		& \textcolor{black}{3240.46 MeV fm$^{-3}$ for $h=1$; 2008.27 MeV fm$^{-3}$ for $h = 1.5$; 1459.81 MeV fm$^{-3}$ for $h = 2$.} \\
		\\
		\hline
	\end{tabular}
\end{table*}

\subsection{\textcolor{black}{Numerical generation of EOS table}}\label{appnumericaleosss}
\textcolor{black}{The EOS used in this work is constructed based on 
	a polytropic relation shown by Eq.~(\ref{polytripiceosss}). The EOS table is 
	generated numerically using a program written in Fortran. In this 
	work, the 
	corresponding polytropic constant $K$ is set to $K =100$ km$^2$, and multiplied by $1.3234 
	\times 10^{-6}$ km$^{-2}$ fm$^3$/MeV (so that it can be in units of fm$^3$/MeV)~\cite{ordaz}. The density is sampled over 
	the range $\rho_{\text{min}} = 10^{-3}$ to $\rho_{\text{max}} = 8 
	\times 10^{3}$ (in the same energy density units), using $N = 300$ 
	grid points distributed logarithmically, i.e., uniformly spaced in 
	$\log_{10}\rho$ between $\log_{10}\rho_{\text{min}}$ and 
	$\log_{10}\rho_{\text{max}}$. This logarithmic spacing is chosen to 
	provide finer resolution at low densities while still adequately 
	covering the high-density regime relevant to compact star-like interiors. 
	For each sampled value of $\rho$, the corresponding pressure $p_r$ is 
	computed directly from the polytropic relation and both values are 
	written to an output data file, which serves 
	as the tabulated EOS used in the subsequent calculations.}

\subsection{\textcolor{black}{TOV equations integration}}\label{tovnumerical}
\textcolor{black}{Using the generated EOS table, the pure polytropic NS configurations are obtained by integrating the standard TOV equations outward from the stellar center using Euler method with a fixed radial step size of 1~m, up to a maximum of $10^5$ steps (corresponding to a maximum radial extent of 100~km). At each step, the EOS is evaluated via linear interpolation between tabulated $(\rho,p)$ points. The integration is terminated once the pressure falls below the minimum tabulated value, at which point the stellar surface radius $R_s$ and total mass $M$ are recorded. For each anisotropy parameter $h$, the reference central energy density $\rho_0$ is determined by repeating this integration over a grid of central densities $\rho_c$ spanning $10^{-4}\leq\rho_c\leq10^4$~MeV~fm$^{-3}$ and identifying the value that yields the maximum stellar mass. Note that the algorithm of calculation follows the one employed in~Ref.\cite{ordaz}. The mass and radius obtained from the TOV integration are then used as references for Approaches~1 and~2. Once these reference values are obtained, the throat radius, ADM mass, and compactness of the CSSTW are calculated for each value of $\rho_c$.}

\subsection{\textcolor{black}{Echo time integration}}\label{appechonum}
\textcolor{black}{The echo time integrals, Eq.~(\ref{echotime}), are evaluated using Mathematica's \texttt{NIntegrate} function with the global adaptive quadrature method in all cases considered in this work. For the non-magnetized configurations (Sec.~\ref{nonmagnetized}), the interior integral ($r_0\leq r<R_s$) is evaluated with \texttt{MaxErrorIncreases} set to 1000 and a working precision of 20 decimal digits to suppress numerical round-off error. For the magnetized configurations (Sec.~\ref{magnetized}), the interior integral is evaluated with \texttt{MaxRecursion} set to 15 and a working precision of 15 decimal digits, given the more complex functional form of $e^{2\nu(r)}$ in this case, ie. Eq.~(\ref{enumagggg}). For the simplified magnetized toy model (Sec.~\ref{magnetized}, Eqs.~(\ref{enutoy}) and (\ref{enuctoyyyy}), the interior integral is evaluated with \texttt{MaxErrorIncreases} set to 10000, \texttt{MaxRecursion} set to 50, and both accuracy and precision goals set to 6 digits, reflecting the additional numerical sensitivity introduced by the $\arctan$ term in Eq.~(\ref{enutoy} over the wide range of magnetic-field strengths considered ($3.5\times10^{13}$--$10^{16}$~G). In all cases, the resulting echo time as a function of the ADM mass (or magnetic field, for the simplified toy model) is subsequently interpolated using a cubic spline over a grid of 50 points to produce smooth curves for the figures of echo times as a function of ADM mass shown in Sec.~\ref{nonmagnetized} and Sec.~\ref{magnetized}.}


{\color{black}\subsection{Radial perturbation eigenvalue calculation (shooting method)}
\label{sec:shooting}

Using the background profiles $\varrho(r)$, $P_r(r)$, $\nu(r)$, and $\lambda(r)$
obtained for each configurations, the fundamental radial
oscillation eigenvalue $\omega^2$ is obtained by numerically integrating the Sturm--Liouville
system, Eqs.~(20)--(21), from the throat $r_0$ outward to the stellar surface $R_s$ using a
shooting method.

The Lagrangian displacement $\zeta(r) = \xi(r)/r$ and the radial-pressure perturbation
$\Delta P_r(r)$ are integrated with Mathematica's \texttt{NDSolve} using
the \texttt{"StiffnessSwitching"} method, with a maximum of $10^{5}$ adaptive steps. To avoid
the coordinate singularities associated with the metric function $\lambda(r)$ at the throat and
with the vanishing pressure at the surface, the integration domain is slightly contracted:
\begin{equation}
	r \in \left[\, r_0(1+10^{-6}),\ R_s(1-10^{-6}) \,\right].
\end{equation}

At the inner boundary, the perturbation is set as
\begin{equation}
	\zeta(r_0) = 10^{-6}, \qquad \Delta P_r(r_0) = 10^{-6},
\end{equation}
approximating the physical condition $\Delta P_r(r_0) \to 0$ and $\zeta(r_0)\to 0$. For a trial eigenvalue $\omega^2$, the
shooting function is defined as the residual radial-pressure perturbation at the (contracted)
surface,
\begin{equation}
	f(\omega^2) \equiv \Delta P_r\!\left(R_s(1-10^{-6});\, \omega^2\right).
\end{equation}

The eigenvalue is located in two stages. First, $f(\omega^2)$ is evaluated on the following discrete grids
of trial values.
\begin{enumerate}
	\item For non-magnetized configurations: 
	\begin{equation}
		\omega^2 \in \left[-0.05,0.80\right],
		\qquad \Delta\omega^2 = 0.00017.\nonumber
	\end{equation}
	\item For magnetized configurations with $B_s=10^{15}$~G and Approach~1: 
	\begin{equation}
		\omega^2 \in \left[-0.05,0.80\right],
		\qquad \Delta\omega^2 = 0.00017.\nonumber
	\end{equation}
	\item For magnetized configurations with $B_s=10^{15}$~G and Approach~2: 
	\begin{eqnarray}
		&\omega^2 \in [-0.05\times10^6,1.10\times10^6],\nonumber\\
		&\nonumber\Delta\omega^2 = 0.00023\times10^6.
	\end{eqnarray}
	\item For magnetized configurations with $B_s=10^{16}$~G and Approach~1: 
	\begin{equation}
		\omega^2 \in \left[-0.05,0.80\right],
		\qquad \Delta\omega^2 = 0.00017.\nonumber
	\end{equation}
	\item For magnetized configurations with $B_s=10^{16}$~G, Approach~1, and $h=1,\: 2$: 
	\begin{eqnarray}
		&\omega^2 \in [-1\times10^4,1.10\times10^4],\nonumber\\
		&\nonumber\Delta\omega^2 = 0.00042\times10^4.
	\end{eqnarray}
	\item For magnetized configurations with $B_s=10^{16}$~G, Approach~1, $h=1.5$, and $R_s\in [0.166,0.339]$~km: 
	\begin{eqnarray}
		&\omega^2 \in [0,2750],\nonumber
		\qquad \Delta\omega^2 = 0.4.
	\end{eqnarray}
	\item For magnetized configurations with $B_s=10^{16}$~G, Approach~1, $h=1.5$, and $R_s\in [0.013,0.166]$~km: 
	\begin{eqnarray}
		&\omega^2 \in [-1\times10^4,1.1\times10^4],\nonumber\\
		&\nonumber\Delta\omega^2 = 0.00042\times10^4.
	\end{eqnarray}
\end{enumerate}
Here, $\Delta\omega^2$ denotes the interval between $\omega^2_i$ and $\omega^2_{i+1}$.
The first bracket $\left[\omega^2_i, \omega^2_{i+1}\right]$ for which $f$ changes sign,
$\mathrm{sign}\,f(\omega^2_i) \neq \mathrm{sign}\,f(\omega^2_{i+1})$, is retained, corresponding
to the fundamental ($n=0$) mode. Second, the bracket is refined to machine precision using the \texttt{FindRoot},
i.e. secant method~\cite{Tan_Ye_Zhang_Zhao_2025},
\begin{equation}
	\omega^2_{n+1} = \omega^2_n - f(\omega^2_n)\,
	\frac{\omega^2_n - \omega^2_{n-1}}{f(\omega^2_n) - f(\omega^2_{n-1})},
\end{equation}
iterated until $|f(\omega^2)|$ falls below the default \texttt{FindRoot} accuracy/precision
tolerance. 
}

\endgroup
\medskip

%
\bibliographystyle{MSP}
\bibliography{sample}

\end{document}